\documentclass[rmp,aps,twocolumn,nofootinbib,floatfix]{revtex4}

\usepackage{graphics}
\usepackage{epsfig}
\usepackage{amssymb}
\usepackage{amsmath}
\usepackage{psfrag}

\usepackage{color}

\newcommand{\eqname}[1]{\label{eq:#1}}
\newcommand{\bgar}{\begin{eqnarray}}
\newcommand{\enar}[1]{\label{eq:#1}\end{eqnarray}}

\newcommand{\ket}[1]{ | #1 \rangle }

\newcommand{\El}{{{\bf E}}}

\newcommand{\kk}{ {\bf k}}
\newcommand{\vv}{ {\bf v}}
\newcommand{\xx}{ {\bf x}}

\newcommand{\rr}{ {\bf r}}
\newcommand{\qq}{ {\bf q}}

\newcommand{\eq}[1]{(\ref{eq:#1})}
\newcommand{\eqs}[2]{(\ref{eq:#1}-\ref{eq:#2})}

\newcommand{\Psihd}{\hat\Psi^\dagger}

\newcommand{\Psih}{\hat\Psi}

\newcommand{\phihd}{\hat\phi^\dagger}
\newcommand{\phih}{\hat\phi}

\newcommand{\Hamilt}{{\mathcal H}}
\newcommand{\ahd}{\hat a^\dagger}
\newcommand{\ah}{\hat a}
\newcommand{\bhd}{\hat b^\dagger}
\newcommand{\bh}{\hat b}
\newcommand{\phd}{\hat p^\dagger}
\newcommand{\ph}{\hat p}
\newcommand{\nh}{\hat n}

\newcommand{\ee}{\mathbf e}
\newcommand{\jj}{\mathbf j}
\newcommand{\AAA}{\mathbf A}

\newcommand{\mume}{\mu\textrm{m}}
\newcommand{\ch}{\hat{c}}
\newcommand{\chd}{\hat{c}^\dagger}

\def\ee{\mathrm{e}}
\def\ii{\mathrm{i}}

\def\wp{\omega_{\text{p}}}
\def\DE{\Delta E}

\begin{document}
\title{Quantum fluids of light}
\author{Iacopo Carusotto}
\email{carusott@science.unitn.it}
\affiliation{INO-CNR BEC Center and Dipartimento di Fisica, Universit\`a di Trento, I-38123 Povo, Italy}
\author{Cristiano Ciuti}
\email{cristiano.ciuti@univ-paris-diderot.fr}
\affiliation{Laboratoire Mat\'eriaux et Ph\'enom\`enes Quantiques, Universit\'e Paris Diderot-Paris 7 et CNRS, B\^atiment Condorcet, 10 rue
Alice Domon et L\'eonie Duquet, 75205 Paris Cedex 13, France}

\begin{abstract}  
This article reviews recent theoretical and experimental advances in the fundamental understanding and active control of quantum fluids of light in nonlinear optical systems. In presence of effective photon-photon interactions induced by the optical nonlinearity of the medium, a many-photon system can behave collectively as a quantum fluid with a number of novel features stemming from its intrinsically non-equilibrium nature. 
We present a rich variety of photon hydrodynamical effects that have been recently observed, from the superfluid flow around a defect at low speeds, to the appearance of a Mach-Cherenkov cone in a supersonic flow, to the hydrodynamic formation of topological excitations such as quantized vortices and dark solitons at the surface of large impenetrable obstacles.
While our review is mostly focused on a class of semiconductor systems that have been extensively studied in recent years (namely planar semiconductor microcavities in the strong light-matter coupling regime having cavity polaritons as elementary excitations), the very concept of quantum fluids of light applies to a broad spectrum of systems, ranging from bulk nonlinear crystals, to atomic clouds embedded in optical fibers and cavities, to photonic crystal cavities, to superconducting quantum circuits based on Josephson junctions.
The conclusive part of our article is devoted to a review of the exciting perspectives to achieve strongly correlated photon gases. In particular, we present different mechanisms to obtain efficient photon blockade, we discuss the novel quantum phases that are expected to appear in arrays of strongly nonlinear cavities, and we point out the rich phenomenology offered by the implementation of artificial gauge fields for photons.
\end{abstract}                                                                 

\maketitle
\tableofcontents
\section{Historical introduction}
\label{sec:intro}
In the last decades, the study of the physics of quantum fluids has attracted a tremendous interest in a variety of different many particle systems, ranging from liquid Helium~\cite{PinesNozieres,Leggett:2004RMP}, electrons in solid-state materials~\cite{Tinkham,Schrieffer,Mahan}, trapped gases of ultracold atoms~\cite{Dalfovo:1999RMP,Giorgini:2008RMP,Bloch:review2008}, quark-gluon plasma in colliders~\cite{quarkgluon1,quarkgluon}, nuclei~\cite{nuclei}.
When the thermal de Broglie wavelength becomes comparable or larger than the average interparticle spacing, the Bose vs. Fermi statistics of the constituent particles starts playing a crucial role in determining the properties of the fluid: 
in non-interacting Fermi gases, the Pauli principle is responsible for the rigidity of the Fermi sphere and the appearance of a Fermi pressure down to zero temperature, while in Bose gases a macroscopic fraction of the particles accumulate into the lowest energy single-particle state, the so-called Bose-Einstein condensate (BEC). The situation is even richer when quantum degeneracy combines with significant inter-particle interactions to produce a variety of spectacular effects such as superconductivity and superfluidity~\cite{Leggett:1999RMP,Tilley} and the fractional quantum Hall effect~\cite{Yoshioka,DasSarma}. 

Historically, most of the theoretical and experimental activities in this field of many-body physics have addressed systems of material particles such as atoms, electrons, nucleons, or quarks. However, in the last decades, a growing community of researchers has started wondering whether in suitable circumstances light can be considered as a fluid composed of a large number of corpuscular photons with sizable photon-photon interactions. Even if this point of view is perfectly legitimate within the wave-particle duality in quantum mechanics, it is somehow at odd with our intuitive picture of light: the historical development of our understandings of matter and light have in fact followed very different paths. 

The idea of matter being formed by a huge number of elementary corpuscles that combine in different ways to form the variety of existing materials dates back to the ancient age with Demokritos'  atomistic hypothesis, while the wavy nature of particles was put forward only in 1924 by de Broglie and experimentally demonstrated by Davisson and Germer in 1927.
On the other hand, the long-standing debate between Newton's corpuscular and Huygens' undulatory theories of light appeared to be solved  in the early nineteenth century with the observation of fringes in Young's double slit experiment and of the remarkable Arago's white spot in the shadow of a circular object. With the microscopic support of Maxwell's theory of electromagnetism, the undulatory thory was able to explain most experimental observations until the beginning of the twentieth century when the corpuscular concept of a photon as a discrete quantum of light was revived by Einstein's theory of the photo-electric effect. Within the wave-particle duality, our standard interpretation of light then consists of a dual wave/particle beam that is emitted by the source and then freely propagates through optical devices until it is absorbed.

While this intuitive picture of light is perfectly sufficient to describe most cases of interest, still it is missing a crucial element, namely the possibility of frequent collisions between photons that allow for collective fluid-like behaviors in the many photon system. While photon-photon interactions have been predicted to occur even in vacuum via virtual excitation of electron-positron pairs~\cite{Heisenberg:1936}, the cross section for such a process is so small that it can hardly be expected to play any role in realistic optical systems.
On the other hand, the nonlinear polarization of nonlinear optical media is able to mediate significant interactions between photons~\cite{Boyd,Butcher}: upon elimination of the matter degrees of freedom, third-order  $\chi^{(3)}$ nonlinearities correspond in the language of Feynman diagrams to four-legged vertices describing, among other, binary collisions between a pair of photons. 

Among the many different configurations that have been studied in the last few decades for nonlinear optical applications, systems in the so-called strong light-matter coupling regime have turned out to be particularly promising in order to obtain the relatively strong nonlinear interactions that are necessary for collective behavior. In this regime, the photon is strongly mixed with matter degrees of freedom, which gives rise to a new mixed quasi-particle, the {\em polariton}~\cite{Hopfield:PR1958}. Pictorially, the polariton can be seen as a photon dressed by a matter excitation: a reinforced optical nonlinearity then appears thanks to the relatively strong interactions between matter excitations. This strong coupling regime can be achieved in a number of material systems, from atomic gases~\cite{Raimond:RMP2001,cavityQED,Fleischhauer:RMP2005} to semiconducting solid state media both in bulk~\cite{yu-cardona,klingshirn} and in cavity~\cite{Weisbuch:PRL1992,Deveaud:2007} geometries, to circuit-QED systems based on superconducting Josephson junctions~\cite{Schoelkopf:Nature2008,You:Nature2011}. 
In the following of the review, we will consider both photon and polariton excitations, depending on the non-resonant or resonant character of the electronic excitation dressing the photon within the material medium.

\begin{figure}[htbp]
\begin{center}
\includegraphics[width=0.8\columnwidth,angle=0,clip]{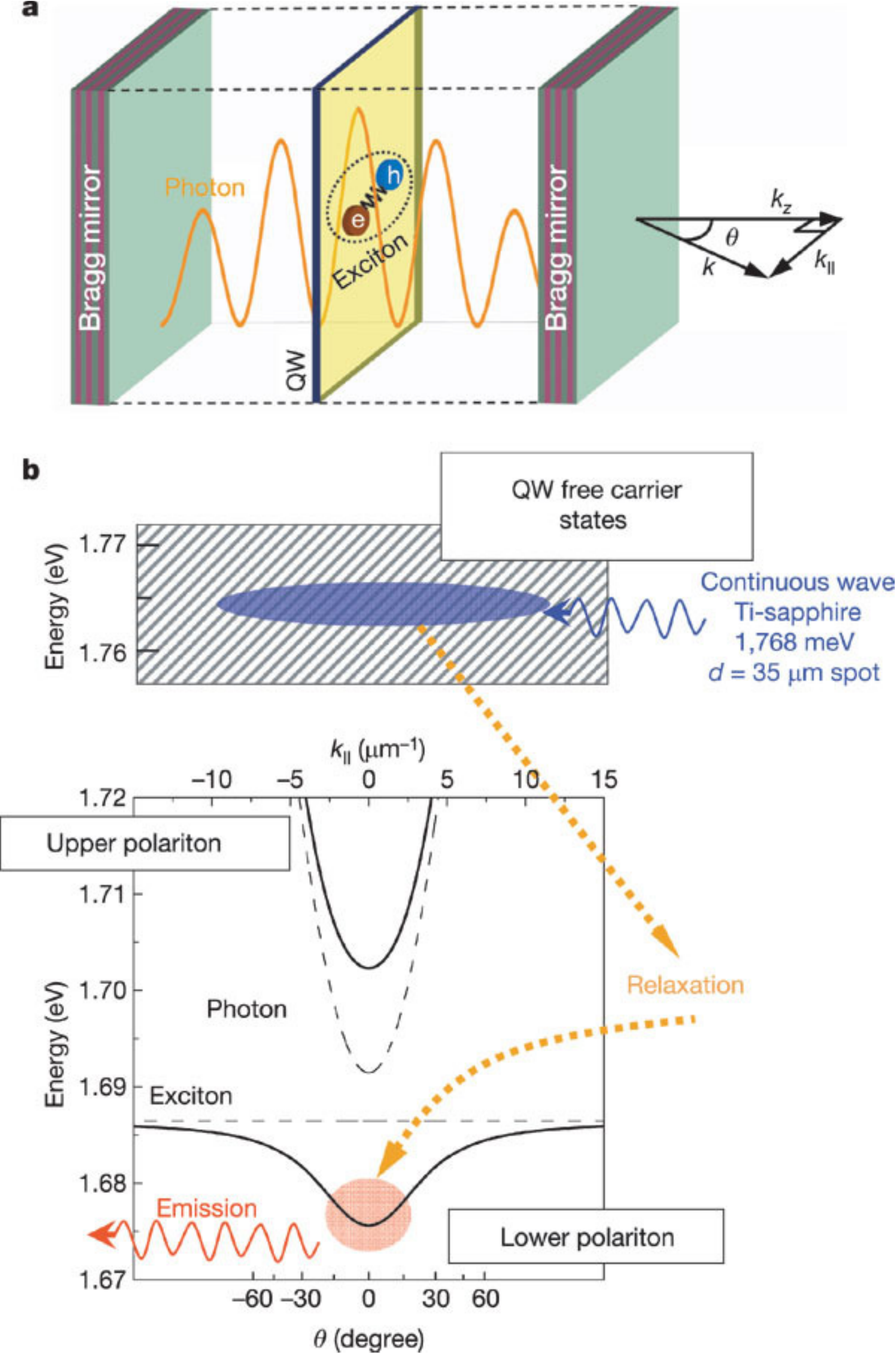}
\includegraphics[width=0.8\columnwidth,angle=0,clip]{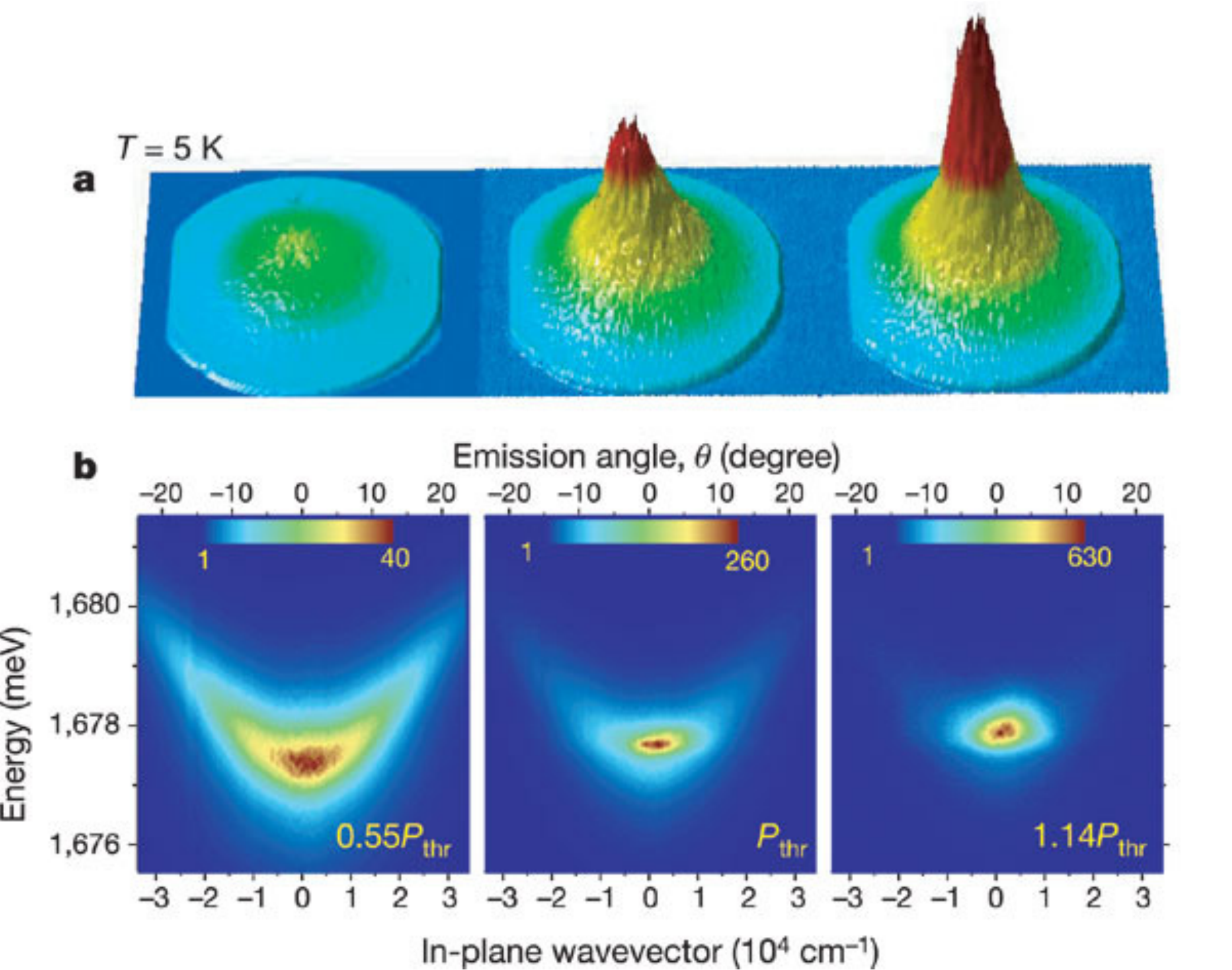}
 \end{center}
\caption{Figure from \onlinecite{Kasprzak:Nature2006}. Upper panel: Sketch of a planar semiconductor microcavity delimited by two Bragg mirrors and embedding a quantum well (QW).  The wavevector in the $z$ direction perpendicular to the cavity plane is quantized, while the in-plane motion is free. The cavity photon mode is strongly coupled to the excitonic transitions in the QWs. A laser beam with incidence angle $\theta$ and frequency $\omega$ can excite a microcavity mode with in-plane wavevector $k_{\parallel} = \frac{\omega}{c} \sin \theta$, while the near-field (far-field) secondary emission from the cavity provides information on the real-space ($\kk$-space) density of excitations. 
Central panel: The energy dispersion of the polariton modes versus in-plane wavevector (angle). The exciton dispersion is negligible, due to the heavy mass of the exciton compared to that of the cavity photon. In the experiments, the system is incoherently excited by a laser beam tuned at a very high energy. Relaxation of the excess energy (via phonon emission, exciton-exciton scattering, etc.) leads to a population of the cavity polariton states and, possibly, Bose-Einstein condensation into the lowest polariton state. Lower panel: Experimental observation of polariton Bose-Einstein condensation obtained by increasing the intensity of the incoherent off-resonant optical pump.}
\label{polBEC}
\end{figure}

To create a stable luminous fluid, it is also crucial to give a finite effective mass to the photon. A simplest strategy to this purpose involves a spatial confinement of the photon by metallic and/or dielectric planar mirrors.
In a planar geometry with a dielectric medium of refractive index $n_0$ and thickness $\ell_z$ enclosed within a pair of  metallic mirrors, the photon motion along the perpendicular $z$ direction is quantized as $q_z=\pi M/ \ell_z$, $M$ being a positive integer. For each longitudinal mode, the frequency dispersion as a function of the in-plane wavevector $\kk$ has the form
\begin{equation}
\omega_{\rm cav}(k) = \frac{c}{n_0} \sqrt{q_z^2 + k^2}\simeq \omega_{\rm cav}^o + \frac{\hbar k^2}{2 m_{\rm cav}},
\eqname{omega_cav}
\end{equation}
where the effective mass $m_{\rm cav}$ of the photon and the cut-off frequency $\omega_{\rm cav}^o$ are related by the relativistic-like expression
\begin{equation}
m_{\rm cav}= \frac{\hbar n_0 q_z}{c}= \frac{\hbar \omega_{\rm cav}^o}{c^2/n_0^2}.
\eqname{m_cav}
\end{equation}
Using suitable values of the effective mass $m_{\rm cav}$ and the cut-off frequency $\omega_{\rm cav}^o$ extracted from microscopic calculations~\cite{savona_cargese}, the generic form \eq{omega_cav} of the dispersion can be  extended to the case of dielectric mirrors. 
In the presence of some electronic excitation resonant with the cavity mode, the elementary excitations of the cavity have a polaritonic character with a peculiar dispersion law that reflects their hybrid light-matter nature. An example of such dispersion is shown in the central panel of Fig.\ref{polBEC}: in spite of the complex light-matter interaction dynamics, the bottom of the lower polariton branch is still well approximated by a parabolic dispersion of the form \eq{omega_cav} with an effective mass $m_{LP}$ and the cut-off frequency $\omega_{LP}^o$.

Historically, the first mention of the concept of photon fluid dates back to the work of~\cite{Brambilla:PRA1991b,Staliunas:PRA1993}, where the time-evolution of the coherent electromagnetic field in a laser cavity with large Fresnel number was reformulated in terms of hydrodynamic equations for the many photon system analogous to the Gross-Pitaevskii equation for the superfluid order parameter. The local light intensity corresponds indeed to the photon density and the spatial gradient of its phase to the local current; the collective behavior originates from the effective photon-photon interactions stemming from the nonlinear refractive index of the medium as well from saturation of gain. In the following years, the transverse dynamics of the electromagnetic field in cavity devices has attracted a lot of attention, in particular the phenomena related to the spontaneous formation of transverse patterns~\cite{Staliunas_book,Denz_book} and to the generation and control of dissipative cavity solitons~\cite{Ackemann:AdvAMOP2009}

In this pioneering literature on hydrodynamics of the photon fluid, a special attention was paid to the phase singularities of the photon field, that were immediately interpreted as quantized vortices~\cite{Coullet:1989}. {After their first experimental observation in bulk nonlinear crystals~\cite{Swartzlander:PRL1992}, most of the following literature addressed the physics of optical vortices in the context of the transverse dynamics of laser or photorefractive oscillators: the drift of vortices under the hydrodynamical effect of the Magnus force due to buoyancy was experimentally studied in~\cite{Vaupel:PRA1995}. The hydrodynamical shedding of vortices in a moving photon fluid hitting a large defect was first predicted in the optical context in~\cite{Staliunas:PRA1993} and a pioneering attempt of experimental investigation of this crucial effect of superfluid hydrodynamics was  reported soon after in~\cite{Vaupel:PRA1996}. }

{From a different perspective, the close analogy between a laser threshold and a second order phase transition was recognized as early as in~\cite{DeGiorgio:PRA1970,Graham:ZPhys1970}: above the laser threshold, the electromagnetic field acquires a well defined phase by spontaneously breaking a $U(1)$ symmetry as it happens to the matter Bose field in a Bose-Einstein condensate of material particles~\cite{Gunton:PR1968,Huang}. This interpretation of lasing as the result of a kind of Bose-Einstein condensation of photons is clearest in spatially extended devices such as vertical cavity surface emitting lasers (VCSELs), where the onset of a coherent laser emission is associated to the appearance of long-range spatial coherence along the cavity plane according to the Penrose-Onsager criterion for off-diagonal long-range order~\cite{BECbook,Huang}
\begin{equation}
\lim_{|\rr-\rr'|\to \infty} \langle \El^\dagger(\rr)\,\El(\rr') \rangle \neq 0.
\end{equation}
} Of course, the non-equilibrium nature of the laser device introduces crucial differences with respect to standard equilibrium BEC as  explained in statistical mechanics textbooks~\cite{Huang}: the steady state of the laser device is in fact not determined by a thermal equilibrium condition, but rather follows from a dynamical balance between the pumping and losses~\cite{Haken:RMP1975}. As we shall see in the following of the review, this feature is responsible for a number of new effects.

From the experimental point of view, the close link between BEC and spontaneous coherence effects in optical systems has been fully recognized only in the last decade, starting with the literature on the so-called BEC of exciton-polaritons in semiconductor microcavities~\cite
{Baumberg:PRB2000,Stevenson:PRL2000,Baas:PRL2006,Kasprzak:Nature2006}. 
In this context, questions related to the analogies and differences between laser operation and photon/polariton BEC have attracted a strong interest from the community, with a special attention to thermalization issues. 
{An experimental observation of microcavity polaritons coherently accumulating in the lowest energy states of a harmonic trap potential according to a Bose distribution was reported in~\cite{Balili:Science2007}: thermalization of the polariton gas was attributed to polariton-polariton collisions within the gas. A similar, apparently thermalized photon distribution was however observed in~\cite{Bajoni:PRB2007} also in a weak-coupling regime where photon-photon interactions are very weak, which raises fundamental questions about the nature of fluctuations on top of a photon/polariton condensate.
These observations are to be contrasted with the strongly non-equilibrium regimes of laser operation observed in a vertical cavity surface emitting laser (VCSEL) device in~\cite{Scheuer:Science1999}: thermalization into the lowest state is completely ineffective and the condensate mode displays a complex structure with array of vortices.}

{In the last years, the quest for condensation effects in photon gases has successfully explored a few other interesting avenues. An early mention of the possibility of a gas of bare photons thermalizing to a Bose condensed state via collisions mediated by the optical nonlinearity is found in~\cite{Navez:PRA2003}.
BEC in a thermalized gas of photons was experimentally observed in~\cite{Klaers:Nature2010} using a macroscopic optical cavity containing a dye solution: because of the rapid decoherence time of the dye molecules, photons are only weakly coupled to the electronic excitations and their thermalization is believed to occur via repeated absorption-emission cycles, which determine the temperature and the chemical potential of the gas in a grand-canonical picture~\cite{Klaers:NatPhys2010,Klaers:PRL2012}.
A kinetic condensation of purely classical light waves was observed in the remarkable experiment of~\cite{Sun:NatPhys2012}: as theoretically discussed in, e.g., ~\cite{Connaughton:PRL2005}, turbulent wave mixing by the optical nonlinearity leads to a redistribution of energy among the different modes and, eventually, to its accumulation into the lowest, condensate mode. A most remarkable feature of this experiment is the complete absence of quantum features, which emphasizes the fundamentally classical origin of the Bose-Einstein condensation phenomenon. More complex condensation phenomena have been reported also in disordered lasers~\cite{Conti:PRL2008} and in actively-mode-locked lasers~\cite{Weill:PRL2010,Weill:OptExpr2010}.}

{Simultaneously to these studies on Bose-Einstein condensation and spontaneous coherence effects, a revived interest has been devoted also to the hydrodynamic properties of the photon gas. The concept of the Bogoliubov dispersion of elementary excitations on top of a photon condensate was first investigated for a planar cavity geometry in the pioneering works~\cite{Chiao:PRA1999,Tanzini:PLA1999}: thanks to the spatial confinement, photons have a massive dispersion of the form \eq{omega_cav} and the photon-photon interactions responsible for the collective behavior of the fluid are provided by the $\chi^{(3)}$ nonlinearity of the cavity medium. None of these works however addressed the crucial role of dissipation in the physics of photon fluids: first steps in this direction appeared in~\cite{Bolda:PRL2001} where the nucleation of vortices in a moving photon fluid past an impenetrable, cylindrical defect in a planar cavity was theoretically investigated. Differently from previous works on oscillators~\cite{Staliunas:PRA1993}, a coherent pumping was proposed as a way to create the photon superfluid in the cavity: this resulted in a different form of the Gross-Pitaevskii equation as originally studied in~\cite{Lugiato:PRL1987} and, more importantly, in the need to switch off the coherent pump to unlock the condensate phase before vortices can appear.}

A completely different approach to the study of superfluidity properties of light involved the paraxial propagation of a light beam through a bulk nonlinear crystal, which can be recast into a superfluid hydrodynamic form under the replacement of the time coordinate with the longitudinal coordinate along the propagation direction. {After the pioneering experiment in~\cite{Swartzlander:PRL1992}, many authors} have theoretically investigated a number of hydrodynamic features in light propagation, from stable liquid-like solitonic structures~\cite{Josserand:PRL1997,Michinel:PRL2006}, to vortices~\cite{Firth:PRL1997,Paz-Alonso:PRL2005}, to the scattering on a defect potential~\cite{Khamis:PRA2008}, to superfluid motion~\cite{Leboeuf:PRL2010}, to dispersive shock waves~\cite{El:PRA2007}, to the generation of optical analogs of acoustic black holes~\cite{Fouxon:EPL2010}. 
In spite of all this theoretical activity, {not many nonlinear optics experiments have specifically addressed hydrodynamic features of light. Among the few exceptions, the experimental study of vortex and soliton stability and interactions~\cite{Mamaev:PRL1996,Mamaev:PRL1996b,Krolikowski:PRL1998}, the generation, propagation and interaction of dispersive shock waves in fluids of light~\cite{Wan:NatPhys2007,Jia:PRL2007}, the propagation of a one-dimensional fluid through a small barrier potential~\cite{Wan:PRL2010}, a study of the generation of vortex pairs in the wake of an obstacle in a two-dimensional geometry~\cite{Wan:FrOpt2008}, a study of the Rayleigh-Taylor instability in stratified fluids~\cite{Jia:NJP2012}, and the realization of a trans-sonic flow of light in a Laval nozzle configuration~\cite{Elazar:arXiv2012}.}
An alternative interesting perspective on many-body physics of photons was developed in the pioneering literature on quantum solitons in nonlinear optical fiber using a quantum nonlinear Schr\"odinger equation as well as Bethe ansatz techniques~\cite{Lai:PRA1989,Lai:PRA1989b,Kaertner:PRA1993,Drummond:Nature1993}.

The research on exciton-polaritons in semiconductor microcavities approached the physics of luminous quantum fluids following a rather different pathway. For many years, an intense activity has been devoted to the quest for Bose-Einstein condensation phenomena in gases of excitons in solid-state materials~\cite{Griffin:book1996}: excitons are neutral electron-hole pairs bound by Coulomb interaction, which behave as (composite) bosons. In spite of the interesting advances in the direction of exciton Bose-Einstein condensation in bulk cuprous oxide and cuprous chloride, bilayer electron systems~\cite{Eisenstein:Nature2004}, and coupled quantum wells~\cite{Butov:JPhysB2007,High:Nano2012}, so far none of these research axes has led to extensive studies of the quantum fluid properties of the alleged exciton condensate. The situation appears to be similar for what concerns condensates of magnons, i.e. magnetic excitations in solid-state materials: Bose-Einstein condensation has been observed~\cite{Demokritov:Nature2006,Giamarchi:Nature2008}, but no quantum hydrodynamic study has been reported yet.

The situation is very different for exciton-polaritons in semiconductor microcavities, that is bosonic quasi-particles resulting from the hybridization of the exciton with a planar cavity photon mode~\cite{Weisbuch:PRL1992}. Following the pioneering proposal by~\onlinecite{Imamoglu:PRA1996}, researchers have successfully explored the physics of Bose-Einstein condensation in these gases of exciton-polaritons. Thanks to the much smaller mass of polaritons, several orders of magnitude smaller than the exciton mass, this system can display Bose degeneracy at much higher temperatures and/or lower densities.

Historically, the first configuration where spontaneous coherence was observed in a polariton system was based on a coherent pumping of the cavity at a finite angle, close to the inflection point of the lower polariton dispersion. As experimentally demonstrated in~\cite{Baumberg:PRB2000,Stevenson:PRL2000}, above a threshold value of the pump intensity a sort of parametric oscillation\cite{Ciuti:PRB2000,Ciuti:PRB2001,Whittaker:PRB2001} occurs in the planar microcavity and the parametric luminescence on the signal and idler modes acquires a long-range coherence in both time and space~\cite{Baas:PRL2006}. As theoretically discussed in~\cite{Carusotto:PRB2005}, the onset of parametric oscillation in these spatially extended planar cavity devices can be interpreted as an example of non-equilibrium Bose-Einstein condensation: the coherence of the signal and idler is not directly inherited from the pump, but appears via the spontaneous breaking of a $U(1)$ phase symmetry. 

The quest for Bose-Einstein condensation in a thermalized polariton gas under incoherent pumping required a few more years to be achieved: after some preliminary claims
~\cite{Dang:PRL1998,Deng:Science02,Richard:PRB2005,Richard:PRL2005}, a conclusive demonstration was reported by ~\onlinecite{Kasprzak:Nature2006}: the onset of Bose-Einstein condensation in a gas of exciton-polaritons was assessed both in $\kk$-space from the macroscopic accumulation of particles into the low-energy states and in real space from the appearance of long-range coherence. The principle of these experiments is illustrated in Fig.\ref{polBEC}: the laser pump injects hot electron-hole pairs, whose excess energy can be dissipated via phonon emission and then Coulomb scattering processes. For sufficiently high pump powers, the density of the incoherently injected polaritons exceeds the critical density for Bose-Einstein condensation and a coherent condensate appears. Even if the system is still a driven-dissipative one, a quasi-thermal equilibrium state {seems to be achieved once} the loss rate is sufficiently slow as compared to the thermalization time of the polariton gas. For a more detailed discussion of the issues related to the relaxation mechanisms and the thermalization of the polariton gases, we refer the reader to the review papers that are already available on the subject, e.g.~\cite{Deng:RMP2010,Keeling:SST2007}. 
While most of this research is being carried out in devices fabricated with inorganic GaAs- or CdTe-based alloys, a first observation of spontaneous polariton coherence in a non-resonantly pumped organic single-crystal microcavity was very recently reported in~\cite{Kena-Cohen:NatPhot2010}.

\begin{figure}[htbp]
\begin{center}
\includegraphics[width=0.95\columnwidth,angle=0,clip]{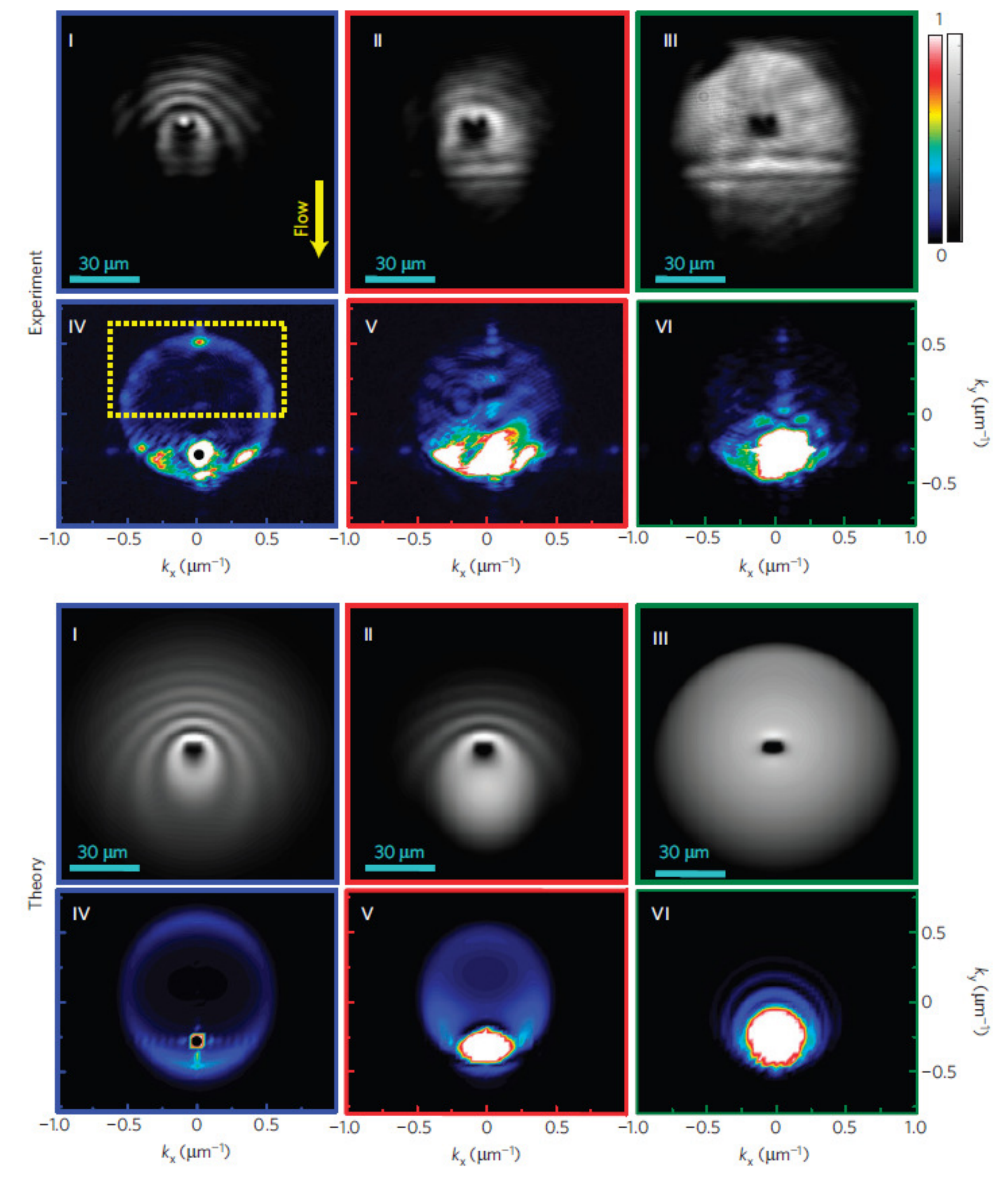}
 \end{center}
\caption{Figures from \onlinecite{Amo:NPhys2009}. Polaritons are coherently injected into the microcavity by a nearly resonant laser field: in contrast to the non-resonant and incoherent pumping scheme of Fig.~\ref{polBEC}, this pumping scheme allows to precisely control the density and the in-plane flow speed of the polariton fluid by changing the parameters (intensity, frequency, incidence angle) of the driving laser. 
Upper group of panels: experimental images of the real- (panels I-III) and momentum- (panels IV-VI) space polariton density extracted from the near-field and far-field emitted light from the cavity. The different columns to increasing values of the polariton density  from left to right refer: for the highest density value, polariton superfluidity is apparent as a suppression of the real-space density modulation (panel III) and the corresponding disappearance of the Rayleigh scattering ring (panel VI). 
Lower group of panels: corresponding theoretical results obtained by numerically solving the non-equilibrium Gross-Pitaevskii equation of~\cite{Carusotto:PRL2004}. }
\label{fig:Amo_superfl}
\end{figure}

The possibility of using exciton-polariton gases for studies of many-body physics and, more precisely, superfluid hydrodynamics was first proposed in~\cite{Carusotto:PRL2004}: a coherent pump configuration as in~\cite{Bolda:PRL2001} was adopted. In contrast to the case of off-resonant and incoherent excitation, in the resonant driving configuration it is possible to perform an {\em ab initio} theoretical description in terms of a generalized Gross-Pitaevskii equation without having to cope with a phenomenological description of complex relaxation processes. In spite of the simplicity of the system, the coherently injected condensate shows peculiar superfluid properties when it hits a defect in the cavity. The shape of the resulting density perturbation could be interpreted in terms of the celebrated Landau criterion of superfluidity using the generalized Bogoliubov dispersion of excitations in the non-equilibrium condensate: at low flow speeds, superfluidity manifests itself in the suppression of the real-space modulation around the defect and, correspondingly, in the disappearance of the Rayleigh scattering ring in $k$-space. For larger flow speeds peculiar patterns appear in both the density and momentum distribution of the condensate polaritons.
Experimental verification of these predictions was reported in~\cite{Amo:NPhys2009} and is summarized in Fig.\ref{fig:Amo_superfl}, where the transition from a dissipative flow (panels I and IV) to a superfluid one (panels III and VI) is apparent in both the real- and the momentum-space images. Remarkably, in the same work it has been directly shown that photon-photon interactions are responsible for the appearance of a sound mode in the polariton fluid, as attested by the appearance of a Cherenkov-Mach cone at ``supersonic'' flow speeds. Following experimental works have extended the study to strong defects, shown the  hydrodynamic nucleation of vortex-antivortex pairs~\cite{Nardin:2011NatPhys,Sanvitto:NPhot2011} and of dark solitons\cite{Amo:2011Science,Grosso:PRL2011} in the flowing superfluid. 
A remarkable theoretical development of the last few years is to use light superfluids in planar geometries to study quantum hydrodynamics effects, in particular the analog Hawking radiation from acoustic black hole configurations~\cite{Marino:PRA2008,Solnyshkov:PRB2011,Gerace:arXiv2012}.

\begin{figure}[t]
\begin{center}
\includegraphics[width=0.65\columnwidth,angle=0,clip]{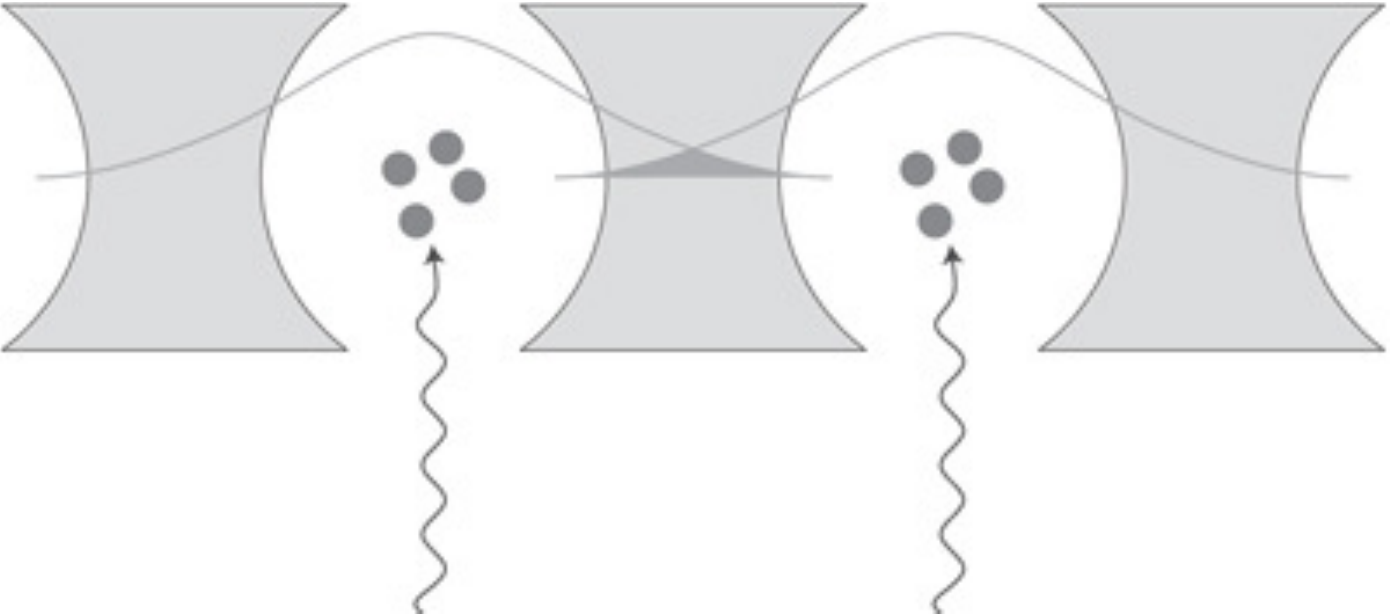}\\
\vspace{0.4cm}
\includegraphics[width=0.7\columnwidth,angle=0,clip]{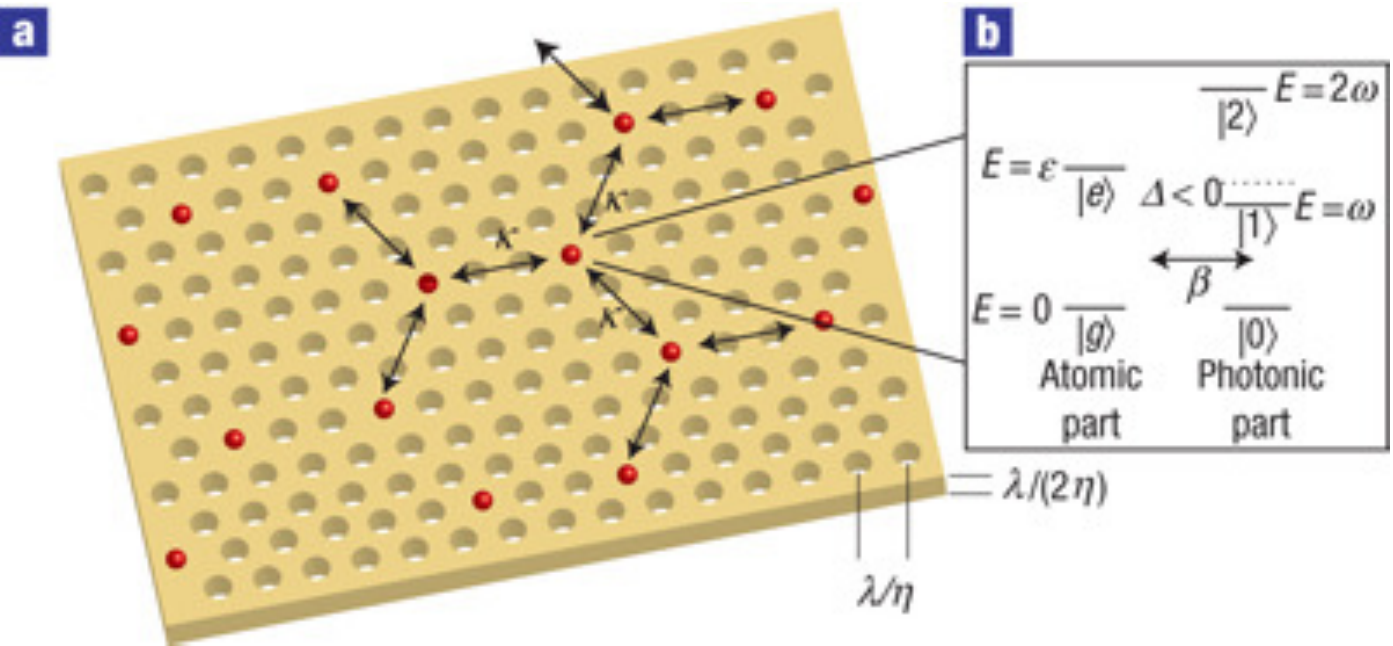}\\
\vspace{0.3cm}
\includegraphics[width=0.75\columnwidth,angle=0,clip]{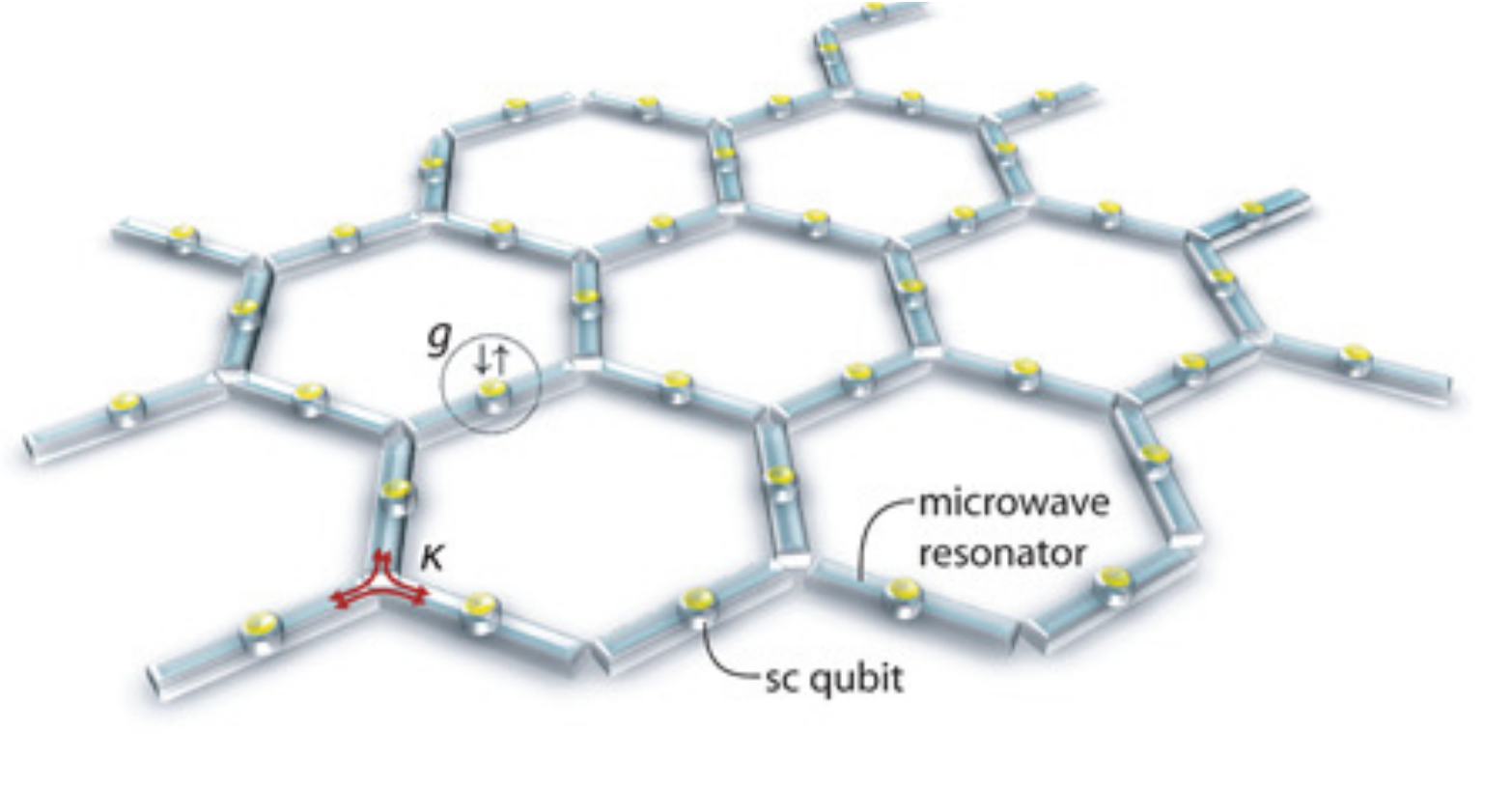}

 \end{center}
\caption{Top panel: figure from \onlinecite{Hartmann:NatPhys2006}.  In an array of coupled optical cavities, photon hopping occurs thanks to the spatial overlap of the photon modes of adjacent cavities. The strong optical nonlinearity is induced by a coherently driven atomic gas present in the cavities. Middle panel: figure from  \onlinecite{Greentree:2006}, showing a schematic diagram of a two-dimensional array of photonic crystal cavities, each cavity containing a single two-level atom. Bottom panel: a Jaynes-Cummings-Hubbard system obtained with superconducting quantum circuits. Each cavity consists of a superconducting transmission line resonator embedding a superconducting qubit, that is an artificial two-level atom. Figure from \onlinecite{Koch:PRA2010}.}
\label{fig:sketch_array}
\end{figure}

In all these works, interactions between single photons are weak and the hydrodynamic behavior of the photon gas originates from the collective interactions of a large number of coherent photons sharing the same orbital wavefunction. Almost in the same period, a series of pioneering theoretical works~\cite{Hartmann:NatPhys2006,Greentree:2006,Angelakis:PRA2007} made the first steps in the theoretical investigation of a completely different regime of {\em strongly interacting photon gases}, where interactions between single photons are large enough to induce sizable quantum correlations in the photon gas.
The starting point of these proposals is the so-called {\em photon blockade} phenomenon predicted in~\cite{Imamoglu:PRL97} and then experimentally observed in~\cite{Birnbaum:Nature2005}: when the optical nonlinearity of a single-mode cavity is large enough for a single photon to shift the resonance frequency by an amount larger than the cavity linewidth, a resonant laser is able to inject photons in the cavity only one at a time and the stream of transmitted photons are strongly antibunched in time. When this strong optical nonlinearity is inserted in a lattice geometry with many coupled cavities, one can expect that the photon gas will show the rich physics of the Bose-Hubbard model~\cite{Fisher:PRB1989}, including the superfluid to Mott-insulator transition recently observed in atomic gases~\cite{Greiner:Nature2002}. The most promising systems to experimentally address this physics are illustrated in the different panels of Fig.\ref{fig:sketch_array}: macroscopic cavities filled by an optically dressed atomic gas; an array of photonic crystal cavities embedding two-level emitters; superconducting circuits embedding artificial two-level atoms based on Josephson junctions. Soon after, these proposals were generalized to other many-body states, in particular Tonks-Girardeau gases of impenetrable photons in a one-dimensional hollow fiber geometry~\cite{Chang:NatPhys2008} and fractional quantum Hall states~\cite{Cho:PRL2008}. 

In contrast to the developments in superfluid hydrodynamics of dilute photon fluids, none of these pioneering works on strongly correlated gases specifically addressed the consequences of the losses that are unavoidably present when dealing with light. Dissipation was in fact considered only as a hindrance limiting the available time for manipulation and observation of the quantum state, with a most detrimental effect on the Mott insulator state. A new perspective on strongly correlated photon gases was introduced in~\cite{Gerace:NatPhys2009} for the case of a two site photonic Josephson system and, soon later, in~\cite{Carusotto:PRL2009} for the case of a Tonks-Girardeau gas: a full inclusion of the interplay of driving, dissipation and strong interactions into the model offers the opportunity to observe novel dynamical features in such a driven-dissipative photon gas, and also suggests new tools for its experimental manipulation. In addition to this, the driven-dissipative regime allows for novel mechanisms of photon blockade to be exploited in coupled cavity systems \cite{Liew:PRL2010}, \cite{Bamba:PRA2011,Bamba:APL2011}.
Very recently, the analogy with on-going developments in ultracold atomic gases~\cite{Dalibard:RMP2011} has opened a new research line in the direction of creating {\em synthethic gauge fields} for photons: with a suitable tailoring of the photonic environment, the motion of a photon can be made to experience an effective vector potential~\cite{Wang:Nature2009,Koch:PRA2010,Hafezi:NatPhys2011,Umucalilar:PRA2011} and novel quantum states of the photon fluid analogous to the fractional quantum Hall effect can be realized~\cite{Cho:PRL2008,Umucalilar:PRL2012}.

This review article is organized as follows. In Sec. \ref{quantumfield}, we briefly summarize the main tools of elementary quantum field theory that are used to theoretically describe the non-equilibrium and quantum physics of the photon fluid in nonlinear optical devices. Even though a particular attention is paid to planar microcavity geometries where most of the recent experimental observations were obtained, most of the concepts are fully general and can can be transferred to other systems. The generalized Gross-Pitaevskii equation describing the dynamics of the photon fluid at mean-field level is introduced in Sec.\ref{GP}: particular emphasis will be devoted to the new features stemming from the non-equilibrium nature of the system. The most significant static and dynamic properties of the coherent photon fluid under the different excitation schemes are reviewed in the following three sections: Sec.\ref{resonant} deals with the case of a coherent and quasi-resonant driving, Sec.\ref{sec:secOPO} deals with the optical parametric oscillation case when the cavity is pumped in the vicinity of the inflection point of the lower polariton branch, and Sec.\ref{nonresonant} deals with the case of a incoherent pumping. We will review the consequences of the non-equilibrium condition on the condensate shape both in real and momentum space as well as its impact on the elementary excitation spectrum. The superfluidity properties of the photon (polariton) fluid are reviewed in Sec.\ref{hydro}: among the several signatures of superfluidity, most emphasis will be devoted to the density modulation induced in the photon fluid by a weak impurity, for which impressive experimental observations have been recently obtained. Subtle issues related to the generalization of the Landau criterion and the role of the interaction-induced speed of sound will be extensively discussed. More complex hydrodynamic effects involving the nucleation of solitons and vortices in the wake of a large and strong defect are reviewed in Sec.\ref{vortices}. The review of the emerging field of strongly correlated photons is the subject of Sec. \ref{strongcorrelation}: we will discuss in detail the different microscopic mechanisms that can be used to achieve an efficient photon blockade regime and the rich new physics that has been predicted for lattices of strongly nonlinear cavities, in particular in the presence of an synthetic gauge field. Finally, conclusions and future perspectives are drawn in Sec.\ref{conclusions}.

\section{Quantum field description of nonlinear planar cavities}
\label{quantumfield}

In this Section,  we will review a second quantization formalism approach to describe the physics of 2D nonlinear cavities. Our goal is to present the essential quantum field theoretical tools that will be useful to understand the discussion of quantum fluid effects in the following sections. While we will give particular emphasis to semiconductor microcavity systems where most of the experimental observations have been carried out, much of the theoretical concepts reviewed here can be applied to arbitrary planar optical resonators embedding a nonlinear slab and are easily generalized to other geometries. We will be careful in pointing out when some properties are specific to a given system or not. Comprehensive introductions to semiconductor microcavity systems can be found in the specific literature, for instance~\cite{Varenna:2003,Deveaud:2007,Varenna:2009,Kavokin_book}.

\subsection{Free cavity fields and input-output formalism}

\subsubsection{The two-dimensional photon field} 
\label{sec:free_field}

Since a  planar cavity is by definition invariant under in-plane translations, the in-plane wavevector $\kk$ is a good quantum number for the free photon dynamics, which can be described by an Hamiltonian of the form
\begin{equation}
\Hamilt_{\rm cav}=\int\!\frac{d^2\kk}{(2\pi)^2}\, \sum_\sigma\,\hbar\omega_{\rm cav}(\kk)\ahd_{C,\sigma}(\kk) \ah_{C,\sigma}(\kk) \eqname{Hamilt_cav}
\end{equation}
where the $\ah_{C,\sigma}(\kk)$ and $\ahd_{C,\sigma}(\kk)$ operators respectively destroy and create a cavity photon with an in-plane wavevector $\kk$ and a polarization state $\sigma$. The creation and destruction operators satisfy standard Bose commutation rules 
\begin{eqnarray}
[\ah_{C,\sigma}(\kk),\ahd_{C,\sigma'}(\kk')]&=&(2\pi)^2\,\delta(\kk-\kk')\,\delta_{\sigma,\sigma'} \eqname{commu_aC1} \\
\big[\ah_{C,\sigma}(\kk),\ah_{C,\sigma'}(\kk')]&=&0. \eqname{commu_aC2}
\end{eqnarray}
Two-dimensional real-space cavity photon field operators $\Psih_{C,\sigma}(\rr)$ and $\Psihd_{C,\sigma}(\rr)$ are defined as the Fourier transform of $\ah_{C,\sigma}(\kk)$,
\begin{equation}
\Psih_{C,\sigma}(\rr)=\int\!\frac{d\kk}{(2\pi)^2}\,\ah_{C,\sigma}(\kk)\,e^{i\kk\rr},
\eqname{psiC}
\end{equation}
and again satisfy Bose commutation rules
\begin{eqnarray}
[\Psih_{C,\sigma}(\rr),\Psihd_{C,\sigma'}(\rr')]&=&\delta(\rr-\rr')\,\delta_{\sigma,\sigma'}, \eqname{commu_psiC1}\\
\big[\Psih_{C,\sigma}(\rr),\Psih_{C,\sigma'}(\rr')]&=&0.
\eqname{commu_psiC2}
\end{eqnarray}
While the $\nh_{C,\sigma}(\rr)=\Psihd_{C,\sigma}(\rr)\,\Psih_{C,\sigma}(\rr)$ operator can be interpreted as a sort of two-dimensional photon density on the cavity plane, the physically observable electric field at the three-dimensional position $(\rr,z)$ is expressed in terms of the field operators by
\begin{equation}
\hat{\El}(\rr,z)=\int\!\frac{d^2\kk}{(2\pi)^2}\,\sum_\sigma\,e^{i\kk\rr}\,\El_\sigma(\kk,z)\,\ah_{C,\sigma}(\kk) + \textrm{h.c.}
\end{equation}
where the $z$-dependence of the photon mode wavefunction $\El_\sigma(\kk,z)$ keeps track of the metallic or dielectric nature of mirrors.
In the simplest case of a planar cavity of refractive index $n_0$ enclosed between metallic mirrors spaced by a distance $\ell_{z}$, the $z$-dependent profile $\El_\sigma(\kk,z)$ of the lowest mode has a sinusoidal shape extending for $0\leq z\leq \ell_{z}$,
\begin{equation}
\mathbf{E}_\sigma(\kk,z)=i\sqrt{\frac{4\pi \hbar \omega_{\rm cav}}{\ell_{z}\,n_0^2}}\,\ee_\sigma\,\sin\left(\frac{\pi\,z }{\ell_{z}}\right):
\eqname{norma_E}
\end{equation}
the boundary conditions at the mirrors fix the position of the nodes, and the global amplitude of the field is determined by the energy of a single photon state $\hbar\omega_{\rm cav}(\kk)$. The unit vectors $\ee_\sigma$ are a basis of light polarizations. A detailed discussion of the more complex case of dielectric cavities enclosed between a pair of Distributed Bragg Reflector (DBR) dielectric mirrors can be found in~\cite{savona_cargese}.

\subsubsection{Polarization effects and effective spin-orbit interaction}
\label{sec:spin}
\begin{figure}[th]
\begin{center}
\includegraphics[width=0.7\columnwidth,angle=0,clip]{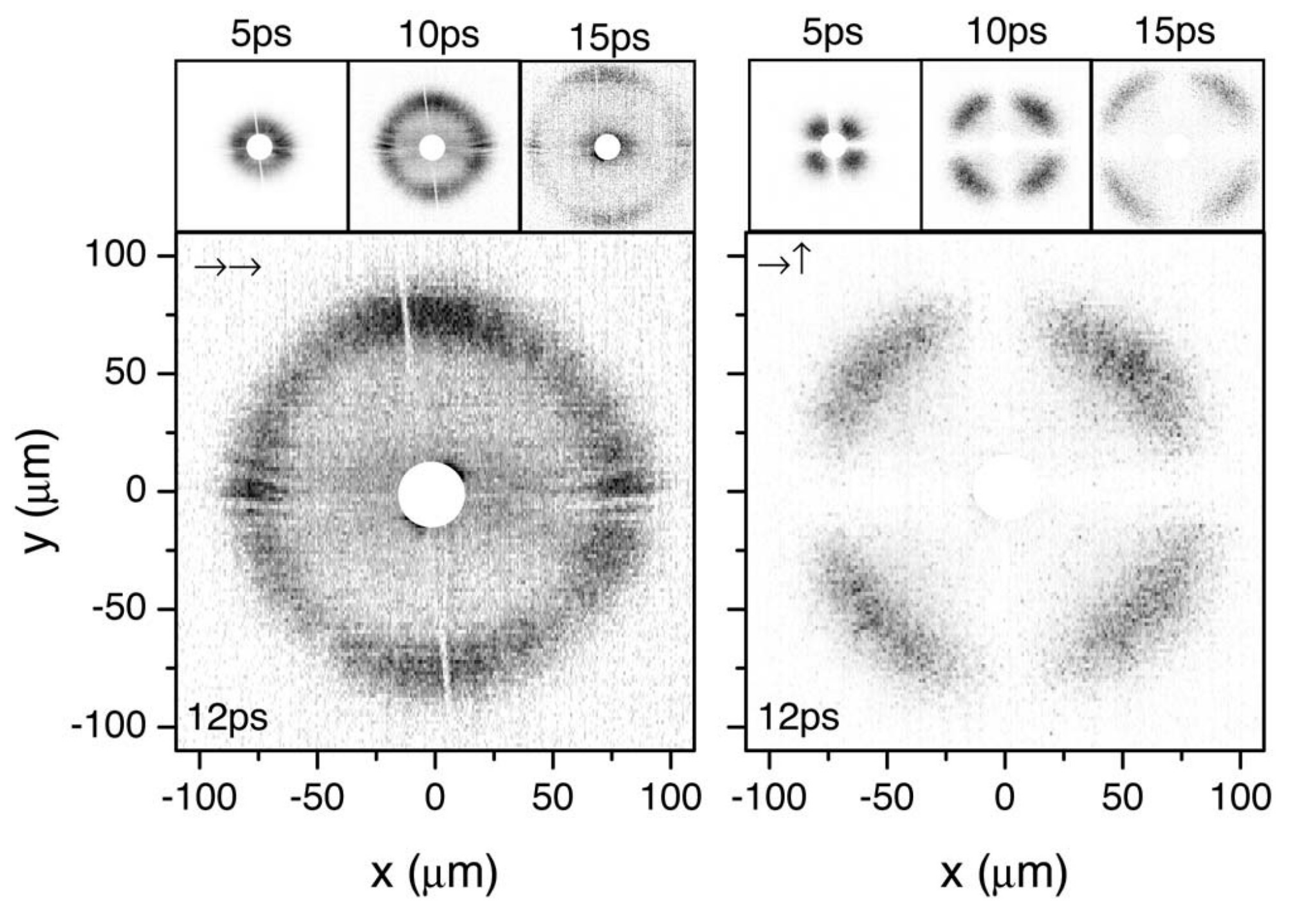} \\
\includegraphics[width=0.6\columnwidth,angle=0,clip]{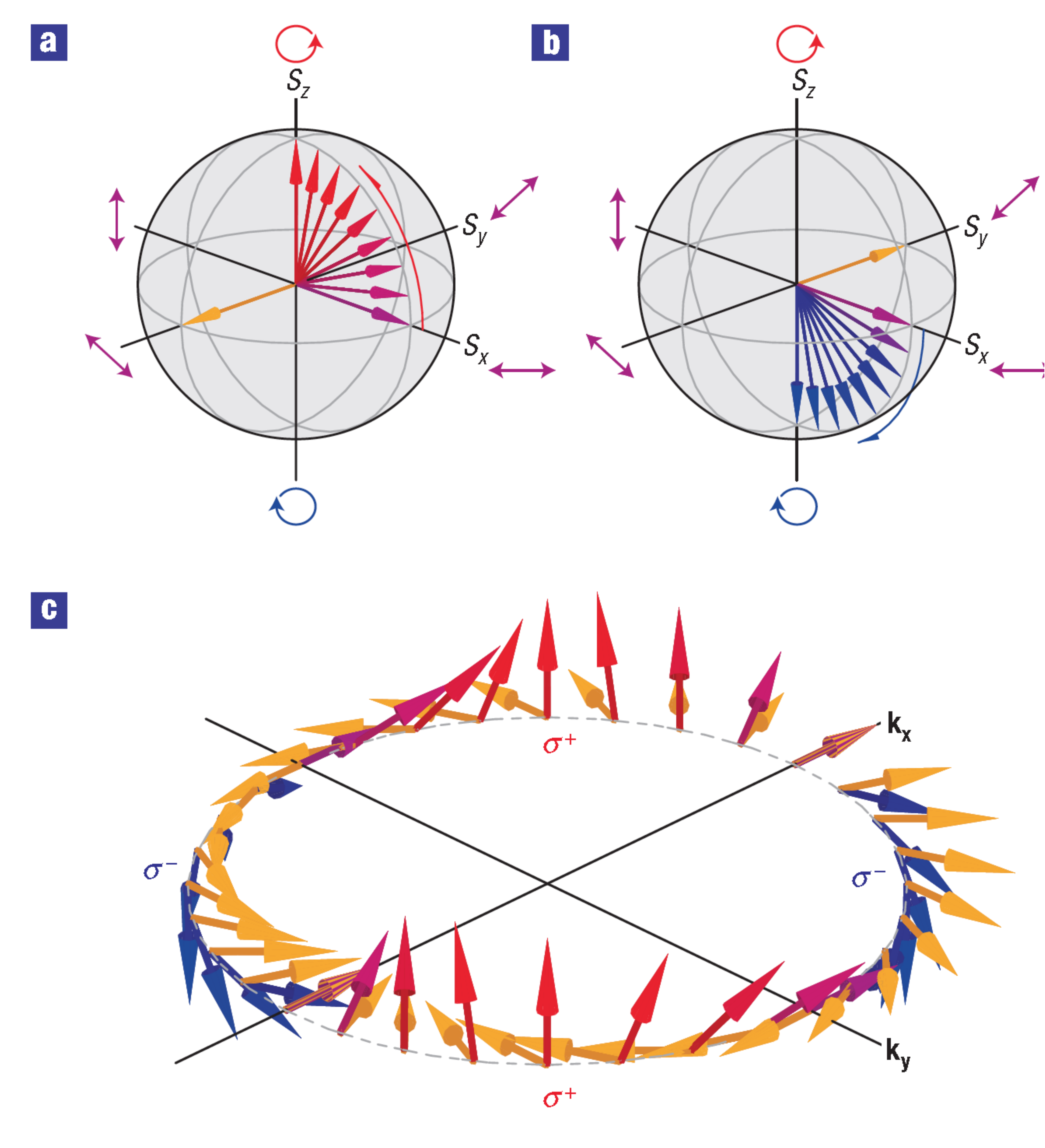} \\
\includegraphics[width=0.4\columnwidth,angle=0,clip]{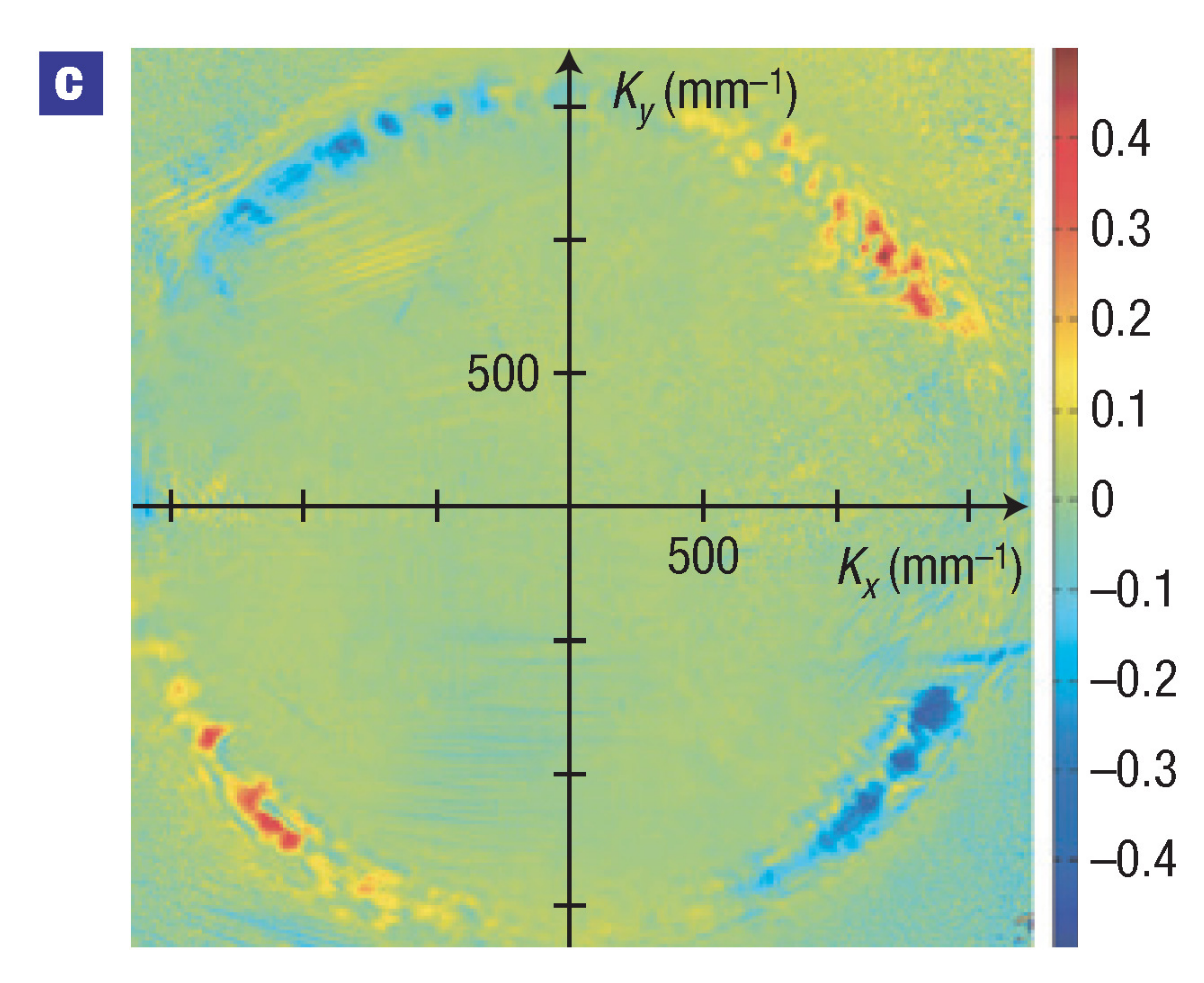} \\
 \end{center}
\caption{Top panel: snapshots of the propagation of cavity excitations created by a tightly focused and linearly polarized laser pulse in a semiconductor planar microcavity. Left and right panels show the optical intensity in the same (left) and in the orthogonal (right) polarization. The polarization-dependent patterns are due to the TE/TM splitting of the planar cavity modes \eq{Hamilt_so}. Top panel taken from \onlinecite{Langbein:PRB2005}. 
{Middle panels labeled as (a-c):} the effective optical spin-orbit interaction \eq{Hamilt_so} is equivalent to a $\kk$-dependent magnetic field (orange arrows) that rotates the pseudo spin associated to the photon polarization state  on the Poincar\'e sphere. {Lower panel}: as anticipated in~\cite{Kavokin:PRL2005}, the combination of the effective spin-orbit interaction with disorder-induced resonant Rayleigh scattering of a linearly polarized pump leads to the optical analogue of the spin Hall effect that is visible in the $\kk$-space pattern of the $z$-polarization shown in the bottom panel.
{Middle and lower panels} taken from \onlinecite{Leyder:NatPhys2007}
\label{fig:spin}}
\end{figure}

In writing the Hamiltonian \eq{Hamilt_cav}, we have implicitly assumed that the polarization states are degenerate.  While this assumption is exact at $k=0$, the reflection amplitudes off a dielectric mirror for the TE (Transverse Electric) and TM (Transverse Magnetic) linear polarization states
are generally different at $\kk\neq 0$, introducing a frequency splitting of the TE/TM modes proportional to $k^2$ in the small $k$ limit~\cite{Panzarini:1999PRB}. 

Physically, the resulting TE/TM splitting can be interpreted as a kind of spin-orbit interaction term coupling the orbital (the wavevector $\kk$) and pseudospin (the polarization) degrees of freedom of the cavity photon and can be described by a Hamiltonian term of the form~\cite{Kavokin:2004PRL,Shelykh:2010SST}
\begin{multline}
\Hamilt_{\rm so}=
\int\!\frac{d^2\kk}{(2\pi)^2}\,\sum_{\sigma,\sigma'}\,\Omega_{\rm so}(k)\,\left[(k_x^2-k_y^2)\,\sigma^{(x)}_{\sigma\sigma'}+2k_x k_y\,\sigma^{(y)}_{\sigma\sigma'}\right]\\ \times \ahd_{C,\sigma}(\kk,t)\, \ah_{C,\sigma'}(\kk,t)= \\
=\int\!\frac{d^2\kk}{(2\pi)^2}\,\sum_{\sigma,\sigma'}\,k^2\,\Omega_{\rm so}(k)\,\left[\cos(2\varphi)\,\sigma^{(x)}_{\sigma\sigma'}+\sin(2\varphi)\,\sigma^{(y)}_{\sigma\sigma'} \right]\\ \times \ahd_{C,\sigma}(\kk,t)\, \ah_{C,\sigma'}(\kk,t) \eqname{Hamilt_so}
\end{multline}
where $\sigma^{(x,y)}$ are the Pauli matrices, the polarization indices $\sigma,\sigma'$ run here over the circular polarizations basis $\sigma_\pm$ and $\Omega_{\rm so}(k)$ quantifies the $k$-dependent magnitude of the TE/TM splitting. In the alternative form shown in the last line, $\phi$ is the angle between the wavevector $\kk$ and the $x$ axis on the plane~\footnote{At a naive glance it may appear that the spin-dependent term \eq{Hamilt_so} breaks rotational symmetry: the point is that the Pauli matrices act here on the space of $\sigma_\pm$ polarized states of a spin-1 (and not $1/2$) object such as the photon, which explains their angular momentum being equal to $2$ and then the factor $2$ in the trigonometric functions on the last line of \eq{Hamilt_so}.} 

If one represents the two $\sigma_\pm$ polarization states of the cavity photon as a $1/2$-pseudospin, then the spin-orbit interaction \eq{Hamilt_so} is equivalent to an effective momentum-dependent magnetic field, which induces a precession of the pseudospins. As shown in Fig. \ref{fig:spin}, this kind of optical spin-orbit interactions can give rise to spectacular effects such as a strongly anisotropic polarization-dependent propagation~\cite{Langbein:PRB2005}. When combined with disorder-induced scattering, this effect is responsible for an optical analogue of the spin Hall effect~\cite{Kavokin:PRL2005,Leyder:NatPhys2007}. More recently, this same physics was investigated in a purely photonic cavity in~\cite{Maragkou:OptLett2011}.

It is worth noting that the term {\em spin Hall effect of light} is sometimes used in the photonic literature to denote a different family of effects stemming from the spin-orbit coupling experienced by photons propagating in bulk optical media with weak spatial inhomogeneities of the refractive index~\cite{Liberman:PRA1992,Bliokh:PRE2004,Onoda:PRL2004,Bliokh:PLA2004,Bliokh:PRL2006}. Recently, this coupling was shown to have remarkable consequences such as a sizable lateral shift of the trajectory of a light beam beyond the geometrical optics prediction~\cite{Hosten:Science2008,Bliokh:NatPhot2008}.

\subsubsection{Pumping and losses: input-output theory and master equation}

A quantum description of the driving of the cavity photon mode by an incident coherent laser beam can be obtained using the {\em input-output} theory of optical cavities. A complete discussion of this theory for single mode cavities can be found in quantum optics textbooks \cite{QuantumNoise,QuantumOptics}. Its extension to the case of planar microcavities with a continuum of in-plane $\kk$ modes can be found in~\cite{Ciuti:2006PRA}: the in-plane translational symmetry of the device guarantees that the in-plane $\kk$ vector is conserved, so that an external radiation of frequency $\omega$ and incident angle $\theta_{\rm inc}$ only couples to the cavity field component of in-plane wavevector $k={\omega}/{c}\;\sin\theta_{\rm inc}$. Correspondingly, the cavity field component of in-plane $\kk$ can only decay into external radiation emitted at an angle $\theta_{\rm out}$ satisfying the analogous condition  $k=\omega/c\,\sin\theta_{\rm out}$: this latter condition is schematically illustrated in the right-most part of the upper panel of Fig.\ref{polBEC}.

The Hamiltonian term describing the external driving of the cavity by a coherent incident field of amplitude  ${E}^{\rm inc}_\sigma(\rr,t)$ can be written in $\kk$-space as
\begin{multline}
\Hamilt_{\rm pump}=i\hbar \int\! \frac{d^2\kk}{(2\pi)^2}\, \sum_\sigma \Big(\eta_\sigma^{\rm fr}(\kk)  \tilde{E}_\sigma^{\rm inc}(\kk,t) \, \ahd_{C,\sigma}(\kk) \\ - \eta_\sigma^{{\rm fr}\,*}(\kk)\,\tilde{E}_\sigma^{{\rm inc}\,*}(\kk,t) \ah_{C,\sigma}(\kk)\Big),
\eqname{Hamilt_p}
\end{multline}
where $\tilde{E}^{\rm inc}_\sigma(\kk,t)$ is the Fourier transform of ${E}^{\rm inc}_\sigma(\rr,t)$ and the $\eta^{\rm fr}_\sigma(\kk)$ coefficient is proportional to the transmission amplitude of the front mirror for light with in-plane wavevector $\kk$. 

The finite transmittivity of the front and the back mirrors of the cavity is responsible for the re-emission of light from the cavity with an amplitude proportional to the in-cavity field operator. Combining this secondary emission with the direct reflection of the coherent laser light off the front mirror, input-output theory leads to the form 
\begin{eqnarray}
\hat{E}^{\rm tr}_\sigma(\kk,t)&=&\kappa^{\rm back}_\sigma(\kk)\, \ah_{C,\sigma}(
\kk,t) \eqname{tr_q} \\ 
\hat{E}^{\rm refl}_\sigma(\kk,t)&=& {E}_\sigma^{\rm inc}(\kk,t) + \kappa^{\rm fr}_\sigma(\kk) \ah_{C,\sigma}(\kk,t) \eqname{refl_q}
\end{eqnarray}
for the quantum operators describing the transmitted and reflected fields, respectively. 
Analogously to the input $\eta^{\rm fr}_\sigma(\kk)$ coefficients in \eq{Hamilt_p}, the output coefficients $\kappa_\sigma^{\rm fr,back}(\kk)$ are proportional to the transmission amplitude of the mirrors and can depend on the in-plane wavevector $\kk$ and the polarization $\sigma$.
An implicit assumption of this formalism is that transmission through the mirrors is almost instantaneous, so that the relations between the external and in-cavity fields \eq{Hamilt_p} are local in time. In frequency space, this means that the $\eta$ and $\kappa$ coefficients do not depend on the frequency $\omega$. The validity of this Markovian assumption is intuitively understood for thin metallic mirrors; a discussion for dielectric DBR mirrors was recently reported~\cite{Sarchi:2010PRB}.

As usual, the emission of light by the cavity is accompanied by a radiative damping of the cavity field at a wavevector- and polarization-dependent rate $\gamma^{\rm rad}_\sigma(\kk)$, proportional to the sum of the mirror transmittivities $|\kappa_\sigma^{\rm fr}(\kk)|^2+|\kappa_\sigma^{\rm back}(\kk)|^2$.
Tracing out the radiative modes of the field outside the cavity, dissipation results in additional terms to be included in the master equation for the evolution of the density matrix of the cavity field~\cite{QuantumOptics,QuantumNoise,Petruccione},
\begin{equation}
\frac{d\rho}{dt}=\frac{1}{i\hbar}[\Hamilt,\rho]+\mathcal{L}[\rho].
\eqname{master}
\end{equation}
Under the assumptions that the temperature of the radiative modes outside the cavity is much lower than the frequency of the cavity mode $k_B T\ll \hbar \omega_{\rm cav}$ and that the radiative coupling is Markovian~\cite{Petruccione}, the super-operator $\mathcal{L}[\rho]$ accounting for the dissipative effects has the zero temperature Lindblad form
\begin{multline}
\mathcal{L}^{\rm rad}[\rho]= \int\frac{d^2\kk}{(2\pi)^2}\,\frac{\gamma^{\rm rad}_{\sigma}(\kk)}{2} \Big[2\ah_{C,\sigma}(\kk)\rho\ahd_{C,\sigma}(\kk)+ \\
-\ahd_{C,\sigma}(\kk)\ah_{C,\sigma}(\kk)\rho-\rho\ahd_{C,\sigma}(\kk)\ah_{C,\sigma}(\kk)\Big].
\eqname{master_loss_rad}
\end{multline}
with a frequency-independent decay rate $\gamma^{\rm rad}_\sigma(\kk)$.
Additional, non-radiative decay channels due, for instance, to absorption in the cavity material can be included in the model by including  into the master equation \eq{master} additional terms $\mathcal{L}^{\rm nrad}[\rho]$ of the same form \eq{master_loss_rad} and proportional to the non-radiative loss rate $\gamma^{\rm nrad}_\sigma(\kk)$. In the following we shall indicate with $\gamma^C_\sigma(\kk)=\gamma^{\rm rad}_\sigma(\kk)+\gamma^{\rm nrad}_\sigma(\kk)$ the total decay rate of a cavity photon.

A complete calculation of the value of the $\eta$, $\kappa$ and $\gamma$ parameters for specific configurations goes beyond the scope of this review and generally requires a microscopic solution of Maxwell equations for the field propagation across the device; in a planar geometry, a tool commonly used for this kind of calculations is the so-called transfer matrix method, reviewed e.g. in~\cite{confined_book,savona_cargese}. Useful general relations can be mentioned for the case when the front and back mirrors of the cavity are identical: under this assumption,  $\eta^{\rm fr}_\sigma(\kk)=\eta^{\rm back}_\sigma(\kk)$, $\kappa^{\rm fr}_\sigma(\kk)=\kappa^{\rm back}_\sigma(\kk)$ and the  radiative decay rate $\gamma^{\rm rad}_\sigma(\kk) =
2\, \eta_\sigma(\kk)\, \kappa_\sigma(\kk)$. Furthermore, explicit expressions for the $\eta_\sigma(\kk)$ and $\kappa_\sigma(\kk)$ coefficients can be given for the case of a planar cavity enclosed by loss-less metallic mirrors of transmittivity $t\ll 1$ separated by a distance $\ell_z$, namely
$ \eta =(c t/2\ell_z) \sqrt{\ell_z/\pi\hbar\omega_c}$ and  $\kappa=-t\,\sqrt{\pi \hbar \omega_c / \ell_z}$.

\subsection{Optical nonlinearities and effective photon-photon interactions}

So far, we have discussed the dynamics of the cavity field at the level of linear optics, where  the quantum dynamics of the non-interacting cavity field reduces to the classical wave equation stemming from Maxwell's electrodynamics in material media.
The situation changes when the cavity layer (or the mirrors) embeds a material with a sizable optical nonlinearity: the non-linear dependence of the matter polarization on the applied electric field is responsible for a number of wave-mixing processes coupling different cavity modes and generating new frequency components~\cite{Butcher,Boyd}. Moreover, it can lead to strong modifications of the quantum fluctuation properties of the cavity field, such as a reduced noise on some field quadrature or even the generation of entangled states for the field~\cite{QuantumOptics}. 
The present review being devoted to the quantum fluid aspects of the photon gas, we will concentrate our attention on third-order nonlinearities proportional to the $\chi^{(3)}$ nonlinear polarizability that can be described in terms of binary interactions between pairs of photons.

Under the standard rotating-wave approximation (that is valid here provided the photon mass is larger than all other energy scales, e.g. losses, kinetic energy, interactions), the total number of photons is conserved and the nonlinear process can be described by a four-operator Hamiltonian term of the form,
\begin{multline}
\Hamilt_{\rm cav-cav}= \\ =
\frac{1}{2}\,\int\!\frac{d^2\kk}{(2\pi)^2}\, \int\!\frac{d^2\kk'}{(2\pi)^2}\,\int\!\frac{d^2\qq}{(2\pi)^2}\,\sum_{\sigma, \sigma'} \,\mathcal{V}^{\rm cav-cav}_{\sigma\sigma'}(\kk,\kk',\qq)\times \\ \times \ahd_{C,\sigma}(\kk+\qq)\, \ahd_{C,\sigma'}(\kk'-\qq)\, \ah_{C,\sigma'}(\kk')\, \ah_{C,\sigma}(\kk):
\eqname{Hamilt_cav-cav}
\end{multline}
where the matter degrees of freedom have been traced out and summarized into the effective photon-photon interaction potential $\mathcal{V}^{\rm cav-cav}_{\sigma, \sigma'}(\kk,\kk',\qq)$. The polarization index runs over the circular polarization states $\sigma=\sigma_\pm$ and spin angular momentum is implicitly assumed to be conserved in the collision process. Total momentum is also conserved in the process of two photons of initial in-plane wavevector $\kk$ and $\kk'$ scattering into the new wavevector states of in-plane momentum $\kk+\qq$ and $\kk'-\qq$. 

As the typical length scale of the electron dynamics in typical bulk material media and in semiconductor heterostructures~\cite{Bastard} used for quantum fluid effects is much shorter than the optical wavelength along the plane, the interaction potential $\mathcal{V}^{\rm cav-cav}_{\sigma, \sigma'}$ can be approximated with its zero-momentum value $\mathcal{V}^{o}_{\sigma\sigma'}$ for $\kk=\kk'=\qq=0$: in real-space, this corresponds to assuming that photon-photon interaction occur via a local potential,
\begin{multline}
\Hamilt_{\rm cav-cav}\simeq \frac{1}{2}\,\int\!d^2\rr\,\mathcal{V}^{o}_{\sigma\sigma'}\,\Psihd_{C,\sigma}(\rr)\,\Psihd_{C,\sigma'}(\rr)\times \\ \times \Psih_{C,\sigma'}(\rr)\,\Psih_{C,\sigma}(\rr).
\eqname{Hamilt_cav-cav-0}
\end{multline}
In two and three dimensions, the use of a strictly local interaction potential beyond the Born approximation often leads to UV divergences: standard techniques to make the theory regular involve renormalization of the interaction potential on a discrete lattice~\cite{Mora:PRA2003} or the use of suitably defined pseudo-potentials~\cite{Huang}.

In the simplest case when the photon frequencies that are involved in the photon fluid dynamics are very far away from electronic resonances in the nonlinear optical material, the optical transitions are virtual and the population of the excited electronic states remains negligible.
In this regime, the photon-photon potential can be expressed as
$\mathcal{V}^{\rm o}_{\sigma\sigma'}=-\frac{3\pi\,n_{nl}\,(\hbar \omega_{\rm cav}^o)^2}{n_o\,\ell_{z}}$
in terms of the nonlinear refractive index $n_{nl}$ normalized in a way such that the effective refractive index is $n_{\rm eff}=n_0+n_{nl}\,|\mathcal{E}|^2$. A more sophisticated theory of photon-photon interactions in a planar cavity device embedding a collection of anharmonic atoms appeared in~\cite{Chiao:PRA2004}. In the case of dielectric cavities, the cavity layer thickness $\ell_z$ has to be replaced by the effective thickness of the cavity mode $\ell_{\rm cav}$~\cite{savona_cargese}.

While sitting far from resonance generally allows to minimize undesired absorption losses in the material, non-resonant optical nonlinearities have the serious drawback of being generally very small and requiring a large number of photons to be observable. While this is not too much a concern for quantum fluid experiments, strong nonlinearities are required to observe photon blockade effects and generate strongly correlated photon fluids. In the next Subsection we shall extend our discussion to systems where the cavity mode is strongly coupled to narrow transitions in the optical medium: in this case, the photon inherits the strong nonlinearity of the matter excitations.

\subsection{Strong light-matter interaction and cavity polaritons}

\label{sec:exciton}

In the last two decades, the physics of strong light-matter interaction has flourished in many interesting domains, including the fields of atomic cavity QED~\cite{Raimond:RMP2001,cavityQED,Fleischhauer:RMP2005}, semiconductor microcavities and superconducting circuit QED~\cite{Schoelkopf:Nature2008,You:Nature2011}. In this section of our review we shall concentrate our attention on the case of planar semiconductor microcavities embedding one or many quantum wells~\cite{Varenna:2003,Deveaud:2007,Varenna:2009,Kavokin_book}. A brief account of other systems will be given later in Sec.\ref{strongcorrelation}.

\subsubsection{Quantum well excitons coupled to the cavity mode}

As detailed in more specialized reviews such as~\onlinecite{Bastard} and \onlinecite{Varenna:2003}, a {\em quantum well} (QW) consists of a thin semiconductor layer (a few nm thick) embedded in a different semiconductor compound acting as 'barrier' material. The chemical composition of the well is chosen to have the bottom of the conduction (the top of the valence) band at a lower (higher) energy than the surrounding material, thus producing quantum confinement of both electrons and holes. The lowest energy optical transition corresponds to the excitation a two-dimensional (hydrogen-like) electron-hole pair confined in the QW layer, the so-called {\em exciton}.
In typical microcavity samples, one or more QWs are embedded in the cavity layer, with their plane parallel to the cavity plane. To have a strong and quasi-resonant coupling of their electronic degrees of freedom with the cavity mode, the QWs are placed at the antinodes of the cavity field and the cavity mode frequency $\omega_{\rm cav}^o$ is tuned in the vicinity of the lowest QW exciton with frequency $\omega_{\rm exc}$.

At the level of linear optics~\cite{klingshirn,Bastard}, the contribution of the quantum well exciton to the optical properties of the cavity can be described in terms of a resonant contribution to the dielectric polarizability of the structure of the form
\begin{equation}
\chi_x(z,\kk,\omega)=\frac{\delta(z-z_{QW})\,f_{2D}}{\omega_{\rm exc}(\kk)-\omega-i\gamma_X/2},
\eqname{exc_classical}
\end{equation}
where the quantum well is approximated as a very thin layer located at $z_{QW}$ and the excitonic transition has an oscillator strength surface density $f_{2D}$.  The exciton resonant frequency is weakly dependent on the in-plane wavevector $\kk$, $\omega_{\rm exc}(\kk)\simeq \omega_{\rm exc}^o+\hbar k^2/2m_{\rm exc}$ with an exciton mass $m_{\rm exc}$ of the order of the electron mass, i.e. orders of magnitude larger than the effective photon mass $m_{\rm cav}$. The linewidth $\gamma_X$ accounts for all non-radiative decay channels of the exciton. 

In a quantum picture, the quantum well exciton can be described in terms of destruction (creation) operators $\ah_{X,\sigma}(\kk)$ ($\ahd_{X,\sigma}(\kk)$) that destroy (create) an exciton with total momentum $\kk$ and their real-space counterparts $\Psih_{X,\sigma}$ and $\Psihd_{X,\sigma}$ defined as their Fourier transform as in \eq{psiC}. In standard QWs~\cite{Bastard}, electrons can have spin projection $\sigma_e=\pm 1/2$ along the growth axis $z$, while holes can have $\sigma_h=\pm 3/2$. Among the four exciton states with spin projection $\sigma= \pm 1, \pm2$ that exist in a QW, here we shall restrict our attention to the $\sigma = \pm 1$ that are coupled to the cavity mode; the ones with $\sigma \pm 2$ are optically inactive.
Provided the interparticle distance remains much larger than their Bohr radius, excitons behave as  bosonic particles, whose creation and destruction operators satisfy Bose commutation rules of the same form as equations \eqs{commu_aC1}{commu_aC2} for cavity photons. The corrections to bosonic behavior due to the composite nature of the exciton have been theoretically addressed in~\cite{Combescot:rev2008}. 
Discussions of exciton physics with a special eye to planar microcavities can be found in the recent review papers by \onlinecite{Varenna:2003}, \onlinecite{Varenna:2009} and \onlinecite{Deng:RMP2010}.

In a second quantized formalism, the Hamiltonian describing in $\kk$ space the exciton dynamics and its coupling to the cavity field can be written as
\begin{multline}
\Hamilt_{\rm exc}=\int\!\frac{d^2\kk}{(2\pi)^2}\, \sum_\sigma\,\Big\{ \hbar\omega_{\rm exc}(\kk)\ahd_{X,\sigma}(\kk) \ah_{X,\sigma}(\kk)+ \\ 
+\hbar\Omega_R\,\Big[\ahd_{X,\sigma}(\kk) \ah_{C,\sigma}(\kk)+\ahd_{C,\sigma}(\kk) \ah_{X,\sigma}(\kk)\Big]\Big\}
\eqname{Hamilt_exc}
\end{multline} 
The last term describes the coherent conversion of an exciton into a cavity photon at a Rabi frequency $\Omega_R$
proportional to the product of the electric dipole of the quantum well exciton transition times the local amplitude of the cavity photon electric field. 
In terms of the classical dielectric properties \eq{exc_classical} of the quantum well, the Rabi frequency $\Omega_R$ in a planar metallic cavity can be related to the exciton oscillator strength $f_{2D}$ by
\begin{equation}
\Omega_R=\left[\frac{4\pi \omega_{\rm cav}^0}{\ell_z}\, f_{2D}\right]^{1/2}\,\frac{E(z)}{E_{\rm max}},
\eqname{Omega_R}
\end{equation}
where the final fraction accounts for the displacement of the quantum well position from an antinode of the cavity mode. In the general case of DBR cavities, the distance $\ell_z$ between the mirrors has to be replaced by the effective thickness $\ell_{\rm cav}$ of the cavity mode~\cite{savona_cargese}. If $N_{QW}$ quantum wells are present in the cavity, $N_{QW}-1$ linear combinations of the exciton states are dark, while the single bright one is coupled to the cavity mode with an enhanced coupling
$
\Omega_R=\sqrt{\sum_{i=1}^{N_{QW}} [\Omega_R^{(i)}]^2}$. When the QWs are located in equivalent positions, the collective coupling enhancement factor is $N_{QW}^{1/2}$.
Note that, in contrast to the case of a quantum well in free space~\cite{Andreani:SSC1991,Tassone:PRB1992}, the presence of the microcavity eliminates the direct coupling of the quantum well exciton to the external radiative modes and radiative decay of the exciton can only take place via the lossy cavity mode.

In addition to these effects stemming from its coupling to radiation, the exciton is also subject to non-radiative recombination processes. In a master equation formalism, these processes can be described by a Lindblad term of the form \eq{master_loss_rad} with a (generally weak) decay rate $\gamma_{X,\sigma}(\kk)$, as well as dephasing processes at a rate $\gamma^{\rm deph}_{X,\sigma}$~\cite{Liew:PRA2011}  due to, e.g., interactions with carriers and spatial inhomogeneities in the quantum well thickness.
The corresponding term in the master equation reads:
\begin{multline}
\mathcal{L}^{\rm deph}_X= \frac{\gamma^{\rm deph}_{X,\sigma}(\kk)}{2}  \Big[ 2 \ahd_{X,\sigma}(\kk)\ah_{X,\sigma}(\kk) \rho \ahd_{X,\sigma}(\kk)\ah_{X,\sigma}(\kk)  + \\
- \ahd_{X,\sigma}(\kk)\ah_{X,\sigma}(\kk) \ahd_{X,\sigma}(\kk)\ah_{X,\sigma}(\kk)  \rho \,+ \\
 - \rho \ahd_{X,\sigma}(\kk)\ah_{X,\sigma}(\kk) \ahd_{X,\sigma}(\kk)\ah_{X,\sigma}(\kk)\Big] \eqname{master_deph_X}
\end{multline}
In view of the on-going theoretical and experimental developments, it is important to remind that the form \eq{Hamilt_exc} of the light-matter coupling term is based on the so-called rotating-wave approximation where anti-resonant terms proportional to $\ah_{X,\sigma}(\kk) \ah_{C,\sigma}(\kk)$ are neglected. 
This approximation is very accurate as long as $\Omega_R\ll \omega_{\rm cav}, \omega_{\rm exc}$, which is generally the case for microcavity systems under examination here where $\Omega_R$ is of the order of $10$~meV for typical III-V based samples and of the order of $20$~meV for II-VI based samples. Significant deviations appear in the opposite {\em ultra-strong coupling} regime where $\Omega_R$ is comparable or larger than $\omega_{\rm cav}, \omega_{\rm exc}$. Such a regime has been recently theoretically predicted and experimentally observed in systems where the cavity mode is strongly coupled to intersubband electronic transitions in doped quantum wells~\cite{Ciuti:PRB2005}, \cite{Anappara:PRB2009,Guenter:Nature09,Todorov:PRL10}, cyclotron transitions of two-dimensional electron gases~\cite{Hagenmuller:PRB2010,Scalari:Science12} and in circuit-QED systems~\cite{Devoret:Annalen2007},~\cite{Bourassa:PRA09,Nataf:PRL10},~\cite{Niemczyk:NaturePhys2010,Fedorov:PRL10}. In addition to going beyond the rotating-wave approximation in the light-matter coupling term in \eq{Hamilt_exc}, a consistent theoretical description of systems in the ultra-strong coupling regime also requires a more sophisticated modeling of the frequency dependence of the dissipative baths~\cite{Ciuti:2006PRA,Carusotto:PRA2012,Ridolfo:arXiv2012}.

\subsubsection{Exciton-polaritons}

The coupled dynamics of photons and excitons in a microcavity is described by the Hamiltonian terms \eq{Hamilt_cav} and \eq{Hamilt_exc} that involve the product of two field operators: as a result, the quantum dynamics exactly recovers the classical Maxwell wave equations inserting a thin layer of material with the susceptibility \eq{exc_classical}. Furthermore, simple linear combinations of the $\ah_{C,\sigma}(\kk)$ and $\ah_{X,\sigma}(\kk)$ operators~\cite{Hopfield:PR1958} can be used to transform the Hamiltonian into a diagonal form
\begin{multline}
\Hamilt_{\rm cav}+\Hamilt_{\rm exc}=\\ =\int\!\frac{d^2\kk}{(2\pi)^2}\,\sum_\sigma\,\hbar\omega_{LP,\sigma}(\kk)\,\ahd_{LP,\sigma}(\kk)\,\ah_{LP,\sigma}(\kk)+ \\ 
+\hbar\omega_{UP,\sigma}(\kk)\,\ahd_{UP,\sigma}(\kk)\,\ah_{UP,\sigma}(\kk),
\end{multline}
where the $\ah_{UP,\sigma}(\kk)$ and $\ah_{LP,\sigma}(\kk)$ operators correspond to hybridized excitations resulting from the linear superposition of exciton and cavity photon modes, the so-called  {\em exciton-polaritons}. Thanks to the translational symmetry along the cavity plane, the $\kk$ wavevector and the spin $\sigma$ remain good quantum numbers for the new quasi-particles. 

The dispersion of the {\em upper} and {\em lower} polariton branches $\omega_{UP,\sigma}(\kk)$ and $\omega_{LP,\sigma}(\kk)$ has the typical anticrossing form
\begin{multline}
\omega_{(UP,LP),\sigma}(\kk)=\frac{\omega_{\rm cav,\sigma}(\kk)+\omega_{\rm exc,\sigma}(\kk)}{2} + \\
\pm\left[\Omega_R^2 + \left(\frac{\omega_{\rm cav,\sigma}(\kk)-\omega_{\rm exc,\sigma}(\kk)}{2}\right)^2 \right]^{1/2}
\eqname{disp_polar}
\end{multline}
that is shown in the middle panel of Fig.\ref{polBEC}: the bare cavity photon and exciton branches (dashed lines) are mixed by the light-matter coupling term into the polariton branches; the minimum splitting $2\Omega_R$ of the two polariton branches is obtained for  $\omega_{\rm cav}(\kk)=\omega_{\rm exc}(\kk)$. The so-called {\em strong light-matter coupling} regime is defined when $\Omega_R>\gamma$, i.e., light-matter coupling exceeding losses.

The photonic and excitonic content of the polariton modes is quantified by the real-valued Hopfield coefficients, which relate the cavity mode and exciton field amplitude to the ones in the lower and upper polariton modes,
\begin{eqnarray}
\ah_{C,\sigma}(\kk)&=&u_c^{LP}(\kk)\,\ah_{LP,\sigma}(\kk)+u_c^{UP}(\kk)\,\ah_{UP,\sigma}(\kk) \eqname{HopfC} \\
\ah_{X,\sigma}(\kk)&=&u_x^{LP}(\kk)\,\ah_{LP,\sigma}(\kk)+u_x^{UP}(\kk)\,\ah_{UP,\sigma}(\kk). \eqname{HopfP}
\end{eqnarray}
Close to the crossing point, both polariton modes have approximately equal photon and exciton content $|u_{c,x}^{LP}|^2\simeq 1/2$, while farther away the two polariton acquire a purely excitonic or photonic character. Differently from the general case of~\cite{Hopfield:PR1958}, the absence of the $\ahd_{(LP,UP),\sigma}$ creation operators in the Bogoliubov transormation \eqs{HopfC}{HopfP} is a consequence of the rotating-wave approximation.

The effective masses $m_{LP,UP}$ of the lower and upper polariton branches are  calculated from the curvature of the dispersion \eq{disp_polar}. In the most remarkable case where $\omega_{\rm cav}^o\approx \omega_{\rm exc}^o$, the condition $m_{\rm cav}\ll m_{\rm exc}$ implies that  $\omega_{LP,UP}^o\simeq \omega_{\rm cav, exc}^o\mp \Omega_R$ and $m_{LP}\simeq m_{UP}\simeq 2m_{\rm cav}$; furthermore, the two polaritons have equal photon and exciton content as long as $\hbar k^2/2m_{\rm cav}\ll \Omega_R$. Most of the experiments that will be discussed in the next sections only involve the lowest states of the lower polariton branch for small wavevectors and a description in terms of a single polariton branch with parabolic dispersion of mass $m_{LP}$ is accurate. More details on the basic properties of microcavity polaritons can be found in~\cite{Varenna:2003,Deveaud:2007,Varenna:2009,Kavokin_book}

\subsubsection{Polariton-polariton interactions}
\label{sec:polariton-polariton}

As it is typically done in the many-body physics, a widespread procedure is to describe the system in terms of an effective model Hamiltonian that is able to reproduce the exciton-exciton interactions without invoking its elementary constituents: the idea is to replace the complex Coulomb interactions between the electron and holes by a simple two-body interaction potential involving the exciton as a whole.
In the simplest version, low-energy scattering at a relative wave vector much smaller than the exciton Bohr radius can be accurately described by a contact two-body interaction potential term,
\begin{multline}
\Hamilt_{XX}=\frac{1}{2}\,\int\!d^2\rr\,\sum_{\sigma\sigma'} \mathcal{V}_{\sigma,\sigma'}^{XX}\,\Psihd_{X, \sigma}(\rr)\,\Psihd_{X,\sigma'}(\rr)\times 
\\ \times
\Psih_{X,\sigma'}(\rr)\,\Psihd_{X,\sigma}(\rr),
\eqname{Hamilt_XX}
\end{multline}
where the spin indices $\sigma,\sigma'$ run over the circular polarization basis $\sigma,\sigma'=\sigma_\pm$. 
Rotational invariance for a contact interaction potential imposes that total exciton spin is conserved and that $\mathcal{V}_{\sigma_+,\sigma_+}^{XX}=\mathcal{V}_{\sigma_-,\sigma_-}^{XX}=\mathcal{V}^{XX}_T$ and $\mathcal{V}_{\sigma_+,\sigma_-}^{XX}=\mathcal{V}_{\sigma_-,\sigma_+}^{XX}=\mathcal{V}^{XX}_S$.

A first estimation of $\mathcal{V}^{XX}$ in terms of a microscopic quantum well electron-hole model of the exciton was provided in~\cite{Ciuti:PRB1998, Tassone:PRB1999} within the Born approximation, giving $\mathcal{V}^{XX}_T={6e^2a_B}/{\epsilon}$ where $\epsilon$ is the dielectric constant of the material and $a_B$ is the exciton Bohr radius. 
This positive and featureless value of the triplet interaction constant appears to be in reasonable agreement with experimental measurements~\cite{Amo:NPhys2009,Ferrier:PRL2011} and provides the most relevant contribution when the microcavity is excited by circularly polarized light.

For other polarization configurations, the much richer physics of the singlet channel needs to be considered: different experiments~\cite{Kavokin:PSSC2005,Amo:PRB2010} have reported values of the singlet interaction constant with opposite signs; first systematic studies of the singlet interaction constant as a function of the microcavity parameters were recently performed in~\cite{Vladimirova:PRB2010,Paraiso:PhD}.
In analogy to the Feshbach resonance phenomenon in atom-atom scattering when the energy of the incident atomic pair is in the vicinity of a molecular intermediate state~\cite{Chin:RMP2010}, several authors~\cite{Savasta:SSC1999,Savasta:PRL2003,Wouters:PRB2007,Carusotto:EPL2010} have suggested the importance of the {\em biexciton} intermediate state (i.e. a two-electron, two-hole bound complex in the singlet scattering channel: in the vicinity of the scattering resonance, the effective interaction constant $\mathcal{V}^{XX}_S$ in the singlet channel is expected to be strongly enhanced and to change sign. More details on the theory of biexciton Feshbach effects will be presented in Sec.\ref{sec:Feshbach}.

In principle, exciton-exciton interactions may also transform a pair of $\sigma_\pm$ bright excitons into a pair of spin $\pm 2$ dark excitons~\cite{Ciuti:PRB1998}. While these processes are important in isolated quantum wells, they no longer conserve energy in microcavities: because of the Rabi coupling between photons and bright excitons, the dark spin $\pm 2$ excitons are at much higher energy than the lower polariton branch and can only play a role as non-resonant intermediate states of high-order processes.

An additional interaction channel originates from saturation of the exciton oscillator strength\cite{Tassone:PRB1999,Rochat:PRB2000,Glazov:PRB2009}: Pauli exclusion principle for electrons and holes forbids that another exciton be created at a distance shorter than the Bohr radius from an existing exciton. 
At the lowest order in the exciton density and retaining only the bright exciton states, this can be modeled as an effective quartic Hamiltonian term of the form
\begin{multline}
\Hamilt_{\rm sat}=\frac{1}{2}\,\int\!d^2\rr\,\sum_{\sigma\sigma'} \mathcal{V}_{\sigma,\sigma'}^{\rm sat}\,\Psihd_{X,\sigma}(\rr)\,\Psihd_{X,\sigma'}(\rr)\times \\
\times \Psih_{X,\sigma'}(\rr)\,\Psih_{C,\sigma}(\rr) + \textrm{h.c.}
\eqname{Hamilt_sat}
\end{multline}
with a saturation potential $\mathcal{V}_{\sigma,\sigma'}^{sat}= \delta_{\sigma,\sigma'}\,{\hbar\Omega_R}/{n_{\rm sat}}$.
As it is discussed in detail in the recent paper by \onlinecite{Glazov:PRB2009}, the exact value of the saturation density $n_{\rm sat} \propto 1/a_B^2$ depends on the specific shape of the wavefunction of the relative motion. For typical values of semiconductor microcavities, the saturation term is generally significantly smaller than the exciton-exciton contribution.

As previously mentioned, in many relevant experimental conditions, one can restrict to the bottom of the lower polariton branch and approximate the dispersion as parabolic. In this regime, we can rewrite the interaction Hamiltonians \eq{Hamilt_XX} and \eq{Hamilt_sat} in terms of polariton operators and keep only the terms involving the lower polariton branch. 
This finally leads to an effective polariton-polariton contact interaction 
\begin{multline}
\Hamilt_{LP-LP}=\frac{1}{2}\,\int\!d^2\rr\,\sum_{\sigma\sigma'} \mathcal{V}_{\sigma,\sigma'}^{LP }\,\Psihd_{LP,\sigma}(\rr)\,
\Psihd_{LP,\sigma'}(\rr)\times \\
\times
\Psih_{LP,\sigma'}(\rr)\,\Psihd_{LP,\sigma}(\rr)
\eqname{Hamilt_LPLP}
\end{multline}
with \begin{equation}
\mathcal{V}_{\sigma,\sigma'}^{LP}= |u_x^{LP}|^4\,\mathcal{V}_{\sigma,\sigma'}^{XX} + 2\,|u_x^{LP}|^2\,u_x^{LP}\, u_c^{LP}\,\mathcal{V}_{\sigma,\sigma'}^{sat}.
\eqname{V_LPLP}
\end{equation}

Recent experiments~\cite{Cristofolini:Science2012} have demonstrated strong coupling of a cavity photon with a hybrid exciton in a double quantum well geometry, where the direct and indirect exciton states are mixed by coherent electron tunneling events across the barrier. As a result, the polariton acquires a finite electric dipole moment, that is expected to strongly reinforce polariton-polariton interactions and, possibly, induce new effects due to long-range dipole form of the interaction potential~\cite{Astrakharchik:PRL2007,Boning:PRB2011}.

\subsection{External potentials affecting the in-plane motion of cavity excitations}
\label{sec:external_pot}

In analogy to magnetic and optical traps for atomic gases~\cite{BECbook}, several strategies have been explored to implement spatially- and spin-dependent external potentials for cavity photons and excitons, creating Hamiltonian terms like
\begin{multline}
\Hamilt_{\rm pot}=\int\!d^2\rr\sum_{\sigma,\sigma'}\,\Big[ V^{(\rm cav)}_{\sigma,\sigma'}(\rr)\, \Psihd_{\rm cav,\sigma}(\rr)\,\Psih_{\rm cav,\sigma'}(\rr) + \\
+V^{(\rm exc)}_{\sigma,\sigma'}(\rr)\, \Psihd_{\rm exc,\sigma}(\rr)\,\Psih_{\rm exc,\sigma'}(\rr) \Big].
\end{multline}
Provided the amplitudes of the potentials $V^{(\rm cav)}$ and $V^{(\rm exc)}$ are much smaller than the Rabi energy $\hbar \Omega_R$ and for a sufficiently smooth spatial variation, the resulting potential acting on lower polaritons can be written as
\begin{equation}
V^{\rm LP}_{\sigma\sigma'}(\rr)= |u_c^{LP}|^2\,V^{(\rm cav)}_{\sigma\sigma'}(\rr) +  |u_x^{LP}|^2\,V^{(\rm exc)}_{\sigma\sigma'}(\rr).
\eqname{V_LP}
\end{equation}
 We will start by considering several experimental protocols to  generate scalar potentials of the form $V^{(\rm cav, exc)}_{\sigma\sigma'}=V^{(\rm cav, exc)}\,\delta_{\sigma\sigma'}$. 
 
The polariton trap used in the Bose-Einstein condensation experiment of~\onlinecite{Balili:Science2007} was created by applying a mechanical stress, resulting in an energy red-shift of the exciton state~\cite{Negoita:APL1999} and thus an attractive potential for excitons ($V^{(\rm exc)}<0$) with an approximately harmonic profile at the bottom.

\begin{figure}[htbp]
\begin{center}
\includegraphics[width=\columnwidth,angle=0,clip]{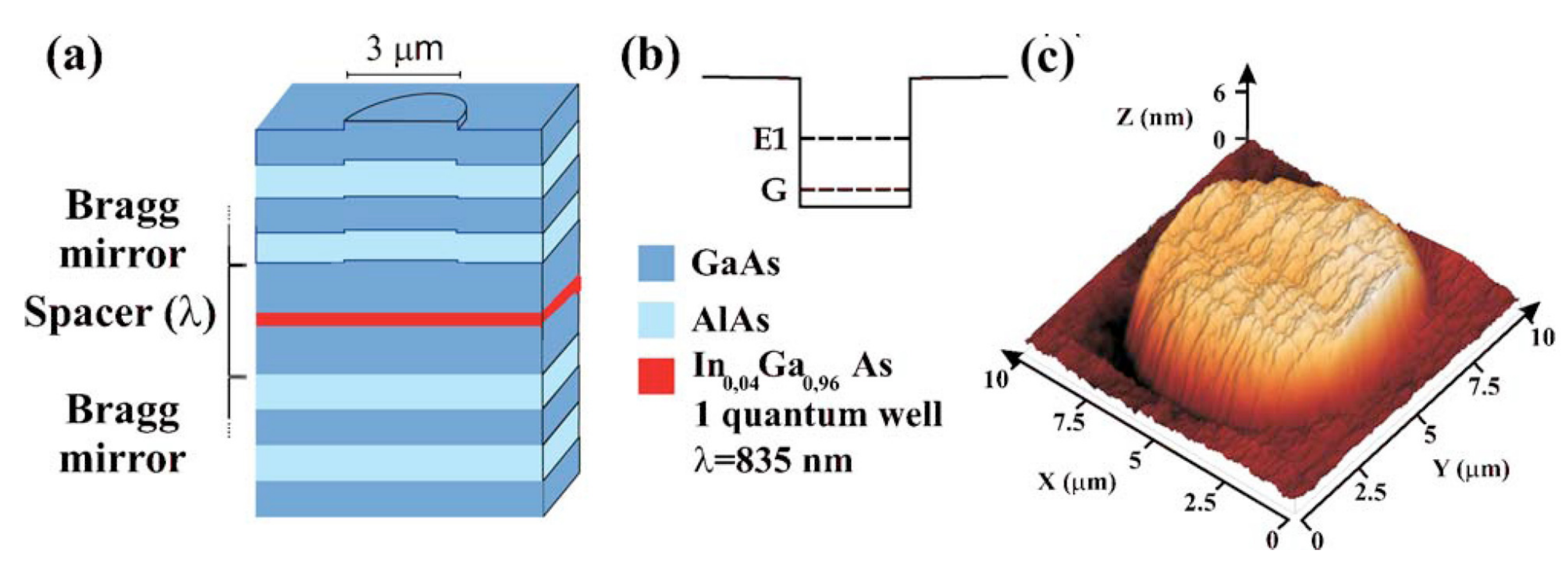}
\includegraphics[width=0.55\columnwidth,angle=0,clip]{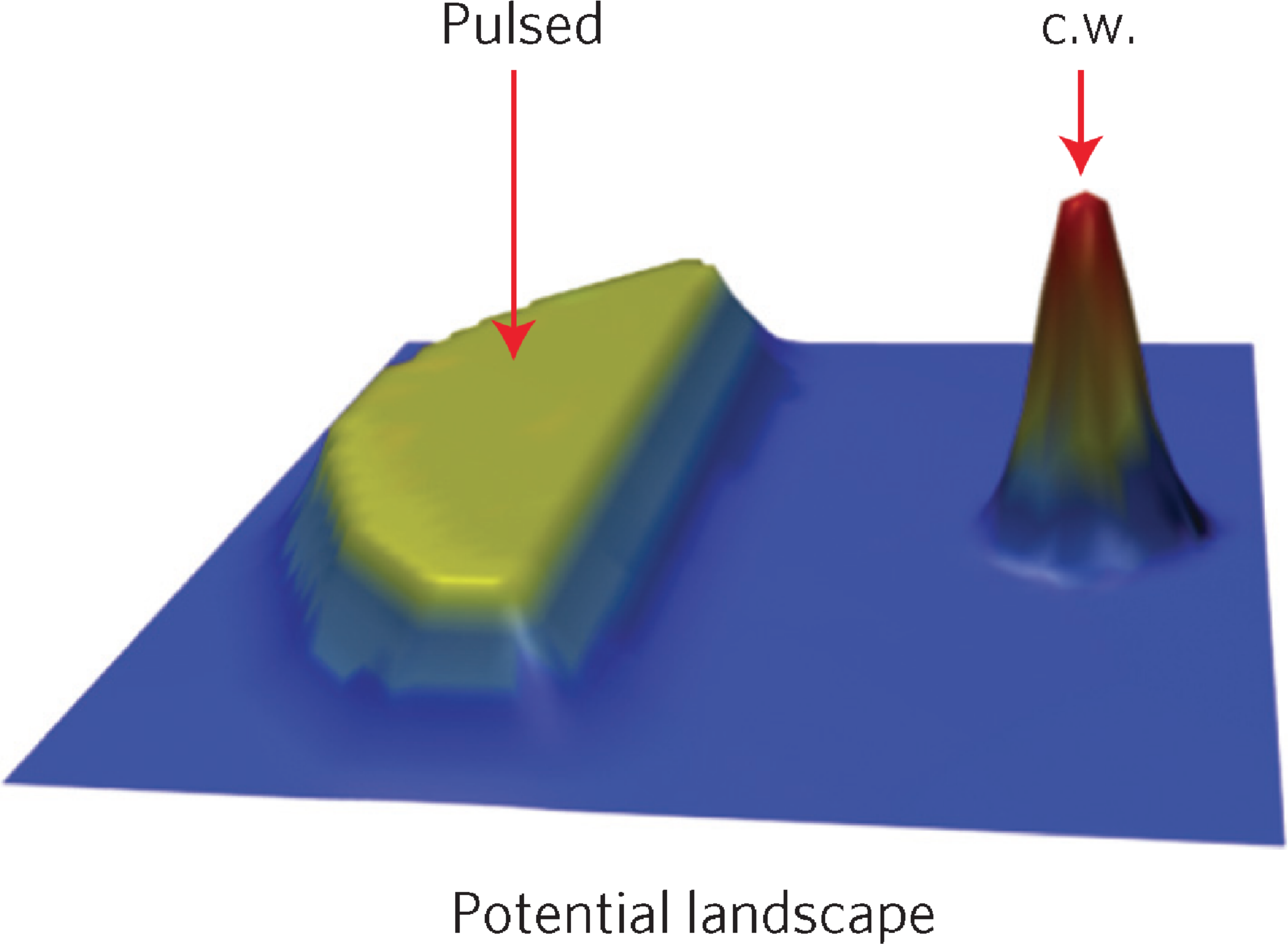}
\end{center}
\caption{Top panel: figure from~\onlinecite{ElDaif:APL2006}. An effective lateral confinement potential for cavity photons (and hence for cavity polaritons) can be obtained with a position-dependent cavity thickness (achievable in semiconductor microcavities via growth, etching and regrowth). Bottom panel: figure from~\onlinecite{Sanvitto:NPhot2011}. An effective lateral confinement potential for cavity photons can be obtained by using control laser spots with different polarizations as first achieved in \cite{Amo:PRB2010}.
}
\label{fig:potential}
\end{figure}
Another concept of polariton trap exploits the dependence of the cavity mode frequency on the thickness of the cavity layer.
This feature is usually exploited to obtain a full scan of the exciton-photon detuning across the resonance on a single microcavity sample by growing a wedge-shaped wafer~\cite{Weisbuch:PRL1992}. In the experiment of~\cite{Sermage:PRB2001}, a polariton acceleration effect was observed as a consequence of the cavity wedge: within a parabolic band approximation, the polariton motion can be described by a Newton-like equation $m_{LP}d^2\rr/dt^2 = -\nabla_\rr \hbar \omega_{LP}(\kk=0)$.

A more sophisticated development of this same idea was implemented in~\cite{Lu:APL2005,ElDaif:APL2006,Kaitouni:PRB2006} to create polariton boxes with an overgrowth technique able to create a position-dependent thickness of the cavity spacer (see top panel of Fig. \ref{fig:potential}). This technique provides a quite flexible tool to design external potentials with arbitrary shapes on characteristic spatial lengths in the few $\mu$m range. 

A more brutal way of confining polaritons is to design a micropillar structure by etching away all the layers forming the top mirror and the cavity layer (and possibly also the lower mirror down to the substrate): in this case, light is confined in the in-plane directions by the large refractive index mismatch at the air-semiconductor interface~\cite{confined_book}. As with the regrowth technique, arbitrary geometries can be realized with full three-dimensional confinement~\cite{Gerard:APL1996,Reithmaier:PRL1997,Ohnesorge:PRB1997} or with free motion along photonic wires~\cite{Zhang:PRL1995,Kuther:PRB1998,Wertz:NatPhys2010}. The direct processing of micropillar devices may come at the price of larger lateral losses at the cavity/air interface. The recent realization of double pillar configurations showing a sizable tunnel coupling of the photonic wavefunctions~\cite{deVasconcellos:APL2011} is very promising in view of creating more complex structures such as cavity arrays.

Another strategy to apply external potentials to exciton-polaritons is to deposit metal films on the surface of the cavity: depending on the geometry, this can either increase the photon energy by squeezing the tails of electric field inside the DBR mirror~\cite{Kim:PSSB2008} or even strongly mix the Tamm plasmon with the exciton and the cavity photon modes as predicted in~\cite{Kaliteevski:APL2009}. Experimental evidence of the strong coupling of a Tamm plasmon with quantum well excitons in a single DBR mirror-metal/air gap microcavity was reported in~\cite{Symonds:APL2009,Grossmann:APL2011}. Using metallic layers also allows for a dynamic tuning of the external potential by applying DC electric fields to the structure: an electric field indeed acts on polaritons by reducing the quantum well oscillator strength by spatially separating the electron/hole pairs and by lowering its energy via the quantum confined Stark effect~\cite{Kim:PSSB2008} and thermal expansion~\cite{Grossmann:APL2011}.

A technique to generate a periodic potential using surface acoustic waves (SAWs) propagating in the semiconductor structure was developed in~\cite{deLima:PRL2006} and then applied to fragment a polariton condensate into an array of strongly elongated gases with reduced coherence~\cite{Cerda:PRL2010}. Microscopically, the SAW strain field simultaneously changes the thickness and the refractive index of the cavity layer, modulating the exciton energy via a deformation potential. Each of these effects is proportional to the real part of the local amplitude of the sound wave. As a result the effective potential moves in space at a speed equal to the SAW phase velocity, on the order of the sound speed in the material ($3\times 10^3$~m/s in~\onlinecite{Cerda:PRL2010}), i.e. much smaller than all other characteristic speeds of the polariton system.

An even more flexible all-optical technique was developed in~\cite{Amo:PRB2010}: a strong $\sigma_-$ laser field is used to inject $\sigma_-$ polaritons into the cavity with a given spatial profile. Assuming that one can neglect spin-flip processes and that a large number of $\sigma_-$ polaritons is injected, polariton-polariton interactions between counter-polarized polaritons create an effective potential for $\sigma_+$ polaritons, whose geometry and time-dependence can be easily controlled via the $\sigma_-$ laser field. As it is shown in the bottom panel of Fig. \ref{fig:potential}, quite localized potentials can be generated by this technique. For instance, they were used in~\cite{Sanvitto:NPhot2011} to study vortex dynamics in polariton superfluids. With respect to SAWs, optical potentials have the advantage that they can be modulated on extremely short time scales and designed with almost any desired spatial shape down to the $\mu$m scale.

Even though the external potential generated by all these methods is mostly a spin-independent one, in some cases it is essential to take into account in the model its spin-dependent component: the different reflection amplitude of TE/TM polarization states at the cavity/air interface in a micropillar device is responsible for a splitting of the two linear polarized states~\cite{Kuther:PRB1998}; an analogous effect may also arise from the coupling of the exciton with mechanical stress in a laterally patterned device~\cite{Dasbach:PRB2002}. In some proposed applications such as~\cite{Umucalilar:PRA2011}, it is essential to be able to impose a specific form of spin-dependent potential to the polaritons. The Zeeman shift under a static magnetic field can be used to split the $\sigma_\pm$ components of the exciton. A (possibly spatially modulated) strain field~\cite{Jansen:APL1996} or a lateral patterning with submicrometer periodicity~\cite{Flanders:APL1983} can be used to generate a linear birefringence in the cavity material, which would then result in a spin-dependent photonic potential of the form
\begin{equation}
V^{(\rm cav)}_{\sigma\sigma'}(\rr)=\Delta(\rr)\,\Big[\cos(2\phi(\rr))\,\sigma^{(x)}_{\sigma\sigma'}+\sin(2\phi(\rr))\,\sigma^{(y)}_{\sigma\sigma'}\Big],
\end{equation}
where the angle $\phi(\rr)$ fixes the orientation of the birefringence axis and the basis of circularly polarized states $\sigma=\sigma_\pm$ is used.

\section{The driven-dissipative Gross-Pitaevskii equation}
\label{GP}

\subsection{The mean-field approximation}

In the previous section we have introduced a quantum mechanical model for the coupled quantum field dynamics of cavity photons and quantum well excitons: the interaction terms describing for exciton-exciton collisions make the dynamics non trivial and are responsible for a number of nonlinear and quantum phenomena. A standard approximation to attack this kind of problems is the so-called {\em mean field approximation}, based on classical evolution equations for the expectation values of the quantum field operators $\mathcal{E}_{\sigma}(\rr)= \langle \Psih_{C,\sigma}(\rr) \rangle$ and $\mathcal{P}_{\sigma}(\rr)= \langle \Psih_{X,\sigma}(\rr) \rangle$: the former is (approximately) proportional to the expectation value of the in-cavity electric field, the latter is proportional to the matter polarization due to the exciton transition. The mean-field equations of motion for $\mathcal{E}_{\sigma}(\rr)$ and $\mathcal{P}_{\sigma}(\rr)$ are obtained from the Heisenberg equations of motion for $\Psih_{C,\sigma}(\rr)$ and $\Psih_{C,\sigma}(\rr)$ by replacing every instance of an operator with the corresponding expectation value.

Of course, this approach is exact as long as we are restricting ourselves to terms in the Hamiltonian that involve at most two operators, e.g. the free dynamics of the coupled cavity photons and excitons, their mutual interconversion, as well as the one-body pumping and loss terms discussed in the previous section. An approximation is instead made on the interactions terms, for which one is assuming,
\begin{multline}
\langle \Psihd_{X,\sigma}(\rr)\, \Psih_{X,\sigma}(\rr) \,\Psih_{X,\sigma}(\rr) \rangle \approx \\ \approx \langle \Psihd_{X,\sigma}(\rr) \rangle\,\langle \Psih_{X,\sigma}(\rr) \rangle\,\langle \Psih_{X,\sigma}(\rr) \rangle.
\end{multline}

In nonlinear optics, a mean-field approximation of this kind is implicitly made whenever one writes the polarization of a medium as the product of a nonlinear susceptibility $\chi^{(n)}$ times the $n$-th power of the classical electric field~\cite{Boyd,Butcher}: this approach provides accurate results in most optical media as nonlinear effects require the presence of a large number of photons. If one is interested in the interplay of the optical nonlinearity with quantum fluctuations, a linearized treatment is generally enough to account for the relatively weak quantum fluctuations~\cite{QuantumOptics}. An alternative, semiclassical approach is based on the truncated-Wigner representation of the quantum field, as discussed in the next subsection. The quantitative validity of these approaches has been tested by comparing with exact calculations for the simplest model of single-mode nonlinear cavity~\cite{Carusotto:PRA2001} as well as with multimode Wigner simulations~\cite{Verger:PRB2007}. Of course, this approximation breaks down completely as soon as one enters the photon blockade regime~\cite{Imamoglu:PRL97} where a single photon is able to substantially modify the response of a device and the discrete, quantum nature of photons starts playing a crucial role.

In the context of quantum gases of material particles, a classical partial differential equation for the superfluid order parameter was first written by~\onlinecite{Gross:NC1961} and~\onlinecite{Pitaevskii:1961} to describe quantum vortices in liquid Helium. Starting from modern formulations of the Bogoliubov theory of the dilute Bose gas, a {\em Gross-Pitaevskii equation} (GPE) of the form
\begin{equation}
i\hbar \partial_t \Psi= -\frac{\hbar^2}{2m}\nabla^2\Psi+V_{\rm ext}(\rr)\,\Psi+\frac{4\pi\hbar^2 a}{m} |\Psi|^2\,\Psi
\eqname{atomicGPE}
\end{equation}
can be derived from first principles to describe the dynamics of the condensate wavefunction $\Psi(\rr,t)$: the macroscopic fraction of particles that populate the condensate mode behave in a collective way and the quantum atomic matter field $\Psih(\rr,t)$ behaves as a classical field $\Psi(\rr,t)$~\cite{BECbook,Leggett:2001RMP}. 
Generally speaking, the GPE is quantitatively accurate as long as the occupation of modes other than the condensate one is small: in three dimensions this requires that 
the temperature is much lower than the transition temperature for Bose-Einstein condensation and that the gas is dilute, i.e. the atom-atom scattering length $a$ is much shorter than the mean interparticle spacing $n a^3\ll 1$.

From a physical standpoint, the GPE is then for the matter field what Maxwell equations are for quantum electrodynamics in nonlinear media. An important difference is however worth emphasizing: while the global phase of the electromagnetic field has a direct and observable physical meaning, conservation of the total number of particles implies that the expectation value of the matter field of massive particles exactly vanishes. Even if treatments based on a spontaneous breaking of the phase symmetry are popular in the literature and have a deep physical foundation in terms of the BEC phase transition~\cite{Gunton:PR1968}, some doubts may remain on the consistence of descriptions based on a classical atomic field, especially for finite systems. The picture has been reconciled by recent theoretical works that have developed particle-number conserving versions of the Bogoliubov theory~\cite{Castin:PRA1998,Gardiner:PRA1997} and have investigated in full detail the meaning of the condensate phase~\cite{Leggett_green,Javanainen:PRL1996,Cirac:PRA1996,Castin:PRA1997}.

\subsection{Generalized GPE under coherent driving}
\label{sec:coherent_GPE}

In many experimental circumstances, it is not necessary to work with the pair of equations of motions for the photonic $\mathcal{E}_{\sigma}(\rr)$ and excitonic $\mathcal{P}_{\sigma}(\rr)$ fields and one can restrict to a single classical field describing the lower polariton field $\Psi_{LP}(\rr,t)$ in a single spin state. This simplified description is generally legitimate provided the Rabi frequency $\Omega_R$ is much larger than all other energy scales of the problem, namely the kinetic and interaction energies, the pump detuning from the bottom of the lower polariton, and the loss rates. 
The resulting Gross-Pitaevskii equation for polaritons has the form
\begin{multline}
i\partial_t  \Psi_{LP}(\rr,t)=\Big[\omega_{LP}^o - \frac{\hbar}{2m_{LP}}\nabla^2 +V_{LP}(\rr)\Big]\Psi_{LP}(\rr,t) + \\
+ g_{LP}\,|\Psi_{LP}(\rr,t)|^2\,\Psi_{LP}(\rr,t)\\
-\frac{i\gamma_{LP}}{2}\, \Psi_{LP}(\rr,t) + i\eta_{LP}\,E^{\rm inc}(\rr,t).
\eqname{GPE_LP}
\end{multline}
With respect to the Gross-Pitaevskii equation \eq{atomicGPE} describing atomic Bose-Einstein condensates, Eq.\eq{GPE_LP} includes additional terms to account for the driven-dissipative nature of the polariton gas, namely a loss rate proportional to $\gamma_{LP}$ and the coherent pumping proportional to the incident field $E^{\rm inc}(\rr,t)$. 

The parameters appearing in the polariton GPE \eq{GPE_LP} are defined in terms of the microscopic description of polaritons discussed in the previous section: $m_{LP}=m_{\rm cav}/|u^{LP}_c|^2$ is the lower polariton mass, $\omega_{\rm LP}^o$ is the frequency of the bottom of the lower polariton branch, $V_{LP}(\rr)$ is the external potential felt by polaritons as defined in \eq{V_LP},  $g_{LP}$ is the polariton-polariton interaction constant introduced in \eq{V_LPLP} and the $\eta_{LP}$ parameter quantifies the coupling of the polariton to incident radiation, $\eta_{LP}=\eta^{\rm fr}_\sigma\,u^{LP}_c$.
Of course, an identical equation holds for planar cavities embedding a nonlinear medium: in this case, the polariton reduces to a pure photon and the interaction constant $g_{LP}$ is related to the nonlinear susceptibility $\chi^{(3)}$ of the medium. In the nonlinear optics literature, this equation often goes under the name of Lugiato-Lefever equation~\cite{Lugiato:PRL1987}.

\subsection{Quantum fluctuations and Wigner representation}
\label{sec:Wigner}
Going beyond the Gross-Pitaevskii equation \eq{GPE_LP} and fully include the quantum and thermal fluctuations of the quantum field $\Psih_{LP}$ is a very difficult task in the general case. In the course of the years, a number of different methods have been developed to attack this problem in different regimes, for instance Path Integral Monte Carlo for the thermal equilibrium state~\cite{Ceperley:RMP1995}, diagrammatic techniques~\cite{Prokofev:PRB2008,VanHoucke:PP2010}, density matrix renormalization group techniques for the both the ground state and the temporal dynamics~\cite{Schollwock:RMP2005}.

In this section we will then limit ourselves to a short review of phase space techniques originally developed in the quantum optical context and recently applied with success to the study of quantum fluids of atoms and photons. The basic idea is to represent the state of the quantum field as a (non-necessarily positive) probability distribution function on a suitable classical phase space and describe the time-evolution of the field in terms of a Fokker-Planck-like partial differential equation or a stochastic partial differential equation. A general introduction to the basic concepts of phase space representations of a quantum field can be found in~\cite{QuantumNoise}.

A crucial feature in view of efficient numerical simulations is that the probability distribution be positive-valued and the time-evolution be described by a Fokker-Planck partial differential equation~\cite{RiskenFP}
\begin{multline}
\frac{\partial P(\xx,t)}{\partial t} = - \sum_{i=1}^M \frac{\partial}{\partial x_i} [ F_i(\xx,t) P(\xx,t) ] + \\
+
\frac{1}{2} \sum_{i,j=1}^M \frac{\partial^2}{\partial x_i\,\partial x_j} [ D_{ij}(\xx)\,P(\xx,t)]
\eqname{FP}
\end{multline}
with a drift force $\mathbf{F}$ and a positive-definite diffusion matrix $D$. The probability distribution $P(\xx,t)$ is here defined on a $M$-dimensional space with real coordinates $x_i$, but probability distributions for complex quantities are straightforwardly included by considering the real and the imaginary parts as independent real variables.
Provided the diffusion matrix $D$ is positive-definite, the Fokker-Planck equation \eq{FP} can be mapped onto a system of $M$ Ito stochastic differential equations~\cite{StochasticMethods} of the form
\begin{equation}
dx_i= F_i (\xx,t)\,dt + dW_i
\eqname{SDE}
\end{equation}
with a Wiener noise satisfying
\begin{equation}
dW_i\,dW_j = D_{ij}(\xx,t)\,dt,
\end{equation}
that can be efficiently simulated on a computer by taking the statistical average over many different realizations of the Brownian motion. 

Unfortunately, in most cases of actual interest this is not possible as either the diffusion matrix $D$ is not positive-definite or additional terms with higher order partial derivatives of $P$ are present in the right-hand side of \eq{FP}: for the system of interacting bosons under investigation here, non-positive diffusion terms appear in the time-evolution of the Glauber $P$ and Husimi $Q$ representations, while third order derivative terms appear in the time-evolution of the Wigner $W$~\cite{Vogel:PRA1988}. 

A possible way out based on the so-called Positive-P representation was proposed in~\cite{Drummond:JPhysA1980,Drummond:JPhysA1980b} and transferred to the realm of atomic condensates in~\cite{Steel:PRA1998}. A remarkable application of this representation to multi-mode optical systems addressed squeezing of quantum solitons propagating in nonlinear optical fibers~\cite{Carter:PRL1987}. Unfortunately, in many other relevant cases the stochastic differential equations in the doubled phase space show strongly divergent trajectories that intrinsically undermine the numerical stability of the method~\cite{Gilchrist:PRA1997}. The use of improved schemes based on the so-called gauge-P representation was proposed in~\cite{Deuar:PRA2002,Drummond:PRL2004} and has provided interesting results on one-dimensional interacting Bose gases~\cite{Deuar:PRA2009}. Another exact stochastic approach to the many-body problem was proposed in~\cite{Carusotto:PRA2001b,Carusotto:JPhysB2001} using a Hartree ansatz in the canonical ensemble: while the method could be proven to be mathematically consistent and to have a finite statistical error at all times, useful results could only be obtained in simple geometries with a limited number of particles~\cite{Carusotto:PRL2003}.

In the following of the section we shall focus our attention on the Wigner $W$ representation that has turned out to be most useful in practical calculations. For the sake of simplicity, we restrict to the coherent pumping case for which a self-contained Hamiltonian description is available~\cite{Carusotto:PRB2005}. Extension to the incoherent pumping case requires some modeling of the relaxation mechanisms in the device: first attempts in this direction were recently reported in~\cite{Wouters:PRB2009}. 
Typical implementations of the Wigner representation require discretizing the quantum field on a $d$ dimensional discrete lattice of $\mathcal{N}^d$ points enclosed in a finite integration box of side $L$. In this geometry, the Wigner distribution is a function of the $\mathcal{N}^d$ complex amplitudes $\psi_i=\psi(\rr_i)$ of the field at the lattice positions $\rr_i$. 
Its time-evolution is described by the Fokker-Planck-like equation
\begin{multline}
\frac{\partial W}{\partial t}= -\sum_i \frac{\partial}{\partial \psi_i}[ F_i\{\psi\}\,W\{\psi\} ] 
 -\sum_i \frac{\partial}{\partial \psi_i^*}[ F_i\{\psi\}\,W\{\psi\} ]+ \\ 
 +\frac{\gamma_{LP}}{2\,\Delta V} \frac{\partial^2 W\{\psi\} }{\partial \psi_i^*\,\partial \psi_i} + \\
 +\frac{i g_{LP}}{4 \Delta V^2} \frac{\partial^2}{\partial \psi_i^*\,\partial \psi_i}
 \left[ \frac{\partial}{\partial \psi_i^*} (\psi_i^* W\{\psi\}) -\frac{\partial}{\partial \psi_i} (\psi_i W\{\psi\})\right],
 \eqname{Wigner_time}
\end{multline}
where $\Delta V=(L/\mathcal{N})^d$ is the volume of the elementary cell of the discrete lattice.
The drift force term on the $\rr_i$ site $F_i\{\psi\}=F\{\psi\}(\rr=\rr_i)$ involves a deterministic evolution of the field very similar to the right-hand side of \eq{GPE_LP},
\begin{multline}
F\{\psi\}(\rr)= -i \Big[\omega_{LP}^o - \frac{\hbar}{2m_{LP}}\nabla^2 +V_{LP}(\rr) -\frac{i\gamma_{LP}}{2} + \\
+ g_{LP}\,\big( |\psi(\rr,t)|^2 -\frac{1}{\Delta V} \big)\Big]\psi(\rr,t)+\\
+ i\eta_{LP}\,E^{\rm inc}(\rr,t).
\end{multline}
The derivatives with respect to the complex variable $\psi$ are defined  as
$\frac{\partial}{\partial \psi}= \frac{1}{2} \left[\frac{\partial}{\partial \textrm{Re}[\psi]} -i \frac{\partial}{\partial \textrm{Im}[\psi]}\right]$,
$\frac{\partial}{\partial \psi^*}=\frac{1}{2} \left[ \frac{\partial}{\partial \textrm{Re}[\psi]} +i \frac{\partial}{\partial \textrm{Im}[\psi]}\right] $.
The second-order derivative term is always positive and can be straightforwardly mapped onto a noise term with local correlations in space. 
On the other hand, the third derivative terms can not be included in a standard stochastic differential equation of the form \eq{SDE}.
Recent works~\cite{Plimak:EPL2001} have tried to solve this issues by generalizing the concept of stochastic differential equation to the case of stochastic {\em difference} equations with discrete time steps of finite size $\Delta t$: provided one does not attempt to take the limit $\Delta t\to 0$, a stochastic process can be found whose average over the noise recovers all terms of the Fokker-Planck-like equation \eq{Wigner_time}. Given the large statistical error of the method, the application of this {\em Positive-Wigner method} has been so far limited to very simple, few-mode models. Another strategy was proposed in \cite{Polkovnikov:PRA2003} in terms of a perturbative expansion in the number of quantum scattering events accounting for the third-order derivative terms

The main practical interest of the Wigner representation follows from the different scaling of the various terms of \eq{Wigner_time} in the dilute gas limit $\psi\to \infty$, $g_{LP}\to 0$ at a constant interaction energy $g_{LP}|\psi|^2$. In this limit, the noise is responsible for a statistical fluctuation of the field $\psi$ around its mean-field value on the order of $\Delta \psi \propto \Delta V^{-1/2}$. Correspondingly, the characteristic magnitude of the third-order derivative term is roughly estimated on the order of $g_{LP}/\Delta V$ to be compared to the diffusion term which is of the order of $\gamma_{LP}$. Provided 
\begin{equation}
\gamma_{LP} \gg g_{LP}/\Delta V,
\eqname{Wigner_condition}
\end{equation}
one may then expect that accurate results for a non-equilibrium gas of photons can be obtained by means of {\em truncated Wigner} calculations where the third-order derivative term is completely neglected and the stochastic partial differential equation has the form:
\begin{equation}
d\psi(\rr,t)= 
F\{\psi\}(\rr)\,dt+
\sqrt{\frac{\gamma_{LP}}{4\,\Delta V}}\,dW(\rr,t)
\end{equation}
with a zero-mean, complex Gaussian noise term $dW$ satisfying
\begin{eqnarray}
dW(\rr,t)\,dW(\rr',t) &=& 0 \\
dW^*(\rr,t)\,dW(\rr',t)&=& 2 \,dt\,\delta_{\rr,\rr'}.
\end{eqnarray}
Under the condition \eq{Wigner_condition}, the non-classical correlations that are introduced by the third-order derivative term are quickly washed away by the losses and the classical noise associated to $\gamma_{LP}$: in particular, the magnitude of the classical noise is the ``right'' one to simulate in the most accurate way the quantum dynamics on a classical computer. In~\cite{Sinatra:JPhysB2002}, it was shown how the truncated Wigner approach is able to capture the quantum fluctuations at least at the level of Bogoliubov theory. 

For typical experimental parameters of dilute photon gases in planar devices, the condition imposed by \eq{Wigner_condition} on the lattice spacing appears to be fully compatible with the characteristic range of wavevectors that are involved in the physics of the system: a first application of the truncated Wigner method to the critical fluctuations across the parametric oscillation threshold of a planar semiconductor microcavity was reported in~\cite{Carusotto:PRB2005} and is reviewed in Sec.\ref{sub:OPOWigner}. More recently, the truncated Wigner method has provided the first quantitative evidence of analog Hawking radiation from acoustic black holes in flowing polariton fluids~\cite{Gerace:arXiv2012}. Remarkably, a careful application of the truncated Wigner method provides useful results also in the conservative case of atomic gases for which $\gamma_{LP}=0$ and the (sufficient) validity condition \eq{Wigner_condition} is not fulfilled. In this case, the truncated Wigner equation  is a purely deterministic one and the quantum fluctuations enter the model via the randomness of the field $\psi$ at the initial time of the simulation. A detailed characterization of the power and the difficulties of the truncated Wigner method to study the thermal equilibrium state and the dynamical properties of atomic gases can be found in~\cite{Sinatra:JPhysB2002}. {A general review of a wider class of C-field methods for simulating the non-equilibrium dynamics of degenerate Bose gases at zero or finite temperature can be found in~\cite{Davis_FINESS2012}; an interesting application of the stochastic GPE~\cite{Gardiner:JPhysB2003} to the formation dynamics of an atomic condensate was discussed in~\cite{Weiler:Nature2008}.}

\subsection{Incoherent and saturable pumping}
\label{sec:incoh_GPE}

The discussion in the previous subsections addressed the case where the microcavity is driven by a coherent, quasi-resonant pump: in this case, the microscopic details of the system are under control and one can develop an {\em ab initio} description of the system. Many recent experiments were performed under different pumping schemes where the polariton gas does not inherit any coherence from the pump and/or coherence is quickly lost during the relaxation process towards the bottom of the lower polariton branch. This is the case of the incoherent and/or far blue-detuned laser pump used in~\cite{Richard:PRL2005,Richard:PRB2005,Kasprzak:Nature2006}, the large angle optical drive of~\cite{Deng:PNAS2003}, as well as, more recently, the electrical injection of polaritons of~\cite{Khalifa:APL2008,Bajoni:PRB2008,Tsintzos:APL2009}. 

A detailed study of the microscopic mechanisms involved in the kinetics of incoherent pumping schemes can be found in~\cite{Porras:PRB2002}: Typically, all these techniques end up accumulating a quite significant density of incoherent polaritons in the so-called bottleneck region in $\kk$-space in the vicinity of the inflection point of the lower polariton dispersion: further relaxation to the bottom of the lower polariton band via phonon-polariton scattering is in fact slowed down by the reduced density of states of the final states (the use of a finite density of electrons in the quantum well to overcome this issue was theoretically proposed by~\onlinecite{Malpuech:PRB2002}). At high enough polariton densities (i.e. for strong enough pumping intensities), another relaxation mechanism based on polariton-polariton collisions becomes active: a pair of polaritons in the bottleneck region collide and are respectively scattered to the bottom of the LP branch and to the large wavevector region where polaritons have a mostly excitonic nature with a very large density of states.  Given the bosonic statistics of polaritons, this relaxation process turns out to be stimulated as soon as the phase-space density of polaritons already present at the bottom of the lower polariton starts being of order one: when stimulation overcomes losses, a macroscopic coherent population of polaritons accumulates in the final state and a condensate appears. On the other hand, the phase-space density on the excitonic branch always remains much smaller than one, which guarantees effective irreversibility of the scattering process. 

A phenomenological description of the condensate dynamics under an incoherent pumping inspired to the semi-classical theory of laser~\cite{Lamb:PR1964} was introduced in~\cite{Wouters:PRL2007}: stimulated scattering into the condensate is described by an amplification term in the field equation of motion, with an amplification rate rate which is a monotonically growing function $R[n_R]$ of the local density of the polariton reservoir $n_R(\rr)$ in the bottleneck region. In practice, this corresponds to adding one more term to the polariton GPE \eq{GPE_LP} of the form 
\begin{equation}
i\partial_t \Psi_{LP}(\rr,t)=\ldots + \frac{i}{2} R[n_R(\rr,t)]\,\Psi_{LP}(\rr,t)
\end{equation}
and describing the polariton reservoir density via a rate equation of the form 
\begin{equation}
\partial_t n_R(\rr,t)=P-R[n_R(\rr,t)]\,|\Psi_{LP}(\rr,t)|^2 -\gamma_R\,n_R(\rr,t).
\eqname{n_R}
\end{equation}
The intensity of the incoherent pumping is phenomenologically described by the pump term $P(\rr)$.
In the typical case where the characteristic relaxation rate $\gamma_R$ of the reservoir is much faster than all other scales~\footnote{Note that the relaxation rate $\gamma_R$ does not need to correspond to the actual decay rate of reservoir polaritons, but may simply account for their redistribution among the different states of the reservoir.}, one can adiabatically eliminate $n_R$.
In the simplest case where $R[n_R]=R\,n_R$, the amplification term then reduces to the form
\begin{multline}
i\partial_t \Psi_{LP}(\rr,t)=\Big[\omega_{LP}^o - \frac{\hbar}{2m_{LP}}\nabla^2 +V_{LP}(\rr)\Big]\Psi_{LP}(\rr,t) + \\
+ g_{LP}\,|\Psi_{LP}(\rr,t)|^2\,\Psi_{LP}(\rr,t)-\frac{i\gamma_{LP}}{2}\, \Psi_{LP}(\rr,t) + \\+ \frac{i P(\rr,t)/2}{\gamma_R+R\,|\Psi_{LP}(\rr,t)|^2}\,\Psi_{LP}(\rr,t)
\eqname{CGLE}
\end{multline}
that closely resembles the complex Ginzburg-Landau equation currently used in the theory of pattern formation in nonlinear dynamical systems~\cite{Cross:RMP1993,Aranson:RMP2002}. {For this reason, some authors prefer to directly call it the {\it complex Ginzburg-Landau equation (cGLE)} for the polariton condensate.} A slightly different, but almost equivalent form of \eq{CGLE} was independently introduced in~\cite{Keeling:PRL2008}. A generalization of the \eq{CGLE} including the energy-dependence of the amplification term was proposed in~\cite{Wouters:PRL2010} to account for the stronger scattering into the lowest polariton states~\cite{Sarchi:PRB2008b,Wouters:PRB2009}.

The phenomenological nature of this model allows for its direct application to a wide class of incoherent pumping schemes: the details of the specific configuration enter via the functional form of the scattering rate $R$ and the other parameters of the theory: the only requirement is the presence of an irreversible process that is able to inject extra polaritons in the lowest energy states and that can be stimulated by pre-existing polaritons. It is worth mentioning that a very similar model was used in~\cite{Kneer:PRA1998} in the completely different context of the theory of atom lasers, still-to-come devices that should produce a coherent atomic matter wave as an output~\cite{Bloch:PRL1999,Chikkatur:Science2002,Lahaye:PRL2004}.

Extension of this model beyond the mean-field approximation to include quantum fluctuations of the polariton field requires a more sophisticated model of the amplification mechanism where the reservoir degrees of freedom are consistently taken into account. A first attempt in this direction based on the Wigner representation of the polariton field was reported in~\cite{Wouters:PRB2009}. A completely different approach based on a Keldysh diagram solution to a non-equilibrium Fermi-Bose model was proposed in~\cite{Szymanska:PRL2006} and provided interesting predictions for the coherence properties of a polariton condensate.

\section{Polariton condensates under quasi-resonant excitation}
\label{resonant}

In the previous section we have introduced the generalized Gross-Pitaevskii equation that describes the dynamics of the polariton condensate at the mean-field level. Now we shall proceed to review its application to derive the stationary state and the spectrum of the elementary excitations of the condensate under the different pumping schemes: The present section is devoted to the case of a coherent pumping at the bottom of the lower polariton branch, the next section is devoted to the so-called parametric regime with a coherent pump in the vicinity of the inflection point of the lower polariton branch, while the following section is devoted to the case of a generic incoherent pumping. To facilitate the reader in appreciating the novel features that stem from the non-equilibrium, driven-dissipative nature of the polariton condensate, we shall start with a short survey of the basic properties of standard condensates at equilibrium, e.g. in ultracold atomic gases.

\subsection{Equilibrium condensates}

\subsubsection{The condensate wavefunction}

The textbook definition of Bose-Einstein condensation in an ideal Bose gas is that a macroscopic fraction of the particles occupy the lowest single-particle energy state: in free space, this is the $\kk=0$ plane wave, while in a trap it is the lowest eigenstate of the trap potential. In presence of weak particle-particle interactions, the condensate wavefunction $\phi_o(\rr)$ is obtained as the lowest $\mu$ eigensolution of the time-independent Gross-Pitaevskii equation~\cite{BECbook}
\begin{equation}
\mu\phi_o(\rr)=-\frac{\hbar^2}{2m}\nabla^2\phi_o(\rr)+V_{\rm trap}(\rr)\,\phi_o(\rr)+g|\phi_o(\rr)|^2\phi_o(\rr),
\eqname{staticGPE}
\end{equation}
where $m$ is the mass of the particles, $V_{\rm trap}(\rr)$ is the trap potential, and $g=4\pi\hbar^2 a/m$ is the interaction constant in terms of the two-body scattering length $a$. A few general features of this equation are worth mentioning.

In agreement with gauge symmetry, the global phase of the wavefunction $\phi_o(\rr)$ is not fixed. As it has no physical consequence, it can be arbitrarily chosen. 
The eigenvalue $\mu$ gives the oscillation frequency of the condensate phase under the time-dependent GPE \eq{atomicGPE}. In a thermodynamical context, it corresponds to the chemical potential of the condensate, i.e. the energy that is required to add one more particle to the gas.
In free space ($V_{\rm trap}=0$), the equation of state relating the chemical potential and the (spatially uniform) density $|\phi_o|^2$ has the simple form
\begin{equation}
\eqname{EOS}
\mu=g\,|\phi_o|^2.
\end{equation}

To the best of our knowledge, the local phase of $\phi(\rr)$ is constant throughout space in all known cases even in the presence of a trap potential $V_{\rm trap}$, which means that $\phi_o(\rr)$ can be reduced to a purely real function~\footnote{This result follows from the unicity of the lowest energy solution $\phi(\rr)$ combined with time-reversal symmetry: if $\phi_o(\rr)$ is an eigensolution of \eq{staticGPE} at $\mu$, also its complex conjugate $\phi_o^*(\rr)$ is an eigensolution with the same $\mu$.}. Physically, having a purely real $\phi_o(\rr)$ means that no macroscopic current is flowing across the condensate in its ground state,
\begin{equation}
\mathbf{J}=\frac{1}{2mi}\left[\phi_o^*(\rr)\nabla \phi_o-\textrm{h.c.} \right]=0
\end{equation}

As we shall see in what follows, polariton condensate show very different features: the system is far from a thermodynamical equilibrium state and the macroscopic occupation of a single state $\Psi_{LP}(\rr)$ can not be derived from a free energy minimization argument. As a result, the shape of $\Psi_{LP}(\rr)$ is determined by a complex dynamical balance of pumping and losses: depending on the geometrical configuration, stationary states with a spatially dependent phase can be found, which physically corresponds to a non-vanishing polariton current $J_{LP}$ along the cavity plane. For instance, polariton fluids with a well-defined and non-zero in-plane momentum $\kk$ can be created using a coherent pump with a non-vanishing incidence angle $\theta_{\rm inc}$. A radial flow in a polariton condensate under a spatially localized incoherent pump was invoked in~\cite{Wouters:PRB2008} to explain the ring-shaped momentum distribution observed in~\cite{Richard:PRL2005}; a similar mechanism due to polariton-polariton interactions underlies the macroscopic flow in a one-dimensional geometry in photonic wires~\cite{Wertz:NatPhys2010}. Quantized vortices are spontaneously present in the steady state of a polariton condensate under a incoherent pump in the presence of disorder~\cite{Lagoudakis:NatPhys2008}. 

\subsubsection{Bogoliubov dispersion of elementary excitations}

The dynamics of weak perturbations on top of a dilute Bose-Einstein condensate can be described within the Bogoliubov theory~\cite{BECbook,CastinLectures}: weakly excited states of the condensate are characterized by bosonic excitation modes, whose frequency and spatial profile can be obtained by linearizing the Gross-Pitaevskii equation around the equilibrium state. 

A weak modulation of the condensate wavefunction can be represented as $\phi(\rr,t)=\left[\phi_o(\rr,t)+\delta\phi(\rr,t)\right]\,e^{-i\mu t/\hbar}$ with small $\delta\phi(\rr,t)$. Inserting this ansatz into the GPE \eq{atomicGPE} an linearizing around the stationary solution leads to a pair of evolution equations
\begin{equation}
i\hbar\partial_t \left(
\begin{array}{c}
\delta\phi(\rr,t) \\ \delta\phi^*(\rr,t)
\end{array} \right) = \mathcal{L}_{\rm Bog}
\left(\begin{array}{c}
\delta\phi(\rr,t) \\ \delta\phi^*(\rr,t)
\end{array} \right)
\eqname{Bogol_L0}
\end{equation}
with the Bogoliubov operator defined by
\begin{widetext}
\begin{equation}
\mathcal{L}_{\rm Bog}=
 \left(
\begin{array}{cc}
-\frac{\hbar^2\nabla^2}{2m} + V_{\rm ext}(\rr) + 2 g n - \mu & g n \\
- g n & \frac{\hbar^2\nabla^2}{2m} - V_{\rm ext}(\rr) - 2 g n + \mu
\end{array} 
\right).
\eqname{Bogol_L}
\end{equation}
\end{widetext}
In the simplest case of a spatially homogeneous system, the equilibrium condensate wavefunction is constant in space $\phi_o(\rr,t)=\sqrt{n}$ and the Bogoliubov modes can be chosen in a plane-wave form of wavevector $\kk$. In this case, the Bogoliubov operator reduces to a matrix and its eigenvalues give the so-called Bogoliubov dispersion of excitations,
\begin{equation}
\eqname{Bogo_omega}
\hbar\omega_{\rm Bog}(\kk)={\pm}\sqrt{\frac{\hbar^2 k^2}{2m}\left(\frac{\hbar^2 k^2}{2m}+2\mu \right) },
\end{equation}
{the $+$ ($-$) sign referring to the so-called positive (negative) Bogoliubov branch, i.e. the modes with positive (negative) Bogoliubov norm. As it is discussed in full detail in textbooks on BEC, the energy of the physical elementary excitations is determined by the positive branch only~\cite{BECbook,CastinLectures}.}

Depending on the value of the wavevector $k$ as compared to the so-called {\em healing length} $\xi=\sqrt{\hbar^2/m\mu}$, two regimes can be identified in the Bogoliubov dispersions: for large momenta $k\xi \gg 1$, the Bogoliubov dispersion recovers the parabolic shape of single particles with a global energy shift due to the Hartree interaction energy,
\begin{equation}
\omega_{\rm Bog}(\kk) \simeq \frac{\hbar^2 k^2}{2m}+\mu. 
\eqname{single_part}
\end{equation}
On the other hand, small wavevector $k\xi\ll 1$ excitations have instead a phononic nature with a sonic dispersion 
\begin{equation}
\omega_{\rm Bog}(\kk)\simeq c_s k
\eqname{sonic}
\end{equation}
with a sound speed $c_s=\sqrt{\mu/m}$. 
The presence of a sharp corner at $\kk=0$ crucially depends on the relative value of the non-diagonal elements $\pm gn$ of $\mathcal{L}_{\rm Bog}$ and the Hartree shift $\pm (2 gn-\mu)=\pm gn$ in the diagonal ones: physically, it can be seen as a remarkable consequence of the dynamical stability of the system combined with the global gauge symmetry under rotations of the global phase of the wavefunction. Experimental studies of the elementary excitation spectrum of liquid Helium showing both the phonon and the roton branches can be found via neutron scattering experiments~\cite{Palevsky:PR1958,Yarnell:PR1959}. Recent studies of the Bogoliubov dispersion in dilute atomic condensates using two-photon Bragg scattering are reviewed in~\cite{Ozeri:RMP2005}.

As we shall see in Sec.\ref{sec:coh_excit}, breaking this condition in the driven-dissipative case  leads to several novel phenomena in the excitation spectrum of polariton condensates.

\subsection{Stationary state under coherent pumping}

In this subsection, we review the steady-state properties of  polariton condensates  under a coherent continuous-wave pump at frequency $\omega_{\rm inc}$ with different spot geometries and, for the sake of simplicity, considering regimes where the form \eq{GPE_LP} of the generalized polariton GPE can be applied. 

\subsubsection{Optical limiting and optical bistability}
\label{sec:coh_lim_bist}
When the cavity is driven by a coherent, continuous-wave pump with a plane-wave spatial profile,
\begin{equation}
{E}^{\rm inc}(\rr,t)=E^{\rm inc}_0\,e^{i\kk_{\rm inc}\rr}\,e^{-i\omega_{\rm inc}t}.
\eqname{planewave}
\end{equation}
we can look for solutions like $\Psi_{LP}(\rr,t)= \Psi_{LP}^0\,e^{i\kk_{\rm inc}\rr}\,e^{-i\omega_{\rm inc}t}$.
Inserting this ansatz into the generalized GPE \eq{GPE_LP}, we obtain the polaritonic analog of the equation of state,
\begin{equation}
\Big[\omega_{\rm inc}- \omega_{LP}^o-\frac{\hbar k_{\rm inc}^2}{2m_{LP}} -g_{LP} |\Psi_{LP}^0|^2 + \frac{i\gamma_{LP}}{2} \Big] \Psi_{LP}^0 = i \eta_{LP}\,E^{\rm inc}_0,
\eqname{EOS_pol}
\end{equation}
relating the polariton field amplitude $\Psi_{LP}^0$ to the pump parameters. Comparing this equation with the standard equation of state \eq{EOS}, we notice that  
the gauge symmetry under global rotations of the condensate phase is explicitly broken by the coherent pumping term on the right-hand side of \eq{EOS_pol}. The consequences of this feature on the elementary excitation spectrum will be discussed in the Sec.\ref{sec:coh_excit}. 

\begin{figure*}[htbp]
\begin{center}
\includegraphics[width=0.9\columnwidth,angle=0,clip]{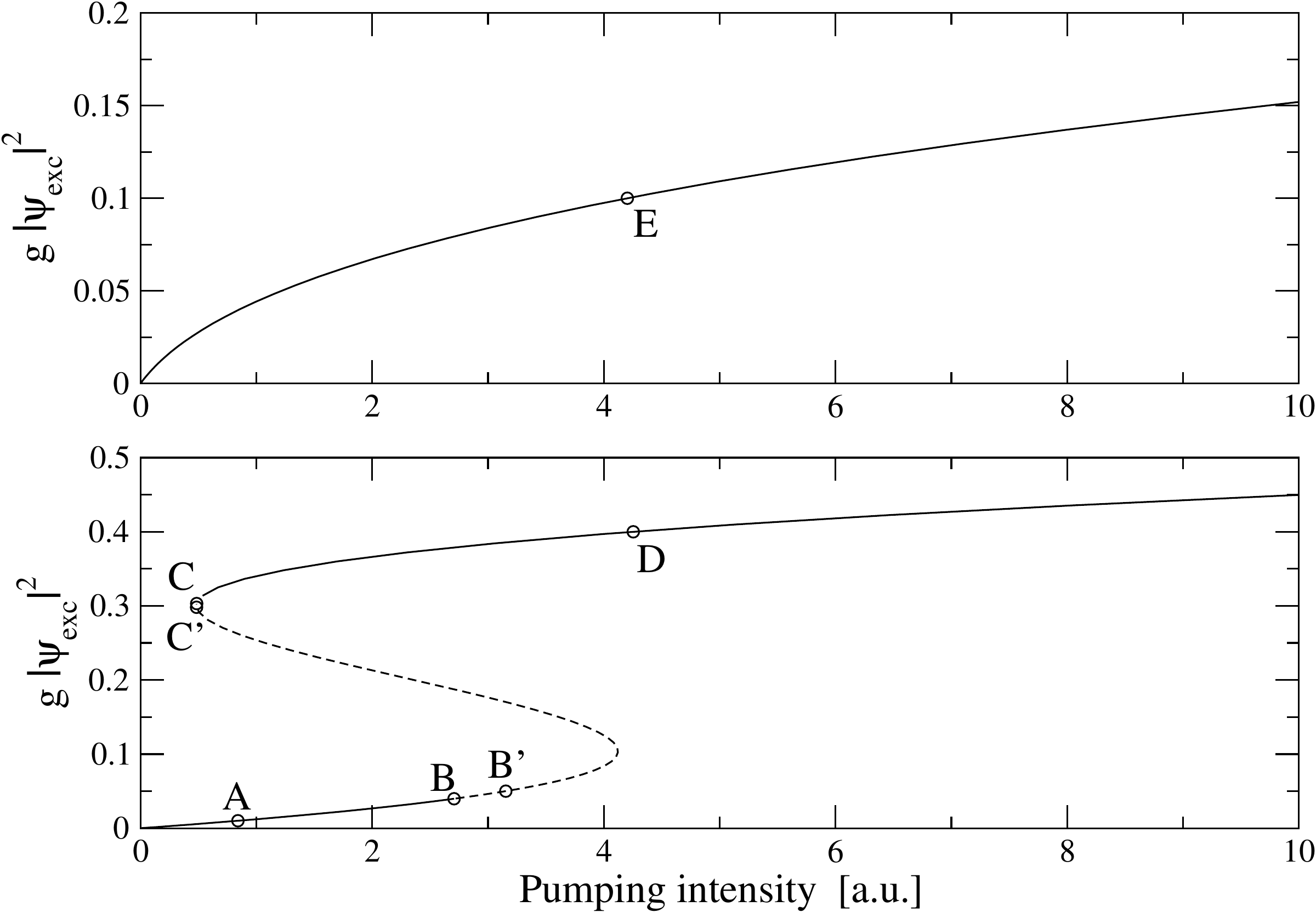}
\hspace{0.1\columnwidth}
\includegraphics[width=0.9\columnwidth,angle=0,clip]{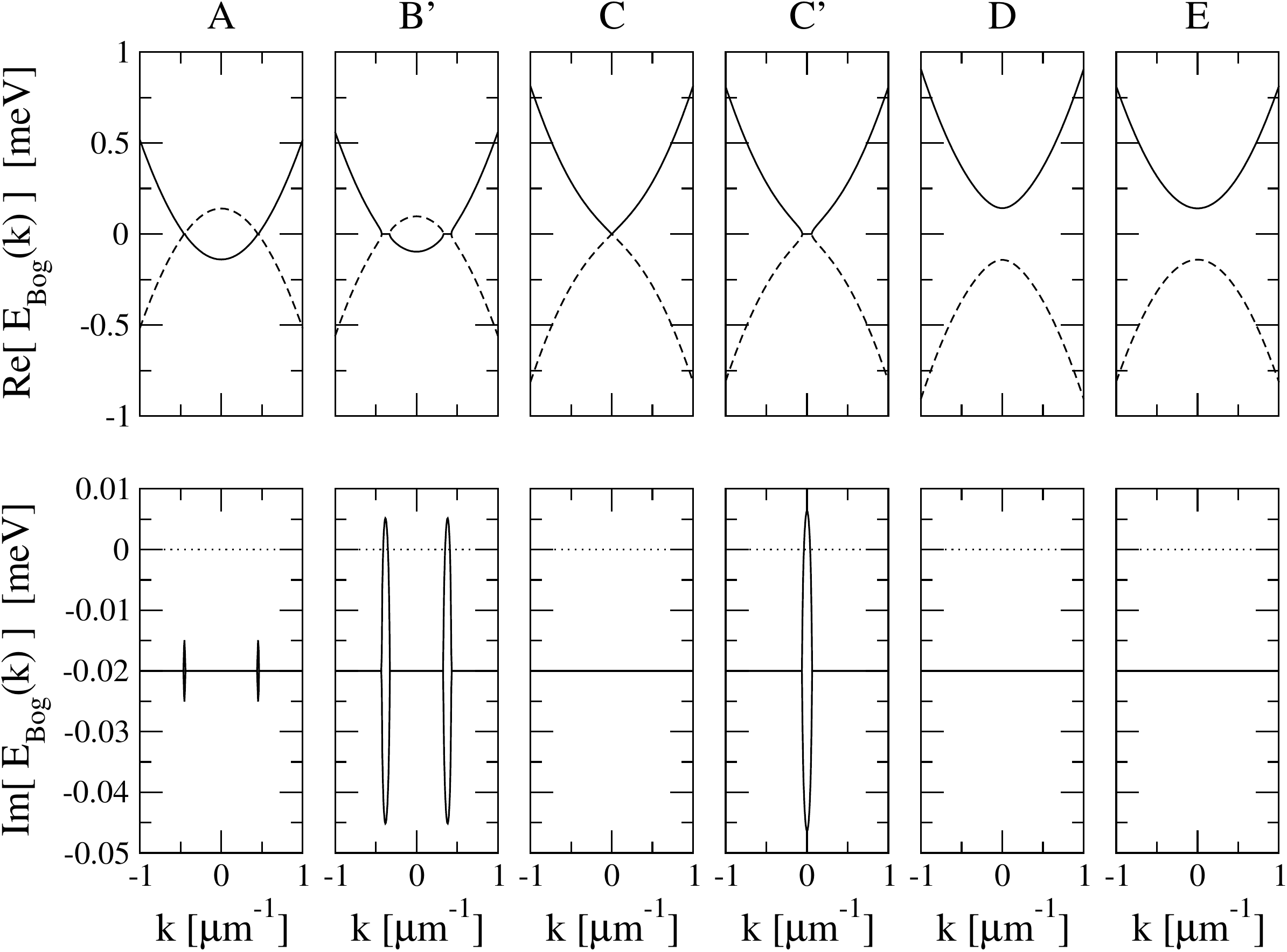}
\end{center}
\caption{Left panel: mean-field energy (meV) for the exciton field versus pump intensity in the optical limiter regime (top) and bistable regime (bottom). The dashed line indicates the unstable regions. 
Right panel: real part (top) and imaginary part (bottom) of the excitation frequencies for the non-equilibrium Bogoliubov modes corresponding to the points A,B,C,C',D,E, indicated in the left panel. 
}
\label{fig:RMP_cohpump}
\end{figure*}
Furthermore, while the oscillation frequency of the matter field in standard equilibrium condensates is related to the density via the chemical potential, the oscillation frequency of the polariton condensate wavefunction $\Psi_{LP}$ is fixed by the pump frequency $\omega_{\rm inc}$. On the other hand, the dependence of the polariton density $n_{LP}=|\Psi_{LP}|^2$ on the pump parameters is strongly affected by the resonance condition between the pump frequency $\omega_{\rm inc}$ and the polariton frequency $\omega_{LP}(\kk_{\rm inc})\simeq \omega_{LP}^o+\hbar k_{\rm inc}^2/2m_{LP}$, possibly shifted by the interaction energy $g_{LP} |\Psi_{LP}^0|^2$.
As a result, the behavior of the density $n_{LP}$ as a function of the pump intensity $I_{LP}=|E_{\rm inc}^0|^2$ is very different depending on whether $\omega_{\rm inc}$ is on the blue or on the red side of $\omega_{LP}(\kk_{\rm inc})$. This is a well-known fact of nonlinear optics~\cite{Boyd} and is illustrated in the left panels of Fig.\ref{fig:RMP_cohpump}.

The $\omega_{\rm inc}<\omega_{LP}(\kk_{\rm inc})$ case is shown in the upper-left panel: the density $n_{LP}$ is a continuous, monotonically growing function of the pump intensity $I_{LP}$. As the nonlinear term tends to shift the polariton frequency further away from resonance with the pump, the growth of the density is sub-linear, a behavior known under the name of {\em optical limiter}.

The $\omega_{\rm inc}>\omega_{LP}(\kk_{\rm inc})$ case
is shown in the lower-left panel: the density $n_{LP}$ is still a growing function of $I_{LP}$ but shows a hysteretic behavior that goes under the name of {\em optical bistability} in the nonlinear optics literature. 
The central branch of the hysteresis loop with a negative slope is always dynamically unstable. Depending on the specific values of pump parameters, also other regions may be dynamically unstable towards the parametric generation of polaritons into other modes at different $\kk\neq \kk_{\rm inc}$. These stability issues will be addressed in full detail in Sec.\ref{sec:coh_excit} using the polariton version of the Bogoliubov theory.

\subsubsection{Local density approximation and experiments} 
\label{sec:coh_lda}

Optical bistability behaviors have been observed in several experiments with polaritons in microcavities~\cite{Baas:PRA2004}: while there is good qualitative agreement with the theory discussed in the previous subsection, a quantitative understanding of the peculiar geometrical features observed in the experiments requires a more refined theoretical treatment accounting for spatial inhomogeneities. Yet, the equation of state \eq{EOS_pol} is a good starting point to describe the polariton wavefunction even in this case. Provided the spatial dependence of the pump and of the external potential is slow enough, one can approximate the local polariton field at each spatial position with the homogeneous system prediction \eq{EOS_pol} using the local value of the pump amplitude and inserting the external potential $V_{LP}(\rr)$ as a shift of  $\omega_{LP}^o$. This procedure of neglecting the kinetic energy term in the GPE \eq{GPE_LP} provides an accurate approximation of the time-independent GPE  of trapped atomic gases \eq{staticGPE} and leads to the so-called Thomas-Fermi profile~\cite{BECbook}.

In the polariton case, one has to consider the additional features due to the optical bistable behavior: for a given value of the pump intensity, several solutions for the polariton density may be available. An intuitive criterion to choose the relevant solution was introduced in~\cite{Shelykh:PRL2008} and successufully compared to complete simulations of the GPE: while moving in space along a streamline (i.e. along the direction of the wavevector $\kk_{\rm inc}$), one has to continuously follow a given branch of the hysteresis loop. Jumps to other branches are only allowed when one reaches the end point of the branch. Optical multistability effects in the presence of spin degrees of freedom were discussed in~\cite{Gippius:PRL2007} and experimentally investigated in~\cite{Paraiso:NatMat2010}. Interesting spatio-temporal switching effects when polarization multistability is combined with a non-trivial spatial dynamics have been theoretically studied in~\cite{Liew:PRL2008} and experimentally demonstrated in~\cite{Amo:NatPhot2010}.

\subsubsection{Spatially localized coherent pump}
\label{sec:local_coh_parag}

A completely different regime is realized when the monochromatic pump is focused onto a spatially small region, with a e.g. Gaussian circular spot of peak amplitude $E^{\rm inc}_0$, radius $\sigma$ and carrier wavevector $\kk_{\rm inc}=k_{\rm inc}\ee_x$.
In the regime of weak pump intensity where nonlinear interactions are negligible, analytic calculations show that the steady-state polariton field
\begin{multline}
\Psi_{LP}(\rr,t)=\sqrt{2\pi}\,\frac{\sigma}{v_g}\,\eta^{\rm fr}\,E^{\rm inc}_0\,\Theta(x)\,e^{-y^2/2\sigma^2}\,e^{-\gamma x/2v_g}\\e^{i\bar{\kk}\cdot\rr}\,e^{-i\omega_{\rm inc}t}:
\eqname{small_spot}
\end{multline}
extends in the direction of the pump wavevector for the distance $\ell=v_g/\gamma$ that cavity photons ballistically cover in a time $\gamma^{-1}$. On the other hand, the polariton field follows the profile of the pump spot along the orthogonal direction $y$ and keeps a narrow width $\sigma$. The small spot assumption is consistent if $\sigma\ll \ell$, while the wavepacket expansion due to the curvature of $\omega_{\rm cav}(\kk)$ is safely negligible provided $\sigma\gg \sqrt{\hbar/2m\gamma}$.

Remarkably, the wavevector $\bar{\kk}$ of the ballistically moving polaritons in \eq{small_spot} is not fixed by the incident laser wavevector $\kk_{\rm inc}$, rather it is determined by the laser frequency $\omega_{\rm inc}$ via the energy conservation condition $\omega_{LP}(\bar{\kk})=\omega_{\rm inc}$. Correspondingly, the group velocity $\vv_g=\nabla_\kk \omega_{LP}$ has to be evaluated at $\kk=\bar{\kk}$, which gives $\vv_g\simeq \hbar \bar{\kk} / m_{LP}$. The missing momentum $\bar{\kk}-\kk_{\rm inc}$ is provided by the spatial inhomogeneity of the pump spot. A similar physics in the presence of interactions for a incoherent pump with a small spatial spot will be discussed in Sec.\ref{sec:smallspot}.

As we shall review in Sec.\ref{vortices}, the fact that the phase of the polariton field in the region outside the pump spot is not fixed by the incident laser is crucial to observe topological excitations in the polariton fluid past an obstacle: this feature was theoretically put forward in~\cite{Pigeon:PRB2011} and experimentally observed in~\cite{Amo:2011Science,Nardin:NPhys2011,Sanvitto:NPhot2011}. A closely related configuration was considered in a recent proposal of analog black holes based on polariton condensates~\cite{Solnyshkov:PRB2011}, \cite{Gerace:arXiv2012}.

\subsection{Elementary excitations and dynamical stability}
\label{sec:coh_excit}

Once the steady state polariton field under a continuous wave pump has been determined, we can proceed with the study of small fluctuations around the steady state. 
This can be done by extending the Bogoliubov theory of dilute Bose gases to the present non-equilibrium context of polaritons using the generalized GPE \eq{GPE_LP}. A pioneering mention of the Bogoliubov dispersion of quasi-particles on top of a luminous fluid appeared in~\onlinecite{Chiao:PRA1999,Tanzini:PLA1999}, where an equilibrium assumption was implicitly made. A first quantitative study taking into account the driven-dissipative nature of the polariton fluid was carried out for the coherent pumping case in~\cite{Carusotto:PRL2004,Ciuti:PSSB2005}. In the following of the discussion we shall closely follow the discussion in these works.

For the sake of simplicity, we shall restrict ourselves here to the illustrative case of a spatially homogeneous system $V_{LP}(\rr)=0$ under a coherent pump with $\kk_{\rm inc}=0$; other  cases will be reviewed in Sec. \ref{goldstone} and \ref{elementary} and \ref{sec:coh_superfl}. Under these assumptions, the steady state has the form $\Psi_{LP}(\rr,t)=\sqrt{n_{LP}}\,\exp(-i\omega_{\rm inc}t)$ and the linearized GPE for the polariton field modulation $\delta\Psi_{LP}(\rr,t)$ reads
\begin{widetext}
\begin{equation}
i\partial_t \left(
\begin{array}{c}
\delta\Psi_{LP}(\rr) \\ \delta\Psi^*_{LP}(\rr)
\end{array} \right)  =
\left(
\begin{array}{cc}
\omega_{LP}(\kk) + 2 g_{LP} n_{LP} - \omega_{\rm inc} -i\gamma_{LP}/2 & g_{LP} n_{LP}  \\
- g_{LP} n_{LP} & -\omega_{LP}(\kk) - 2 g_{LP} n_{LP} + \omega_{\rm inc} -i\gamma_{LP}/2
\end{array} 
\right)
\left(
\begin{array}{c}
\delta\Psi_{LP}(\rr) \\ \delta\Psi^*_{LP}(\rr)
\end{array} \right)
\eqname{Bogol_LP}
\end{equation}
being  reminiscent of the equilibrium Bogoliubov equations \eqs{Bogol_L0}{Bogol_L}. 
The additional terms are the loss rate  $\gamma_{LP}$ and pump frequency $\omega_{\rm inc}$ next to the Hartree energy\cite{Ciuti:PRB2001}: in particular, the real part $\omega_{LP}^0+2 g_{LP} n_{LP}-\omega_{\rm inc}$ is no longer equal to the non-diagonal term $g_{LP} n_{LP}$, leading to a wealth of new behaviors (see the plots in the right panels of Fig. \ref{fig:RMP_cohpump}) described by the analytic formula 
\begin{equation}
\omega_{\rm Bog}(\kk)= \pm \Big[\Big(\omega_{LP}^o+\frac{\hbar k^2}{2m}+2 g_{LP} n_{LP} -\omega_{\rm inc} \Big)^2  - (g_{LP} n_{LP})^2\Big]^{1/2}-i\,\frac{\gamma_{LP}}{2}.
\eqname{Bogo_neq_coh}
\end{equation}
\end{widetext}
As expected, the presence of the coherent pump locking the phase of the condensate directly reflects in the absence of a Goldstone branch whose frequency $\omega$ tends to zero for $\kk\to 0$. 

The optical limiter case $\omega_{\rm inc}<\omega_{LP}^0$ is illustrated in panel (E): the two Bogoliubov branches are split in frequency and the sonic behavior at small $\kk$ \eq{sonic} disappears. The stronger the pump intensity, the wider the gap. The imaginary parts of the two branches are flat and equal to $-\gamma_{LP}$.

In the optical bistability case $\omega_{\rm inc}>\omega_{LP}^0$, several regimes have to be distinguished. On the upper branch of the bistability loop [panel (D)], a behavior similar to the optical limiter case is recovered as interactions have effectively shifted the polariton branch $\omega_{LP}(\kk)+ 2 g_{LP} n_{LP}$ to frequencies larger than $\omega_{\rm inc}$: the Bogoliubov branches are split in energy by a gap that increases for growing pump intensity. A sonic dispersion of the form \eq{sonic} is recovered only at the end-point C of the upper branch, which satisfies the effective resonance $\omega_{\rm inc}= \omega_{LP}^0+g_{LP} n_{LP}$ condition. Only at this point the oscillation frequency $\omega_{\rm inc}$ happens to fulfill the analogous of the  equilibrium equation of state \eq{EOS}.

The dynamical instability on intermediate branch (including point C') is the standard Kerr single-mode instability of optical bistability~\cite{Drummond:JPhysA1980,Ciuti:PSSB2005}, signaled by the imaginary part becoming positive in a neighborhood of the pump wavevector $\kk_{\rm inc}=0$.
 
The physics is more interesting for pump values on the lower branch: the two Bogoliubov branches cross on a ring of wavevectors [panels (A,B,B')]. Because of the anti-Hermitian non-diagonal terms in the matrix \eq{Bogol_LP} that provide an effective ``attraction" between modes~\cite{Ciuti:PRB2001,Savvidis:PRB2001}, the branches stick in the vicinity of the crossing point giving rise to a flat region where $\textrm{Re}[\omega_{\rm Bog}]=0$. On the other hand, the imaginary parts $\textrm{Im}[\omega_{\rm Bog}]$ are split: as soon as one of them becomes positive, the system becomes dynamically unstable (as at point B') and a parametric oscillation sets in, as discussed in Sec.\ref{sec:secOPO}.
It is worth noting the qualitative analogy of this behavior with the soft mode of an open Dicke model~\cite{Dimer:PRA07} when the critical point of the Dicke transition is approached~\cite{Emary:PRE03,Nagy:PRA11,Nagy:PRL10}; experimental investigations of the Dicke physics using an optically driven atomic condensate embedded in an optical cavity have recently appeared in~\cite{Baumann:Nature10,Baumann:PRL11}.

In Sec.\ref{sec:coh_superfl} we shall discuss in detail the consequences of the form \eq{Bogo_neq_coh} of the Bogoliubov dispersion onto the superfluidity properties of the polariton fluid and we shall review the experiments that have been performed in the different regimes. A direct study of the Bogoliubov dispersion was performed in~\cite{Kohnle:PRL2011} by means of an angle-resolved four-wave mixing technique {as proposed in~\cite{Wouters:PRB2009b}}. In this way, the transition from the parabolic single-particle dispersion at low densities to the sonic dispersion at high densities was observed: under the pulsed pump configuration used in this experiment the polariton Bogoliubov dispersion recovers in fact the standard one \eq{Bogo_omega} for equilibrium condensates, the only difference being the global $-i\gamma_{LP}/2$ decay rate. {In contrast to previous studies of the Bogoliubov dispersion with photoluminescence tecnhiques~\cite{Utsonomiya:NatPhys2008}, the improved sensitivity of four-wave mixing allowed to detect a signal also from the negative Bogoliubov branch.}

\section{Parametric oscillation regime}
\label{sec:secOPO}

Historically, the first configuration that was used to experimentally study the spontaneous onset of macroscopic coherence in the polariton gas and the collective dynamics of coherent polariton fluids in planar microcavity systems was in the so-called optical parametric oscillation (OPO) regime. In this section, we shall review the main properties of the OPO condensate from the point of view of hydrodynamics. An early review of this physics can be found in~\cite{Ciuti:SST2003}.

\begin{figure}[t]
\begin{center}
\includegraphics[width=0.55\columnwidth,angle=0,clip]{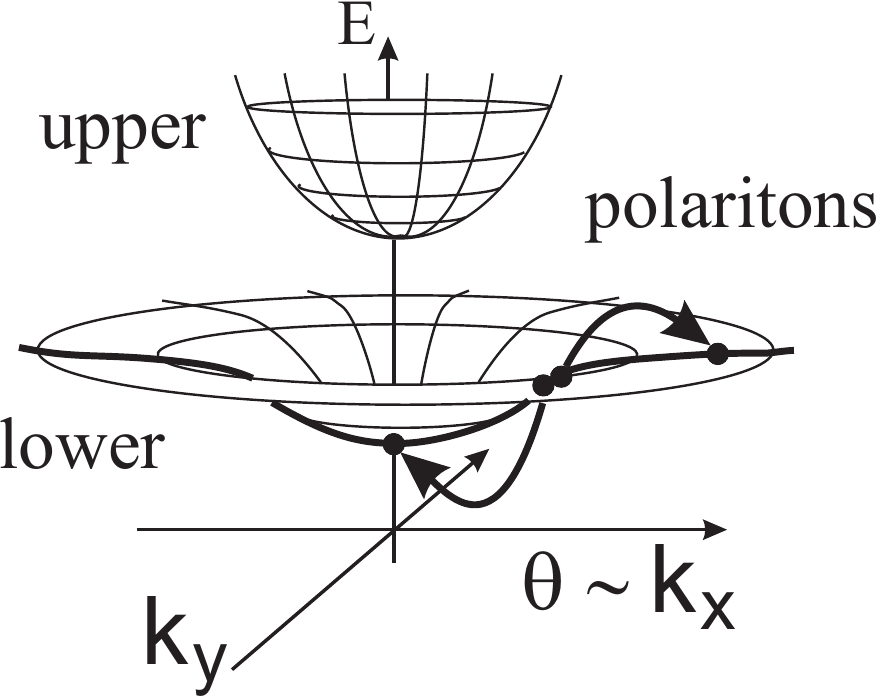} \\
\includegraphics[width=0.7\columnwidth,angle=0,clip]{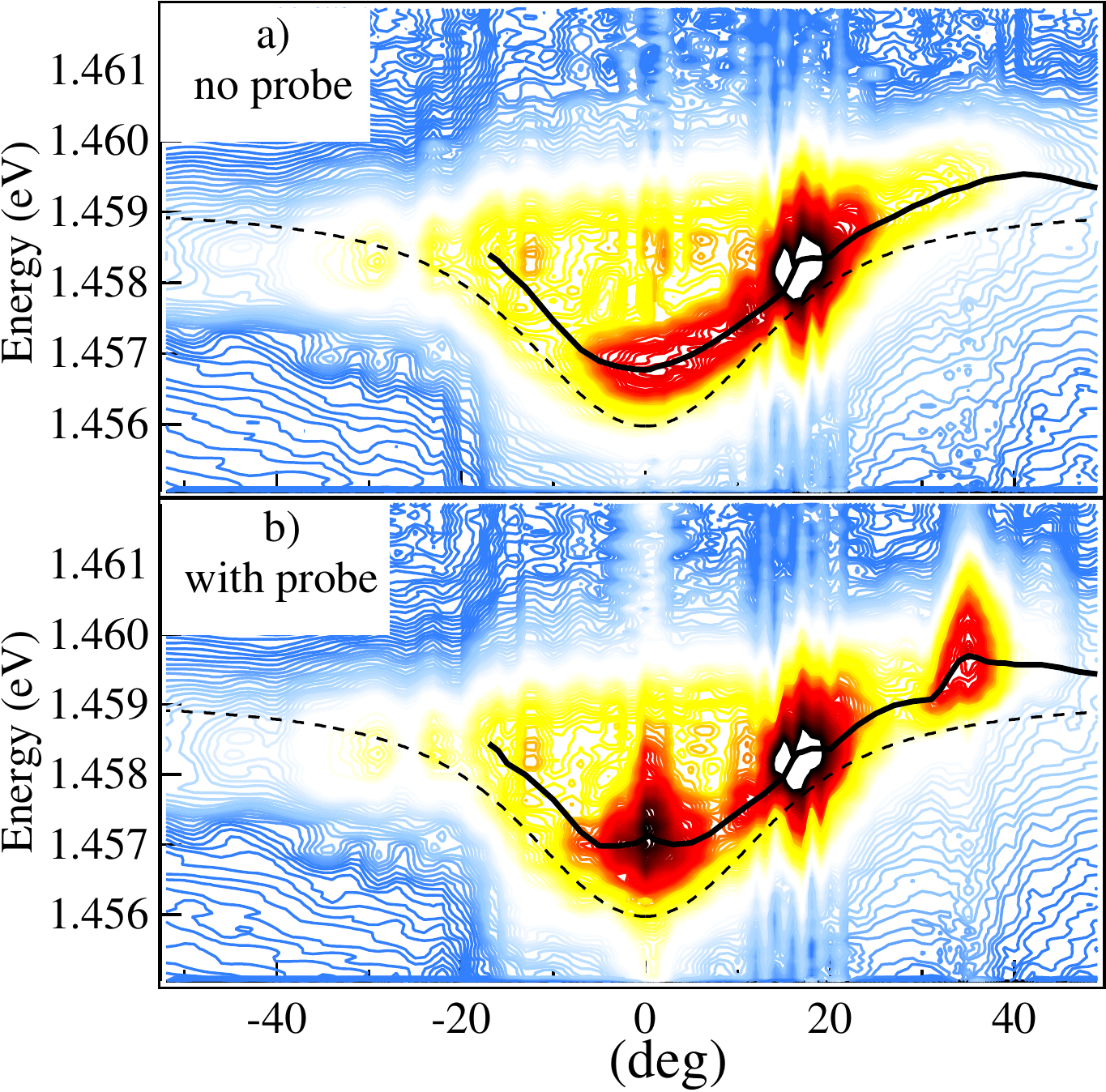} \\
\includegraphics[width=0.6\columnwidth,angle=0,clip]{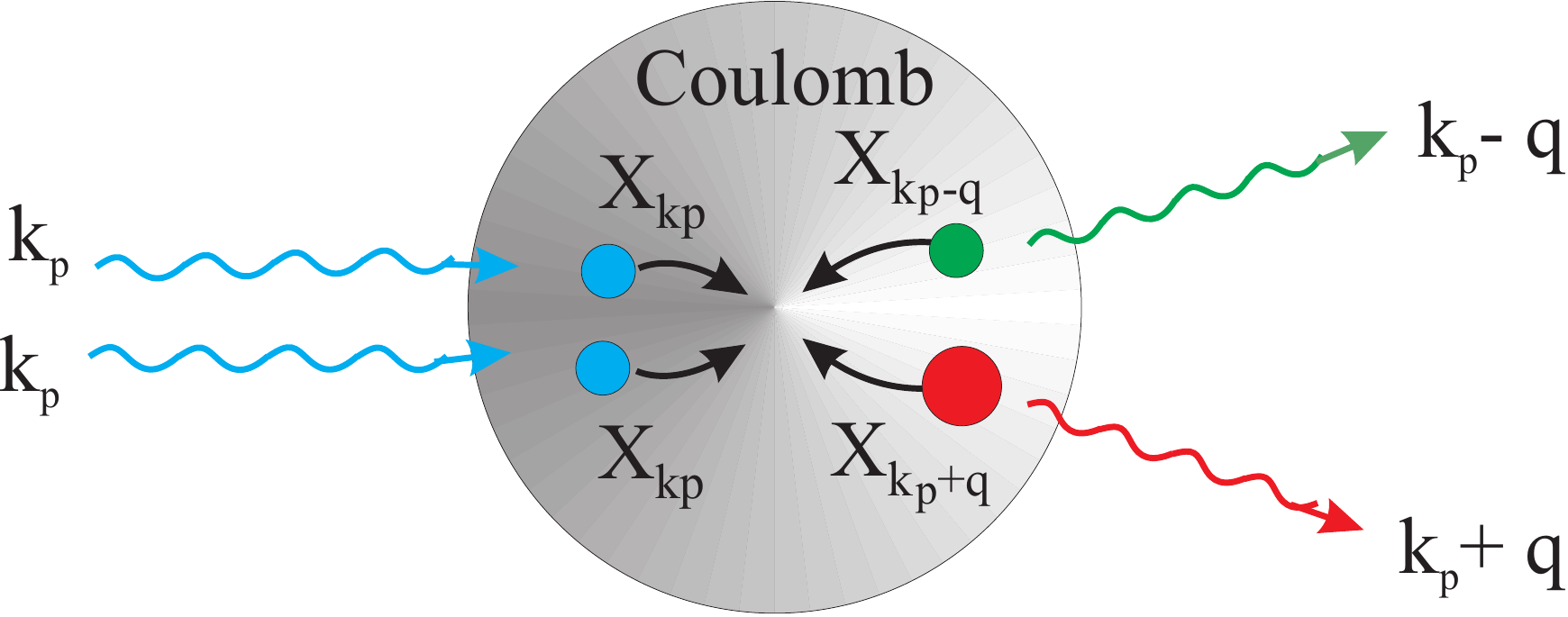} 
 \end{center}
\caption{(Color online) Top panel: sketch of a polariton-polariton scattering process conserving both total energy and momentum. The first experimental observation of such process was reported by \onlinecite{Savvidis:PRL2000}. Middle: contour plots of angle- and frequency-resolved polariton parametric scattering in the spontaneous (top) and stimulated (bottom) regime. Figure from \onlinecite{Savvidis:PRB2000}. Bottom panel: in the coherent regime, such scattering process can be seen as a polariton four-wave-mixing process \cite{Ciuti:PRB2000,Ciuti:PRB2001} due to the nonlinearity of the excitonic component, in particular via the Coulomb interaction. Figures are courtesy of P. G. Savvidis.
}
\label{fig:OPOsketch}
\end{figure}

A schematic sketch of the pumping scheme is shown in the upper panel of Fig.\ref{fig:OPOsketch}: a coherent pump is shined on the microcavity at a finite incidence angle such that the wavevector $\kk_p$ lies in the vicinity of the inflection point of the lower polariton dispersion. Because of collisions, polaritons are then scattered into a pair of other modes at different wavevectors $\kk_s$ and $\kk_i$. In nonlinear optical terms, the modes at $\kk_s$ and $\kk_i$ represent the {\em signal} and {\em idler} modes of an optical parametric amplifier/oscillator, whereas the $\kk_p$ mode is the pump mode: the $p+p\rightarrow s+i$ parametric scattering is mediated by the strong $\chi^{(3)}$ optical nonlinearity resulting from polariton-polariton interactions as shown in the lower panel. Differently from most other parametric devices, the peculiar shape of the lower polariton dispersion and the choice of the pump wavevector $\kk_{p}$ in the vicinity of the inflection point allows for the parametric process to occur in a remarkable triply-resonant way with conservation of both total energy and momentum.

\subsection{Parametric amplifier and parametric luminescence}

\begin{figure}[t]
\begin{center}
\includegraphics[width=0.45\columnwidth,angle=0,clip]{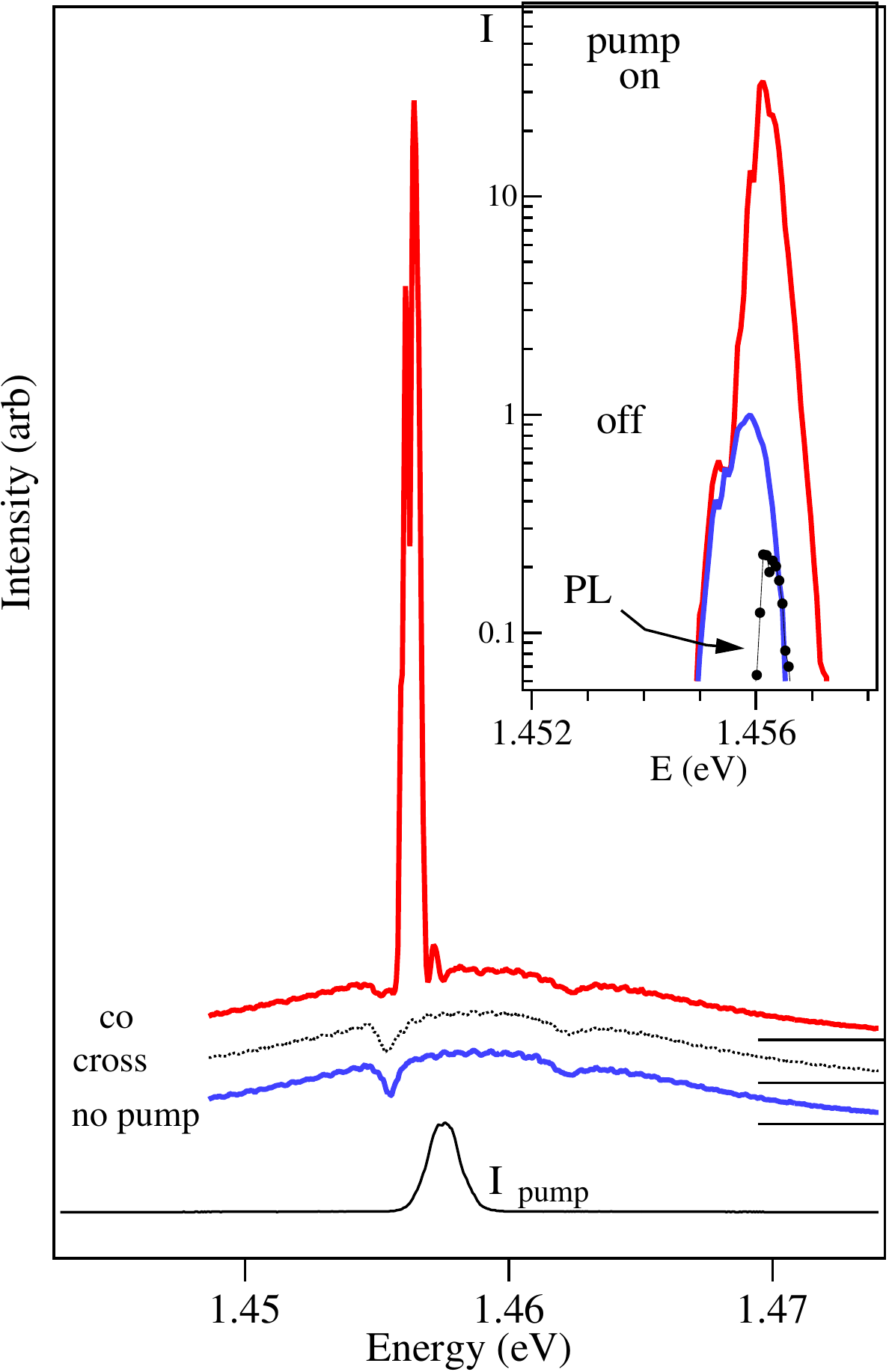} 
\includegraphics[width=0.45\columnwidth,angle=0,clip]{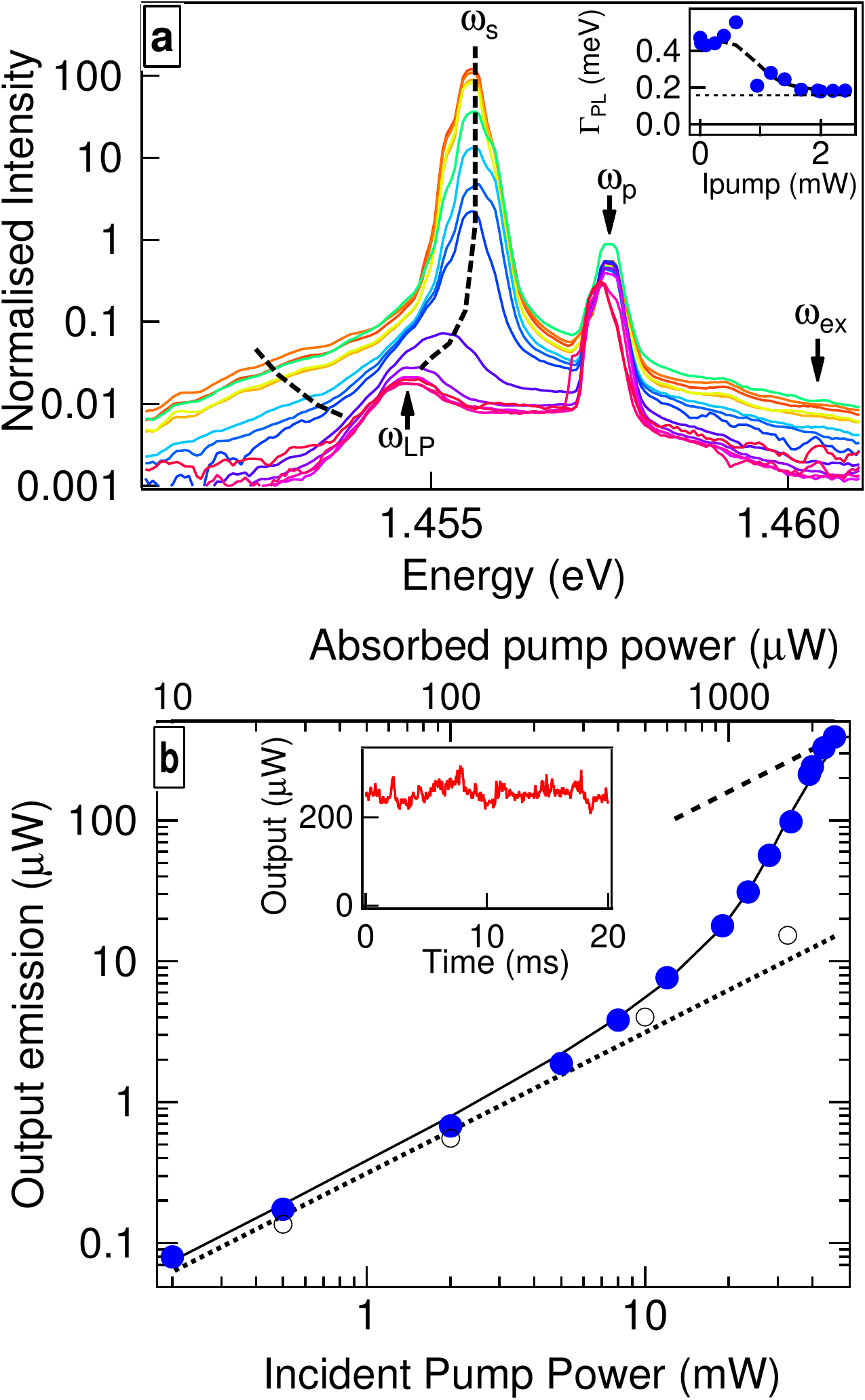}
 \end{center}
\caption{Left: amplification spectra due to stimulated polariton scattering \cite{Savvidis:PRL2000}. If the pump is circularly polarized, due to the scattering selection rules, the amplification (responsible for the very sharp peak in the red curve) occurs for co-circularly polarized seeding, while no effect is visible for counter-circular polarization. Right: power-dependance of the emission under continuous wave pump excitation. Figures from \cite{Baumberg:PRB2000}}
\label{fig:threshold}
\end{figure}

For low pump intensities, the parametric scattering process takes place in a spontaneous way and produces an incoherent luminescence into the signal and idler modes, characterized by a broad spectral distribution in both energy and wavevector (i.e. angle, see middle panels of Fig.\ref{fig:OPOsketch}). If a additional seed beam is incident on the microcavity around the signal (idler) wavevectors, this gets coherently amplified \cite{Savvidis:PRL2000} (see Fig.\ref{fig:threshold}) and in turn generates a four-wave mixed beam at the idler (signal) wavevector~\cite{Ciuti:PRB2000}.  Following the first observations, many authors have experimentally unveiled different aspects of this sort of parametric amplifier physics~\cite{Messin:PRL2001,Erland:PRL2001,Saba:Nature2001}, with a special attention to its coherent and quantum aspects \cite{Huynh:PRL2003,Kundermann:PRL2003,Savasta:PRL2005}.

\subsection{Optical parametric oscillator}

As usual in bosonic systems, scattering processes can be stimulated by an existing population in the final states, here the signal and idler modes. 
When the pump intensity is strong enough for the stimulated scattering rate to overcome losses, a new stationary regime with a macroscopic occupation of single signal and idler modes is reached and a pair of coherent signal and idler beams are emitted with a narrow distribution in the energy and in-plane wavevector~\cite{Baumberg:PRB2000,Stevenson:PRL2000}. Other general aspects of the parametric oscillation in microcavity polariton systems were reported in~\cite{Savvidis:PRB2000,Houdre:PRL2000,Tartakovskii:PRB2002,Butte:PRB2003}.

From a fundamental point of view, the parametric oscillation in spatially extended geometries such as planar microcavities is an interesting example of non-equilibrium phase transition~\cite{Haken:RMP1975}. The symmetry that is spontaneously broken at the critical point is the $U(1)$ symmetry corresponding to the simultaneous and opposite rotation of the global signal and idler phases by the arbitrary angle $\varphi$,
\begin{equation}
S\rightarrow S\,e^{i\varphi} \hspace{1cm} I\rightarrow I\,e^{-i\varphi}:
\end{equation}
even though the mean-field equations are (approximately~\cite{Wouters:PRA2007}) invariant under the $U(1)$ symmetry, a specific value of the signal and idler phases is chosen at every instance of the experiment. 
The strong analogies between the  parametric oscillation operation and a Bose-Einstein condensation transition are apparent in the Penrose-Onsager criterion 
\begin{equation}
\lim_{|\rr-\rr'|\rightarrow \infty} \langle \El^\dagger_{s}(\rr)\,\El_{s}(\rr') \rangle \neq 0
\end{equation}
that was used in the experiment~\cite{Baas:PRL2006} and in the numerical simulations of~\cite{Carusotto:PRB2005} to assess the appearance of long-range order in the signal (or equivalently the idler) emission. In both theory and experiments, the signal field $\El_{s}(\rr)$ was isolated from the pump and idler ones by an angular filtering in $\kk$-space. 

\subsection{The signal/idler condensate} 

Early theoretical modelling of the polariton parametric effects in planar microcavities was based on a three-mode approximation and consisted in a set of coupled nonlinear differential equations for the time-evolution of the lower polariton field $\mathbb{C}$-number amplitudes in the pump $\kk_p$, signal $\kk_s$, and idler $\kk_i$ modes~\cite{Ciuti:PRB2000,Whittaker:PRB2001,Gippius:EPL2004,Whittaker:PRB2005}.
This corresponds to making a three-mode ansatz for the lower polariton field:
\begin{multline}
\psi_{LP}(\kk,t) =S(t)\,\delta_{\kk,\kk_s}\, e^{-i\omega _{s}t}
+P(t)\, \delta_{\kk,\kk_p}\, e^{-i\omega_{p}t} + \\
+I(t)\, \delta_{\kk,\kk_i}\,e^{-i \omega_i t},
\eqname{3-mode}
\end{multline}
where the signal/idler frequencies and wave vectors are related by $\omega_i=2\omega_p-\omega_s$ and $\kk_i=2 \kk_p-\kk_s$.
Below the pump oscillation threshold, the pump-only solution with ${S}={I}=0$ is dynamically stable. 
Above threshold, the pump-only solution becomes dynamically unstable (as shown in the B' panel of Fig.\ref{fig:RMP_cohpump}) and a new steady-state is reached with finite signal and idler amplitudes that spontaneously break the $U(1)$ phase symmetry.
Depending on the specific  pump parameters, a variety of bistable and reentrant behaviors as a function of pump intensity were anticipated; in particular, the parametric threshold can be a second-order one with a continuous growth of the signal/idler intensity past the critical point, or a first-order one with a sudden jump to the OPO regime~\cite{Whittaker:PRB2005,Wouters:PRB2007b}. Experimental investigations of this rich nonlinear dynamics have appeared in~\cite{Baas:PRB2004,Demenev:PRB2009}.

\subsubsection{Spatial patterns}

A main drawback of the three-mode approximation based on the ansatz \eq{3-mode} is that it is limited to spatially homogeneous geometries and does not provide a criterion to determine the specific modes $\kk_{s,i}$ into which OPO operation will take place. Naively,  one could consider the  $\kk_{s,i}$ modes for which parametric gain is the strongest and then perform a local density approximation. However, $\kk_{s,i}$ themselves strongly depend on the pump intensity and the theory has to be solved self-consistently. Some insight can be obtained by applying techniques borrowed from pattern formation theories in nonlinear dynamical systems~\cite{Cross:RMP1993}.
\begin{figure}[t]
\begin{center}
\includegraphics[width=\columnwidth,angle=0,clip]{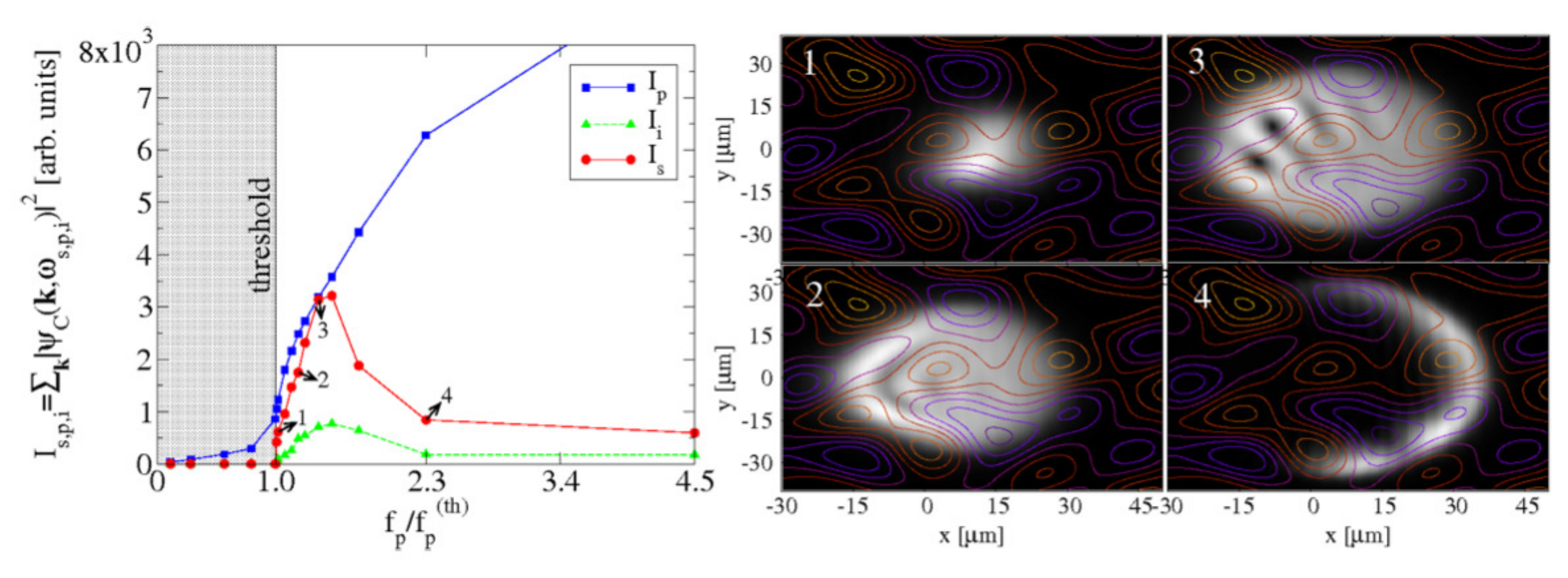}
\includegraphics[width=\columnwidth,angle=0,clip]{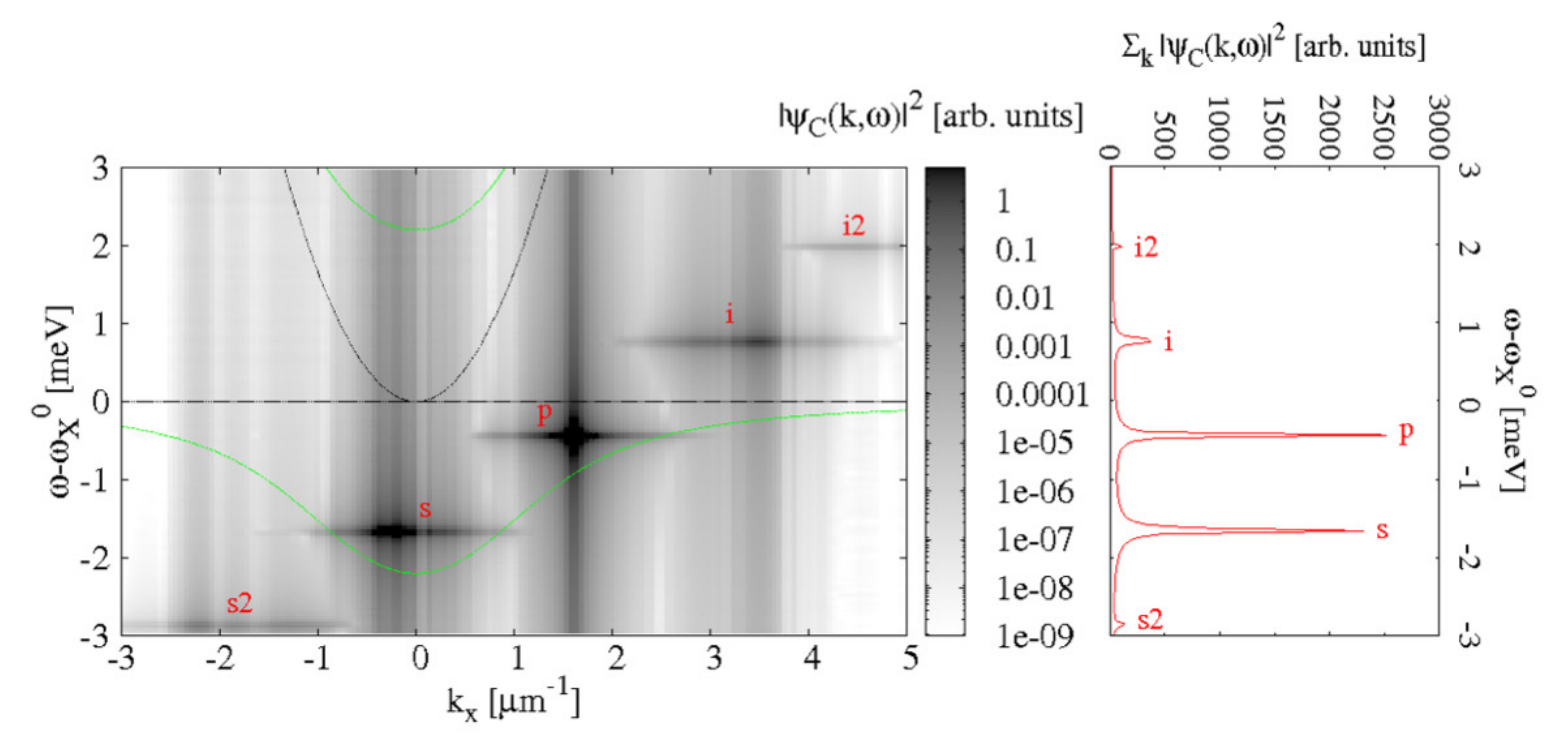}
\includegraphics[width=\columnwidth,angle=0,clip]{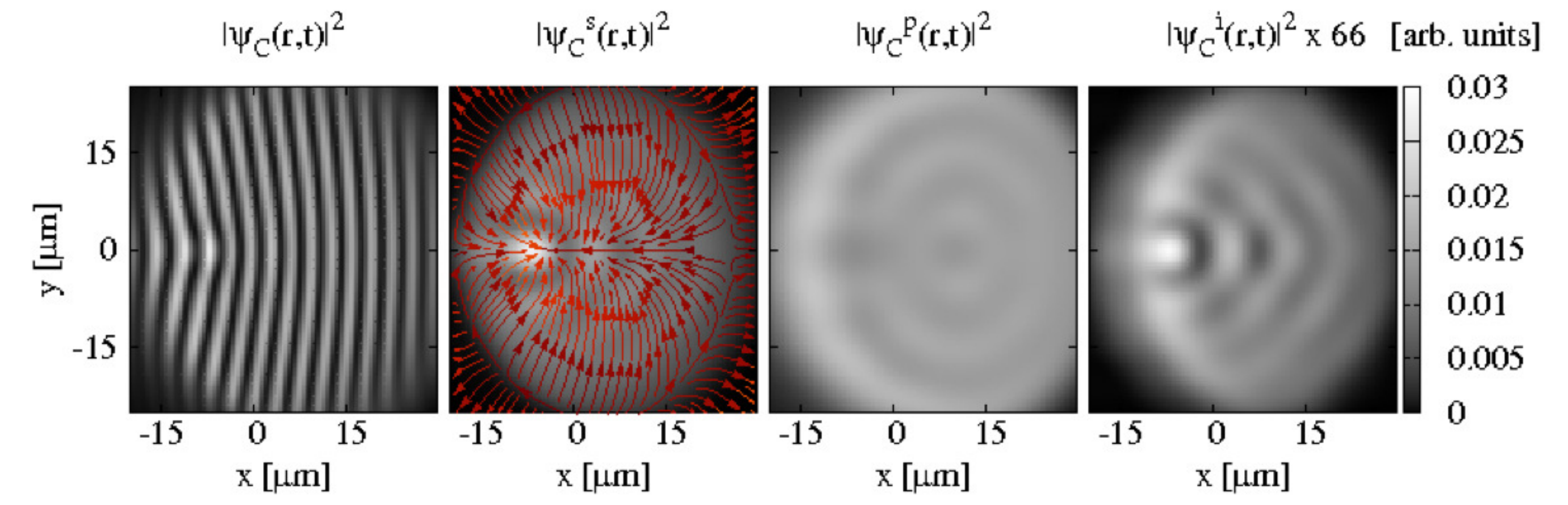}
 \end{center}
\caption{Figures from~\cite{Marchetti:chapter2012}. Top panels: calculated evolution of the signal, idler and pump state intensities as a function of the renormalized pump intensity (left) and space profiles of the filtered signal at different values of the pump intensity in the presence of some photonic disorder potential (right). Middle panel: photonic density spectra in momentum-frequency space. Bottom panel: full emission (first panel) and filtered emission of signal (second panel), pump (third panel)  and idler (fourth panel). The arrows in the second panel indicate the current pattern that is present in the steady state of the signal field.}
\label{fig:marchetti:OPO}
\end{figure}

Alternatively, the spatial pattern can obtained by numerically integrating the non-equilibrium Gross-Pitaevskii equation. This was first done by~\onlinecite{Whittaker:PSSC2005}. A discussion along these lines recently appeared in~\cite{Marchetti:chapter2012} for a flat-top pump spot geometry (see Fig.\ref{fig:marchetti:OPO}). As a function of pump intensity, the parametric oscillation operation first appears in the center of the spot  and then extends to the whole pumped region (top-right panels). More details on the spectral shape of the polariton wavefunction are shown in the middle panels of Fig.\ref{fig:marchetti:OPO}: as we are dealing with a steady-state, the pump, signal and idler emissions are monochromatic at $\omega_{p,s,i}$. On the other hand, the finite spatial size of the emission is visible in their significant broadening in $\kk$ space around their central wavevectors $\kk_{p,s,i}$. The weaker satellite emissions that are visible at frequencies $\omega_{s2}=2\omega_s-\omega_p$, $\omega_{i2}=2\omega_i-\omega_p$ are due to secondary parametric scattering processes; experimentally, they were first observed in~\cite{Savvidis:PRB2001,Tartakovskii:PRB2002}.
 
The real-space shape of the polariton wavefunction in the parametric regime is illustrated in the lower left panel of the same figure. As it was mentioned in~\cite{Carusotto:PRB2005}, interference of the signal, idler and pump beams is responsible for moving fringes in the polariton density pattern: from this point of view, the OPO phase transition can be interpreted as the spontaneous formation of a stripe pattern in the intensity profile of the cavity field, as discussed at length in the nonlinear optics literature~\cite{Oppo:PRA1994,Staliunas_book,Vaupel:PRL1999,Lugiato:JMO1997}.

In the finite pump spot geometry under consideration here, the interplay of nonlinearities, parametric gain and losses is responsible for the appearance of a complex structure of macroscopic currents also in the the filtered signal emission (second panel from the left). In~\cite{Marchetti:PRL2010}, it was shown that the behavior can be even richer when the pump beam is spatially narrow and the configuration with spatially smooth signal and idler can become dynamically unstable towards the spontaneous appearance of quantized vortices. Once the vortex has reached its equilibrium position, the system remains stably in the new steady-state with a macroscopic current flowing around the cloud.

As originally pointed out in~\cite{Whittaker:SM2007}, vortices can also be forced into the system by seeding the signal (or idler) modes with a temporally short pulse with a Laguerre-Gauss spot profile. Such a triggered parametric device is often indicated by the acronym TOPO. For suitably chosen pump conditions, the orbital angular momentum imprinted by the seed can remain in the signal polaritons for macroscopically long times until the pump beam is switched off. As we shall see better in Sec.\ref{sec:superfluid_metastable}, the presence of several metastable states  with non-trivial macrocopic current patterns is one characteristic signature of superfluid behavior. 

\subsubsection{Coherence of the signal/idler condensate}
\label{sub:OPOWigner}

The Gross-Pitaevskii description reviewed in the previous section is able to capture a good deal of the mean-field physics of the signal/idler emission well above the threshold. As usual in statistical mechanics, fluctuations play a crucial role in the critical region in the neighborhood of the transition point, which calls for more sophisticated theoretical approaches to sort out the analogies and differences between the parametric critical point and the standard Bose-Einstein condensation phase transition in equilibrium statistical mechanics.

A numerical study of the coherence properties of the signal emission across the OPO threshold was performed in~\cite{Carusotto:PRB2005} using the Wigner Monte Carlo technique of Sec.\ref{sec:Wigner} to include quantum fluctuations. 
Below threshold, it results from numerical calculations that the signal emission has a incoherent, thermal nature with a short-ranged first-order coherence and a significant spatial bunching on a distance scale set by the coherence length. As the threshold is approached, the coherence length becomes of the order of the system size. Related features in the correlation functions were addressed in the nonlinear optical literature in~\cite{Gatti:PRA1995,Zambrini:PRA2000} and often go under the name of ``quantum images''~\cite{Lugiato:JMO1997}. A recent work on the universal critical fluctuations around the optical parametric threshold appeared in~\cite{Drummond:PRL2005}. Above threshold, 
first-order coherence extends to the whole system and intensity fluctuations are almost completely suppressed $g^{(2)}\approx 1$. These numerical observations are in agreement with the usual picture of second order phase transitions and with the experimental measurement of coherence by~\onlinecite{Baas:PRL2006}.

As the polariton system is naturally a two-dimensional one, one may expect that fluctuations have a major impact on the long-distance coherence of the signal emission. At equilibrium, the Bose-Einstein condensation phase transition in two-dimensions is replaced by a Berezinski-Kosterlitz-Thouless (BKT) transition to a superfluid but still non Bose-condensed state~\cite{Minnhagen:RMP1987} and true long-range order of a Bose condensate is only observable at strictly zero temperature. So far, the spatial size of the systems considered in numerical calculations of two-dimensional OPO systems was not large enough to numerically solve this issue. Indirect evidence based on analytical calculations of~\cite{Szymanska:PRL2006} for a different but related model suggests that also at non-equilibrium the transition is of the BKT type.

Quasi-condensation behavior is instead clearly visible in calculations for a one-dimensional photonic wire geometry: even well above threshold, coherence has an exponential decay in space~\cite{Carusotto:PRB2005,Wouters:PRB2006}. At the threshold, the coherence length shows a clear increase and grows to macrocopic values, still it does not diverge to infinity. On the other hand, the suppression of intensity fluctuations is almost unaffected by the reduced dimensionality. Pioneering experimental observation of suppressed coherence in one-dimensional parametric oscillators appeared in~\cite{Cerda:PRL2010} {and then in~\cite{Spano:NJP2012}.} An experimental study of the BKT-like power-law decrease of coherence in a two-dimensional polariton gas under incoherent pumping has recently appeared in~\cite{Roumpos:PNAS2012}. {Following the theoretical proposal in~\cite{Small:PRA2011}, pioneering experimental studies of the BKT transition of classical nonlinear waves have recently been presented~\cite{Situ:APS2012} using an optical configuration similar to the kinetic BEC experiment of~\cite{Sun:NatPhys2012}.}

Experimental investigations of the temporal coherence of the signal emission were reported in~\cite{Krizhanovskii:PRL2006}. 
On one hand, the interplay of the OPO operation with spatial inhomogeneities due to sample disorder (mainly of photonic origin) and the pump beam profile was responsible for the simultaneous presence of several coherent signal beams at slightly different frequencies with different spatial distributions. 
On the other hand, the temporal coherence of each frequency component showed a coherence time in the few 100~ps range, almost a couple of orders of magnitude longer than the polariton lifetime. 
Even longer coherence times in the ns range was reported in~\cite{Spano:arXiv2011} using a high-quality sample where the OPO emission enjoys a remarkably good spatial uniformity and spatial coherence: the experimentally observed trends suggest the combined effect of number fluctuations
and inter-particle interactions in the condensate as the main source of decoherence according to the single mode model of~\cite{Whittaker:EPL2009}.

\subsubsection{Elementary excitations: the Goldstone mode}
\label{goldstone}
As usual in the statistical mechanics of phase transitions, the spontaneously broken continuous $U(1)$ symmetry is responsible for the  appearance of a soft Goldstone mode in the dispersion of the elementary excitations of the system: in the long-wavelength limit $k\rightarrow 0$, its frequency dispersion $\omega_G(k)$ tends to zero in both the real and imaginary parts. Physically, the Goldstone mode corresponds to a spatially slow twist of the order parameter, which feels a vanishingly small restoring force in the long-wavelength limit.

\begin{figure}[t]
\begin{center}
\includegraphics[width=0.9\columnwidth,angle=0,clip]{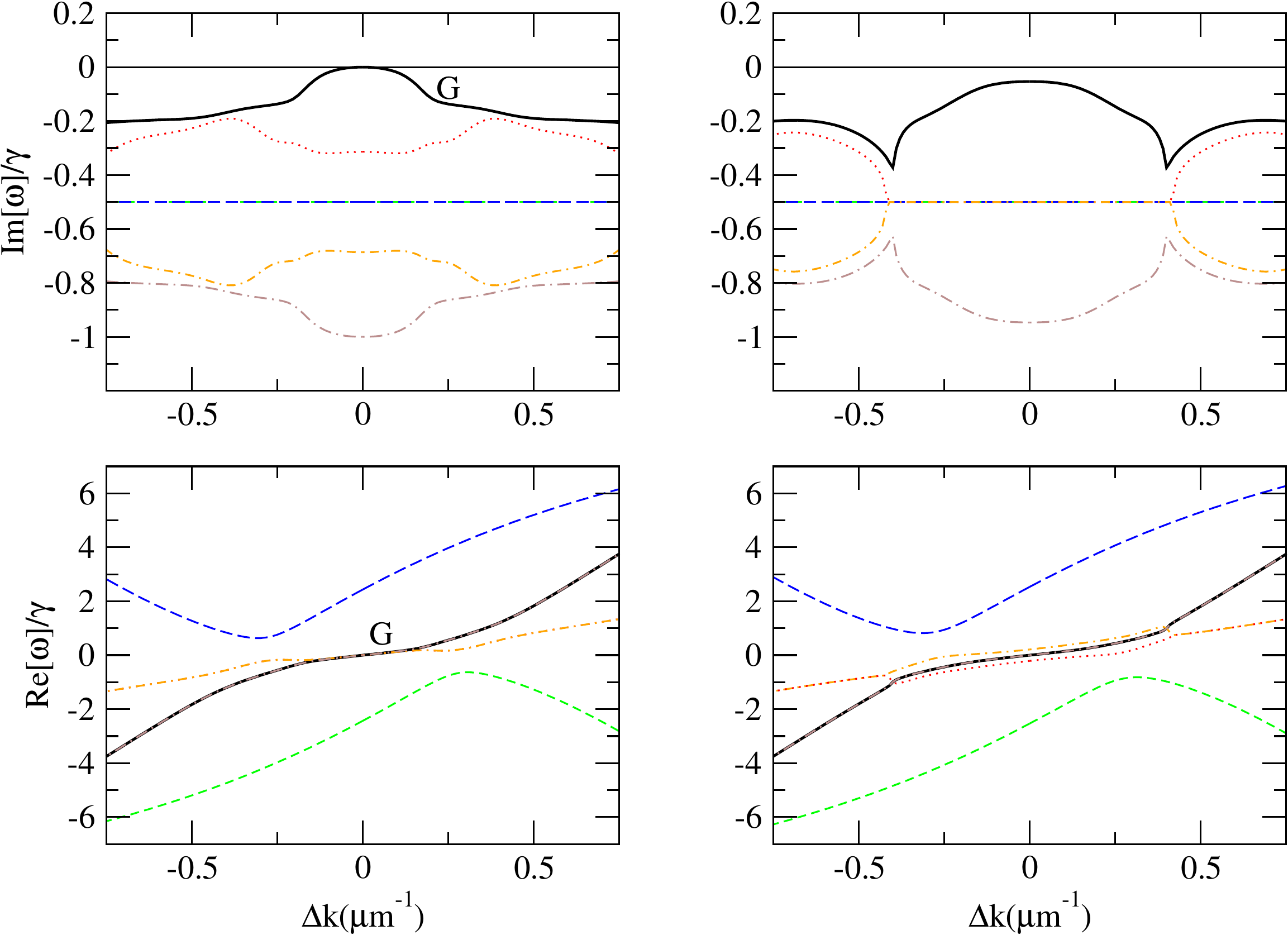}
 \end{center}
\caption{(Color online) 
Figure from~\cite{Wouters:PRA2007}. Top panel: imaginary part of the energy of the elementary excitations without (left) and with (right) symmetry-breaking probe. Bottom panel: same but for the real parts. The letter G indicates
the Goldstone mode.}
\label{fig:Goldstone}
\end{figure}

Among the most celebrated examples of Goldstone modes in condensed matter physics we can mention the zero-sound mode of superfluid Helium 4 or dilute Bose-Einstein condensates and the magnon excitations in ferromagnets: zero-sound is related to the spontaneous breaking of the $U(1)$ gauge symmetry of the quantum Bose field below the Bose-Einstein condensation temperature~\cite{Huang,Forster,PinesNozieres}, while the magnon branch is related to the spontaneous breaking of the rotational symmetry of the magnetic moment orientation below the Curie temperature~\cite{LandauStat2}.

Calculations of the elementary excitation spectrum on top of an OPO coherent state were performed in~\cite{Wouters:PRB2006,Wouters:PRA2007} starting from a three-mode model of OPO operation. Small spatial fluctuations of the mode amplitudes are included in the form 
\begin{eqnarray}
S &\rightarrow& S + u_s\, e^{i(\Delta k\, x-\omega\, t)}
+v^*_s\, e^{-i(\Delta k\, x-\Delta \omega\, t)} \label{S_lin}  \\
P &\rightarrow& P + u_p\, e^{i(\Delta k\, x-\omega\, t)}
+v^*_p\, e^{-i(\Delta k\, x-\Delta \omega\, t)} \label{P_lin} \\
I &\rightarrow& I + u_i\, e^{i(\Delta k\, x-\omega \,t)}
+v^*_i\, e^{-i(\Delta k\, x-\Delta \omega\, t)}, \label{I_lin}
\end{eqnarray}
and the dispersion $\Delta\omega(\Delta k)$ of the elementary excitations is obtained by plugging this ansatz into the full wave equation \eq{GPE_LP}. {A pioneering experimental investigation of these excitations was reported in~\cite{Savvidis:PRB2001}, including the remarkable observation of a significant {\em off-branch} luminescence from the negative Bogoliubov modes.}

An example of {theoretical dispersion taken from~\cite{Wouters:PRA2007}} is shown in the left panels of Fig.\ref{fig:Goldstone}: note in particular the Goldstone mode (indicated by the "G" label) whose frequency tends to zero (in both real and imaginary parts) as $\Delta k\rightarrow 0$. Its functional dependence on $\Delta k$ is however very different from the usual one \eq{Bogo_omega} of equilibrium Bose systems  which starts as $\omega(k)\simeq c_s\,|k|$ for small $k$'s with a singularity at $k=0$.

Here, no singularity appears in the dispersion relation $\omega_G(\Delta k)$ of the Goldstone mode around $\Delta k=0$. 
The continuous and non-vanishing slope of the real part $\textrm{Re}[\omega(\Delta k)]\propto \Delta k$ is due to the flow of the 
pump polaritons which are injected with a finite wave vector $k_p$ and are then able to drag the elementary excitations.
On the other hand, the imaginary part of $\omega_G(\Delta k)$ has the low-$\Delta k$ form $\textrm{Im}[\omega(\Delta k)]\approx -\alpha\,(\Delta k)^2$ with a positive $\alpha>0$.
As a result, the Goldstone mode of a planar OPO consists of a spatially slowly varying twist of the signal and idler phases: a localized perturbation of the signal/idler phase will not propagate as a sound wave, but rather relax back to the equilibrium state while being dragged by the pump polariton flow. Almost simultaneously, a similar diffusive behavior of the Goldstone mode was obtained in a completely different model of polariton condensation under a incoherent pump in~\cite{Szymanska:PRL2006}; a review of these results will be given in Sec.\ref{nonresonant}.

The soft nature of the Goldstone mode is destroyed in the presence of external fields that fix the direction of the order parameter.
While this can appear as quite artificial for Bose-Einstein condensation of material particles~\cite{Huang,Gunton:PR1968}, it naturally happens in ferromagnets in the presence of an external magnetic field that imposes the direction of magnetization. 
As it was discussed in~\cite{Wouters:PRA2007}, the signal/idler phases of an OPO can be fixed by an additional, weak laser field that resonantly drives the signal (or the idler) mode. 
The effect of this additional weak laser field on the elementary excitation spectrum is illustrated in the right panels of Fig.\ref{fig:Goldstone}: now the imaginary part of the dispersion does not tend to zero for $\Delta k\rightarrow 0$, which signals the presence of a restoring force than tends to bring the signal/idler phase back to their steady-state determined.

\section{Condensates under incoherent pumping}
\label{nonresonant}

In Sec. \ref{sec:incoh_GPE} we have introduced the generalized Gross-Pitaevskii equation \eq{CGLE} describing the mean-field dynamics of a driven-dissipative polariton condensate under an incoherent pumping. In this section we shall review its application to specific configurations of actual experimental interest and we shall compare the theoretical predictions with observations. Even though our focus will be concentrated on microcavity polariton systems for which the analogy with BEC of material particles is the clearest, most of the features are expected a much wider application range and can be observed to any spatially extended laser oscillator, for instance vertical cavity surface emitting lasers, macroscopic laser cavities at large Fresnel number, or even laser oscillation in disordered media, the so-called {\em random lasing} (see e.g. \onlinecite{Tureci:Science2008}).

\subsection{The condensate shape}
\begin{figure}[t]
\begin{center}
\includegraphics[width=\columnwidth,angle=0,clip]{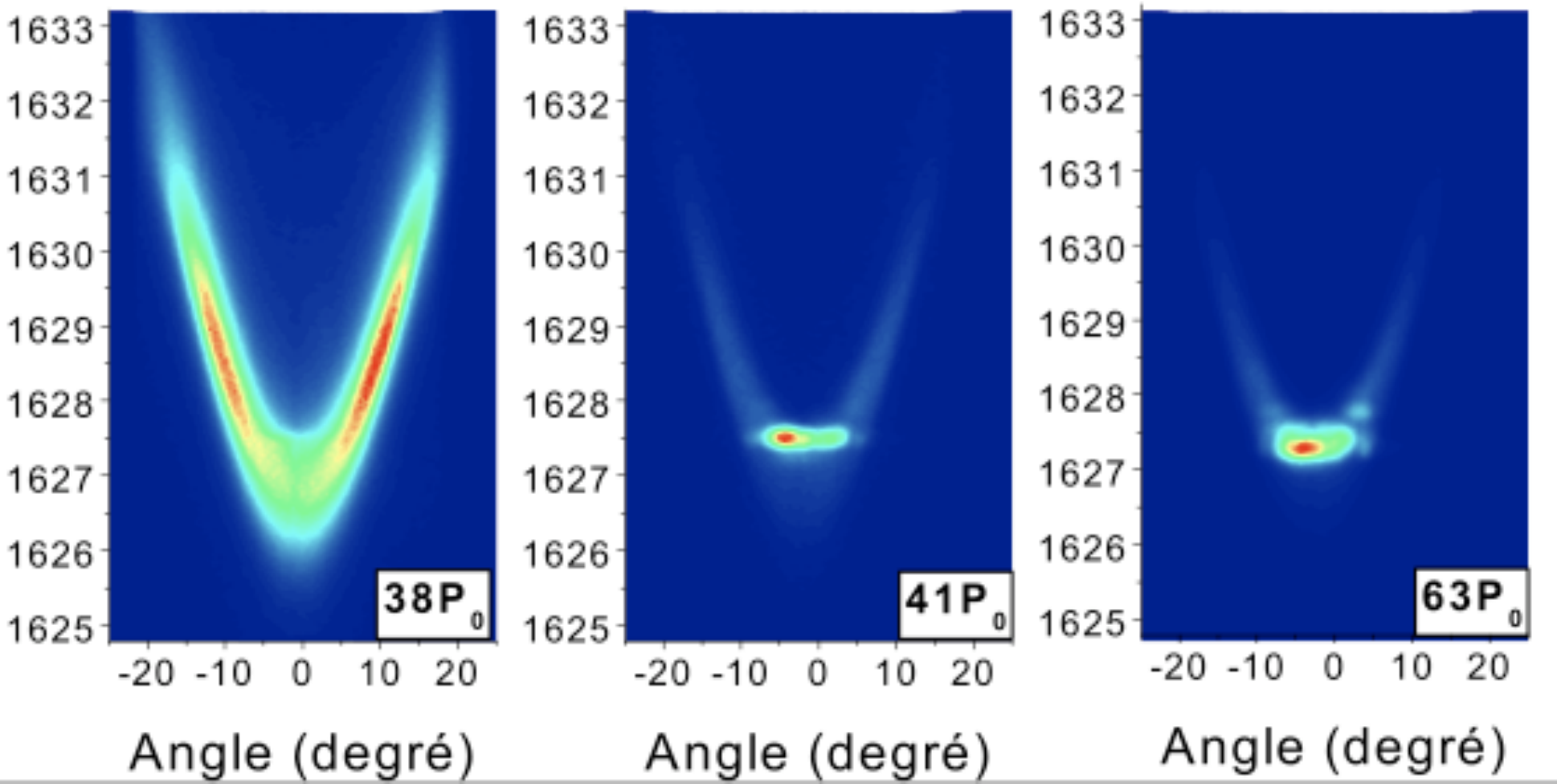}
 \end{center}
\caption{(Color online) Experimental observations of polariton condensation under incoherent pumping with a large spot of diameter $25\,\mu\textrm{m}$. The different panels show the emission pattern in the $(k,E)$ plane for different values of the pump intensity. The condensation threshold is situated at a pump intensity value between the second and the third panel. Figure from~\cite{Richard:PhD}.}
\label{fig:Richard_wide}
\end{figure}
\begin{figure}[t]
\begin{center}
\includegraphics[width=1.\columnwidth,angle=0,clip]{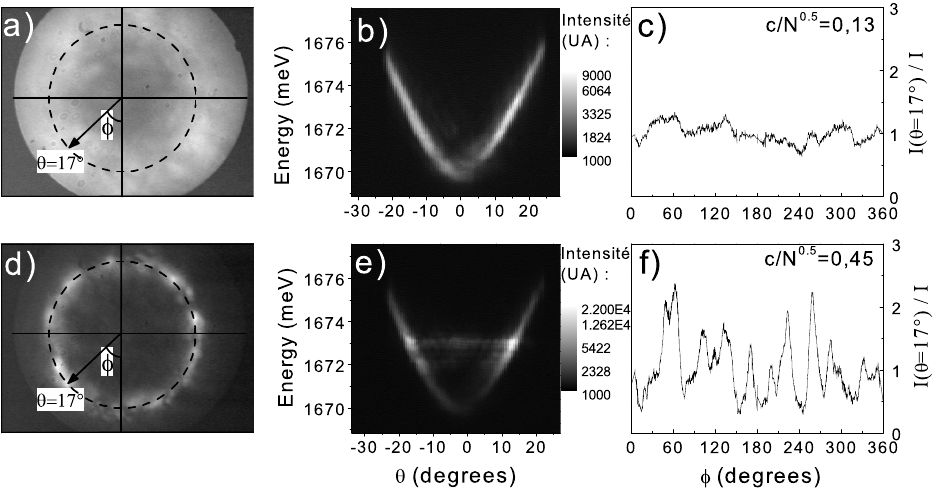}
 \end{center}
\caption{Experimental observations of polariton condensation under incoherent pumping with a small spot of diameter $3\,\mu\textrm{m}$. The first and second rows are obtained below and above the
stimulation threshold, respectively. Left column: far-field emission. The middle column:  emission pattern in the $(k,E)$ plane. Right panels: intensity profile along the azimuth angle on the $\theta=17^o$ ring indicated as a dashed line in (a,d).
Figure from~\cite{Richard:PRL2005}.}
\label{fig:Richard_ring}
\end{figure}
The first quantity to consider is the shape of the condensate in both real space and in momentum space: an unexpected feature of the first, pioneering experiments of polariton condensation under an incoherent pumping was in fact the strong dependence of the momentum distribution of the coherent polaritons on the size of the pump spot~\cite{Richard:PRL2005,Richard:PRB2005}. For a large pump spot ($\sigma\approx 20~\mu$m), the momentum distribution was wide but centered at $\kk=0$ as shown in the two rightmost panels Fig.\ref{fig:Richard_wide}. On the other hand, for a small pump spot ($\sigma\approx 3~\mu$m) the momentum distribution was centered on a narrow ring in $\kk$-space, as shown in Fig.\ref{fig:Richard_ring}(d). 

\subsubsection{Spatially homogeneous system}

A physical explanation of these observations can be obtained in terms of the generalized GPE {that in the present case of an incoherent pumping has the complex Ginzburg-Landau equation form \eq{CGLE}. As a first step, let us} consider the spatially homogeneous case with a spatially constant pumping $P(\rr)=P$ and without any trapping potential~\cite{Wouters:PRL2007}. 

For low pumping $P<P_c=\gamma_R\, \gamma_{LP}$, the only dynamically stable mean-field solution is $\Psi_{LP}=0$. For a stronger pumping $P>P_c$, the stimulated scattering overcomes losses and the zero mean-field solution becomes dynamically unstable. Depending on the initial fluctuation that seeds the condensation process, a stable condensate can appear in any momentum mode,
\begin{equation}
\Psi_{LP}(\rr,t)=\Psi_{LP}^0\,e^{i\kk \rr}\,e^{-i\omega t}
\end{equation}
with a condensate density
\begin{equation}
n_{LP}=|\Psi_{LP}^0|^2=\frac{\gamma_R}{R}\left(\frac{P}{P_c}-1\right).
\eqname{P_n}
\end{equation}
Given the $U(1)$ symmetry of the GPE equation under global rotations of the condensate phase, this latter is arbitrarily chosen at every instance of the experiment according to the same spontaneous symmetry breaking mechanism as in the equilibrium BEC phase transition~\cite{Gunton:PR1968}. A completely different approach to non-equilibrium condensation based on a Bose-Fermi model was proposed in~\cite{Szymanska:PRL2006}, giving the same qualitative features for the phase transition.

The condensate frequency is determined by the pumping intensity via the equation of state
\begin{equation}
\omega=\omega_{LP}(\kk)+g_{LP}\,|\Psi_{LP}^0|^2+\Delta \omega_{\rm res}(P):
\eqname{EOS_P}
\end{equation}
the second erm on the right-hand side accounts for the interactions between the condensate polaritons with an interaction constant $g_{LP}$ already discussed in \eq{V_LPLP}. In contrast, an explicit expression for the last term $\Delta\omega_{\rm res}(P)$ describing the (generally repulsive) interactions with the reservoir polaritons requires a microscopic modeling that goes beyond the scope of the present review. In recent experiments~\cite{Wertz:NatPhys2010}, it appears that this latter term is the dominating one in the pumped region. The fact that in such a model a stable condensate can appear in any $\kk$ mode is of course an artifact of the assumption of a spatially homogeneous pump $P(\rr)=P$ and disappears in more sophisticated calculations including in the generalized GPE \eq{CGLE} the spatial profile of the (unavoidably finite-size) pump spot~\cite{Wouters:PRB2008}: the drift of polaritons outside the pumped region reduces the effective amplification of $\kk\neq 0$ moving condensates and provides a criterion to select the mode into which condensation does occur. The next two subsections describe the steady-state condensate shape for simple geometries.
\subsubsection{Condensation under a large pump spot}
\label{sec:largespot}
The condensate shape in the case of a large pump spot with a smooth spatial profile was theoretically studied within the local density approximation in~\cite{Wouters:PRB2008}. While the local density $n_{LP}(\rr)=|\Psi_{LP}(\rr)|^2$ is univocally determined by the local pumping intensity via \eq{P_n}, more care has to be paid to the phase: in contrast to the standard GPE \eq{staticGPE}, breaking of time reversal symmetry by the pumping and loss terms of \eq{CGLE} allows for the steady-state wavefunction to have a spatially varying phase, corresponding to macroscopic currents flowing across the condensate cloud. 

\begin{figure}[h]
\begin{center}
\includegraphics[width=1.\columnwidth,angle=0,clip]{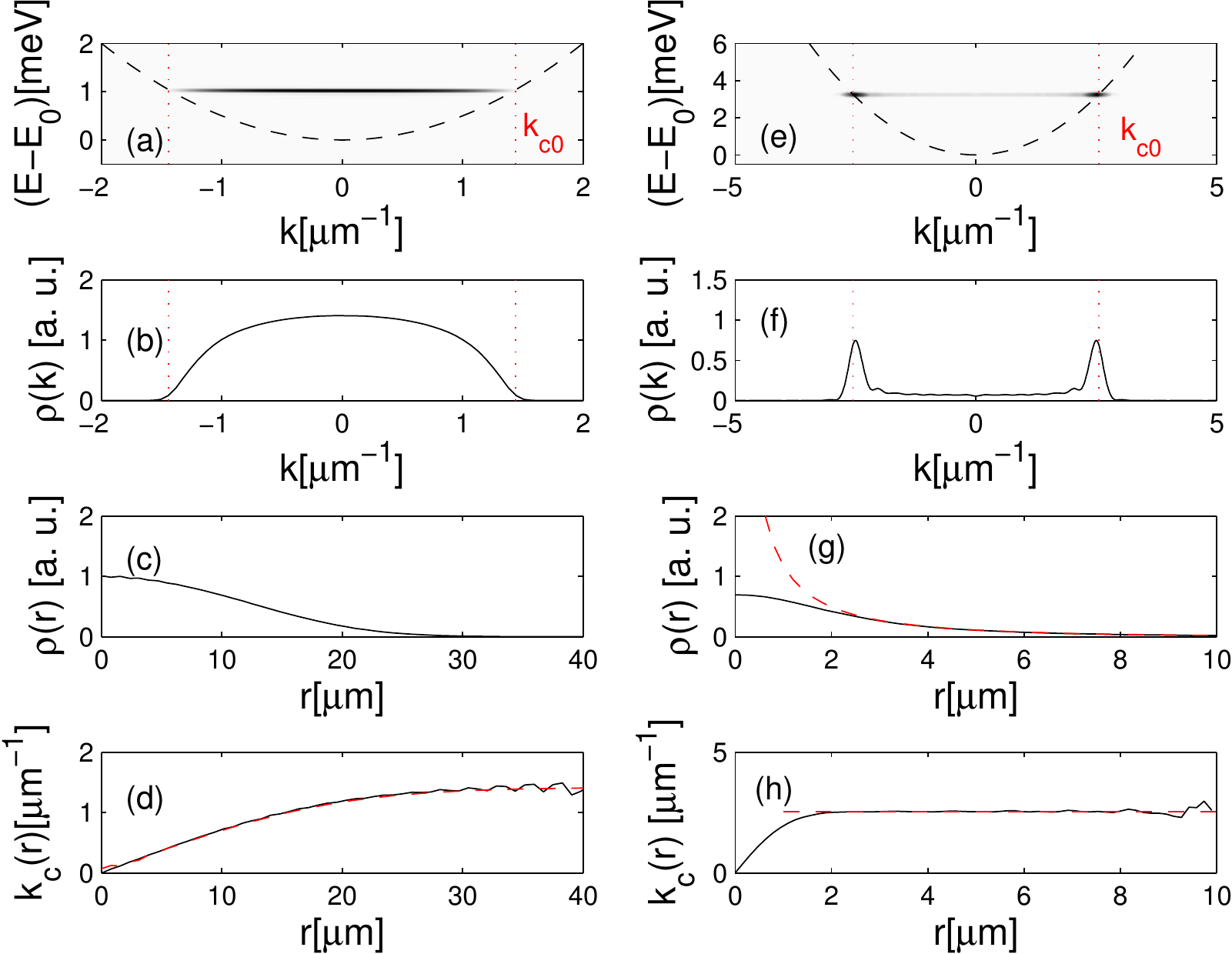} 
\end{center}
\caption{(Color online) Numerical results of generalized GPE simulations in the absence of disorder for respectively a large $\sigma_p=20\,\mume$ (a-d) and a small $\sigma_p=2\,\mume$ (e-h) circular excitation pump spot. Panels (a,e) give the $(k,E)$ emission pattern, (b,f) the polariton distribution in momentum space, (c,g) the polariton distribution in real space and (d,h) the local wave vector $k_c(r)$. The wave vector $k_{c0}$ of a free polariton at  $\omega_c$ is indicated by the dotted lines in (a,b,e,f); the dashed line in (g) is the analytical approximations to the density tail, and the dashed line in (d,h) are the LDA predictions to the local wave vector. 
All quantities in real (momentum) space depend only on the radial coordinate $r=|\rr|$ ($k=|\kk|$). Figure from~\cite{Wouters:PRB2008}.}
\label{fig:ring_th}
\end{figure}

As the frequency $\omega$ of a condensate is constant throughout the cloud, the spatial variation of the pumping intensity $P(\rr)$ and of the condensate density $n_{LP}(\rr)$ has to be compensated in \eq{EOS_P} by a non-trivial spatial dependence of the local wavevector $\kk(\rr)=\nabla[\textrm{Arg}[\Psi_{LP}(\rr)]$. 
The geometric details of the resulting polariton current are easily understood in the simplest case of cylindrically symmetric pump with a monotonically decreasing pump intensity along the radial direction, see Fig.\ref{fig:ring_th}(a-d); an example of the complex current pattern that appears in a polariton condensate under a OPO pumping for was shown on the bottom row in Fig.\ref{fig:marchetti:OPO}.

Under the reasonable assumption that rotational invariance is not spontaneously broken, the wavevector at the center of the cloud $\kk(\rr=0)=0$ and $\kk(\rr)$ is everywhere in the outward radial direction, $\kk(\rr)=k(r)\,\ee_\rr$. From the explicit forms of \eq{EOS_P} and \eq{P_n}, the flow wavevector $k(r)$ is found to be an increasing function of the radial coordinate $r$ and gets to its maximum value $k_{\rm max}$ at the external edge of the condensate cloud where the condensate density vanishes ($n(r_{\rm max})=0$). Within the local density approximation, the radius $r_{\rm max}$ of the condensate cloud coincides with the point where the local pump intensity equals the critical pump value, $P(r_{\rm max})=P_c$. Most remarkably, the maximum momentum $k_{\rm max}$ corresponds to the value of wavevector at which the free polariton dispersion equals the condensate frequency $\omega_{LP}(k_{\rm max})=\omega$. 

Even though no local measurement of the condensate phase has yet directly confirmed this prediction for the spatial dependence of $\kk(\rr)$ in wide condensate clouds, several other experimental observations~\cite{Richard:PRB2005} are in agreement with this theory: in particular the emission pattern in the $(k,E)$ plane consists of horizontal segments at a constant $\omega$, whose extremes coincide with the intersection with the free polariton dispersion.  Another indirect evidence of the presence of a non-trivial current pattern in the condensate is visible in the non-centrosymmetric shape of the momentum distribution shown in the right panel of Fig.\ref{fig:Richard_wide}: the Fourier transform $\tilde{f}(\kk)$ of any function $f(\rr)$ with a spatially constant phase has in fact a centro-symmetric modulus $|\tilde{f}(-\kk)|^2=|\tilde{f}(\kk)|^2$.

\subsubsection{Condensation under a small pump spot}
\label{sec:smallspot}

The situation is different when the size $\sigma$ of the pump spot is smaller than the ballistic propagation distance and polaritons are able to travel away from the pump spot before decaying. In this geometry, a quantitative calculation of the frequency $\omega$ and of the peak density requires including the kinetic energy terms of \eq{CGLE} and has to be performed numerically. Nonetheless, important qualitative information on the condensate shape in both real and momentum space can be obtained under the reasonable assumption that rotational invariance is not broken. In particular, an analytical solution for the condensate wavefunction far from the pump spot can be obtained for non-interacting polaritons in the form~\cite{Wouters:PRB2008}
\begin{equation}
\Psi_{LP}(r)\simeq \frac{A}{\sqrt{r}}\,e^{ik_{\rm max} r}\, e^{-r/2\ell}
\eqname{radial}
\end{equation}
with $A$ a normalization factor, $k_{\rm max}$ is the above-defined maximum wavevector and $\ell=v_{\rm max}/\gamma_{LP}$ is the spatial decay rate of the ballistically flowing polaritons at speed $v_{\rm max}=\nabla_\kk\omega(k_{\rm max})$. The additional $1/\sqrt{r}$  dependence is a consequence of particle conservation for a radial flow in two dimensions. The non-interacting polariton assumption is generally valid as soon as the density $n(\rr)$ is low enough for $g_{LP} n(\rr) \ll \omega_{LP}(k_{\rm max})$.

Under the condition $\ell\gg\sigma$, we expect that the total number of polaritons ballistically expanding at $k_{\rm max}$ is larger than the number of polaritons under the pump spot and that the momentum distribution will be approximately given by the Fourier transform of \eq{radial}. As a result, the condensate shape in momentum space consists of a ring in $\kk$ space of radius $k_{\rm max}$ and width proportional to $1/\ell$. This analytic guess was numerically verified by~\onlinecite{Wouters:PRB2008}: the result for the steady-state condensate wavefunction $\Psi_{LP}(\rr)$ is illustrated in Fig.\ref{fig:ring_th}(e-h) and provides a theoretical interpretation to the observations by~\onlinecite{Richard:PRL2005} reported in Fig.\ref{fig:Richard_ring}. 
This picture of ballistic motion of condensed polaritons away from the pumped region has been confirmed by later experiments~\cite{Wertz:NatPhys2010,Christmann:PRB2012} measuring the local wavevector of the expanding polaritons at different positions.

For the sake of completeness, it is worth mentioning that an alternative explanation of the observed ring involves fragmentation of the condensate into several sub-condensates at different $\kk$'s. The conceptual difference between these two states is formally illustrated as
\begin{equation}
\frac{1}{\sqrt{N!}}\left[\frac{1}{\sqrt{2}} \left( \ahd_1 + \ahd_2\right)\right]^N |\textrm{vac}\rangle \neq \frac{1}{(\frac{N}{2})!} (\ahd_1 )^{\frac{N}{2}} ( \ahd_2 )^{\frac{N}{2}} |\textrm{vac}\rangle.
\eqname{BEC_vs_frag}
\end{equation}
For condensates at equilibrium, energy arguments~\cite{nozieres_green,Leggett:2001RMP} favor the coherent superposition state in the LHS of \eq{BEC_vs_frag}, but these arguments no longer apply in the driven-dissipative polariton case. Mutual coherence of the emission in the $\pm \kk$ directions was experimentally measured in~\cite{Richard:PhD,Richard:PRL2005} with a Billet interferometer, which confirmed that one is indeed dealing with a single, non-fragmented condensate with a spatially complex wavefunction $\Psi_{LP}(\rr)$. A direct observation of the mutual coherence of polaritons ballistically moving in opposite directions away from the small pump spot was reported in~\cite{Wertz:NatPhys2010}.

\subsubsection{Effect of disorder}

A complete account of the experimental findings requires including in the model the effect of the disorder that is unavoidably present in the microcavity samples, mostly of photonic origin. 
The main effect of disorder is to introduce a strong modulation of the condensate density, the highest density being concentrated around the bottom of the disorder potential wells. It is then natural to wonder about the mutual coherence of the different regions: in the experiment~\cite{Kasprzak:Nature2006}, the spatial coherence of the different regions was assessed by measuring the contrast of fringes when light from different spatial positions is made to interfere. This observation was a crucial step in establishing the Bose-condensed nature of the polariton system in contrast to the Bose-glass phase described in~\cite{Malpuech:PRL2007}. The temporal dynamics of the onset of long-range coherence was experimentally investigated in~\cite{Nardin:PRL2009}. Remarkably, the problem of the spatial and temporal coherence of the laser emission in disordered systems is one of the key issues of present-day research in random lasers, i.e. devices that exploit disorder to facilitate laser oscillation~\cite{Wiersma:NatPhys2008}.

This physics was further investigated in~\cite{Baas:PRL2008}: when the potential barrier separating the high density spots is too large for tunneling to lock the phase, a synchronized state with a single condensate is replaced by a desynchronized state with multiple independent condensates. A theoretical model of phase locking phenomena in non-equilibrium polariton condensation in multiple-well geometries was put forward in~\cite{Wouters:PRB2008b,Eastham:PRB2008}. 

While all these features have a counterpart in equilibrium BEC in ultracold atomic gases or liquid Helium, novel features stem from the interplay of disorder with the possibility of macroscopic currents in the steady-state wavefunction.  Numerical examples of such patterns for the case of a wide pump spot are presented in~\cite{Wouters:PRB2008}: while the short-distance profile of the current pattern is strongly affected by the disorder, its global shape with a radially outgoing current is robust against disorder.
\begin{figure}[t]
\begin{center}
\includegraphics[width=\columnwidth,angle=0,clip]{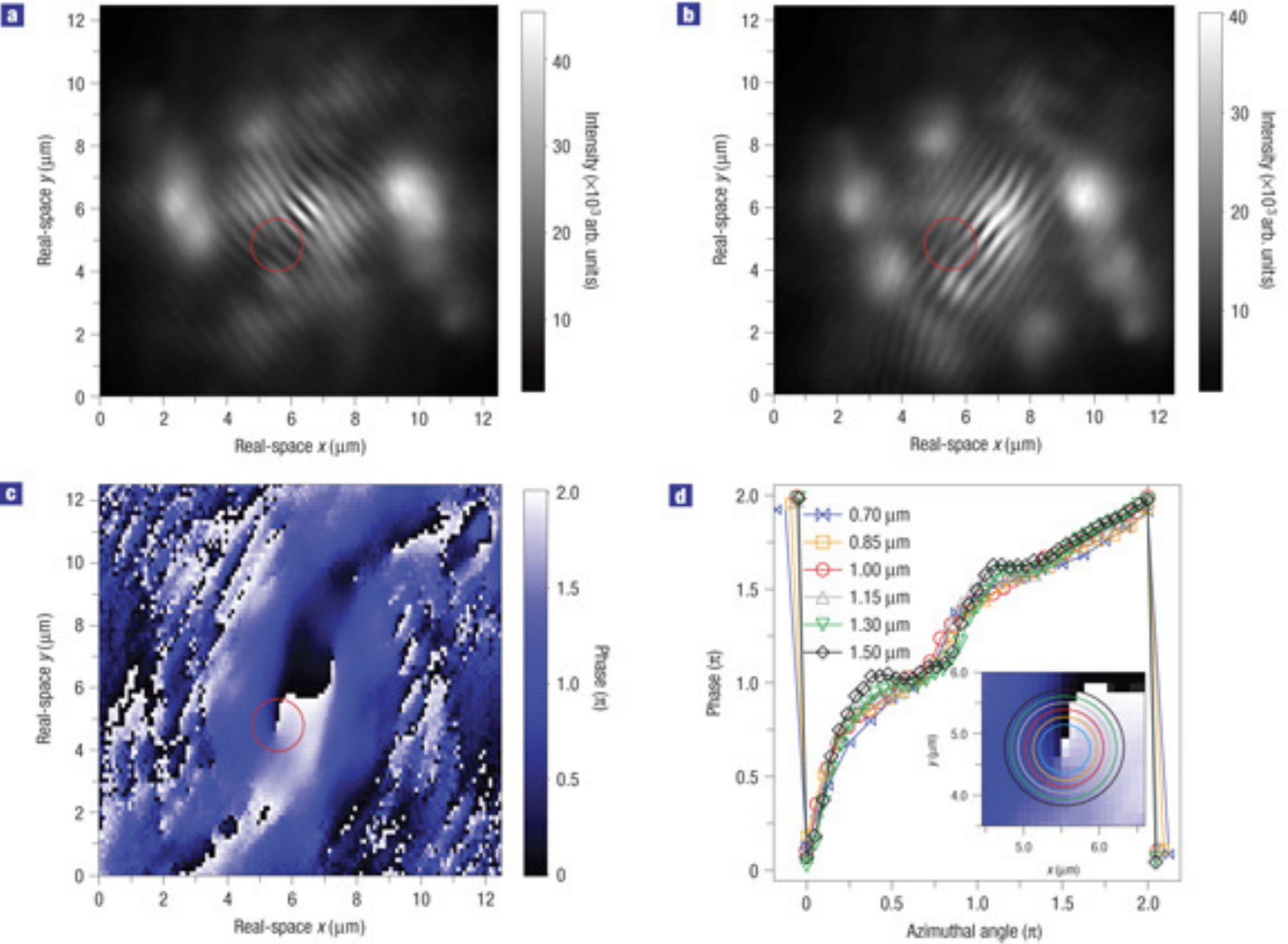}
 \end{center}
\caption{(Color online) 
Figure from~\cite{Lagoudakis:NatPhys2008}. (a) Interferogram with a  vortex: the fork-like dislocation can be seen in the red circle. (b) same interferogram, but this time the vortex is overlapped with a different region of the condensate and a different fringe orientation is selected. The vortex appears to be pinned in the same position. (c) Real-space phase profile calculated from the interferogram in (a). The red circle encloses the vortex (same real-space area as in a,b). (d) Phase as a function of the azimuthal angle for a range of different radii as shown in the inset of (d) (magnification of (c)). }
\label{fig:}
\end{figure}

From a closer look at the real-space experimental images of the non-trivial condensate phase pattern, it was possible~\cite{Lagoudakis:NatPhys2008} to ascertain the presence of quantized vortices in the steady-state condensate. On one hand, the number and the position of the vortices strongly depends on the pumping conditions, e.g. the position of the pump spot with respect to the disorder potential and its intensity. On the other hand, their fixed position for given pumping conditions pinpoints a deterministic mechanism for vortex nucleation and excludes an interpretation in terms of thermal vortices as in the Berezinskii-Kosterlitz-Thouless transition to a non-superfluid state in finite-temperature two-dimensional Bose gases~\cite{Minnhagen:RMP1987}. In the present experiment, vortices spontaneously appear at the interface of spatial regions with macroscopic currents in different directions. With a careful choice of the pump beam profile, vortex-antivortex pairs were experimentally observed~\cite{Roumpos:NatPhys2011}. The spontaneous appearance of half-vortices in spinor condensates via a related mechanism was observed in~\cite{Lagoudakis:Science2009}. The observation of topological defects when several atomic condensates are merged together was reported in~\cite{Scherer:PRL2007}.

A striking self-oscillation effect was observed in the time-resolved data reported in~\cite{Lagoudakis:PRL2010} for a polaritonic Josephson junction. With a careful choice of the position on the sample allowed to identify a spatial region where the spatial dependence of the disorder potential exhibits a sort of double-well configuration and the condensate density is concentrated at the bottom of the two traps. Above a certain value of the pump intensity, time-resolved measurements of the relative population and of the relative phase showed an oscillatory behavior that bears analogies with the ac Josephson effect.

\subsubsection{Condensation in a trap potential}

Differently from the experiments discussed in the previous section where the spatial size of the condensate was mostly determined by the size of the pump spot, the work by \onlinecite{Balili:Science2007} used a mechanical stress to confine polaritons in a harmonic potential as discussed in Sec.\ref{sec:external_pot}. This strategy took inspiration from experiments with atomic condensates and the experimental findings were similar: the condensate appears at the bottom of the harmonic potential, and its size is determined by an interplay of harmonic trapping and repulsive interactions. A similar observation was reported in the photon BEC experiments by~\onlinecite{Klaers:Nature2010}.

Some care has to be paid if one wishes to use the generalized GPE~\eq{CGLE} to model a harmonically trapped polariton condensate under a spatially homogeneous incoherent pumping: as shown in~\cite{Keeling:PRL2008}, a naive application of the theory would in fact predict that the rotationally symmetric steady state is dynamically unstable towards the exponential growth of high angular momentum modes at the surface of the condensate and a new stationary state is eventually reached in the form of a vortex lattice. The discrepancy of this theoretical prediction from the experimental observations can be traced back to the finite pump laser spot used in~\cite{Balili:Science2007}. It is worth mentioning that a regular lattice of vortices was instead observed in~\cite{Scheuer:Science1999} using a VCSEL system electrically driven well above threshold.

\subsubsection{More complex trap potentials}

Investigations of condensation in complex geometries started from the case of a double well potential: early work used a pair of neighboring polariton traps created by disorder. This allowed to observe the transition between independent condensates to a synchronized regime where the condensate phases are locked by tunneling across the separating barrier~\cite{Baas:PRL2008}. Theoretical models accounting for the observed featured appeared in~\cite{Eastham:PRB2008,Wouters:PRB2008b}. More recently, a similar physics was investigated in an artificially fabricated double micropillar system~\cite{Galbiati:arXiv2011}. A time-dependent regime showing spontaneous Josephson oscillations between the two wells were observed in~\cite{Lagoudakis:PRL2010}. Theoretical studies of Josephson physics in coherently driven polariton systems appeared in~\cite{Sarchi:PRB2008} and first experimental investigations were reported in~\cite{Bloch:private2012}.

Polariton condensation into several trapped states of complex wavefunction has been observed in~\cite{Wertz:NatPhys2010} using a combination of the repulsive effect of the pump and of the geometrical end of the polariton wire. Later measurement of the emission coherence in~\cite{Tosi:NatPhys2012} suggests that condensation is occurring into a time-dependent polariton wavepacket that bounces back and forth within the trap potential and gets amplified each time it hits the pumped region.

\subsubsection{Multiply connected geometries}
\label{sec:ring}

An interesting physics can be observed when the geometry of the condensate is multiply connected. In the simplest case of a ring, the pump intensity is maximum along a circle and vanishes at the center. As a result, several metastable steady states exist for the condensate wavefunction, 
\begin{equation}
\Psi_{LP}(\rr)=f(r)\,e^{i L\theta}.
\end{equation}
labeled by the value of the quantized angular momentum $L$~\cite{Leggett:1999RMP,Mueller:PRA1998}. Differently from simply-connected finite-spot geometries where the drift of polaritons outside the pump spot suppresses the effective amplification rate and makes a moving  condensate to be quickly replaced by a condensate at rest, finite super-current states in multiply-connected geometries can live for macroscopically long times: unwinding the condensate phase requires in fact creating a node in the condensate wavefunction, which is a quite unlikely event. The strict conceptual link between metastability of super-currents and superfluidity effects will be the subject of Sec.\ref{sec:superfluid_metastable}.

\subsection{Stability and the elementary excitation spectrum}
\label{elementary}

After having determined the shape of the condensate, the next logical step is to investigate the spectrum of elementary excitations on top of the condensate. For the case of an incoherent pumping, this problem was first attacked in~\cite{Szymanska:PRL2006} using Keldysh diagram techniques applied to a non-equilibrium Bose-Fermi model of condensation. Here we shall follow a different route initiated in~\cite{Wouters:PRL2007} and based on the non-equilibrium GPE \eq{CGLE}, which leads to qualitatively similar predictions in a perhaps more transparent way. The main result of these works is the diffusive behavior of the Bogoliubov dispersion at low $\kk$'s, which recovers the almost simultaneous work~\cite{Wouters:PRB2006,Wouters:PRA2007} for the planar optical parametric oscillator reviewed in Sec.\ref{goldstone}. The fact that the same behavior is found in several microscopically very different models suggests that it must be a generic feature of non-equilibrium phase transitions.

\begin{figure}[t]
\begin{center}
\includegraphics[width=0.8\columnwidth,angle=0,clip]{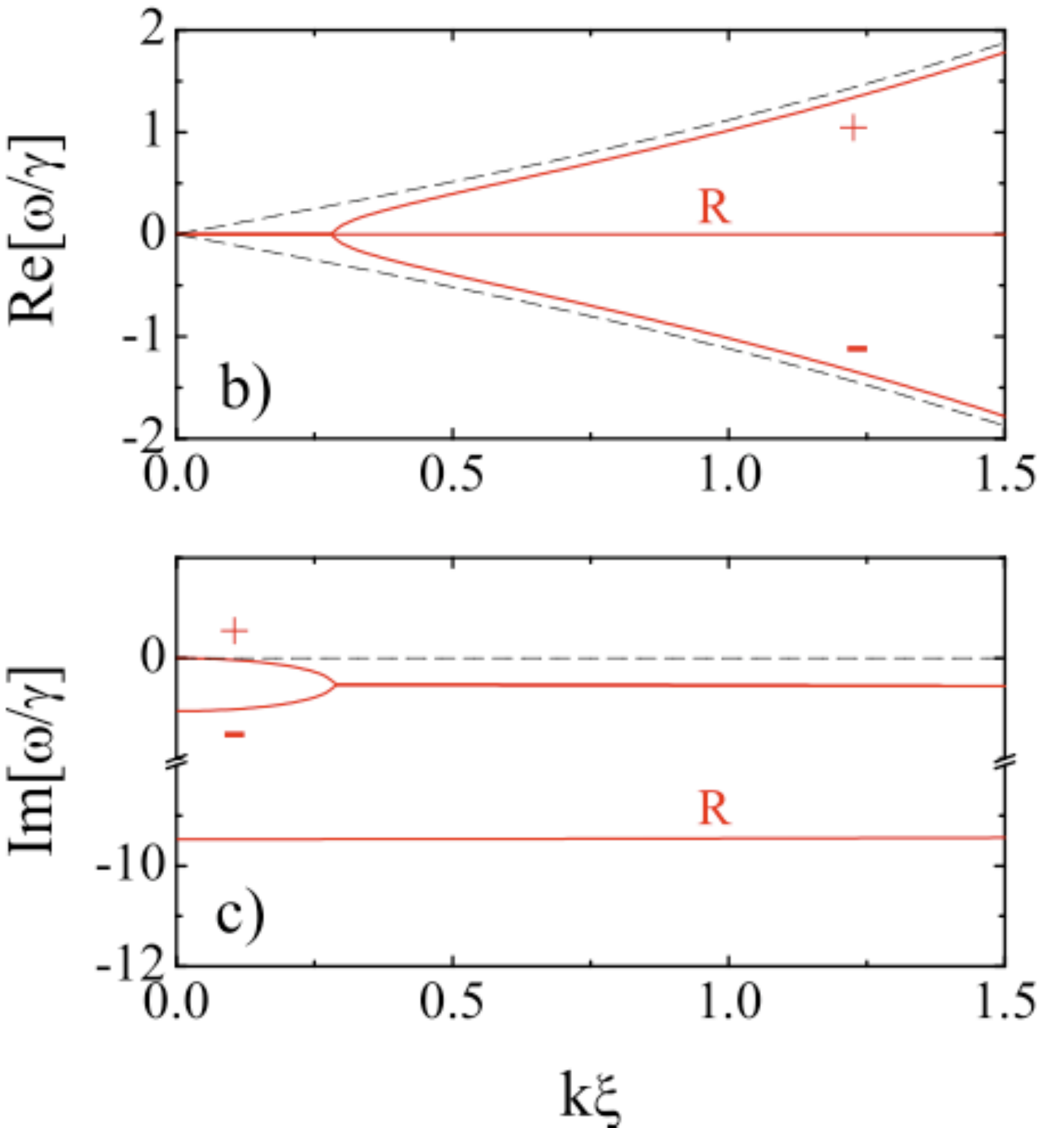}
 \end{center}
\caption{(Color online) Dispersion of the Bogoliubov modes of an incoherently pumped condensate. The $\pm$ branches show the prediction \eq{omega_bog_nonres} for the equilibrium Bogoliubov modes. The $R$ branch indicates the fast relaxation of the reservoir at rate $\gamma_R$ according to \eq{n_R}. The diffusive behavior of $\omega_{\rm Bog}(k)$ is clearly visible at low $\kk$'s.
Figure adapted from~\cite{Wouters:PRL2007}.}
\label{fig:Bogo_incoh}
\end{figure}

In the simplest case of a spatially homogeneous condensate at rest $\Psi_{LP}(\rr,t)=\sqrt{n_{LP}}\,e^{-i\omega t}$, the Bogoliubov equations describing the dynamics of small fluctuations have the form
\begin{widetext}
\begin{equation}
i\partial_t \left(
\begin{array}{c}
\delta\Psi_{LP}(\rr) \\ \delta\Psi^*_{LP}(\rr)
\end{array} \right) =
\left(
\begin{array}{cc}
\omega_{LP}^o-\frac{\hbar \nabla^2}{2m_{LP}} + 2 g_{LP} n_{LP} - \frac{i \Gamma}{2} - \omega & g_{LP}n_{LP} - i \Gamma/2   \\
- gn  - i \Gamma/2  & -\omega_{LP}^o+\frac{\hbar \nabla^2}{2m_{LP}} - 2 g_{LP}n_{LP} - \frac{i \Gamma}{2}  + \omega 
\end{array} 
\right)
\left(
\begin{array}{c}
\delta\Psi_{LP}(\rr) \\ \delta\Psi^*_{LP}(\rr)
\end{array} \right).
\eqname{Bogol_LP_incoh}
\end{equation}
\end{widetext}
with an effective damping rate
\begin{equation}
\Gamma=\frac{P R n}{(\gamma_R + R n)^2}={\gamma}\,\frac{P/P_c - 1}{ P/P_c}.
\end{equation}
Within the usual parabolic band approximation, the eigenmodes of \eq{Bogol_LP_incoh} give the dispersion
\begin{equation}
\omega_{\rm Bog}(\kk)=-\frac{i\Gamma}{2}\pm\sqrt{\frac{\hbar k^2}{2m_{LP}}\left(\frac{\hbar k^2}{2m_{LP}} + 2 g_{LP} n\right)-\frac{\Gamma^2}{4}}
\eqname{omega_bog_nonres}
\end{equation}
for the elementary excitations. An example of such a dispersion is plotted in Fig.\ref{fig:Bogo_incoh}: two regimes are clearly identified depending whether the argument of the square root in \eq{omega_bog_nonres} is positive or negative. The transition wavevector $k_o$ corresponds to the point where the equilibrium Bogoliubov dispersion crosses the effective damping $\Gamma$: as expected, for smaller loss rates $\gamma$, the effective decay rate $\Gamma$ decreases and the system behavior recovers the equilibrium case.

For low values of the wavevector $k<k_o$, the argument of the square root is negative and the real parts of the two branches stick to the common value $\textrm{Re}[\omega_{\rm Bog}(\kk)]=0$, while the imaginary parts are split. Stability of the condensate is guaranteed by the fact that for all $\kk$'s one has $\textrm{Im}[\omega_{\rm Bog}(\kk)]\leq 0$. 

As $k\rightarrow 0$, one of the two branches tends to a finite, purely imaginary value $-2i\Gamma$: from an explicit calculation of the eigenvector, one sees that it corresponds to a spatially homogeneous fluctuation of the condensate density and does not affect the condensate phase. When approaching the critical point from above $P\rightarrow P_c^+$, its decay rate $\Gamma$ tends to zero. In the laser literature~\cite{MandelWolf}, such a mode is involved in the intensity dynamics of the device at switch-on and determines the temporal correlation function of intensity fluctuations~\footnote{In standard lasers, the decay rate $\Gamma$ of intensity fluctuations remain finite even at threshold because of the relatively large fluctuations in a single mode~\cite{MandelWolf}. Even though such a prediction $\Gamma(P=P_c)=0$ was obtained within mean-field theory, this result is expected to be robust against fluctuations provided a spatially extended system is considered. The linearized fluctuation theory of~\cite{Szymanska:PRL2006} appears to confirm this expectation.}. Of course, a more complex dynamics than simple relaxation is observed as soon as the gain medium shows a significant frequency dependence of the amplification rate.

A similar {\em amplitude mode} is encountered in many contexts of condensed-matter physics when a continuous symmetry is spontaneously broken at a thermal or quantum phase transition. The simplest case is the superfluid to Mott-insulator transition for bosons in a lattice. Close to the critical point on the superfluid side, a second branch of excitations appears in addition to the Bogoliubov sound. Physically, this new Higgs-like mode can be understood in terms of an oscillating amplitude of the superfluid order parameter, that is a local interconversion of the condensed and non-condensed fractions~\cite{Altman:PRL2002,Huber:PRB2007,Menotti:PRB2008}. A summary of physical systems where this physics may be observed (atomic gases, antiferromagnets, charge-density-wave systems, superconductors) is reported in~\cite{Podolsky:PRB2011}: very recently, experimental evidence of the amplitude mode was reported for ultracold atoms in an optical lattice~\cite{Bissbort:PRL2011}. A related physics was investigated for a Dicke model of polariton condensation in~\cite{Brierley:PRL2011}.

The other mode which tends to $0$ in both real and imaginary parts as $k\to 0$ is the Goldstone mode corresponding to the $U(1)$ spontaneous symmetry breaking associated to the BEC transition: given the underlying $U(1)$ phase rotation symmetry of the problem, the phase of the condensate is randomly chosen at the phase transition and a global rotation of the phase can not experience any restoring force. This interpretation is confirmed by a calculation of the corresponding $\kk=0$ eigenmode of \eq{Bogol_LP_incoh} that indeed describes a pure global phase rotation. 
{Most remarkably, the so-called Schawlow-Townes linewidth of a standard laser~\cite{QuantumOptics,Schawlow:PR1958} can be seen as arising from the excitation of this soft mode by spontaneous emission events.}

For finite $0<k<k_o$, the amplitude and phase modes are mixed by the kinetic energy term in \eq{Bogol_LP_incoh} but maintain a diffusive nature with a decay rate tending to zero in the long wavelength limit. A perturbation imprinted at the initial time $t=0$ on the condensate phase $\varphi(\rr)$ will not propagate through the system, rather it will diffusively expand in space according to a heat equation
\begin{equation}
\frac{\partial \varphi(\rr,t)}{\partial t}= \frac{c_s^2}{\Gamma}\,\nabla^2 \varphi(\rr,t),
\eqname{heat}
\end{equation}
$c_s=\sqrt{g_{LP} n_{LP} /m}$ being defined as the usual sound velocity at equilibrium.

For $k\gtrsim k_o$, the imaginary parts collapse to a single value $-\Gamma/2$, while the real part starts from zero proportionally to $\sqrt{k-k_o}$. At larger values of $k$, the dispersion recovers the standard Bogoliubov dispersion of equilibrium systems (black dashed line in Fig.\ref{fig:Bogo_incoh}). As usual in Bogoliubov theory, the off-diagonal terms in the matrix in \eq{Bogol_LP_incoh} are responsible for a mixing of the $\delta\Psi_{LP}(\rr)$ and $\delta\Psi_{LP}^*(\rr)$ field, i.e. particle and holes operators in a quantum framework. As a result, both the positive and the negative branches should be observable in luminescence experiment~\cite{Keeling:PRL2005,Marchetti:PRB2007,Byrnes:arXiv2011}. A related prediction was put forward for atom laser beams extracted from atomic condensates in~\cite{Japha:PRL1999}.

Even though such negative Bogoliubov branches (also called ghost branches) have been observed with polariton systems in several {other pumping regimes}~\cite{Savvidis:PRB2001,Kohnle:PRL2011}, no trace of it was visible in the recent experimental study of the Bogoliubov dispersion in an incoherently pumped condensate in a planar geometry~\cite{Utsonomiya:NatPhys2008}: Most likely, this is due to the fact that the emission from this negative branch is easily masked by the much stronger background of the condensate emission. On the other hand, a spectral feature that may be attributed to the negative branch was observed in a pillar geometry by~\onlinecite{Lagoudakis:private}.

Extension of this theory to the supercurrent states in multiply-connected geometries described in Sec.\ref{sec:ring} requires more attention. A naive application of \eq{Bogol_LP_incoh} to a moving condensate at speed $\vv_0=\hbar \kk_0/m$ would predict (within the parabolic approximation for the lower polariton branch) a mere Doppler shift of the real part of the Bogoliubov dispersion,
\begin{equation}
\omega^{(\vv_0)}_{\rm Bog}(\kk)=\omega_{\rm Bog}(\kk)+\kk\cdot \vv_0,
\end{equation}
while the imaginary part of the dispersion remains completely unaffected.
This prediction that moving condensates in ring geometries are dynamically stable up to arbitrarily high values of the flow speed $v_0$ is at odd with intuitive expectations that condensation into fast moving states should be unfavored. A more refined treatment including an energy-dependent amplification mechanism was developed in~\cite{Wouters:PRL2010}: as expected, it turns out that the region of stability extends up to a maximum speed value $v_0<v_{\rm stab}$. For higher speeds $v_0>v_{\rm stab}$, the instability of the condensate is due to the birth of another condensate around $\kk=0$ which eventually replaces the one at $\kk_0$.

\section{Superfluid hydrodynamics of the photon fluid}
\label{hydro}

Superfluid behaviors were first observed in 1938 almost simultaneously by Allen and Miesner in Cambridge~\cite{Allen:Nature1938} and Kapitsa in Moscow~\cite{Kapitsa:Nature1938} in liquid Helium-4 cooled at temperatures below the so-called $\Lambda$ point at $T_\Lambda=2.17$~K. The first experimental signature was a sudden drop of mechanical viscosity when the Helium fluid was flowing along a narrow channel.
Since then, a number of other fascinating features of superfluid Helium-4 have been unveiled~\cite{Leggett:1999RMP,Pitaevskii:1992JLT}, from {\em metastability of supercurrents} (with interesting applications as a gyroscope for rotation sensing~\cite{Sato:RPP2012}), to novel excitation branches with out-of-phase oscillations of the normal and superfluid components (the so-called {\em second sound}~\cite{Peshkov:JETP1944,Peshkov:JETP1946}), to the reduced moment of inertia \cite{Andronikashvili:RMP1966} and the Hess-Fairbank effects~\cite{Hess:PRL1967} in rotating bucket experiments, to the {\em fountain effect}~\cite{Allen:Nature1938b}, to the frictionless motion of impurities through the fluid when the speed is below some critical value~\cite{Rayfield:PRL1966,Phillips:PRL1974}. 
Even though our conceptual understanding of superfluidity in liquid Helium is now based on London's intuition that superfluidity originates from Bose-Einstein condensation of the constituent quantum particles and on Tisza and Landau's two fluid hydrodynamics, the experimental evidence of the presence of a Bose-Einstein condensate in superfluid Helium is still quite elusive and based on indirect neutron scattering data~\cite{Sosnick:PRB1990,Sokol:1995,Azuah:PRB1997,Glyde:PRB2000,Glyde:PRB2011}.

The interest in the physics of superfluids was suddently revived in 1995 when Cornell and Wieman at JILA~\cite{Anderson:Science1995} and soon after Ketterle at MIT~\cite{Davis:PRL1995} observed Bose-Einstein condensation in a gas of laser- and then evaporatively cooled atoms stored in a magnetic trap at nanokelvin temperatures. In contrast to liquid Helium where the condensate fraction is limited to around 10\%, atomic gases have a low enough density to be in the dilute regime where the condensate fraction is large and clearly visible in the experimental data.

In the immediate wake of the first evidence of condensation, researchers started investigating the superfluid properties of the ultra-cold atomic clouds. Thanks to the dilute nature of the atomic gas and the flexibility of atomic manipulation and diagnostic techniques, a direct comparison with microscopic calculation was possible, which allowed to get a deeper insight into the basic mechanisms underlying, e.g. the critical velocity for frictionless motion of weak impurities across the cloud~\cite{Raman:PRL1999}, the nucleation of quantized vortices at the surface of a large defects~\cite{Neely:PRL2010}, the effect of a reduced moment of inertia onto the collective scissor mode of an anisotropic cloud~\cite{Marago:PRL2000}, the metastability of supercurrents in multiply-connected geometries~\cite{Ryu:PRL2007}. In spite of these exciting advances, many subtle features of superfluidity of ultracold atomic clouds are still waiting for experimental confirmation, in particular for what concerns the conceptual distinction between the superfluid and the condensed fractions, in particular in low-dimensional geometries~\cite{Carusotto:Physics2010}.

In this section we shall give an overview of the recent theoretical and experimental developments in the superfluid hydrodynamics of light. Along the same lines of classical studies of superfluidity in condensed-matter systems~\cite{Leggett:1999RMP}, different aspects of superfluid behavior will be addressed: in contrast to standard equilibrium systems, the different aspects of superfluidity can appear separately in driven-dissipative systems, which gives rise to a richer variety of effects. 

\subsection{The Landau criterion}

A most celebrated textbook presentation of superfluidity~\cite{BECbook} is based on the so-called Landau criterion for superfluidity, which determines the maximum speed at which a weak impurity can travel across a fluid without experiencing any friction force. This criterion was originally proposed by Landau to explain superfluidity in liquid Helium and led to the prediction of the roton minimum in the elementary excitation spectrum of the fluid. 

In terms of the dispersion $\omega(k)$ of the elementary excitations in a generic fluid at rest, the Landau critical velocity has the form
\begin{equation}
v_c=\min_k\left[\frac{\omega(k)}{k}\right].
\eqname{landau_critical}
\end{equation}
In its simplest formulation, the idea underlying the Landau criterion can be summarized as follows. We consider a fluid uniformly moving at speed $\vv$: as seen from the laboratory reference frame, the elementary excitation spectrum in the fluid has the Galilean-transformed form
\begin{equation}
\omega'(\kk)=\omega(\kk)+\kk\cdot\vv.
\end{equation}
For slow flow speeds $v<v_c$, a finite positive amount of free energy is always required to create excitations in the fluid at any value of the momentum $\kk$.
For faster flows $v>v_c$, there exist values of $\kk$ for which $\omega'(\kk)<0$, so it is thermodynamically favorable to create a large number of excitations in the moving fluid. As a result, superfluidity is broken and an effective friction force appears. Of course, the validity of the Landau criterion is restricted to impurity  potentials (due, e.g., to the roughness of the container walls) that are weak enough to modify the fluid density only in a perturbative way. As we shall see in the following of this section, non-perturbative processes in the presence of stronger impurities can create bound complexes of many elementary excitations (e.g. vortices) even at $v<v_c$; experimentally, this appears to be the case in most experiments with macroscopic Helium samples.

\subsubsection{Bogoliubov formulation of the Landau criterion for weak impurities: the equilibrium case}
\label{sec:BogoLandau}

Landau's thermodynamical argument can be made complete and quantitative in the framework of the Bogoliubov theory of dilute gases in the limiting case of a weak impurity at rest in a uniformly moving homogeneous condensate~\cite{Astrakharchik:PRA2004,Carusotto:PRL2006}. A unified discussion of the wave emission by uniformly moving sources in the different contexts of superfluid hydrodynamics, \u Cerenkov effect in classical electromagnetism, and classical hydrodynamics of surface waves can be found in~\cite{Carusotto:arXiv2012}.

Introducing the compact notation $\delta{\vec \phi}(\rr,t)=(\delta\phi(\rr,t),\delta\phi^*(\rr,t) )^T$, the weak perturbation created by the impurity in the moving condensate at speed $\vv=\hbar \kk_0 /m$ can be straightforwardly obtained within Bogoliubov theory by the linear equation
\begin{equation}
  \label{eq:Bogo}
  i\hbar\frac{\partial}{\partial t}\delta{\vec \psi}={\mathcal L}_\vv \cdot \delta{\vec
  \psi}+{\vec F}_d,
\end{equation}
where the source ${\vec F}_d$ is proportional to the defect potential $V_d$ and the unperturbed condensate wavefunction $\phi_o(\rr)=\sqrt{n}\,\exp(i\kk_0\cdot \rr)$,
\begin{equation}
  \label{eq:F_d}
{\vec F}_d(\xx)=
V_d(\xx)\,\left(
\begin{array}{c}
\phi_o(\xx) \\
-\phi_o^*(\xx)
\end{array}
\right).
\end{equation}
The Bogoliubov operator ${\mathcal L}_{\vv}$ for the moving fluid has the form:
\begin{equation}
  \label{eq:BogoL}
{\mathcal L}_\vv=
\left(
\begin{array}{cc}
-\frac{\hbar^2}{2m}\nabla^2+ g n & g n \,e^{2i\kk_0 \xx} \\
-g n\,e^{-2i\kk_0 \xx} & +\frac{\hbar^2}{2m}\nabla^2- g n
\end{array}
\right),
\end{equation}
where the condensate motion is clearly visible as a phase factor in the off-diagonal matrix elements. 
As usual, the dispersion of the Bogoliubov modes in the moving fluid is obtained from the eigenvalues of $\mathcal{L}_\vv$,
\begin{multline}
  \label{eq:BogoEnerg}
  \omega_{\rm Bog}^{(\vv)}(\kk)=\vv\cdot(\kk-\kk_0) \\
  \pm
\sqrt{\frac{\hbar(\kk-\kk_0)^2}{2m}
\Big(\frac{\hbar(\kk-\kk_0)^2}{2m}+2g\rho_0\Big)}:
\end{multline}
the finite velocity of the condensate is apparent in the Doppler shift term $(\kk-\kk_0)\cdot \vv$.

\begin{figure}[t]
    \begin{center}
\includegraphics[width=0.95\columnwidth,clip]{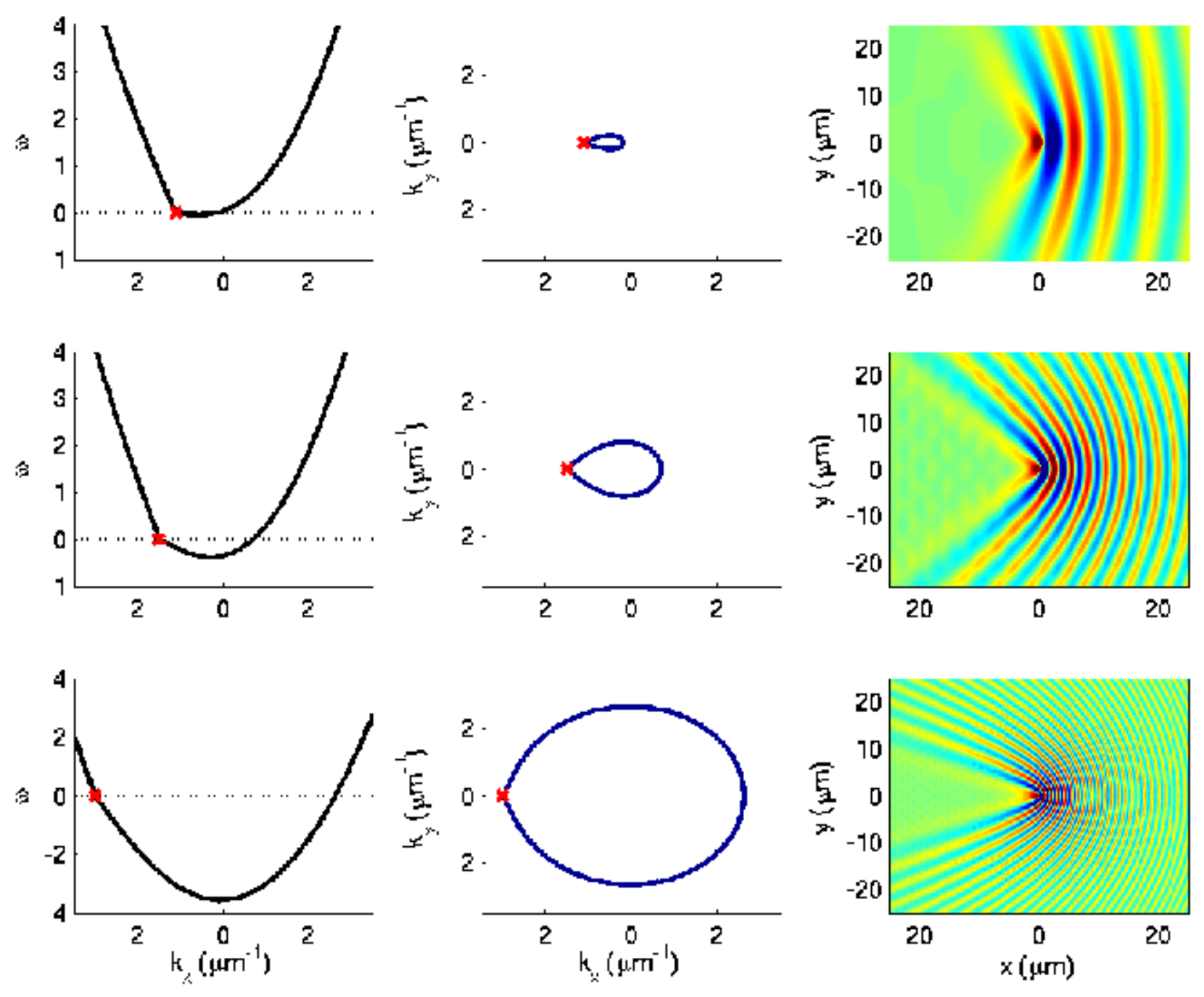}
      \caption{Dispersion of Bogoliubov modes in the flowing condensate (left column), $\kk$-space  locus $\Sigma$ of the 
  resonantly excited Bogoliubov modes (center  column) and spatial profile of the density wake around an impurity at rest (right column).
 The different rows correspond to different values of the (supersonic) flow speed $v_0/c=1.1, 1.5, 3$. The red crosses in the left and central columns indicate the condensate wavevector $\kk_0$. Figure adapted from~\cite{Carusotto:PRL2006}.
      \label{fig:Supersonic}}
    \end{center}
  \end{figure}

The steady state in the presence of the defect potential $V_d$ is then obtained from the motion equation \eq{Bogo} as:
\begin{equation}
  \label{eq:stationary}
\delta{\vec \psi}_d=-\big({\mathcal L}_\vv-i\,0^+\,{\mathbf 1}\big)^{-1}
\cdot {\vec F}_d,
\end{equation}
where the infinitesimal imaginary term is required to ensure causality of the solution. 
As it can be seen from the resonant denominator, the time-independent defect potential is able to effectively excite only the $\kk\neq \kk_0$ Bogoliubov modes whose energy is $\omega_{\rm Bog}^{(\vv)}(\kk)=0$. The shape of the $\kk$-space locus $\Sigma$ of such modes is shown in the central column of Fig.\ref{fig:Supersonic} for a few values of the flow speed. According to the Landau criterion \eq{landau_critical}, two main regimes can be identified depending on whether this is slower or faster than the speed of sound $c_s$ in the BEC.

In the subsonic regime $v_0<c_s$, the intersection locus $\Sigma$ is empty. As a consequence, no Bogoliubov mode can be resonantly excited by the defect and the fluid is able to flow in a frictionless way around the defect without suffering any dissipation. Nevertheless, the non-resonant excitation of the Bogoliubov modes by the impurity is responsible for a sizable density modulation in the vicinity of the impurity, that quickly decays to zero in space with an exponential law and which results in a sizable renormalization of the mass of the impurity~\cite{Astrakharchik:PRA2004}. Of course, all these conclusions are based on a mean-field description of the condensate that neglects quantum fluctuations: more sophisticated Bethe ansatz calculations for a strongly interacting one-dimensional Bose gas~\cite{Astrakharchik:PRA2004} have anticipated the appearance of a finite drag force also at sub-sonic speed. A calculation including higher order terms of the Bogoliubov theory led the authors of~\cite{Roberts:PRL2005} to a similar claim for a three-dimensional condensate; this conclusion however appears at odd with classical literature on liquid Helium~\cite{Iordanskii:JETP1978}.

In the supersonic regime $v_0>c$, the intersection locus $\Sigma$ is not empty and consists of the closed curve $\Sigma$ shown in the central column of Fig.\ref{fig:Supersonic}. Resonant excitation of these Bogoliubov modes makes the kinetic energy of the flow to be partially dissipated into Bogoliubov phonons and is associated to a sizable drag force exerted by the fluid onto the defect. Some characteristic features can be identified in the corresponding density modulation pattern:

(i) The emission of low-wavevector Bogoliubov modes in the sonic region is responsible for a Mach cone located downstream of the defect with the usual aperture $\theta$ such that $\sin \theta=c_s/v_0$. This emission mechanism is closely analogous to the \u Cerenkov emission of electromagnetic waves by charged particles in super-luminal motion across a dielectric, $v>v_{\rm ph}=c_0/n$~\cite{Jelley}.

(ii) A series of curved precursors located upstream and laterally to the defect that result from the single-particle interference between the incident plane wave and the spherical scattered wave. An analytical calculation of their shape is given in~\cite{Gladush:PRA2007}; a simplified approximate model is discussed in~\cite{Carusotto:arXiv2012}; their one-dimensional analog was studied in~\cite{Leboeuf:PRA2001,Pavloff:PRA2002}.

\subsubsection{Experiments with liquid Helium and atomic condensates}

 \begin{figure}[t]
    \begin{center}
\includegraphics[width=0.46\columnwidth]{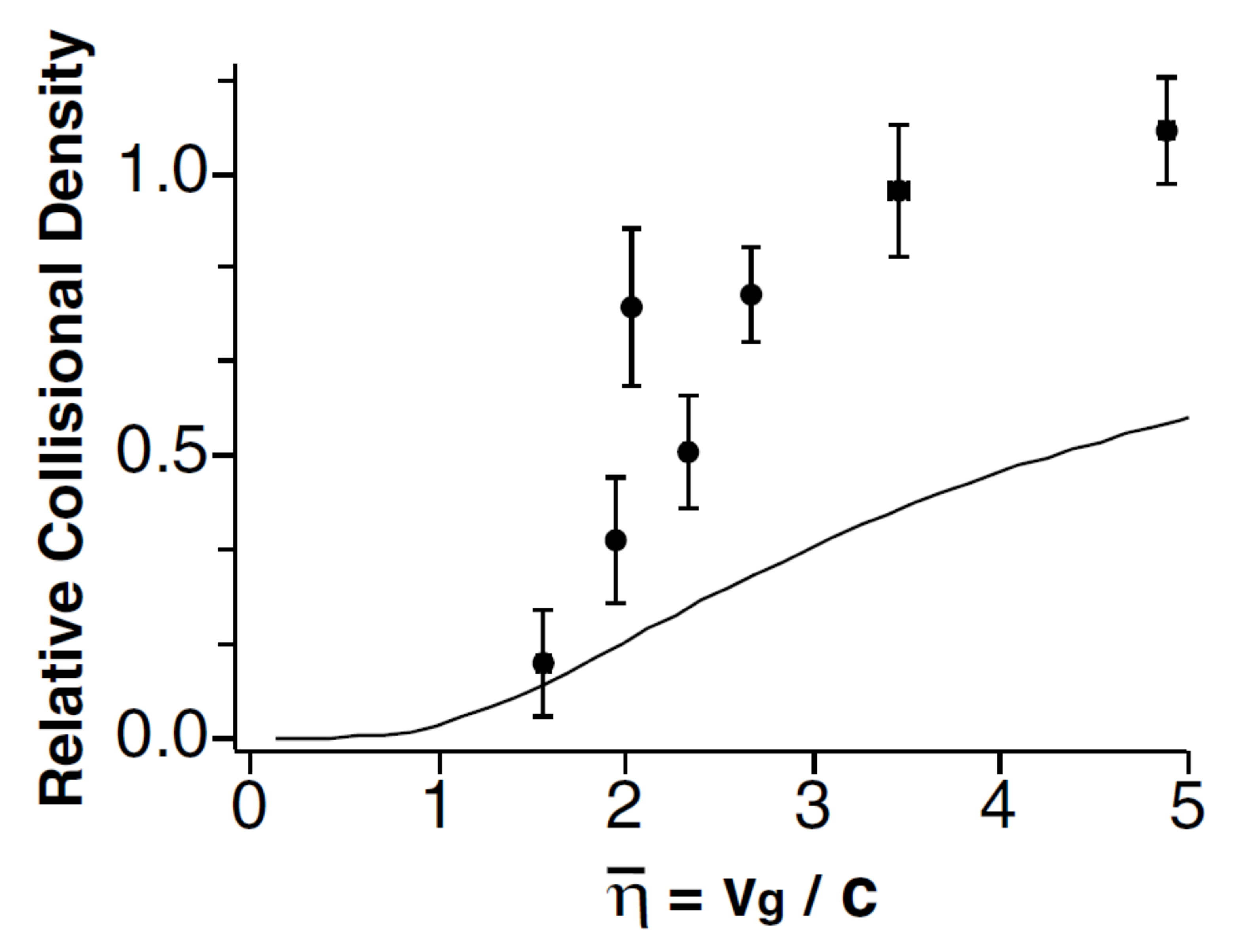}
\includegraphics[width=0.27\columnwidth,clip]{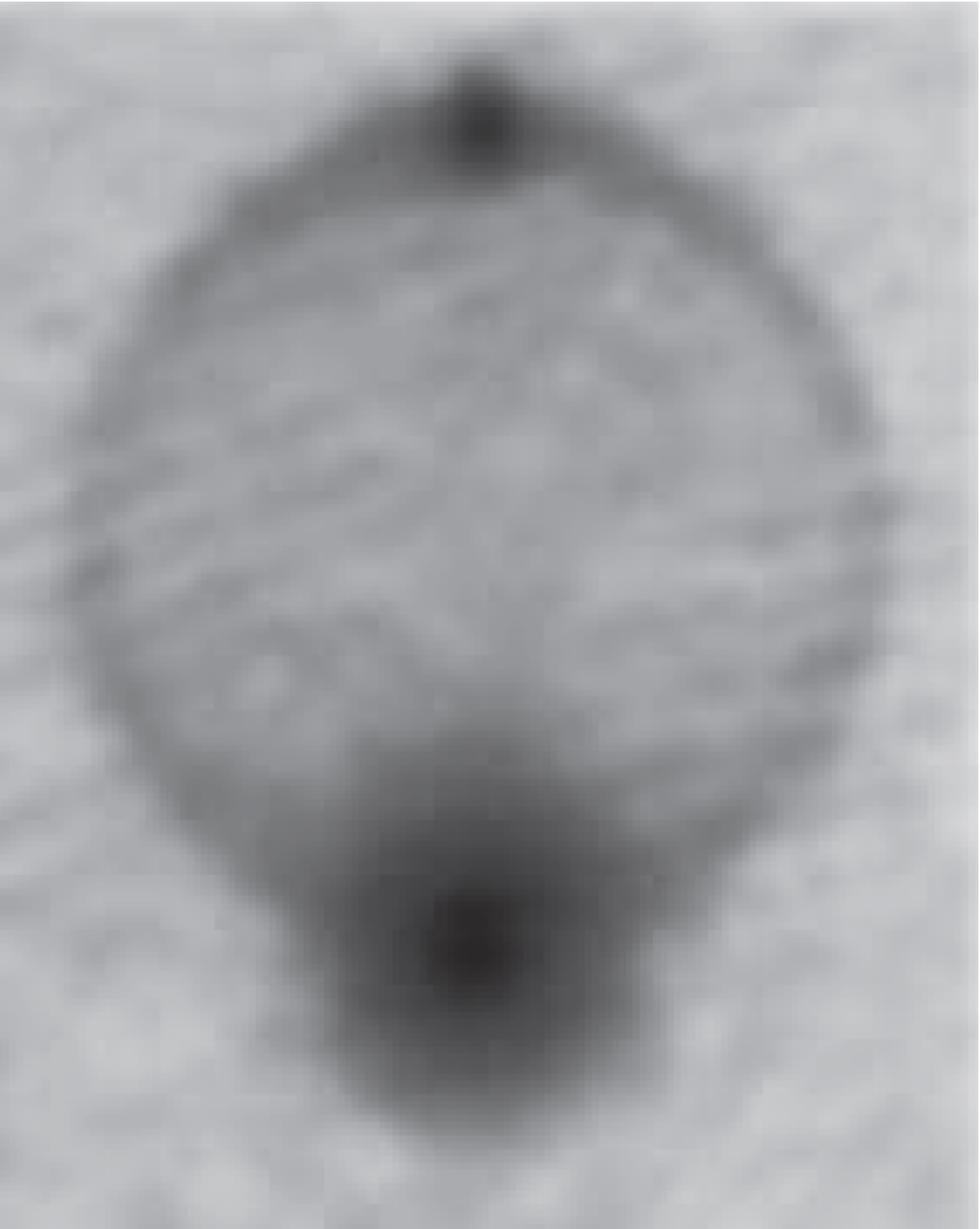}
\includegraphics[width=0.20\columnwidth,clip]{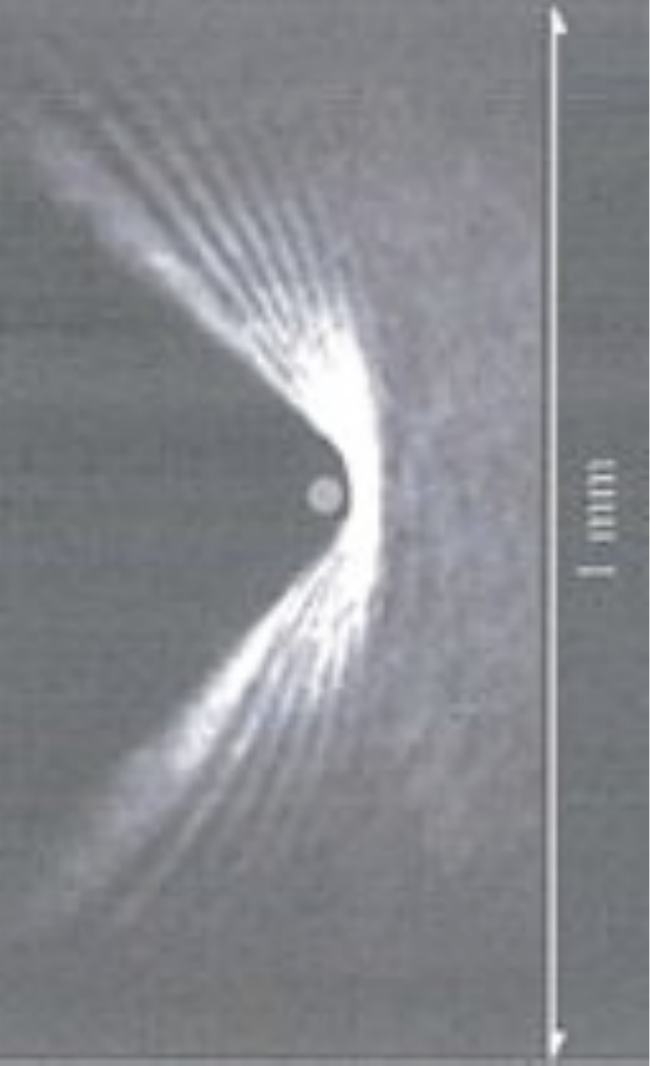} \\
\includegraphics[width=0.8\columnwidth,clip]{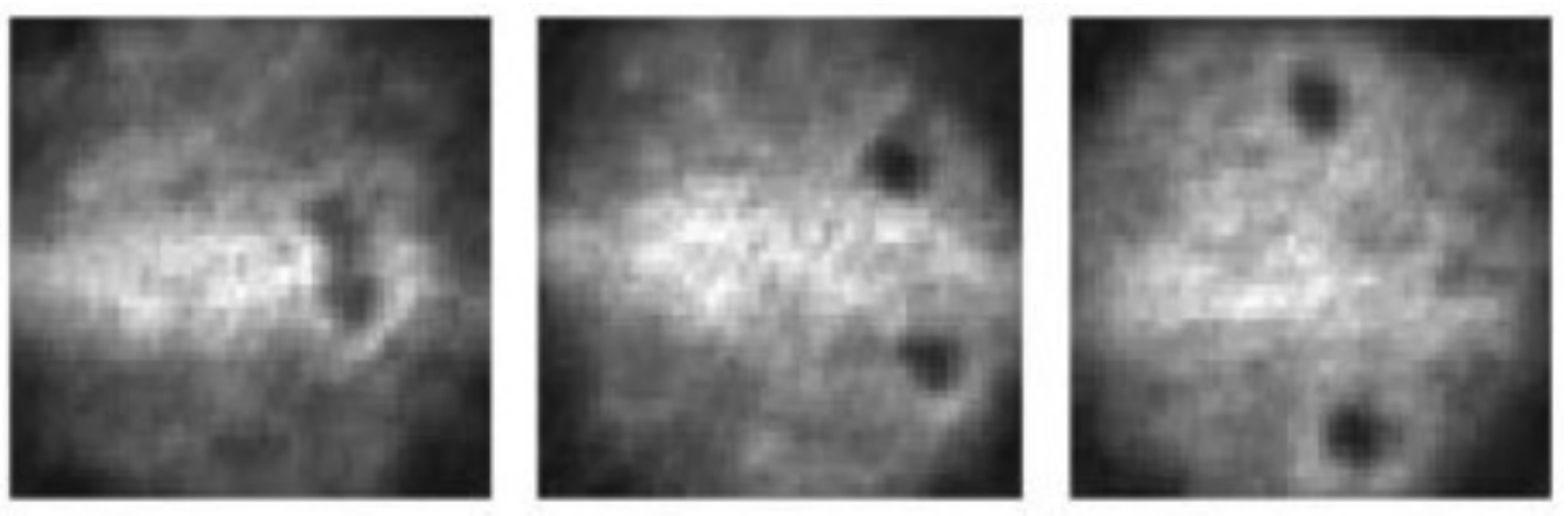} 
      \caption{Upper left and middle panels: experimental studies of weak impurities travelling across dilute atomic condensates. Left: relative density of scattered atoms as a function of the impurity speed $v_0/c_s$, showing a marked threshold at $v_0=c_s$. Figure from~\onlinecite{Chikkatur:PRL2000}. Middle: atomic momentum distribution after elastic scattering between the condensate (lower black dot) and super-sonically moving impurity atoms (upper spot); figure from~\cite{Ketterle:Lecture2000}.
Upper right panel: real-space density profile of a BEC hitting an obstacle at supersonic velocity $v_0/c_s= 13$ in the left direction. Figure courtesy of P. Engels and E. A. Cornell~\onlinecite{Cornell:UCSB}. Lower panels: sequences of images showing the dynamics of vortices nucleated by the repulsive potential of a blue-detuned laser scanned across a pancake-shaped condensate in the rightwards direction. Figure taken from~\onlinecite{Neely:PRL2010}.
}
 \label{fig:superfl_exp_atoms}
    \end{center}
  \end{figure}

As originally anticipated by Landau, the critical velocity \eq{landau_critical} is determined in liquid Helium by the roton branch of excitations. 
Experimental studies of the drag force as a function of the velocity were reported in~\cite{Rayfield:PRL1966,Phillips:PRL1974} using ions as microscopic impurities traveling across a macroscopic Helium sample. Interpretation of the experimental data was made difficult by the number of competing phenomena that may be simultaneously taking place, e.g. generation of charged vortex rings.

In  recent years, a much more detailed experimental access to the microscopic physics underlying the Landau criterion has been provided by experiments with ultracold atomic clouds. Some remarkable results are illustrated in Fig.\ref{fig:superfl_exp_atoms}. 
In the upper-left and middle ones, the impurity consists of atoms of the same species crossing the condensate cloud at finite speed after a Bragg scattering process. The onset of drag as soon as the speed of the impurity exceeds the speed of sound is clearly visible in the right panel. The ring-shaped momentum distribution of scattered atoms shown in the middle panel provides information on the Bogoliubov modes that are excited by the moving impurity. 
The right panel shows the real-space wake created in an expanding condensate by the strong repulsive potential of a blue detuned laser beam: while the \u Cerenkov cone in the downstream region is strongly modified by a trivial shadow effect, the curved precursors upstream of the defect closely resemble theoretical prediction in Fig.\ref{fig:Supersonic}. Experimental studies of this physics in polariton fluids were reported in Sec.\cite{Amo:NPhys2009} and are reviewed in Sec.\ref{sec:polar_SF_exp} and Figs.\ref{fig:Amo_superfl} and \ref{fig:Amo_cer}.

\subsubsection{Beyond the Landau criterion: strong defects}

In most experiments involving macroscopic bodies moving in liquid Helium or strong potentials in atomic condensates, the critical velocity for the onset of drag turns out to be much lower than the prediction of the Landau criterion \eq{landau_critical}. A possible explanation for this behavior was proposed in~\cite{Frisch:PRL1992} in terms of the strong modification of the velocity pattern around a large and strong defect: as a result of the spatial compression of the stream lines, the local velocity on the surface of the defect can largely exceed the flow speed at infinity. As soon as the critical velocity \eq{landau_critical} is locally exceeded, quantized vortices start being nucleated at the surface of the object and give rise to a drag force. 
According to the first calculations in~\cite{Frisch:PRL1992}, this happens when the asymptotic speed far from the defect exceeds a critical value of $\sqrt{2/11}=0.43$ times the speed of sound. Subsequent, more refined calculations in~\cite{Berloff:JPhysA2000,Berloff:JPhysA2001,Rica:PhysicaD2001} predict a slightly different value $0.37$ for the critical Mach number, in agreement with numerical calculations. Further numerical work for penetrable objects was reported in~\cite{Winiecki:JPHYSB2000}. This dissipation mechanism based on vortex nucleation was experimentally confirmed in~\cite{Neely:PRL2010} using a strong blue-detuned beam crossing a pancake-shaped atomic condensate: examples of experimental snapshots of the density pattern are reproduced in the bottom panel of Fig.\ref{fig:superfl_exp_atoms}. Experimental investigations of this physics with polariton condensates were reported in~\cite{Nardin:2011NatPhys,Sanvitto:NPhot2011} and are reviewed in Sec.\ref{vortices} and Fig.\ref{fig:nardin_vort}.

As originally predicted in~\cite{El:PRL2006}, at flow speeds larger than the speed of sound, vortex nucleation is replaced by the appearance of one or several pairs of oblique solitons  in the wake of the large defect. Stabilization of the dark soliton against the snake instability and the subsequent decay into a string of vortices  is prevented by the large flow speed as explained in~\cite{Kamchatnov:PRL2008,Kamchatnov:ARXIV2011}. While no evidence of this physics has yet been obtained yet in liquid Helium nor in atomic clouds, experimental studies of hydrodynamic generation of oblique dark solitons in polariton fluids were reported in~\cite{Amo:2011Science} and soon after in~\cite{Grosso:PRL2011}. This physics will be reviewed in Sec.\ref{vortices} and illustrated in Fig.\ref{fig:Amo_solit}.

\subsection{Coherently pumped polariton condensate flowing against a weak defect: superfluidity effects}
\label{sec:coh_superfl}
The first experiments unambiguously showing superfluidity effects in luminous systems~\cite{Amo:NPhys2009} were performed under a coherent pump with a quasi-plane-wave spatial profile, along the lines of the proposal in~\cite{Carusotto:PRL2004}. In this subsection we shall review the theoretical concepts underlying the experiment and then we shall illustrate the experimental data.

\subsubsection{Theory}

The value of the in-plane pump wavevector $\kk_{\rm inc}$ and of the pump frequency $\omega_{\rm inc}$ are chosen in the parabolic region at the bottom of the lower polariton branch, so to avoid the parametric instabilities discussed in Sec.\ref{sec:secOPO} and have, in the absence of defects, a spatially uniform polariton condensate steadily flowing along the microcavity plane. However, as we have discussed in Sec.\ref{sec:coh_lim_bist}, the non-equilibrium nature of the polariton system eliminates the one-to-one relation between the field oscillation frequency $\omega_{\rm inc}$ and the density $n_{LP}$ and is responsibile for the larger variety of Bogoliubov dispersions discussed in Sec.\ref{sec:coh_excit}.

\begin{figure}[t]
\begin{center}
\includegraphics[width=0.95\columnwidth,angle=0,clip]{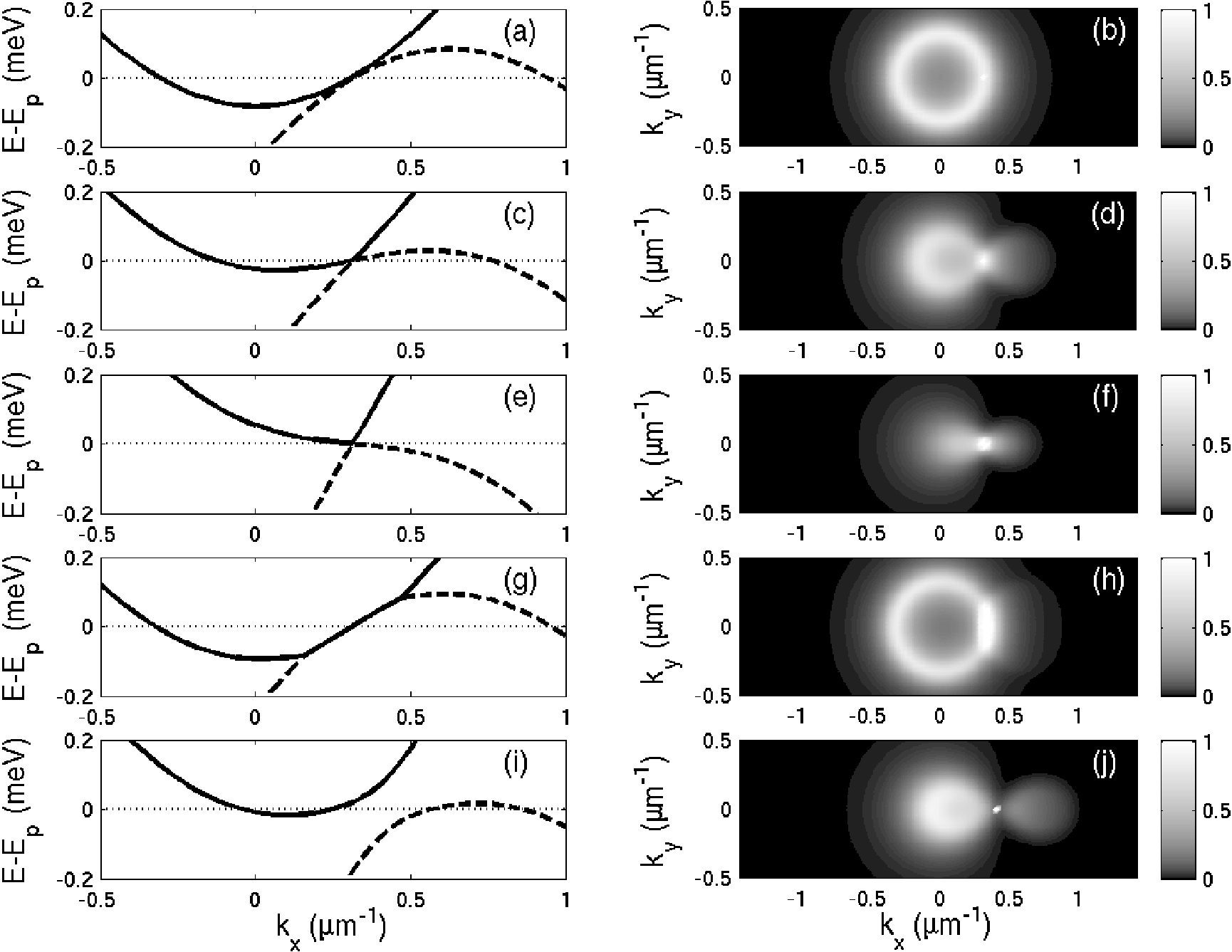}
 \end{center}
\caption{Left panels: Bogoliubov dispersion of elementary excitations in a coherently pumped driven-dissipative polariton condensate flowing in the rightward direction. Right panels: corresponding far-field emission patterns in the presence of a weak defect at rest. The pumping regimes considered in the different rows are explained in the text. 
Figure taken from~\cite{Carusotto:PRL2004}. 
}
\label{fig:PRLsuperfl_k}
\end{figure}

The Bogoliubov matrix for a coherently pumped polariton condensate flowing at speed $\vv=\hbar \kk_{\rm inc}/m$ was first derived in~\cite{Carusotto:PRL2004,Ciuti:PSSB2005}. Its form is closely similar to the matrix \eq{Bogol_LP} for the polariton condensate at rest except for the phase factors $e^{\pm 2i\kk_{\rm inc}}$ in the off-diagonal terms as in the Bogoliubov matrix \eq{BogoL} for an equilibrium condensate in motion.
Its eigenvalues give the dispersion of elementary excitations. The most relevant examples are shown in the left panels of Fig.\ref{fig:PRLsuperfl_k}: the effect of the finite $\kk_{\rm inc}$ is to Doppler shift the branches by $(\kk-\kk_{\rm inc})\cdot \vv$. As usual, the frequency and wavevector of the Bogoliubov modes is defined with respect to the oscillation frequency $\omega_{\rm inc}$ and the wavevector $\kk_{\rm inc}$ of the unperturbed condensate.

Superfluidity of the polariton fluid in the spirit of the Landau criterion can be investigated by looking at the response of the flowing polariton fluid to a local and time-independent impurity potential modeling, e.g., a fabrication defect in the microcavity sample or an optically generated potential. The perturbation induced in the fluid by the weak defect is straightforwardly obtained using the formalism review in Sec.\ref{sec:BogoLandau} with the appropriate form \eq{eq:Lmoving} of the Bogoliubov operator: as the defect potential is a time-independent one, the emission of Bogoliubov modes is concentrated on those modes for which the real part of the frequency $\textrm{Re}[\omega(\kk)]$ is zero; the effect of the finite lifetime of the Bogoliubov excitation on the broadening of the emission as well as on the drag force was discussed in~\cite{Cancellieri:PRB2010}. The peculiar shape of the $\kk$-space locus of excited Bogoliubov modes directly reflects on the shape of the $\kk$-space density pattern, which is concentrated on the resonant modes: a few examples of such patterns are shown in the right column of Fig.\ref{fig:PRLsuperfl_k}.

Panels (a,b) on the top row of Fig.\ref{fig:PRLsuperfl_k} illustrate the case of low-intensity resonant pump, where polariton-polariton interactions are negligible. In this case, the locus of resonant modes has a circular shape and the $\kk$-space emission pattern is characterized by the so-called resonant Rayleigh scattering ring~\cite{Freixanet:PRB1999,Houdre:PRB2000,Langbein:PRL2002}. An analogous feature was illustrated in the left panel of Fig.\ref{fig:superfl_exp_atoms} in the case of ultracold atomic gas.

Panels (c,d) on the second row correspond to the case of an interacting polariton gas in supersonic motion: the pump frequency and intensity is adjusted at the end-point C of the bistability loop of Fig.\ref{fig:RMP_cohpump}, so that the low-$\kk$ dispersion has a sonic shape with a sound speed lower than the speed of flow. This results into a strong deformation of the $\kk$-space  resonant Rayleigh scattering ring into an eight-shaped pattern. 

Panels (e,f) illustrate the case of a subsonic motion, where the polariton gas behaves as a superfluid: no Bogoliubov mode can be any longer resonanly excited and the resonant Rayleigh scattering ring disappears. The weak emission that is still visible in panel (f) corresponds to non-resonant scattering events favored by the finite linewidth of the polariton modes.

All these features are in close parallel to what is predicted for dilute atomic gases. Novel behaviors that stem from the non-equilibrium nature of the polariton fluid are illustrated in the two lowest rows of Fig.\ref{fig:PRLsuperfl_k}.
Panels (g,h) correspond to the case where the pump intensity is on the lower part of the bistability loop (i.e. in the vicinity of point A on Fig.\ref{fig:RMP_cohpump}) and the Bogoliubov branches stick within a finite region in $\kk$ space. As a consequence of the reduced linewidth of the Bogoliubov modes, the $\kk$-space emission is reinforced on a short segment parallel to the $k_y$ axis corresponding to the sticking region.

Panels (i,j) correspond to a optical limiter case or to the upper branch of the bistability loop where the Bogoliubov branches are separated in energy. In this case, the $\kk$ locus still has a ring-like shape but no longer intersects the pump wavevector. Of course, for larger separation of the Bogoliubov branches one may even recover a superfluid behavior without any resonant scattering channel available.

\begin{figure}[t]
\begin{center}
\includegraphics[width=0.95\columnwidth,angle=0,clip]{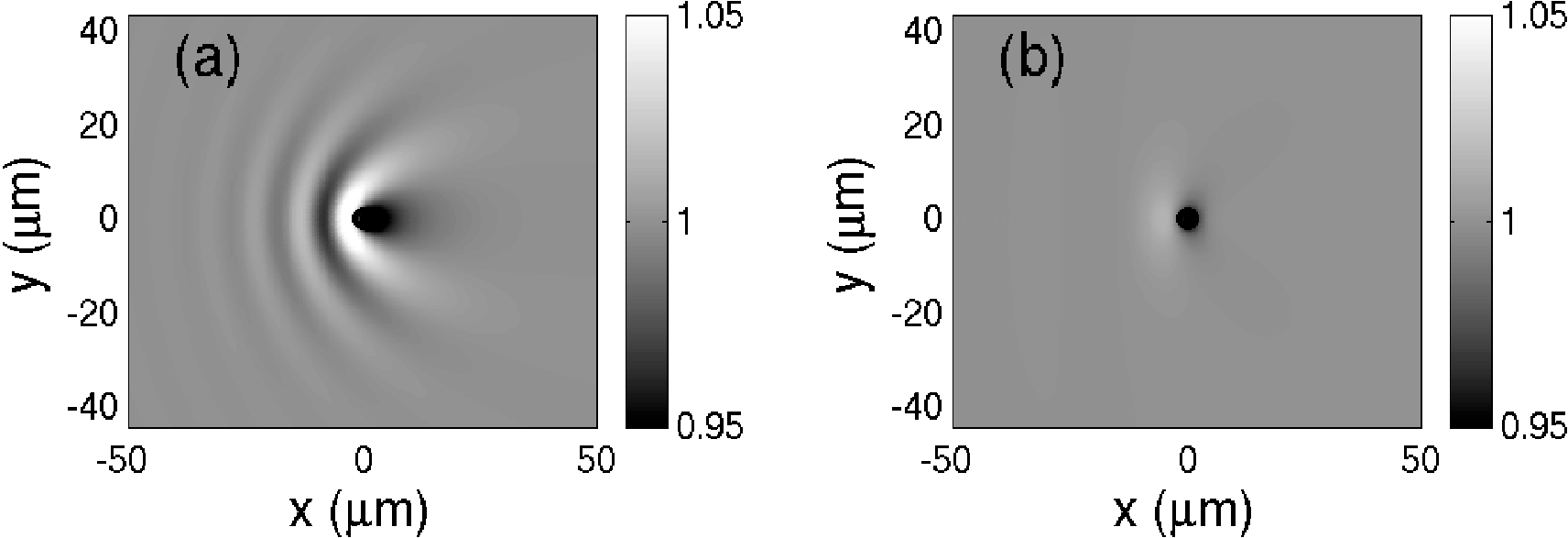}\\
\includegraphics[width=0.45\columnwidth,angle=0,clip]{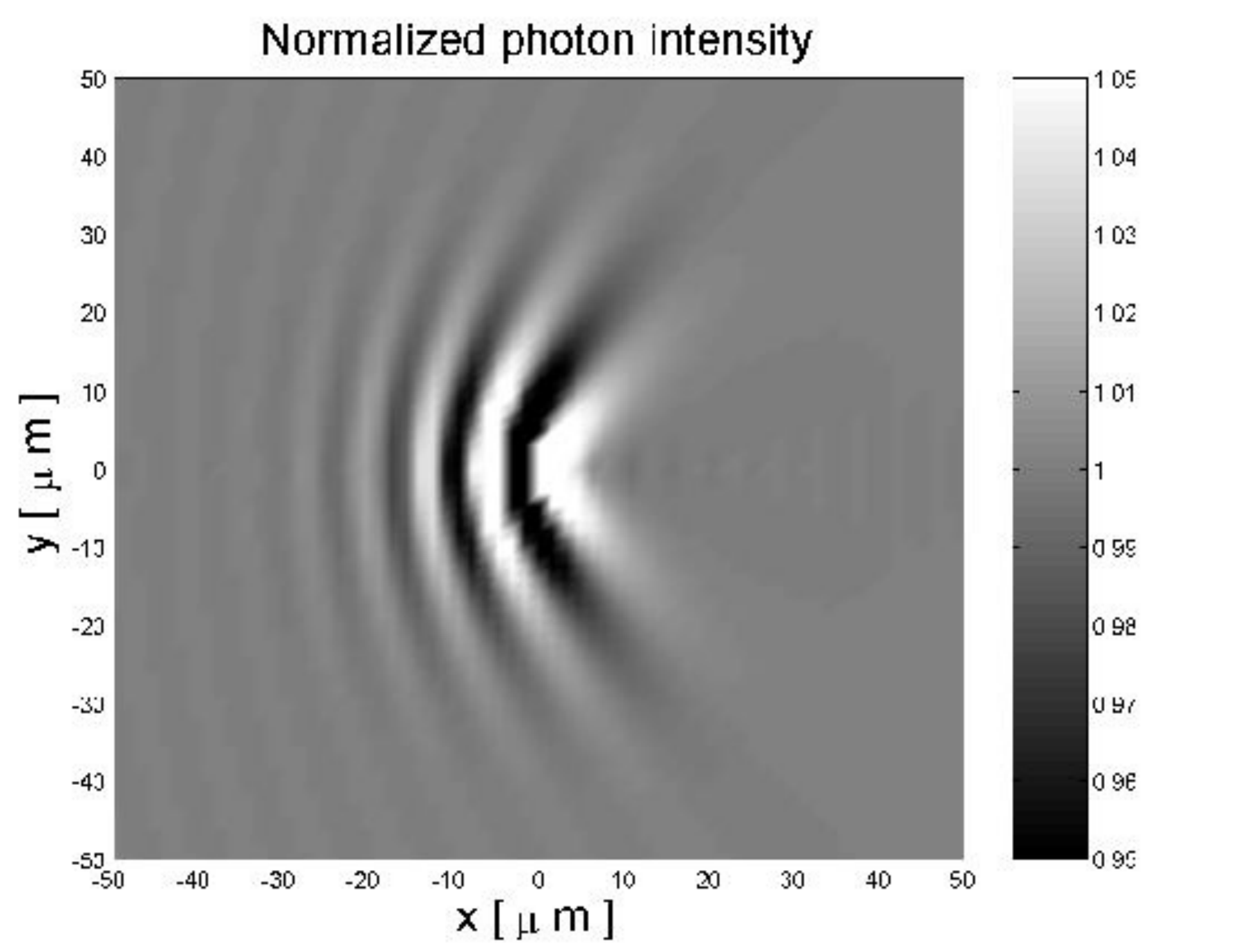}
\includegraphics[width=0.41\columnwidth,angle=0,clip]{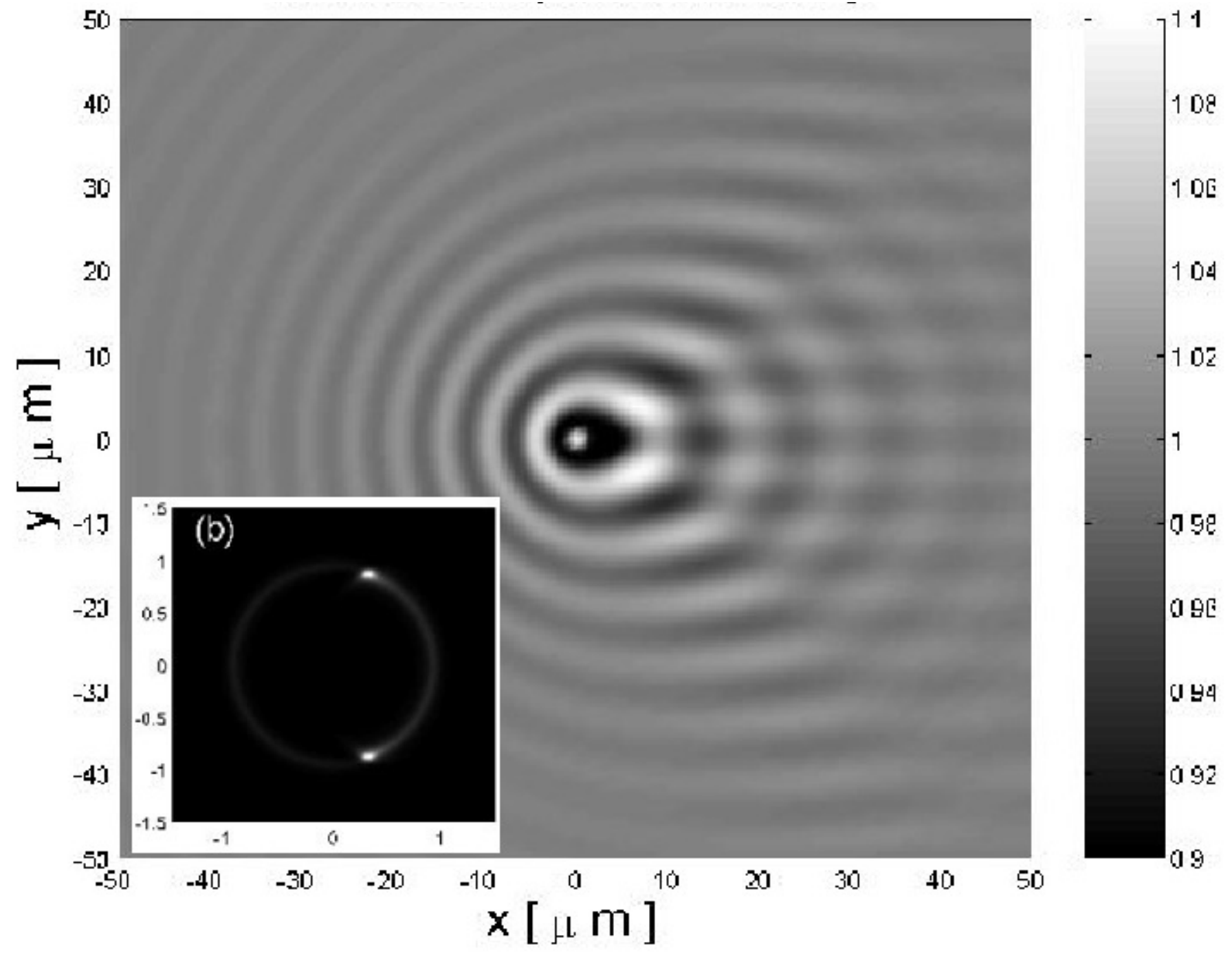}
 \end{center}
\caption{Examples of real-space polariton density patterns for coherently pumped polariton condensates flowing in the rightward direction against a defect at rest. The defect is located at the center of each panel. Upper-left panel is for the non-interacting polariton regime of Fig.\ref{fig:PRLsuperfl_k}(a,b). Upper-right panel is for the superfluid regime of Fig.\ref{fig:PRLsuperfl_k}(e,f). Panels from~\cite{Carusotto:PRL2004}. The lower-left panel is for a super-sonic flow regime of Fig.\ref{fig:PRLsuperfl_k}(c,d). The lower-right panel illustrates the {\em Zebra-\u Cerenkov} effect in the vicinity of a parametric instability; the corresponding $\kk$-space emission pattern is shown in the inset. Panels from~\cite{Ciuti:PSSB2005}. 
}
\label{fig:PRLsuperfl_x}
\end{figure}

Examples of real space density patterns for the different cases are shown in Fig.\ref{fig:PRLsuperfl_x}.
The upper-left panel refers to the linear regime of non-interacting polaritons whose $\kk$-space features are shown in Fig.\ref{fig:PRLsuperfl_k}(a,b). The density pattern consists of a series of parabolic wavefronts extending in the upstream direction of the defect and originating from the interference of the plane-wave incident condensate with the spherical wave of scattered polaritons. Of course, the finite polariton lifetime makes the perturbation disappear at large distances from the impurity.
The upper-right panel corresponds to the superfluid regime of Fig.\ref{fig:PRLsuperfl_k}(e,f) where no Bogoliubov mode is resonantly excited and the density perturbation remains localized in a very small region in the neighborhood of the impurity. The two panels on the lower row correspond to a polariton fluid in supersonic motion: the deformed shape of the resonant Rayleigh ring shown in Fig.\ref{fig:PRLsuperfl_k}(c,d) is responsible for the rich structure of the pattern with curved precursors upstream of the defect and a Mach cone extending in the downstream direction. 

The lower-right panel of fig.\ref{fig:PRLsuperfl_x} illustrates a novel regime of {\em zebra-\u Cerenkov} effect that is only possible far from equilibrium when the condensate is close to a parametric instability: the $\kk$-space emission is concentrated around two symmetric spots where the linewidth of the Bogoliubov modes is strongly quenched as a precursor of the instability; correspondingly, the real-space image is characterized by a horizontal stripe pattern downstream of the defect.

\subsubsection{Experiments}
\label{sec:polar_SF_exp}

\begin{figure}[t]
    \begin{center}
\includegraphics[width=0.95\columnwidth]{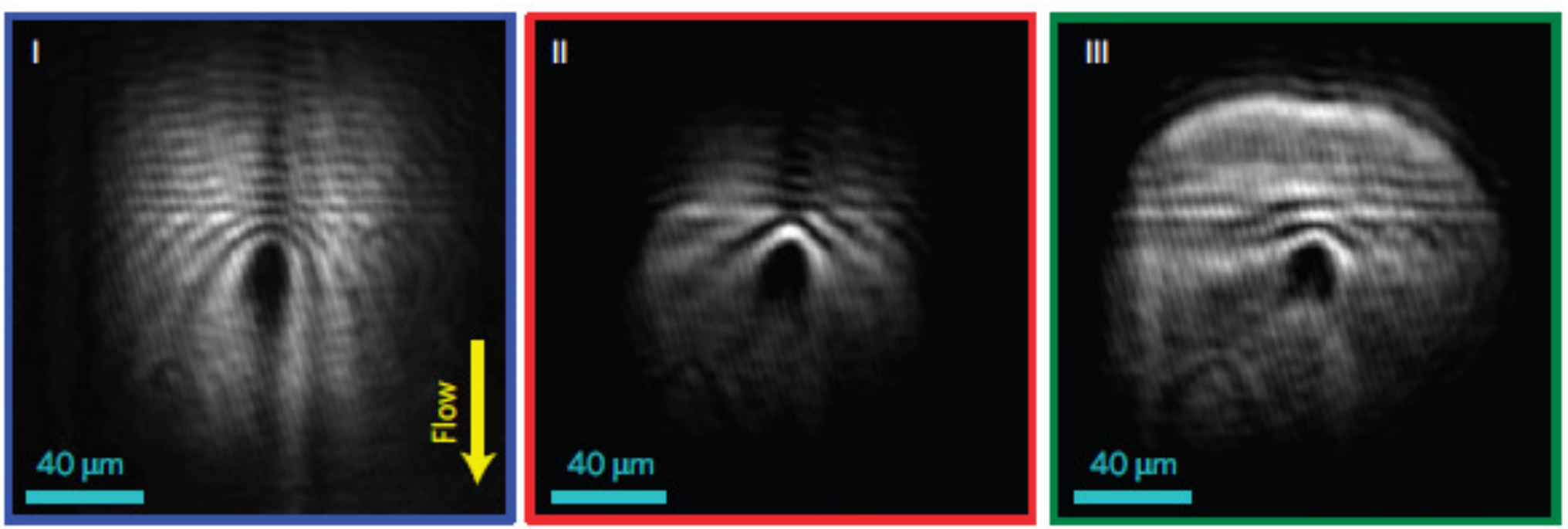}
\includegraphics[width=0.95\columnwidth]{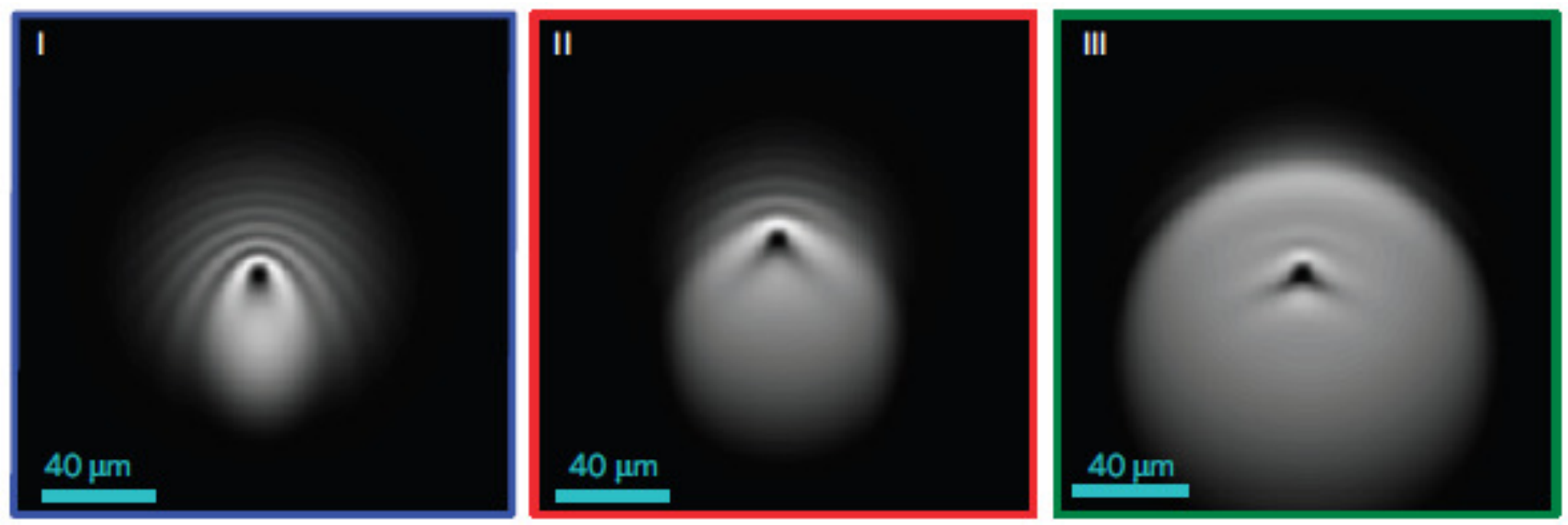}
      \caption{Experimental (top) and theoretical (bottom) real-space images of a fast moving polariton condensate hitting a point defect. The different columns correspond to increasing values of the polariton density (from left to right). Figure taken from~\cite{Amo:NPhys2009}.
}
 \label{fig:Amo_cer}
    \end{center}
  \end{figure}

The theoretical predictions reviewed in the previous subsection have been experimentally confirmed in~\cite{Amo:NPhys2009}. The most significant images are reproduced in Figs.\ref{fig:Amo_superfl} and \ref{fig:Amo_cer}. In this experiment, polaritons are coherently injected by a laser beam with a wide spot incident on the microcavity sample at a finite angle so to generate a polariton directed in the downwards direction. The static impurity consists of a localized fabrication defect naturally present in the microcavity structure. 

The first and second rows of Fig.\ref{fig:Amo_superfl} show the experimental images for the real and $\kk$-space emission patterns. The different columns refer to increasing polariton densities from left to right.
The left column correspond to a non-interacting polariton fluid: the parabolic wavefronts in real space and the resonant Rayleigh scattering ring in $\kk$-space are clearly visible. The weaker spot that is visible at the upmost point on panel (iv) can be attributed to a coherent multiple scattering events on cavity disorder~\cite{MuellerLectures,Langbein:PRL2002b}.

The right column correspond to a superfluid regime where the $\kk$-space emission is quenched and the disturbance created by the impurity remains spatially localized in its close vicinity. Together with the experiment~\cite{Amo:Nature2009} that will be reviewed later on in Sec.\ref{sec:Amobullet}, these observations provide the first experimental evidence of superfluid behaviors in a luminous fluid.
Theoretical images obtained by solving the non-equilibrium Gross-Pitaevskii equation are shown in the lower rows. Exception made for the additional microcavity disorder that was not implemented in the calculations, the agreement with experimental data is very good and confirms that the main source of broadening of the $\kk$ space images is the finite spatial size of the polariton fluid.

Fig.\ref{fig:Amo_cer} illustrates the case of a coherently pumped polariton fluid moving along the plane at a faster speed as a result of a larger incident wavevector $\kk_{\rm inc}$. In this case, the available laser power was not sufficient to penetrate a superfluid regime. The supersonic flow regime is visible in the real-space patterns shown in the central and right columns: the precursors in the upstream direction are always visible and the aperture of the \u Cerenkov cone increases for increasing polariton density (i.e. decreasing Mach number).

Soon after, a similar experiment was performed in~\cite{Amo:PRB2010} using another laser beam to create the defect potential by injecting control polaritons with counter-polarized $\sigma_-$ polarization with a suitably designed spatial distribution The all-optical nature of this method appears extremely promising in view of generating more complex potential geometries where to trap polaritons: linear-shaped barriers were explored in the same work~\cite{Amo:PRB2010}, while other configurations were considered in~\cite{Sanvitto:NPhot2011}.

\subsection{Ballistically moving polariton fluid}

The main drawback of the coherent pumping configuration is that the condensate phase is locked by the pump laser, apparently restricting the range of hydrodynamic phenomena that can occur. In particular, topological features such as solitons and vortices are forbidden, as they involve the spontaneous development of non-trivial phase patterns in the fluid. This issue was first pointed out in~\cite{Bolda:PRL2001}, who proposed to use a temporally pulsed coherent pump to inject polaritons and then follow in time their free evolution when hitting a defect. Such a pump configuration was adopted in~\cite{Sermage:PRB2001} to study the in-plane propagation of polariton bullets in presence of a finite acceleration.

Another strategy to study hydrodynamical effects in fluids of polaritons is inspired to the experiment of~\cite{Richard:PRL2005} that we have reviewed in Sec.\ref{sec:local_coh_parag}: the condensate of polaritons is generated in a spatially restricted region from which it propagates away in a ballistic way before decaying on a length scale set by $\ell=v_{\rm gr}/\gamma$. As theoretically explained in~\cite{Wouters:PRB2008}, the group velocity $v_{\rm gr}$ is determined by the repulsive potential felt by polaritons in the pumped region, while $\gamma$ is the inverse polariton lifetime. A more complete study of the density and flow profile of such a ballistically propagating polariton condensate can be found in~\cite{Kamchatnov:ARXIV2011}.

Focussing our attention onto a small enough spatial region where the condensate density $\bar{n}_{LP}$ and local momentum $\bar{\kk}$ can be taken as approximately constant, the Bogoliubov matrix has the form
\begin{equation}
  \eqname{eq:Lmoving}
{\mathcal L}=
\left(
\begin{array}{cc}
-\frac{\hbar^2\nabla^2}{2m_{LP}}+ g_{LP}\bar{n}_{LP} -\frac{i\gamma_{LP}}{2} & g_{LP}\bar{n}_{LP} \,e^{2i\bar{\kk} \rr} \\
-g_{LP}\bar{n}_{LP} \,e^{-2i\bar{\kk} \rr} & +\frac{\hbar^2\nabla^2}{2m_{LP}}-g_{LP}\bar{n}_{LP} -\frac{i\gamma_{LP}}{2}
\end{array}
\right):
\end{equation}
exception made for the linewidth $\gamma_{LP}$ of the modes, the resulting spectrum is then the standard sonic one of equilibrium condensates \eq{BogoL}. Note how a similar spectrum is obtained for condensates generated by a spatially extended, but temporally pulsed coherent pump; an experiment addressing this last case was reported in~\cite{Kohnle:PRL2011}.

A direct proof of coherent ballistic polariton propagation was provided by the experiment in~\cite{Wertz:NatPhys2010} by comparing the momentum distribution in the different spatial regions with theoretical predictions and then by measuring the interference pattern between polaritons located on opposite sides of the pumped region. These observations complements the $\kk$-space Billet interferometer experiment of~\cite{Richard:PhD,Richard:PRL2005}.
Intriguing observation of back-scattering suppression in supersonically flowing condensates have been recently reported in~\cite{Tanese:arXiv2011}: the experimentally observed superfluid flow appears at odd with the expectation based on the naive Landau criterion used so far. In the same work, a novel possible mechanism for creating an effectively superfluid state via parametric instabilities is proposed.
\begin{figure}[t]
    \begin{center}
\includegraphics[width=0.95\columnwidth]{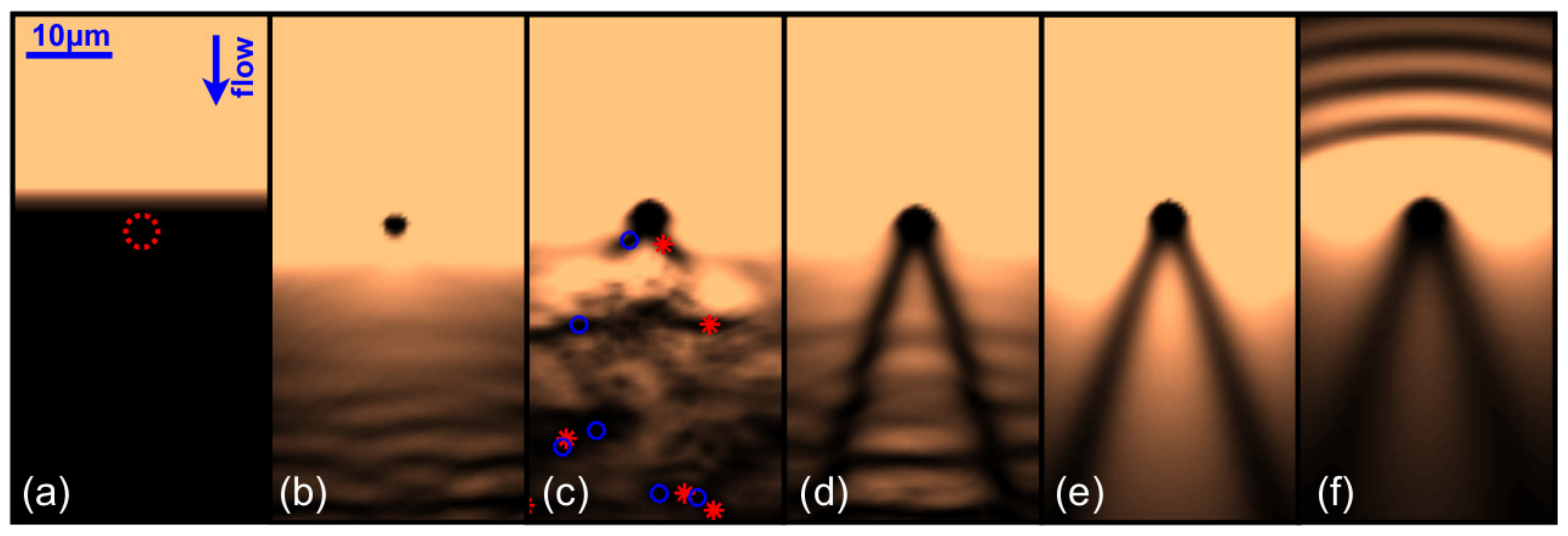}
      \caption{Theoretical images of the ballistic propagation of a polariton condensate against a large defect. 
      Panel (a): spatial profile of the pump spot localized in the top half-plane, just above a point defect. Panels (b-f) show the different flow behaviors: (b) superfluid propagation around the defect; (c) turbulent creation of vortex-antivortex pairs (unlike the other images, this one is not stationary in time);  (d-f) generation of a pair of dark oblique solitons. Figure from~\cite{Pigeon:PRB2011}.}
 \label{fig:vortex_th}
    \end{center}
  \end{figure}
  
The potential of ballistic propagation schemes for studying the hydrodynamic nucleation of topological excitations in a condensate was pointed out in~\cite{Pigeon:PRB2011} using a configuration with a coherent pump in a half-plane geometry as shown in Fig.\ref{fig:vortex_th}(a). Panels (b-f) show snapshots of the polariton density for increasing values of the Mach number $v/c_s$. For intermediate values of the Mach number, turbulent vortex nucleation is observed and then a pair of oblique solitons are formed. Qualitatively, this phenomenology is reminiscent of the theoretical works in the context of liquid Helium~\cite{Frisch:PRL1992} and ultracold atoms~\cite{El:PRL2006, Kamchatnov:PRL2008}. Experiments along these lines will be reviewed in Sec.\ref{vortices}.
\subsection{Spontaneous symmetry breaking and non-equilibrium polariton superfluidity}

\label{sec:Amobullet}

Historically, the first experiment that has addressed superfluidity properties of polariton fluids was reported by~\onlinecite{Amo:Nature2009}. The chosen configuration is very different from the one of the experiments discussed so far, and raises many intriguing questions about its theoretical understanding.

\begin{figure}[t]
    \begin{center}
 \includegraphics[width=0.99\columnwidth,clip]{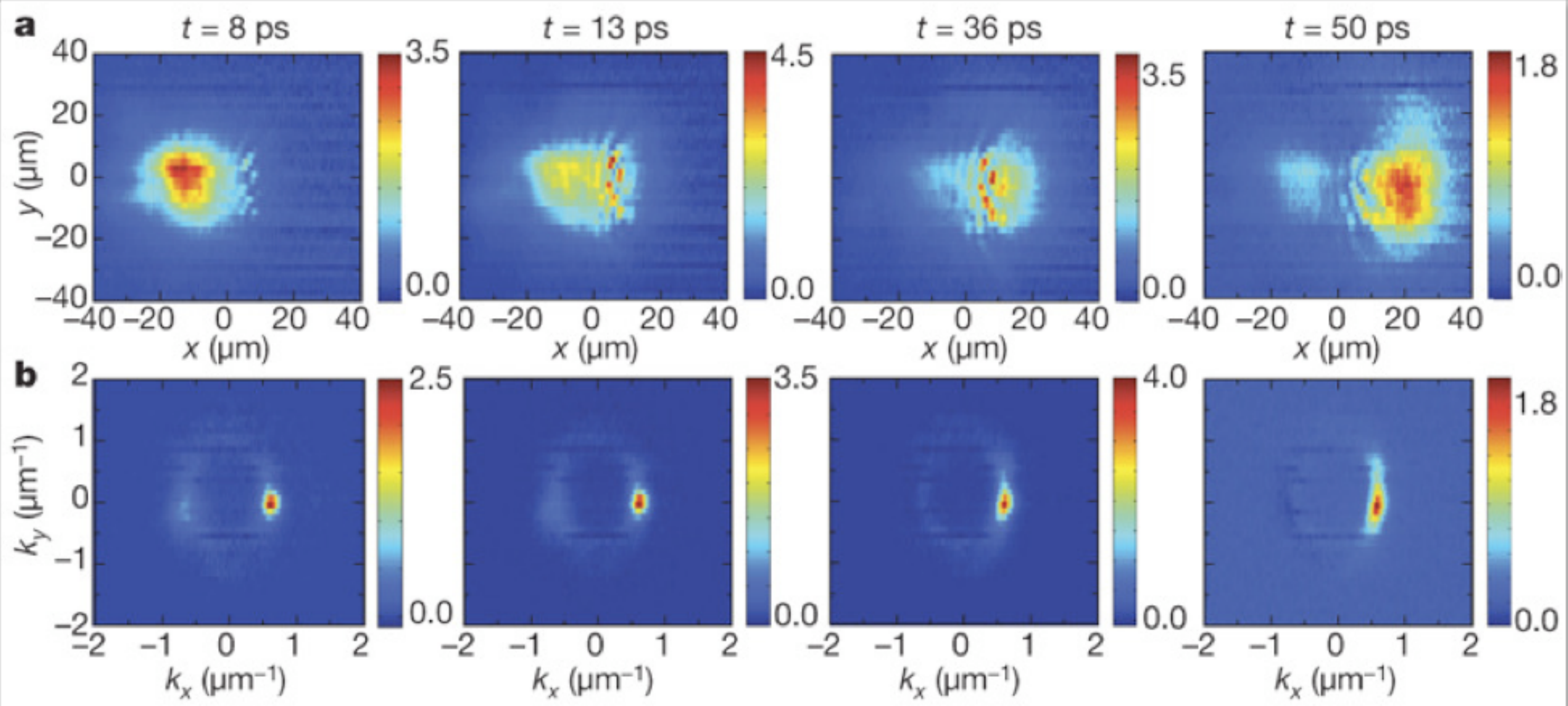}
      \caption{Figure taken from~\cite{Amo:Nature2009}. Top panels: real space images of a ``bullet'' of signal polaritons hitting a natural defect of the microcavity. 
      The bullet of signal polaritons has a very long lifetime thanks to the parametric gain induced by the continuous-wave pump. Bottom panel: corresponding $\kk$-space space images.
}
 \label{fig:Amo_superfl_OPO}
    \end{center}
  \end{figure}

A spatially wide, continuous-wave pump is incident on the cavity at a finite $\kk$-vector close to the inflection point of the polariton dispersion with an intensity slightly below the parametric threshold discussed in Sec.\ref{sec:secOPO}. A second probe pulse excites the cavity with a small wavevector, creating a slowly moving polariton bullet that propagates within the pumped region. Parametric scattering processes from the pump strongly enhance the lifetime of the signal polariton bullet~\cite{Ballarini:PRL2009}. The remarkable experimental observation is that this polariton bullet does not spread in space during propagation and is not destroyed by the collision against a defect: snapshots of real and momentum space distributions illustrating this superfluid-like behavior are shown in Fig.\ref{fig:Amo_superfl_OPO}.

Since the experimental results were published, several mechanisms have been invoked to explain the observations. As -to the best of our knowledge- no consensus has been yet reached in the community, we limit ourselves here to an overview of some most significant features and open questions. An alternative point of view on this experiment can be found in the review~\cite{Marchetti:chapter2012}.

A main difficulty of the theoretical description is that the experiment is carried out in the vicinity of the OPO critical point, where nonlinear effects play a dominant role in determining the so-called TOPO (triggered OPO) spatio-temporal dynamics. With respect to the already complex discussion of the OPO equation of state in~\cite{Wouters:PRB2007b}, the temporal and spatial evolution of the signal/idler wavepackets has now to be fully taken into account. 

Slowly decaying and non-spreading polariton wavepacket propagation has been numerically studied in~\cite{Szymanska:PRL2010} by solving the full OPO wave equations: the polariton bullet consists of spatially superposed signal and idler beams; parametric scattering from the pump is able to compensate losses and keep the spatial shape constant.
For a simplified model with a spatially infinite pump beam, non-spreading solitonic solution have been numerically found and characterized in~\cite{Egorov:PRL2010,Egorov:PRB2011}. A dedicated experimental study of bright polariton solitons in microcavities under a cw-pump in a OPO-like configuration recently appeared in~\cite{Sich:arXiv2011}. We expect that these studies of soliton propagation may provide useful insight in the mechanism underlying the polariton bullets in the TOPO configuration of~\cite{Amo:Nature2009}.

Another, even more challenging feature of this experiment is the robust shape of the polariton bullet after hitting the defect, see the right panel of Fig.\ref{fig:Amo_superfl_OPO}. The spatial modulation of the density profile that is visible in the two central panels is attributed to the Cerenkov wake that is imprinted by the defect onto the pump beam. 
A naive reasoning based on the Landau criterion applied to the elementary excitations spectrum of an OPO condensate shown in Fig.\ref{fig:Goldstone} does not provide an explaination of this experimental observation: because of the diffusive nature of the Goldstone mode, a number of $\kk$-modes are in fact available into which polaritons can be scattered by the defect. On this basis, one would rather expect the signal polariton bullet to be immediately destroyed after hitting the defect.

\begin{figure}[t]
    \begin{center}
\includegraphics[width=0.99\columnwidth,clip]{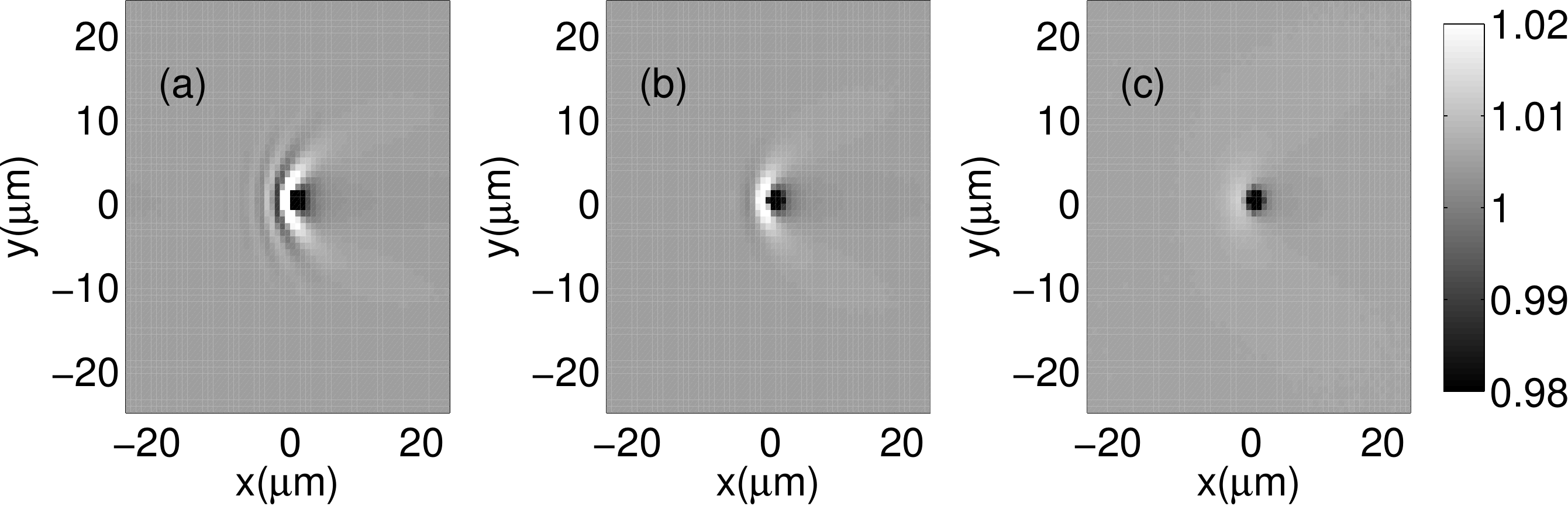}
\includegraphics[width=0.99\columnwidth,clip]{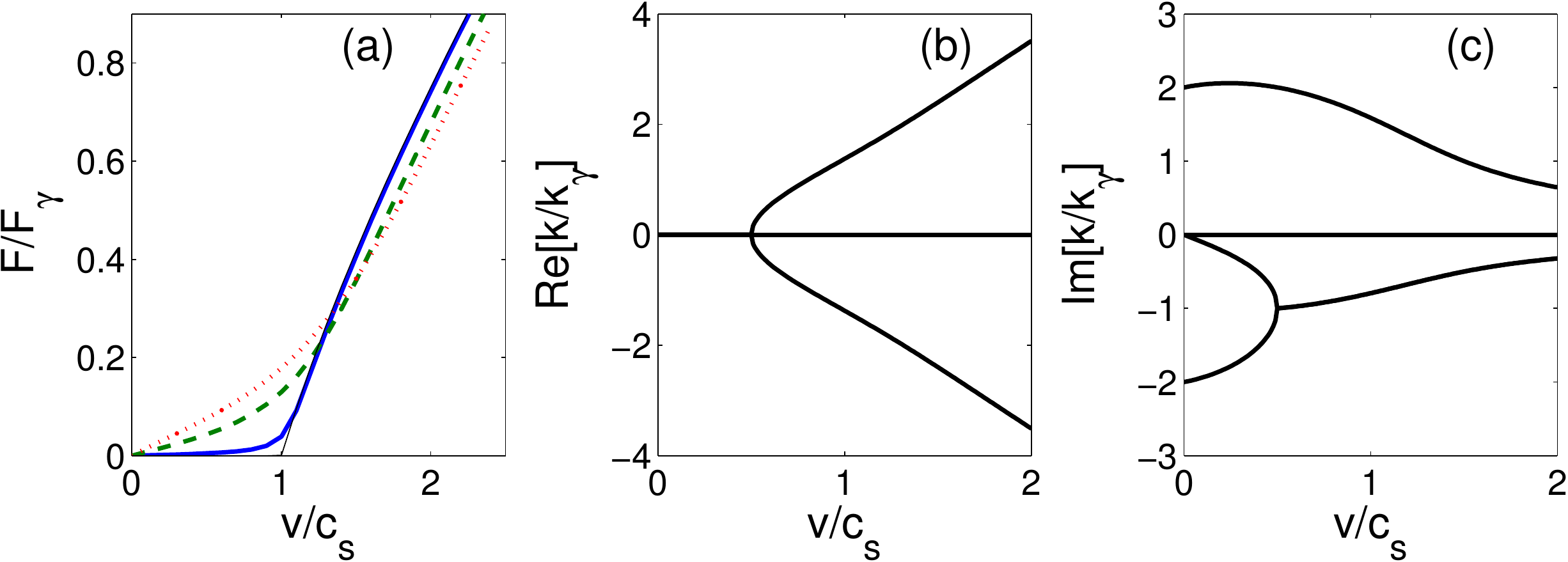}
      \caption{
Upper panels: Generalized GPE simulation of an incoherently pumped condensate hitting a weak and stationary defect at three different speeds $v/c_s=1.5,1,0.4$ across the critical value for superfluidity (from left to right). 
Lower left panel: force exerted on the defect by the moving condensate as a function of the speed $v$. The different (thin black solid, blue solid, green dashed, red dotted lines) curves correspond to growing values of the non-equilibrium parameter. Lower center and right panels: Real and imaginary parts of the complex wavevector $\tilde{k}$ of the zero-frequency Bogoliubov mode as a function of $v$. 
Figure taken from~\cite{Wouters:PRL2010}.
}
 \label{fig:GPE_noneq_nonres}
    \end{center}
  \end{figure}

The unexpected outcome of numerical simulations of the Gross-Pitaevskii equation \eq{CGLE} for polariton condensate under incoherent pumping in~\cite{Wouters:SLMC2008,Wouters:PRL2010} has suggested a possible path to reconcile the picture. A naive application of the Landau criterion \eq{landau_critical} to the real part of the diffusive spectrum of elementary excitations \eq{omega_bog_nonres} shown in Fig.\ref{fig:Bogo_incoh} would predict that the critical speed vanishes and an impurity is able to emit phonons in the fluid independently of the value of the flow speed. The numerical results shown in the upper panels of Fig.\ref{fig:GPE_noneq_nonres} strongly disagree with this expectation and indicate that at low speeds the incoherently pumped condensate indeed behaves as a superfluid and is almost unaffected by the defect. The critical speed for the onset of the usual wake is of the order of the (equilibrium) speed of sound $c_s=\sqrt{g_{LP}n_{LP}/m_{LP}}$. These numerical observations are confirmed by the clear threshold that is visible close to $c_s$ in the velocity dependence of the friction force shown in the lower left panel: the weaker the effective loss rate $\Gamma$, the sharper the threshold. 

Taking inspiration from classical work in electrodynamics of absorbing media~\cite{Tait:PRB1972}, an analytical understanding of these results can be obtained by noting that in a stationary state both the condensate wavefunction and the Bogoliubov modes oscillate at the single frequency $\omega$, while the imaginary part of the dispersion \eq{omega_bog_nonres} has to be reabsorbed into the complex wavevector $\tilde{k}$.
Restricting our attention to zero-frequency $\omega_{\rm Bog}=0$ Bogoliubov modes propagating along the velocity axis, the real and imaginary parts of the complex wavevector $\tilde{k}$ are plotted in the bottom central and left panels of Fig.\ref{fig:GPE_noneq_nonres} as a function of the speed $v$: the generalized Landau critical velocity $\tilde{v}_c$ corresponds to the threshold for the appearance of a non-vanishing real part. The weaker the effective decay rate $\Gamma$, the closer $\tilde{v}_c$ to the equilibrium prediction $c_s$.

As we have reviewed in Sec.\ref{sec:secOPO}, polariton condensates in the OPO regime share the same spontaneous $U(1)$ symmetry breaking mechanism and the same diffusive Goldstone mode as the incoherently pumped ones. On this basis, we then expect that the mechanism of superfluidity illustrated in Fig.\ref{fig:Bogo_incoh} for incoherently pumped condensates may provide a physical explanation of the intriguing superfluidity observations in the OPO regime reported in~\cite{Amo:Nature2009}.

\subsection{Metastability of supercurrents}
\label{sec:superfluid_metastable}

A striking manifestation of superfluidity is the metastability of supercurrents~\cite{Leggett:1999RMP}. For simplicity let's consider a multiply connected geometry in the form of a torus: single-valuedness of the condensate wavefunction $\Psi(\rr)$ imposes the so-called {\em Onsager-Feynman quantization condition} to the supercurrent around the torus,
\begin{equation}
\oint \vv_s \cdot d\mathbf{l} = \frac{2\pi \hbar}{m} N_w,
\eqname{O-F-quant} 
\end{equation}
where $\vv_s$ is the superfluid flow velocity and the integer $N_w$ is the so-called {\em winding number} indicating the number of times the phase of the wavefunction winds up in a loop around the torus.

While in classical hydrodynamics any friction process is able to continuously slow down the flow around the torus, the quantization condition \eq{O-F-quant} makes the lifetime of finite $N_w\neq 0$ supercurrents extremely long in repulsively interacting superfluids with no extra internal degree of freedom even in the presence of a significant wall roughness. In order for the winding number $N_w$ to vary, a node has in fact to appear in the condensate wavefunction, which (exception made for the critical region just below the superfluid critical temperature $T_\Lambda$) requires surmounting a very high free-energy barrier. As a result, a number of different metastable supercurrent states exist for the condensate, labelled by the winding number $N_w$.

This physics was first investigated in superfluid liquid $^4$He~\cite{Vinen:PRSL1961,Reppy:PRL1964,Reppy:PRL1965} and, more recently, in ultra-cold atom condensates~\cite{Ryu:PRL2007,Ramanathan:PRL2011}: in this latter case, it was explicitly shown that the toroidal geometry is essential to guarantees metastability of the supercurrent; a vortex in a harmonically trapped condensate can in fact escape from the cloud with no energy cost~\cite{Fetter:JPHYS2001}.

\begin{figure}[htbp]
    \begin{center}
\includegraphics[width=0.99\columnwidth,clip]{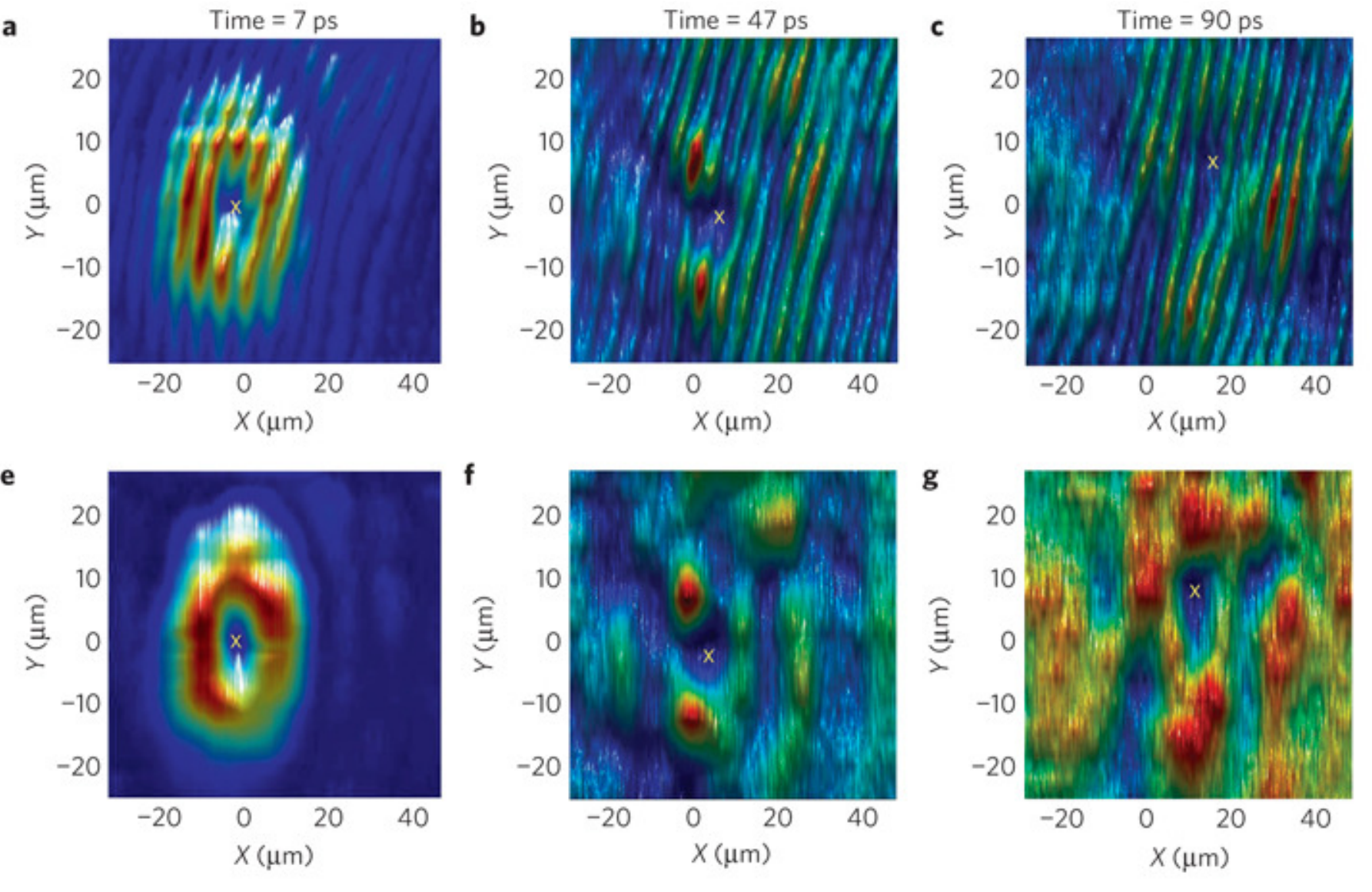}
       \caption{Snapshots of the time evolution of the phase (upper row) and of the density (bottom row) of an OPO polariton condensate triggered by a pulsed probe carrying a $m=1$ vortex. To improve visibility of the vortex dynamics, the contribution of the unperturbed polariton signal (in the absence of the probe pulse) is subtracted from all data. Figure {adapted} from~\cite{Sanvitto:NatPhys2010}.
}
 \label{fig:metastable_vortex}
    \end{center}
  \end{figure}

An experiment along these lines was carried out in polariton condensates in a OPO configuration in~\cite{Sanvitto:NatPhys2010}: a pump beam with a wide spot profile is shone on the microcavity with an intensity slightly above the threshold for OPO operation. A temporally short and spatially narrow probe pulse in a Laguerre-Gauss state with a finite orbital angular momentum is used to force OPO operation to occur into a rotating state: after the extra polaritons injected by probe pulse have disappeared, a vortex remains imprinted into the condensate wavefunction in the signal/idler modes and lasts for macroscopically long times as shown in Fig.\ref{fig:metastable_vortex}. Analogously to the case of harmonically trapped atomic condensates of~\cite{Ryu:PRL2007}, the fate of the vortex is to eventually drift out of the condensate: the time-scale of this decay process is however orders of magnitude longer than the radiative lifetime of a signal polariton.

Further experimental work has investigated the size of the vortex core under a continuous-wave probe beam injecting angular momentum~\cite{Krizhanovskii:PRL2010} and the appearance of vortex-antivortex pairs when the spatial size of the probe pulse is much smaller than the pumped area~\cite{Tosi:PRL2011}. The dynamical stability of multiply-charged vortices was experimentally addressed and numerically confirmed in~\cite{Sanvitto:NatPhys2010}: while in trapped atomic condensates with simply connected geometries, these objects are generally dynamically unstable towards the splitting into several singly-charged vortices~\cite{Fetter:JPHYS2001}, stable $m=2$ vortices could be observed in the polariton case. A related theoretical study of metastable vortices and supercurrents in polariton condensates under an incoherent pumping scheme appeared in~\cite{Wouters:PRB2010b}.

\subsection{Response to transverse vector potentials}

From a formal point of view, the most precise and quantitative definition of superfluid $f_s$ and normal $f_n$ fractions involves the response of the quantum gas to a weak transverse vector potential~\cite{Hohenberg:AnnPhys1965}. The interaction Hamiltonian describing the coupling of the particles to a vector potential has the form
\begin{equation}
V=-\int\!d^2\rr\,\jj(\rr)\cdot \AAA(\rr)
\eqname{A-V}
\end{equation}
with the current operator defined as usual as
\begin{equation}
\jj(\rr)=\frac{\hbar}{2im}\left[\phihd(\rr)\nabla\phih(\rr) - \textrm{h.c.} \right].
\end{equation}
For a spatially homogeneous system of density $n$, the susceptibility tensor relating the average current to the applied vector field can be written in Fourier space as
\begin{equation}
\langle \jj \rangle (\qq,\omega) =\chi(\qq,\omega)\,\AAA(\qq,\omega).
\eqname{chi}
\end{equation}
and the normal fraction $f_n$ is defined as the low-momentum, low-frequency limit of the transverse susceptibility
\begin{equation}
f_n=\lim_{q\rightarrow 0} \lim_{\omega\rightarrow 0} \frac{m}{n}\chi_{T}(\qq,\omega).
\eqname{f_n}
\end{equation}
This definition of the normal fraction is widely used in numerical quantum Monte Carlo calculations~\cite{Pollock:PRB1987} to assess the transition to a superfluid state and then quantitatively evaluate the superfluid fraction: a most interesting feature in view of studies of exotic superconductor and superfluid states is that it does not require {\em a priori} knowledge of the microscopic nature of the order parameter nor of its symmetry.

To get a more intuitive physical understanding of the definition \eq{f_n}, one may note that a purely transverse vector potential $\AAA(\rr)$ naturally appears in the description of a mechanical system in a rotating reference frame at angular velocity $\mathbf{\Omega}$~\cite{CCT:CdF},
\begin{equation}
\AAA_{\rm rot}(\rr)=\mathbf{\Omega}\times \rr:
\eqname{Arot}
\end{equation}
the current pattern induced by the vector potential \eq{Arot} via the Hamiltonian \eq{A-V} then has a rigid-body spatial shape proportional to the normal fraction $f_n$.
As in the rotating bucket gedanken experiment~\cite{Leggett:1999RMP}, the superfluid fraction stays at rest in the fixed star reference frame, while the normal fraction is dragged into rotation by the vessel. The Andronikashvili experiment~\cite{Andronikashvili:RMP1966} showing superfluidity of liquid Helium samples was based on an implementation of the rotating bucket idea using a torsional oscillator. 

Recent developments in the manipulation of ultracold atomic clouds have demonstrated the possibility of generating artificial gauge fields coupled to the motion degrees of freedom of neutral atoms~\cite{Dum:PRL1996,Lin:Nature2009,Dalibard:RMP2011}. The measurement of the response of the atomic cloud to a suitably taylored artificial gauge fields was proposed in~\cite{Cooper:PRL2010,John:PRA2011,Carusotto:PRA2011} as a way to measure the normal and superfluid fractions of an ultracold atomic cloud independently from Bose-Einstein condensation.

A related theoretical proposal for the case of non-equilibrium polariton condensates appeared in~\cite{Keeling:PRL2011}. 
A many-body calculation using the Schwinger-Keldysh diagrammatic approach for non-equilibrium systems has shown that the superfluid fraction of the polariton gas  according to the definition \eq{f_n} remains finite in spite of the vanishing Landau critical velocity \eq{landau_critical} and the diffusive nature of the Goldstone mode. In contrast to the atomic and liquid Helium cases, the non-equilibrium nature of the polariton condensate makes the normal fraction to remain finite even at $T=0$. The artificial gauge field for polaritons can be generated by applying a real magnetic field to the microcavity in a suitable imbalanced anti-Helmholtz configuration. Other proposals to implement artificial gauge fields for polaritons will be reviewed in Sec.\ref{sec:gauge}.

\section{Hydrodynamic formation of dark solitons and vortices}
\label{vortices}

\begin{figure*}[htbp]
    \begin{center}
 \includegraphics[width=1\columnwidth]{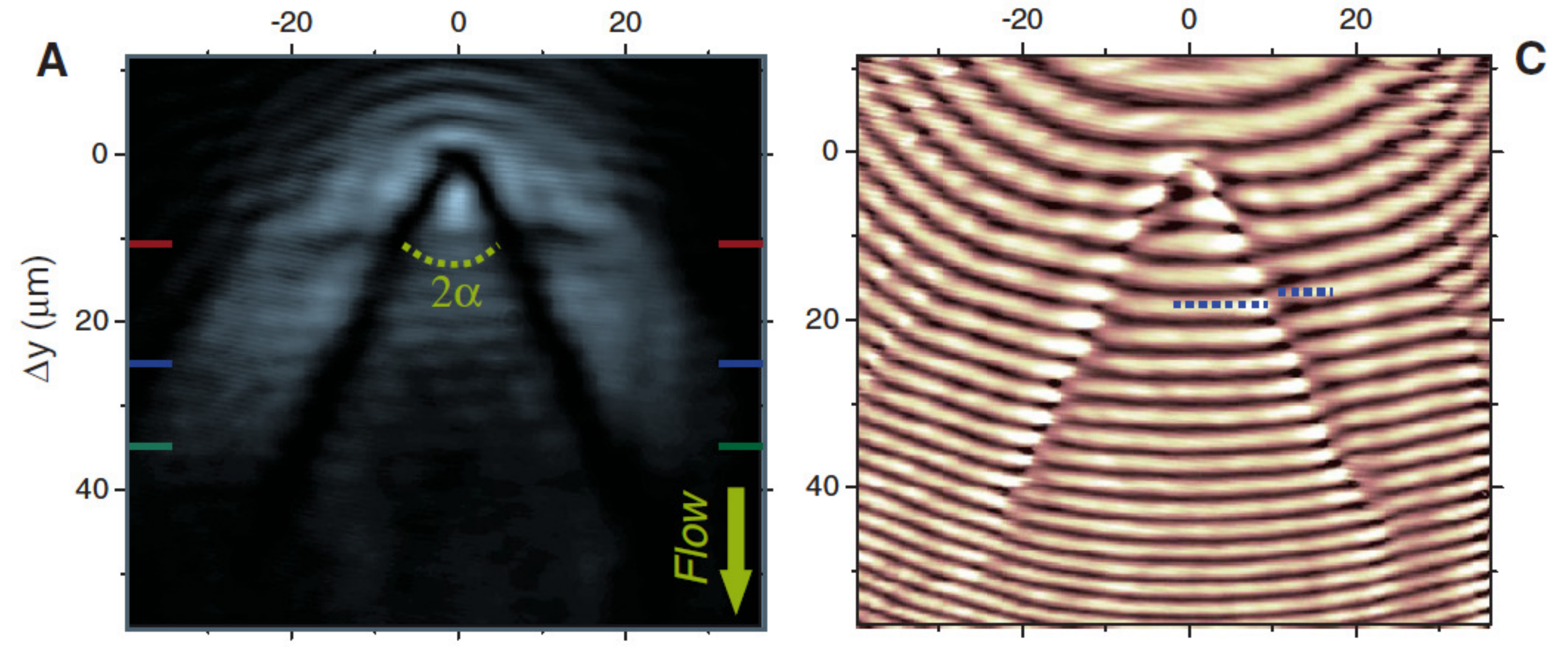}
 \includegraphics[width=0.4\columnwidth,clip]{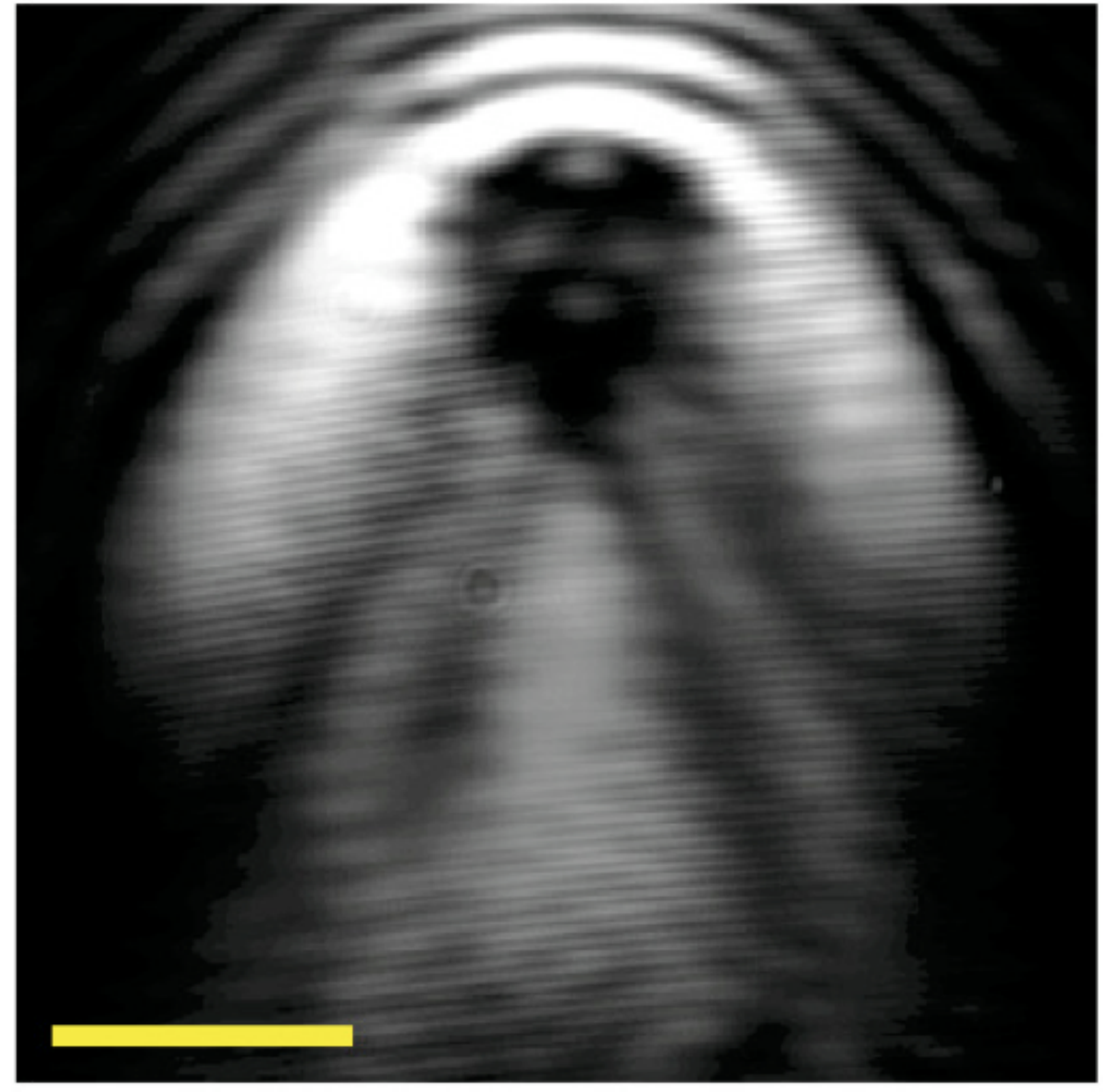}
      \caption{Figure taken from~\cite{Amo:2011Science}. Experimental observation of hydrodynamically generated oblique dark solitons in a polariton superfluid (left). The phase interferogram (middle panel) clearly reveals a phase jump across the soliton lines. With bigger defects, a quadruplet of dark solitons is observed (right panel).
}
 \label{fig:Amo_solit}
    \end{center}
  \end{figure*}

{As we have reviewed in the introductory section of this article, a great deal of the early literature on superfluid photon hydrodynamics dealt with the physics of quantized vortices in the coherent photon fluid. However, in spite of a number of theoretical proposals~\cite{Staliunas:PRA1993,Bolda:PRL2001} and experimental developments using photorefractive oscillators~\cite{Vaupel:PRA1996} and nonlinear optical crystals in a propagating geometry~\cite{Wan:FrOpt2008}, no complete study of the fundamental hydrodynamic processes of soliton and vortex nucleation at the surface of a large defect was reported. Even more surprisingly, even though this mechanism was predicted to play a crucial role in determining the actual critical speed of superfluid liquid Helium \cite{Frisch:PRL1992}, its first direct observation in a condensed matter context appeared only very recently using an atomic condensate in~\cite{Neely:PRL2010}.}

{In this section we shall review the recent experimental investigations of the physics of a superfluid of light hitting a large defect. Thanks to their flexibility, these experiments using microcavity polaritons have provided the first complete characterization of the different behaviors as a function of the flow parameters.}

\subsection{{Oblique dark solitons}}

The first experimental investigation of a flowing polariton condensate hitting a large defect appeared in~\cite{Amo:2011Science}: with respect to the earlier demonstration by the same group of polariton superfluidity~\cite{Amo:NPhys2009}, a different fabrication defect in the same microcavity sample was chosen, with a suitably larger spatial extension. Another crucial difference was the implementation of the proposal by~\onlinecite{Pigeon:PRB2011} to use a pump spot spatially restricted to the upper half space upstream of the defect. Without this judicious trick, the phase of the condensate would be locked to the pump laser and no topological excitation could appear in the fluid. Experimental images of the density and phase profile for a large flow speed well above the speed of sound $c_s$ are shown in Fig.\ref{fig:Amo_solit}. In addition to the curved precursors located upstream and laterally to the defect, a pair of oblique {dark} solitons is apparent in the wake of the defect. Their solitonic nature is confirmed by the large phase jump $\theta\simeq \pi$ across the dark region that is visible in the phase pattern. {In contrast to the moving dark solitons that appear past a strong obstacle in a one-dimensional flow~\cite{Hakim:PRE1997,Engels:PRL2007,Wan:PRL2010}, the tip of the oblique solitons shown in Fig.\ref{fig:Amo_solit} is pinned at the defect position.}

The aperture angle $\alpha$ can be related to the Mach number and to the phase jump across the soliton using the theory of dark solitons~\cite{BECbook}. 
This predicts a soliton speed equal to $v_{\rm sol}=c_s\,\cos(\theta/2)$, where the phase jump $\theta$ is related to the relative value of the minimum density $n_{\rm min}$ by $\cos^2(\theta/2)={n_{\rm min}}/{n}$.
In order to be at rest in the laboratory frame, the soliton speed has to be equal to $v_{\rm sol}=v\,\sin\alpha$, where $\alpha$ is the aperture angle of the soliton as indicated in the left panel of Fig.\ref{fig:Amo_solit}.
Combining these relations, one obtains for the aperture $\alpha$ the expression 
\begin{equation}
\sin \alpha=\frac{c_s}{v}\,\cos\left(\frac{\theta}{2}\right):
\eqname{alpha}
\end{equation}
a direct consequence of the $\cos(\theta/2)$ factor in \eq{alpha} is that the aperture of the oblique soliton pair is always narrower than the one of the Mach cone of linear waves~\cite{El:PRL2006,Gladush:PRA2007}. 

Remarkably, the same $\cos(\theta/2)$ factor \eq{alpha} predicts that solitons are kinematically allowed also for subsonic speeds $v<c_s$; however, theoretical work for lossless systems\cite{Zakharov:JETP1974,Brand:PRA2002} has shown that such sub-sonic solitons should be dynamically unstable against the decay into vortices via the so-called snake instability~\cite{Anderson:PRL2001} unless the flow is supersonic~\cite{Kamchatnov:PRL2008}. The experiment~\cite{Amo:2011Science} appears in partial disagreement with this prediction, as stable solitons are observed even for sub-sonic speeds: according to~\onlinecite{Kamchatnov:ARXIV2011}, the discrepancy may be explained as a result of the finite lifetime of polaritons. The quite different physics of solitons under a pulsed excitation was investigated in~\cite{Grosso:PRL2011}: within a few picoseconds, solitons decay into vortex streets. Numerical simulations in the same work point out the crucial role of cavity disorder in triggering this instability mechanism. The rightmost panel in Fig.\ref{fig:Amo_solit} confirms the prediction by~\onlinecite{El:PRL2006} that for very large defect sizes, the single oblique solitons should be replaced by a fan structure of many oblique solitons. Very recently, the study of oblique solitons was extended to the case of half-solitons in spinor condensates, where the interplay of the flow and the spin-orbit coupling effects discussed in Sec.\ref{sec:spin} are responsible for the appearance of peculiar spin structures around the soliton axis~\cite{Flayac:PRB2011,Hivet:arXiv2012}.

\subsection{Turbulent behavior}

\begin{figure}[t]
    \begin{center}
\includegraphics[width=0.95\columnwidth]{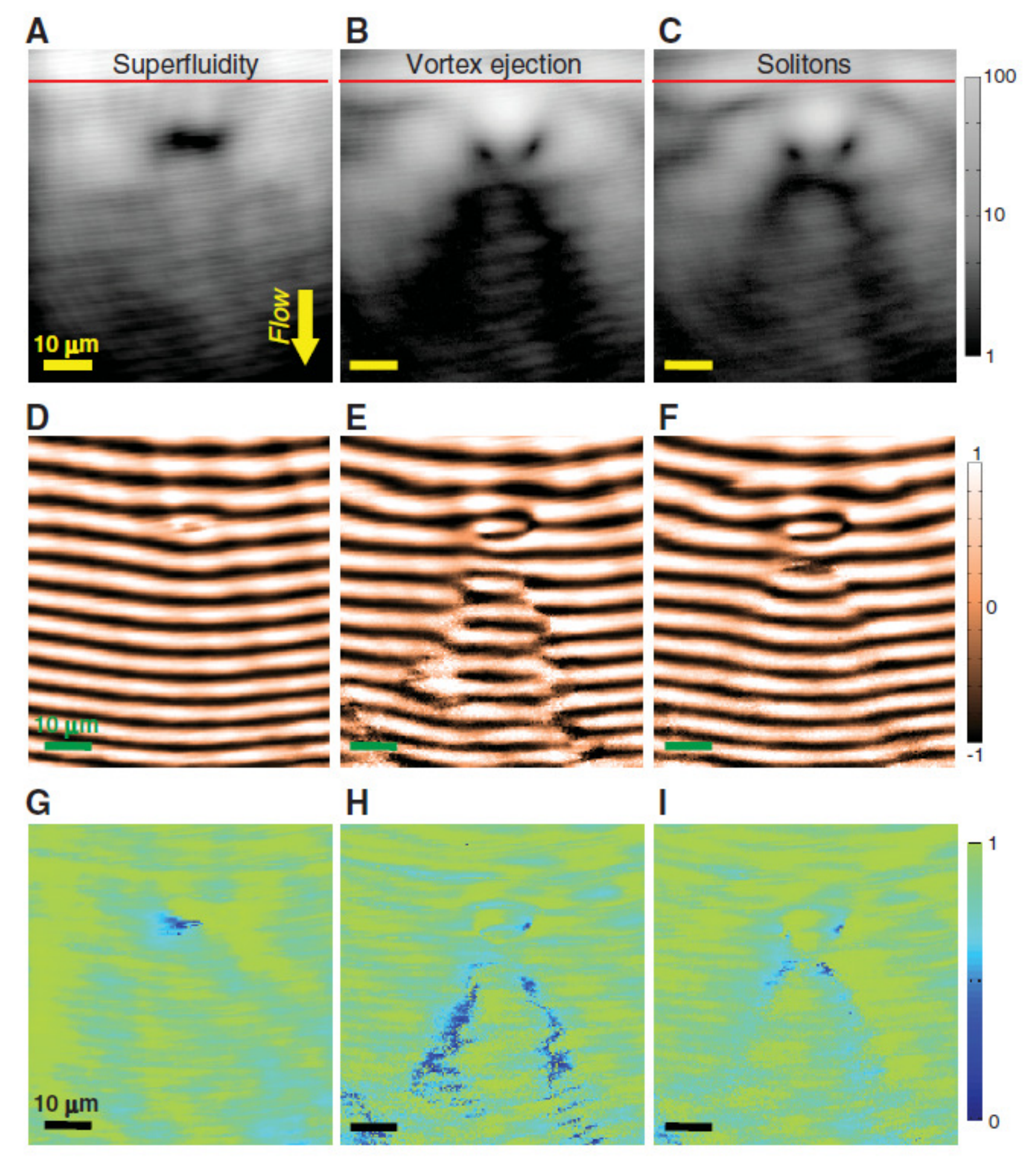}
       \caption{Different regimes of flow in the presence of a large fabrication defect in the microcavity. From left to right, the different columns correspond to growing values of the Mach number $v/c_s$. For each value, the top panels show the real space density profile. The middle panels show the interferograms giving the phase. The bottom panels show the first-order coherence function $g^{(1)}$ of the emission.
Figure taken from~\cite{Amo:2011Science}. }
 \label{fig:Amo_turb}
    \end{center}
  \end{figure}

The outcome of this same experiment for different values of the Mach number $v/c_s$ is summarized in Fig.\ref{fig:Amo_turb}. The left panels correspond to a higher polariton density and lower $v/c_s$: a superfluid behavior is visible where the perturbation induced by the defect remains spatially localized (upper row) and the condensate phase keeps a plane wave form (central row). The right panel shows again the fast flow case discussed in the previous sub-section, where a pair of dark lines are visible in the density pattern and are associated to a sudden jump in the phase pattern. In both these cases, the coherence function $g^{(1)}$ remains close to unity at all positions, which signals that the condensate is fully coherent.

As it was expected from the theoretical calculations by~\onlinecite{Frisch:PRL1992,Winiecki:JPHYSB2000,Pigeon:PRB2011}, the situation is completely different for the intermediate value of the speed that is illustrated in the central column: while the density profile exhibits a significant decrease behind the defect, the phase pattern shows strong irregularities in the same region with a number of vortex singularities distributed with no apparent order. An interpretation of these observations in term of a turbulent behavior is confirmed by the spatial behavior of the  $g^{(1)}$ coherence function: the regions of suppressed coherence in the wake of the defect signal the presence of moving vortices with a complex nucleation and drift process.

{Before proceeding, it is interesting to note that the corresponding regime in a one-dimensional geometry consists of a time-dependent nucleation of solitons past the obstacle, as originally predicted in~\cite{Hakim:PRE1997} and experimentally observed with atomic condensates in~\cite{Engels:PRL2007}. Some first evidence of this effect in a fluid of light was reported in~\cite{Wan:PRL2010} using a nonlinear crystal in a propagation geometry.}

\subsection{Vortices}

The experiments reviewed in the previous subsection were performed under a spatially localized, but temporally continuous-wave coherent pump. This pumping regime is ideal to study the stationary state of the polariton fluid, but has a limited access to time-dependent phenomena such as the vortex nucleation process: as this effect relies on a dynamical instability of the stationary flow at the surface of the defect, the exact nucleation time is strongly affected by any experimental disturbance or system imperfection. To experimentally observe quantized vortices in the wake of the defect as in the theoretical image of Fig.\ref{fig:vortex_th}(c), one then needs either a very high temporal resolution of the imaging apparatus or the possibility of taking a single-shot image of the fluid. While the former issue can be solved using streak-camera devices with almost ps temporal resolution, the latter one still appears very challenging with present-day technology. 

\begin{figure}[t]
    \begin{center}
  \includegraphics[width=1\columnwidth]{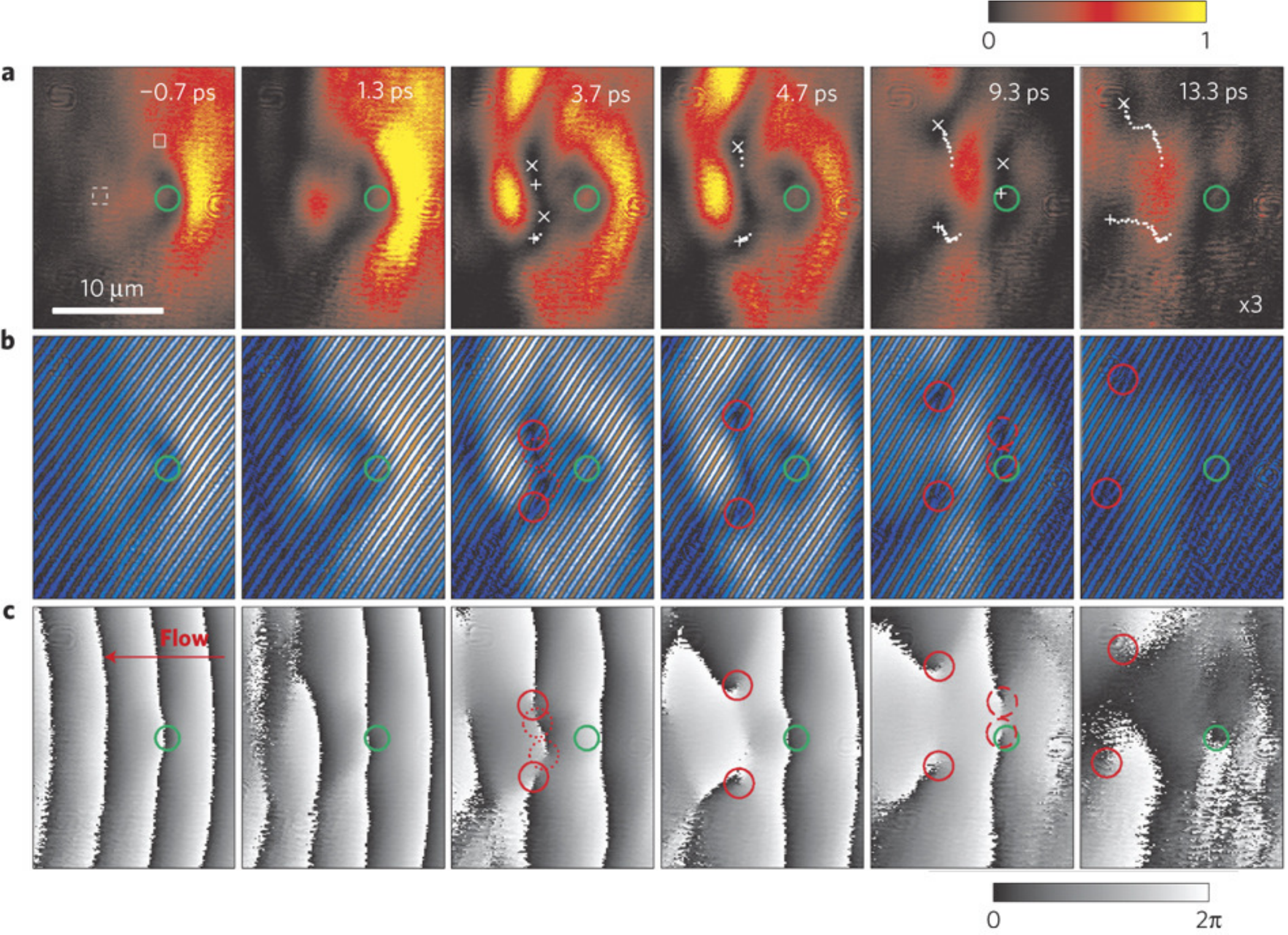}
      \caption{Temporal sequence of experimental images of the scattering of a leftward moving polariton cloud on a structural defect. The three rows show the polariton density (a), the fringes of the measured interferogram (b) and the polariton phase (c). The defect position is indicated by the green circles. Vortices start being visible in the third column and are indicated by white markers ($\times$ for vortex, $+$ for anti-vortex) on the density plot and are circled in red on the fringes and phase plots. Dotted circles indicate short-lived vortices. On the later columns, the previous motion of the long-lived vortex pair is indicated by white dots on the density plots.
Figure taken from~\cite{Nardin:NPhys2011}.
}
 \label{fig:nardin_vort}
    \end{center}
  \end{figure}

In the experiments of~\cite{Nardin:NPhys2011} and soon after of~\cite{Sanvitto:NPhot2011}, this problem was circunvented using a train of identical short pump pulses. The spacing of the pulses is chosen much longer than the lifetime of excitations in the sample, so that each realization of the experiment can be considered as independent. In particular, the exact nucleation time of the vortices is deterministically pinned to the temporal profile of the pump pulse, which allows to repeat the experiment a huge number of times and accumulate good statistics on the polariton density and phase pattern. As suggested in~\cite{Bolda:PRL2001}, short pump pulse durations guarantee that the condensate phase is free to evolve even for pump pulses with a plane wave spatial shape.
Fig.\ref{fig:nardin_vort} shows a series of snapshots at different times before and after the arrival of the coherent pump pulse (of duration 3~ps). The polariton flow is from the right to the left and hits a structural defect in the microcavity sample: the nucleation of pairs of vortices with opposite circulation is apparent in the snapshots for $t=4.7$~ps shown in the third column; after being nucleated, the vortices are dragged downstream by the polariton flow. Images from an analogous experiment carried out using a gas of ultracold atoms by~\onlinecite{Neely:PRL2010} are shown in the lower panel of Fig.~\ref{fig:superfl_exp_atoms}.

\begin{figure}[t]
    \begin{center}
 \includegraphics[width=1\columnwidth]{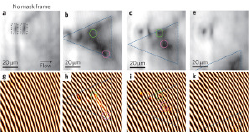}
      \caption{
Vortex storage in a triangular trap. Real-space images (top row) and corresponding interferograms (bottom row) for polaritons injected by a continuous-wave laser beam and flowing in the rightward direction against a natural defect in the microcavity (marked by rectangles in a). The different columns correspond to the no-mask case (a,g) and then to different positions of the triangular mask, indicated by the blue dotted triangle. Figure {adapted} from~\cite{Sanvitto:NPhot2011}.
}
 \label{fig:Sanvitto_mask}
    \end{center}
  \end{figure}

A similar experiment was reported in~\cite{Sanvitto:NPhot2011} using an optically generated defect potential~\cite{Amo:PRB2010} instead of a structural defect of the microcavity. This apparently minor difference was responsible for some qualitative differences in the process of vortex nucleation, in particular because of the soft nature of the optical potential. This same work also reported implementation of a proposal in~\cite{Pigeon:PRB2011} to permanently trap and store the quantum vortices that are hydrodynamically generated in the wake of a natural defect. 
The most significant results are summarized in Fig.\ref{fig:Sanvitto_mask}: the plane-wave, continuous-wave coherent pump is blocked by a triangular-shaped mask that suppresses the coherent pumping in the region inside the dashed triangle. If the mask is located right behind the defect (b-d), the vortices that are hydrodynamically generated by the defect remain permanently stored within the un-pumped region where the condensate phase is free. Their escape is indeed forbidden by the pump beam that fixes the phase in the external region. If the mask is too far away (e) or is not present (a), no vortex can be created at all because the condensate phase in the neighborhood of the defect is fixed by the pump laser. As these figures are obtained under a continuous-wave pump with a long integration time, the only vortices that are visible in the figure are the ones whose position is stationary in time.

\section{Strongly correlated photons}
\label{strongcorrelation}

Most of the physics reviewed so far originates from the collective behavior of a large number of interacting photons or polaritons coherently sharing the same wavefunction. As a result, the system can be  described in terms of a mean-field theory based on a generalized non-equilibrium Gross-Pitaevskii equation, where the driven-dissipative nature of the photon fluid is accounted for. 
On the other hand, the mean-field theory is inaccurate when classical and/or quantum fluctuations become important and the state of the many-particle system is no longer described by a single macroscopic wavefunction. A first example of situation where fluctuations play a central role around the critical threshold point (see Sec.\ref{sec:secOPO} for the case of the optical parametric oscillation).  As long as interactions between polaritons are weak enough, a theoretical description of the coherence properties of the strongly fluctuation polariton field across the OPO threshold in~\cite{Carusotto:PRB2005} could be obtained in a semiclassical way in terms of the stochastic Gross-Pitaevskii equation within the truncated Wigner representation reviewed in Sec. \ref{sec:Wigner}.

The physics is much richer when interactions between polaritons are strong and induce sizable quantum correlations in the fluid.  A simple illustrative example of this phenomenon is provided by the so-called superfluid to Mott-insulator transition predicted for Bose particles trapped at the minima of a periodic potential~\cite{Fisher:PRB1989,Jaksch:PRL98} and experimentally observed with ultra-cold atoms in optical lattices~\cite{Greiner:Nature2002}. Depending on the ratio of the hopping energy $J$ between neighboring sites  and the on-site interaction energy $U$, the ground state of the $N$-particle system at integer filling $\nu=N/M$ ranges from a coherent state describing the superfluid (for  $J/U \gg 1$)
\begin{equation}
\vert \Psi_{SF} \rangle = \left(\sum_{j=1}^M \frac{\ahd_j}{\sqrt{M}} \right)^N\,\vert\textrm{vac}\rangle
\eqname{SF}
\end{equation}
to a Mott insulator state (for $J/U \ll 1$)
\begin{equation}
\vert \Psi_{MI} \rangle = \prod_{j=1}^{M} \left(\ahd_j\right)^\nu \vert\textrm{vac}\rangle
\eqname{MI}
\end{equation}
 While the former state is well captured by a mean-field theory based on the GPE, the Wigner representation of the latter would involve phase space regions where the Wigner function is negative, signaling the importance of quantum correlations and preventing the use of semiclassical stochastic approaches.

Another illustrative example of strongly correlated state that appears in the theory of strongly interacting Bose fluids is the so-called Tonks-Girardeau gas of impenetrable bosons in a one-dimensional geometry~\cite{Olshanii:PRL1998}. In stark contrast with mean-field states where the many-body wavefunction is the product of single particle states, the many-body wavefunction of a Tonks-Girardeau gas is characterized by zeros whenever two particles approach to each other. A remarkably elegant exact solution to this problem was obtained in~\cite{Girardeau:1960} using a rigorous mapping between the one-dimensional gas of impenetrable bosons and a gas of non-interacting spinless fermions: the energy spectra of the two systems are identical, as are all configurational probability distributions in real space. On the other hand, the signature of the strong correlations existing between the bosons are clearly visible in the momentum distributions, which remain instead quite different from the fermionic ones. 
Experimental evidence of Tonks-Girardeau gases of ultracold atoms trapped in one-dimensional geometries was reported in~\cite{Paredes:Nature2004,Kinoshita:Science2004}: the strongly-correlated, fermionized nature of the strongly interacting Bose gas was assessed from macroscopic observables such as the cloud size, its internal energy and the momentum distribution of the atoms.
We refer to~\onlinecite{Cazalilla:RMP2011} for a recent review of the physics of one-dimensional Bose gases in both the condensed-matter and the atomic contexts.

In this section we will review the recent theoretical advances towards the realization of strongly correlated gases of photons or polaritons. A complementary point of view on this same physics can be found in~\cite{Tureci:NPhys2011}. 
The basic block in this direction is the so-called {\em photon blockade} effect first proposed in~\cite{Imamoglu:PRL97}: in the presence of a strong enough optical nonlinearity, photons in single-mode cavities behave as effectively impenetrable particles~\cite{Birnbaum:Nature2005,Faraon:NatPhys2008,Reinhard:NatPhot2011,Lang:PRL2011}. 
When a system of many cavities arranged in a lattice structure is considered, photon-photon interactions are expected to drive the photon gas from a coherent, superfluid state to a correlated Mott-insulator state. The observability of this phase transition in realistic devices was first theoretically investigated in~\cite{Hartmann:NatPhys2006,Hartmann:Laser2008,Greentree:2006,Angelakis:PRA2007} following the analogy with material particles, namely assuming a quasi-equilibrium condition and neglecting the driven-dissipative nature of the photon fluid. A related theoretical proposal to generate a Tonks-Girardeau gas of fermionized photons in an optical fiber appeared in~\cite{Chang:NatPhys2008}. 
The next crucial step was to include in the model photon losses and the pump mechanism that is used to replenish the photon gas. While these issues were well known in the literature on photon hydrodynamics reviewed in the previous sections, the first works addressing and possibly taking advantage of the non-equilibrium nature in the context of strongly correlated photon gases were~\cite{Gerace:NatPhys2009} and soon after~\cite{Carusotto:PRL2009}.

A first experimental realization of large optical nonlinearities at the single photon level leading to a strongly correlated stream of photons was reported very recently in~\cite{Peyronel:Nature2012} using the dramatically reinforced optical nonlinearity of coherently dressed atoms in a Rydberg EIT configuration~\cite{Sevincli:PRL2011,Shahmoon:PRA2011,Gorshkov:PRL2011,Petrosyan:JPhys2012,Dudin:Science2012}.

 \subsection{Non-equilibrium Bose-Hubbard-like systems}
\label{sec:S-I-model}

Many of the recent studies of the physics of strongly interacting photon gases are based on lattice models where photons are trapped in an array of optical cavities. A few examples of such configurations in a two-dimensional geometry are schematically illustrated in Fig.\ref{fig:sketch_array} using an array of photonic crystal cavities (central panel) or an array of superconducting transmission line resonators (lower panel). At a simplest level of description, the system can be modelled by a generalized driven-dissipative Bose-Hubbard Hamiltonian of the form
\begin{multline}
\mathcal{H}=\mathcal{H}_{BH}+\mathcal{H}_F+\mathcal{H}_{env}= \\ 
= \sum_i \hbar \omega_0 \chd_i \ch_i + \hbar U\,\chd_i\chd_i\ch_i\ch_i 
-
\sum_{\langle i,j \rangle} \hbar J \, \chd_{i} \ch_{j} + \\
+\sum_i \left[ F_i(t)\,\chd_i + F_i^*(t)\,\ch_i \right]  + \mathcal{H}_{env},
\eqname{BH}
\end{multline}
where each cavity is assumed to support a single photon mode whose destruction (creation) operator is $\ch_i$ ($\chd_i$). 
Photon tunneling through the non-perfectly reflecting cavity mirrors are responsible for the hopping amplitude between neighboring cavities with an amplitude $J$. The photon-photon interactions are included by a two-particle term of proportional to the on-site energy $\hbar U$ proportional to the $\chi^{(3)}$ third-order optical nonlinearity of the cavity medium~\cite{Drummond:JPhysA1980,Ferretti:PRB2012},
\begin{equation}
U\simeq \frac{3(\hbar\omega_0)^2}{4\epsilon_0V_{\rm eff}}
\frac{\bar{\chi}^{(3)}}{\bar{\epsilon}_r^2}
\eqname{U_nonresonant}
\end{equation}
where $V_{\rm eff}$ is the effective volume of the cavity mode, $\bar{\chi}^{(3)}$ and $\bar{\epsilon}$ are the spatially averaged relative dielectric constant and optical nonlinearity of the cavity material.

The driven-dissipative nature of the photon system is accounted for by the terms on the third line of \eq{BH}: coherent excitation by a pump laser field is described by the site- and time-dependent external field $F_i(t)$, while coupling with the environment is described by $\mathcal{H}_{env}$. Most works have considered an environment consisting of a zero-thermal radiative bath into which photons from each site can escape at a rate $\Gamma$. Once the environment degrees of freedom have been traced out, the corresponding term in the master equation for the reduced density matrix of the system has the Lindblad form
\begin{equation}
\frac{d\rho}{dt}=\frac{1}{i\hbar}[\mathcal{H}_0,\rho] + \frac{\Gamma}{2}\sum_i\left[
2 \ch_i \rho \chd_i- \chd_i \ch_i \rho - \rho \chd_i \ch_i 
\right]
\eqname{mastereq}
\end{equation}
A different but related form of Lindblad term in the master equation \eq{mastereq} was used in~\cite{Hoffmann:PRL2011} to describe the driving of the cavity array by a broadband incident radiation. Of course, incoherent pumping of the cavities by an amplifying bath as discussed in Sec.\ref{sec:incoh_GPE} may be another interesting option, but so far it has not been investigated yet in the context of strongly interacting photons.
 
The driven-dissipative Bose-Hubbard model of \eq{BH} is just an example of a wide class of models describing systems of strongly interacting photons. 
Another prominent model is obtained by replacing on each site the two-particle interaction term with a Jaynes-Cummings interaction term with a two-level emitter of transition frequency $\omega_{\rm eg}$ strongly coupled to the cavity mode. The resulting {\em Jaynes-Cummings-Hubbard} Hamiltonian reads
\begin{multline}
\mathcal{H}_{JCH} = \sum_i \hbar \omega_0 \chd_i \ch_i + \sum_{\langle i,j \rangle} \hbar J \, \chd_{i} \ch_{j} + \\ +  \hbar \omega_{\rm eg} \vert {\rm e}\rangle_i \langle {\rm e}\vert_i  +\hbar g \left(\ch_i \vert {\rm e}\rangle_i \langle {\rm g}\vert_i  + \chd_i \vert {\rm g}\rangle_i \langle {\rm e}\vert_i \right)
\eqname{JC}
\end{multline}
With $\vert {\rm g}\rangle_i$ and $\vert {\rm e}\rangle_i$, we indicate the ground and excited states of the emitter on site $i$. The coupling between the photon mode and the two-level system is quantified by the vacuum Rabi frequency $g$. Hamiltonians of this form can be realized in the physical systems shown in Fig.\ref{fig:sketch_array} by coupling a semiconductor quantum dot to a photonic crystal cavity (middle panel) or a superconducting Josephson artificial two-level atom to the microwave transmission line resonator (lower panel). Overall, most of the predictions of the Jaynes-Cummings Hamiltonian \eq{JC} are in close qualitative agreement with the ones of the simpler Bose-Hubbard one \eq{BH}; some remarkable differences concerning the many-body physics of the photon gas are pointed out in~\cite{Schmidt:PRL2009,Schmidt:PRL2010} and in~\cite{Grujic:arXiv2012}.

Another variation of the generalized Bose-Hubbard model involves a cavity mode strongly coupled to a nonlinear matter excitation, 
\begin{multline}
\mathcal{H}_{PBH} = \sum_i \hbar \omega_0 \chd_i \ch_i +  \hbar \omega_{\rm exc} \bhd_i \bh_i +  \hbar \Omega_R \left (\ch_i \hat{b}^{\dagger}_i + \chd_i \hat{b}_i \right) + \\
+ U  \hat{b}^{\dagger}_i  \hat{b}^{\dagger}_i b_i b_i
+ \sum_{\langle i,j \rangle} \hbar J \, \chd_{i} \ch_{j}. 
\eqname{PBH}
\end{multline}
where  $\hbar \omega_{\rm exc}$ is the energy of the matter excitation created by the bosonic operator $\bhd_i$ and $\Omega_R$ is the vacuum Rabi frequency of the light-matter coupling.  This {\em polariton-Bose-Hubbard} model naturally arises in the description e.g. of a semiconductor micropillar cavity mode strongly coupled to an quantum well exciton transition: as discussed in Sec.\ref{sec:polariton-polariton}, strong interactions arise from the exciton-exciton interactions. In the spatially confined geometry, their strength can be quantified by the single parameter $U$~\cite{Verger:PRB2006}.

Provided the energy scale set by the vacuum Rabi frequency is larger than the nonlinear interaction, the pumping and the losses, one can restrict the description to a single polariton mode and recover a Bose-Hubbard Hamiltonian of the form \eq{BH} with rescaled hopping and nonlinear interaction parameters, $U_{\rm eff}=|X|^4\,U$ and $J_{\rm eff}=|C|^2\,J$.

\subsection{Photon blockade effects}
\label{sec:S-I-blockade}

The first example of nonlinear optical process at the single photon level is the so-called {\em photon blockade} effect, anticipated in~\cite{Carmicheal:PRL1985} and fully elaborated in~\cite{Imamoglu:PRL97} in analogy with the Coulomb blockade effect of electron transport through mesoscopic devices~\cite{Fulton:PRL1987,Kastner:RMP1992}.

The basic idea is to consider a single-mode cavity (i.e. a single site of the Bose-Hubbard Hamiltonian \eq{BH}) with a strong enough optical nonlinearity to be in the $U\gg \gamma$ regime. In this case, the two-photon state is spectrally shifted of an amount $2U$ much larger than its linewidth $2\gamma$ and a laser field resonant with the linear cavity frequency $\omega_0$ will not be able to inject into the cavity more than one photon at a time. In physical terms, the presence of a single photon in the cavity is able to effectively block the entrance of a second one: as a result, the transmitted light across the cavity show a strong antibunching, closely analogous to the one observed in the resonance fluorescence by a single two-level atom~\cite{Kimble:PRL1977}. 
The figure of merit of this antibunching is the ratio $U/\Gamma$, with the zero-delay photon correlation function
$g^{(2)}(0) = {\langle \chd \chd \ch \ch \rangle}/{\langle  \chd \ch \rangle^2}  = {1}/{(1+4 (U/\gamma)^2)}$
quantifying the ratio of having two photons on a single site. Generalization of the photon blockade to nonlinear atom optics using matter waves was proposed in~\cite{Carusotto:PRA2001,Kolomeisky:PRA2004}.

\subsubsection{Jaynes-Cummings nonlinearities}

The main ingredient to observe the traditional photon blockade effect is a large difference between the two-excitation state energy $E_{n=2}$ and $2 E_{n=1}$, i.e., twice the energy of the single-excitation state, regardless of the physical origin of the nonlinearity. In particular, $\vert E_{n=2} - 2 E_{n=1} \vert \gg \hbar \gamma$ is a sufficient photon blockade condition, being $\gamma$ the loss rate.
Indeed, the photon blockade effect turned out to be experimentally easier to observe in Jaynes-Cummings systems where the optical nonlinearity is due to a single two-level emitter strongly coupled to the cavity mode.

\begin{figure}[t]
\includegraphics[width=.33\columnwidth,clip]{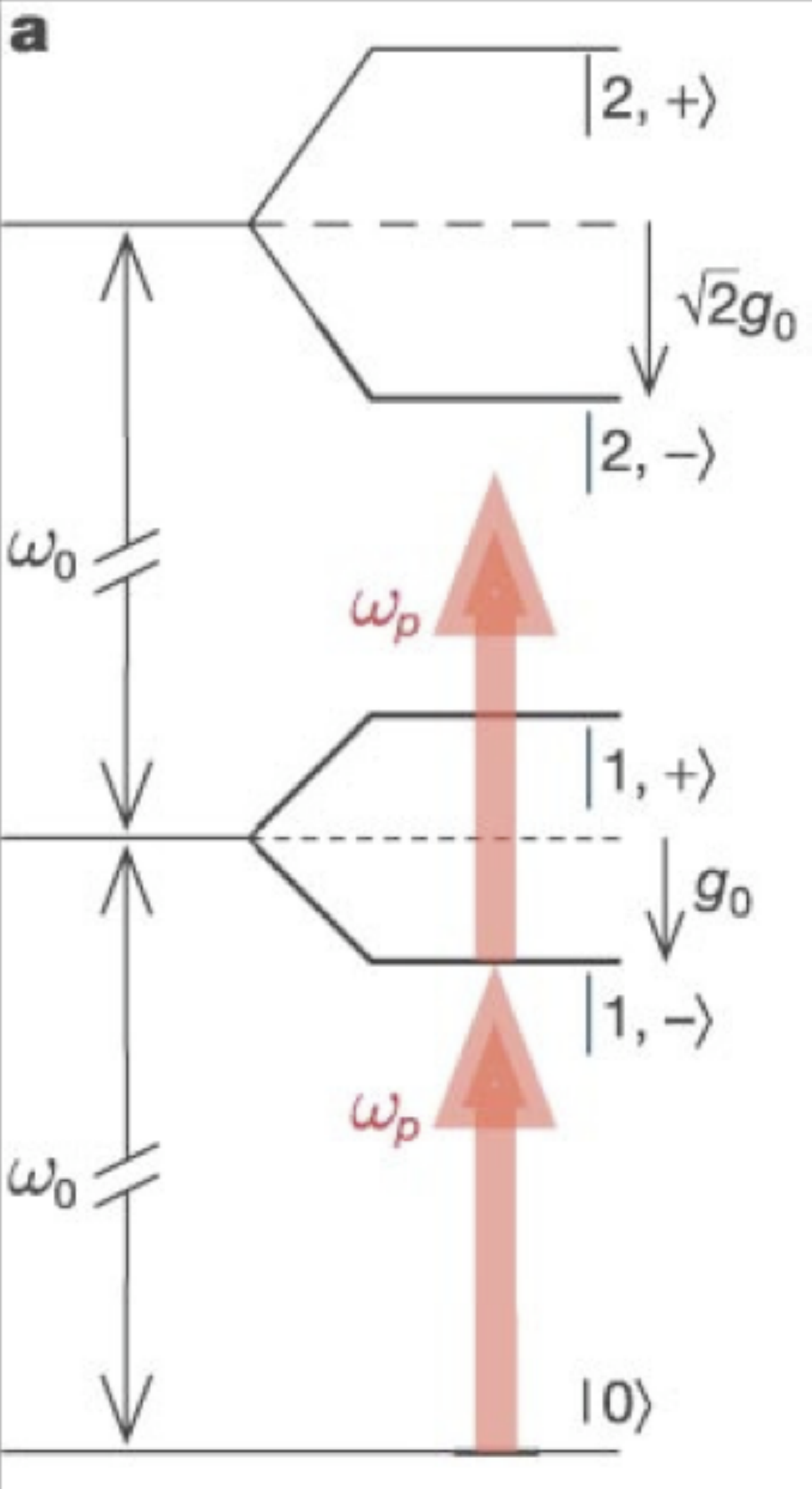}
\hspace{0.02\columnwidth}
\includegraphics[width=0.62\columnwidth,angle=0,clip]{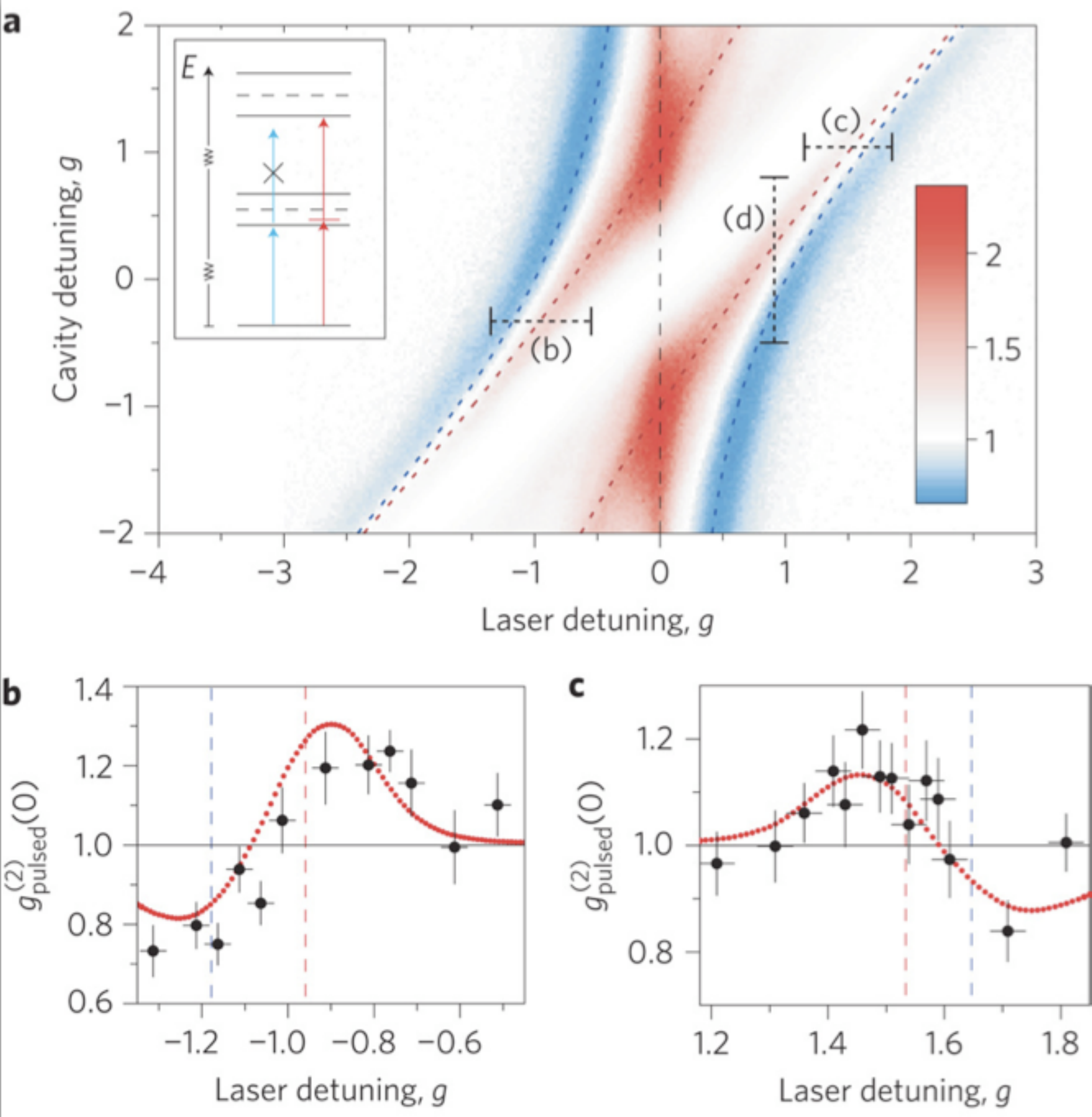}
	\caption{Left panel: Sketch of the level structure of the Jaynes-Cummings model with a laser pump on resonance with a one-photon transition. Figure from~\cite{Birnbaum:Nature2005}.
	Right panels: second-order correlation $g^{(2)}(0)$ for a single quantum dot in a photonic crystal microcavity. Upper panel: Theoretical calculations for $g^{(2)}(0)$ as a function of cavity-exciton and laser detuning. Blue (red) regions correspond to sub- (super-) Poissonian statistics.
Lower panels: Experimental results (points) compared with theoretical curves for specific values of the cavity detuning. Figure from \onlinecite{Reinhard:NatPhot2011}.	
	}
	\label{fig:JC}
\end{figure}
As it is shown in the left panel of Fig.\ref{fig:JC}, the level scheme of the Jaynes-Cummings model consists of an infinite series of doublets above a non-degenerate ground state. When the emitter is exactly on resonance with the cavity mode, the splitting of the $n^{\rm th}$ doublet is equal to $2\sqrt{n}\,g$, $g$ being the emitter-cavity coupling energy.
For a pump field on resonance with, e.g., the lower state of the $n=1$ one-excitation doublet, the closest two-excitation state is then detuned by a frequency amount $(2-\sqrt{2})\,g$. As a result, a sufficient condition to observe significant antibunching in the transmitted light is that the vacuum Rabi frequency $g$ is much larger than the loss rates of both the cavity mode and the emitter.

In experiments, a sizeble antibunching was observed in the optical domain using either a single atom in a macroscopic optical cavity~\cite{Birnbaum:Nature2005}, or -to a slightly less extent- using a single quantum dot in a photonic crystal nanocavity~\cite{Faraon:NatPhys2008,Reinhard:NatPhot2011}. This latter case is illustrated in the right panels of Fig.\ref{fig:JC}. The most spectacular results were obtained in the microwave domain of circuit QED experiments~\cite{Lang:PRL2011}, where the large $g/\Gamma \sim 100$ leads to the very pronounced antibunching illustrated in the right panels of Fig.\ref{fig:circuit_QED}. An experimentally different strategy to assess photon blockade effects was adopted in~\cite{Hoffmann:PRL2011} using a broadband microwave excitation instead of a coherent drive. 

An interesting new perspective on photon blockade effects was put forward by~\onlinecite{Shen:PRL2007}: instead of the usual quantum optics approach of studying the optical response of a coherently driven two-level emitter, they applied Bethe ansatz techniques to solve the quantum dynamics of a two-photon state in terms of the two-photon wavefunction. In particular, they extracted from the scattering $S$ matrix interesting predictions for bunching and antibunching features in the scattered light.

\subsubsection{Resonant optical nonlinearities in multi-level systems}


It is common wisdom in nonlinear optics that the strongest optical nonlinearities are found in the vicinity of optical resonances, where unfortunately also absorption is generally large~\cite{Boyd,Butcher}. The devices in the strong light-matter coupling regime that we have mentioned so far overcome this difficulty using very narrow transitions with weak non-radiative losses. Another, different strategy is to follow the pioneering proposal in~\cite{Harris:PRL1990} based on Electromagnetically Induced transparency (EIT) media~\cite{Arimondo:1996,Fleischhauer:RMP2005}: by suitably dressing an otherwise strongly absorbing three-level medium with a coherent dressing field, a subtle destructive interference effect turns out to suppress one-photon resonant absorption, while keeping a large matrix element for the two-photon transitions that are responsible for the optical nonlinearity.

\begin{figure}
	\centering
	\psfrag{g13}{\hspace{-0.04cm}$g_{13}$}
	\psfrag{g24}{\hspace{-0.16cm}$g_{24}$}
	\psfrag{o}{$\Omega$}
	\psfrag{d}{\hspace{-0.12cm}$\Delta$}
	\psfrag{d2}{\hspace{-0.04cm}$\delta$}
	\psfrag{e}{$\varepsilon$}
	\psfrag{w1}{\hspace{0.02cm}$\omega_C$}
	\psfrag{w2}{\hspace{-0.18cm}$\omega_C$}
	\psfrag{1}{\raisebox{-0.1cm}{$1$}}
	\psfrag{2}{\raisebox{-0.1cm}{$2$}}
	\psfrag{3}{$3$}
	\psfrag{4}{$4$}
	\includegraphics[width=.7\linewidth]{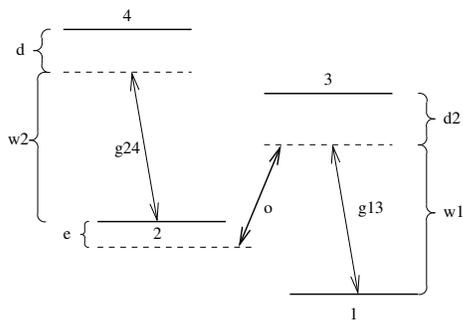}
	\caption{EIT scheme to enhance the strength of optical nonlinearity in a coherently dressed four level atom. The cavity mode at frequency $\omega_C$ is coupled to the $1 \to 3$ and $2\to 4$ atomic transitions, the coherent field dressing the atom on the $2\to 3$ transition has a Rabi frequency $\Omega$ . The parameteres $g_{13}$ and $g_{24}$ quantify the respective dipole couplings and $\delta$, $\Delta$ and $\varepsilon$ are the detunings. Figure from \cite{Hartmann:NatPhys2006}.}
	\label{fig:4level}
\end{figure}

The idea of using a single-mode cavity filled with an EIT medium was at the heart of the first proposal for photon blockade~(\onlinecite{Imamoglu:PRL97} (see also \onlinecite{Grangier:PRL1998} and \onlinecite{Imamoglu:PRL1998}). Under suitable assumptions on the light-matter coupling parameters indicated in level diagram of Fig.\ref{fig:4level}, a large on-site nonlinear interaction energy can be found in the form~\cite{Werner:PRA1999,Hartmann:PRL2007b},
\begin{equation}
U_{EIT}=-\frac{g_{24}^2}{\Delta-i\gamma_4}\frac{N g_{13}^2 \Omega^2}{\left(N g_{13}^2 + \Omega^2\right)^2}
\eqname{U_EIT}
\end{equation}
The real part of \eq{U_EIT} is responsible for the detuning of the two-photon state: the blockade effect sets in as soon as the on-site interaction energy $U$ exceeds the effective decay rate $\gamma_{EIT}$ of the filled cavity.
As it was experimentally demonstrated in~\cite{Laupretre:OptLett2011}, the dramatically reduced group velocity of EIT media is responsible for a corresponding suppression of the cavity decay rate by the same factor\footnote{{Note that the use of detuned EIT media to increase the group velocity above $c$ has a much more complicate effect on the cavity decay dynamics~\cite{Laupretre:NJP2012}.}}
\begin{equation}
\gamma_{EIT}=\frac{\gamma}{1+Ng_{13}^2/\Omega^2}.
\end{equation}

From \eq{U_EIT}, it is immediate to see that a small value of $\Delta$ is favorable to reinforce $U_{EIT}$: a naive upper limit to $U_{EIT}$ is set by the two-photon losses described by the imaginary part of \eq{U_EIT}.
This statement was partially overturned by the recent works in~\cite{Hafezi:PRA2012,Kiffner:PRA2010}: inspired to related work in the atomic context where strong correlations are induced in cold molecular gases by a strong dissipation~\cite{Syassen:Science2008,Durr:PRA2009}, it was realized that an effective blockade effect without significant losses can also be observed via a large $\textrm{Im}[U]$: the underlying mechanism is based on the so-called Zeno effect, which effectively forbids a second particle from entering a site if the two-photon loss rate exceeds tunneling. A possible implementation of this  dissipative blockade idea may involve the strong dissipative nonlinearity \eq{U_EIT} of EIT systems in the $\Delta=0$ regime, as discussed in~\cite{Harris:PRL1998}. An alternative scheme to reinforce the photonic nonlinearity was proposed in~\cite{Brandao:NJP2008}.

Even though EIT allows to suppress absorption losses, the resulting value of the resonant optical nonlinearity remains of the same order as the one of bare two-level atoms. A strategy to further increase the effective photon-photon interactions using EIT with a pair of counter-propagating dressing fields was proposed in~\cite{Andre:PRL2005}. As it was experimentally demonstrated in~\cite{Bajcsy:Nature2003}, this configuration allows one to take advantage of Bragg scattering processes to freeze light as a stationary excitation within the EIT medium while still keeping a sizable photon fraction in the polariton.

Another strategy to enhance photon-photon interactions involves the strong dipole-dipole interactions of Rydberg atoms, i.e. alkali-like atoms with a single electron promoted into a highly excited orbital~\cite{Saffman:RMP2010}: when the third state of the EIT lambda configuration consists of a Rydberg state, the presence of a single photon in the system is able to drive away from the Rydberg-EIT resonance condition all atoms contained in a surrounding volume of mesoscopic size. For large enough optical densities, a second photon traveling across this volume turns out to be absorbed and/or to suffer a sizable phase shift~\cite{Sevincli:PRL2011,Shahmoon:PRA2011,Gorshkov:PRL2011,Petrosyan:JPhys2012}, which can be exploited to obtain an effective blockade effect.

All these schemes to enhance the optical nonlinearity are however not limited to discrete geometries involving cavities: recent theoretical~\cite{Chang:NatPhys2008} and experimental works have investigated fiber geometries with a gas of atoms coupled to strongly confined optical modes propagating along the fiber with mode waists as small as 100~nm: to this purpose, hollow-core photonic crystal fibers have been used~\cite{Bajcsy:PRL2009,Venkatamaran:PRL2011}, as well as standard optical fiber strongly tapered by mechanical stretching~\cite{Vetsch:PRL2010}. 
In both cases, the nonlinear optical medium consists of atoms that are injected in the hollow core of the fiber or are optically trapped at its surface. The quite high optical depths achieved in this way have led to efficient all-optical switching at low powers using a EIT scheme~\cite{Bajcsy:PRL2009} or a two-photon absorption one~\cite{Venkatamaran:PRL2011}: the transmission of a probe beam is significantly affected by the presence of a control beam containing a number of photons on the order of a thousand (for the EIT scheme) and few tens (for the TPA scheme). 

Very recently, the large optical nonlinearity in an optically dressed dense atomic cloud in a Rydberg-EIT scheme has allowed for the first experimental claim of quantum nonlinear optics at single photon level in a propagating, cavity-less geometry: the signature of strong photon correlation is the conversion of an incident coherent laser light beam into a non-classical train of single-photon pulses~\cite{Peyronel:Nature2012}. 

\subsubsection{Feshbach blockade effect}
\label{sec:Feshbach}

\begin{figure}[t]
\begin{center}
\hspace*{0.cm} \includegraphics[height=3cm,clip]{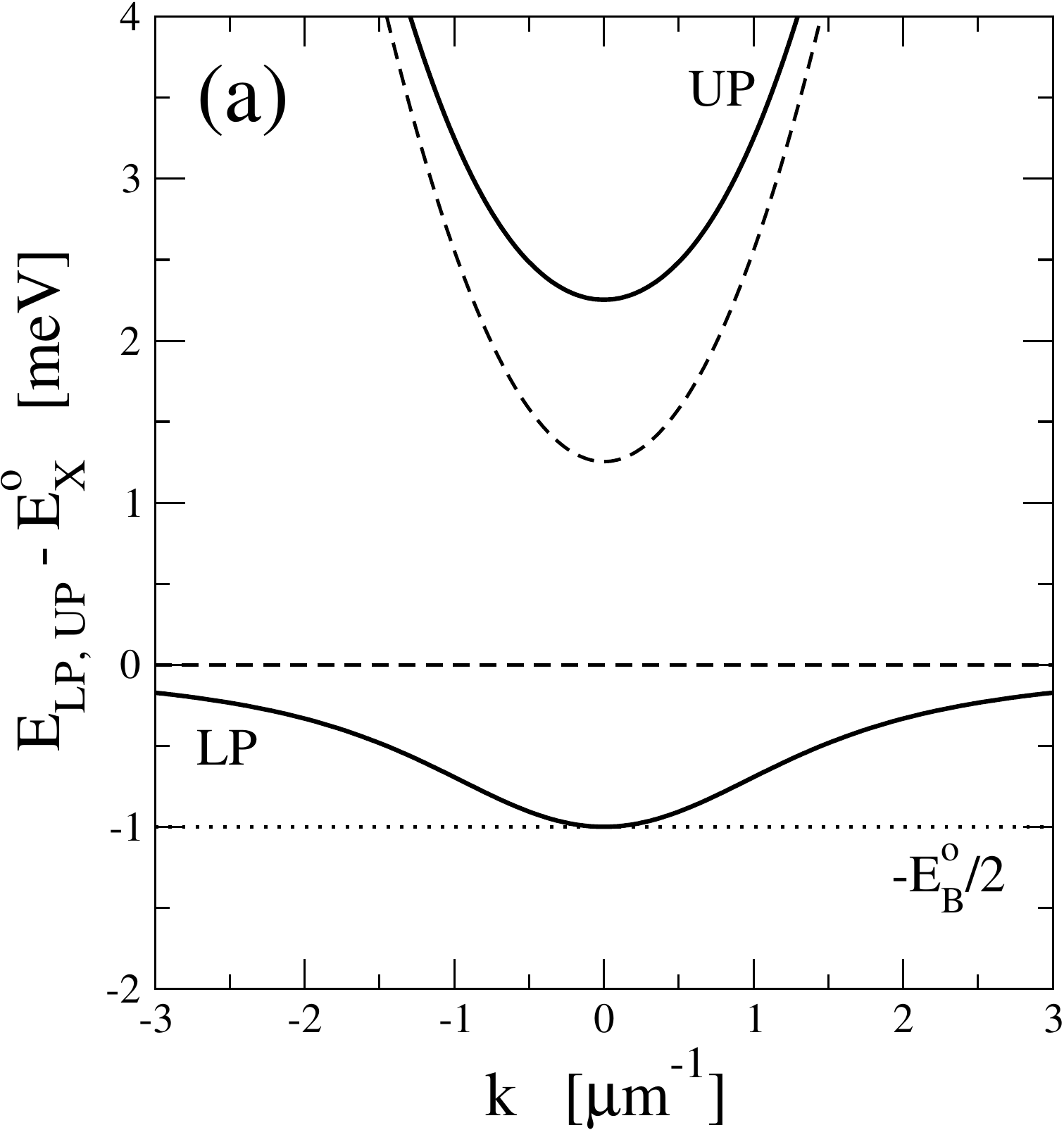}
\hspace{0.2cm} \includegraphics[height=3cm,clip]{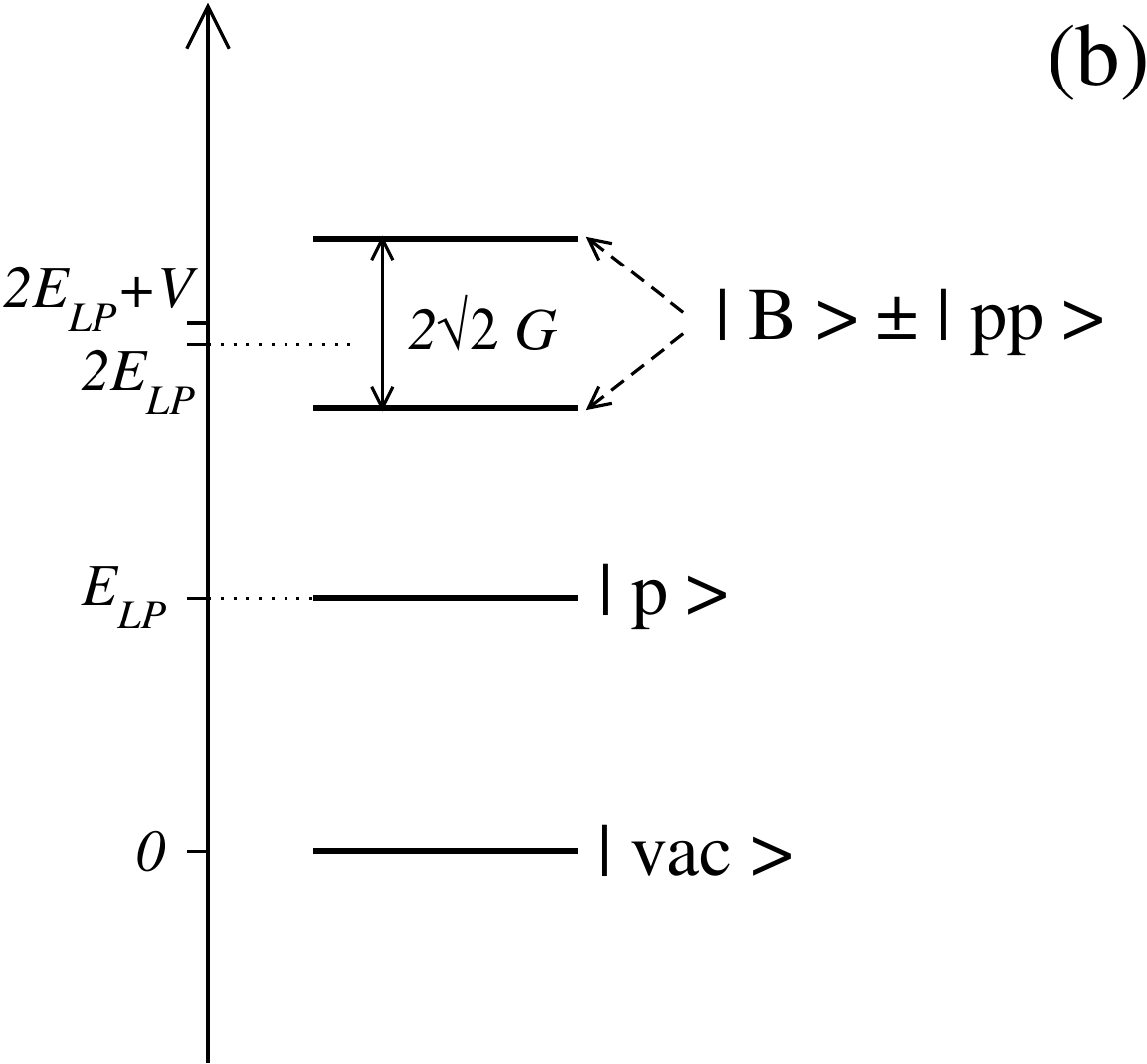}\\
\vspace*{0.2cm}
\hspace{0.2cm} \includegraphics[width=8cm,clip]{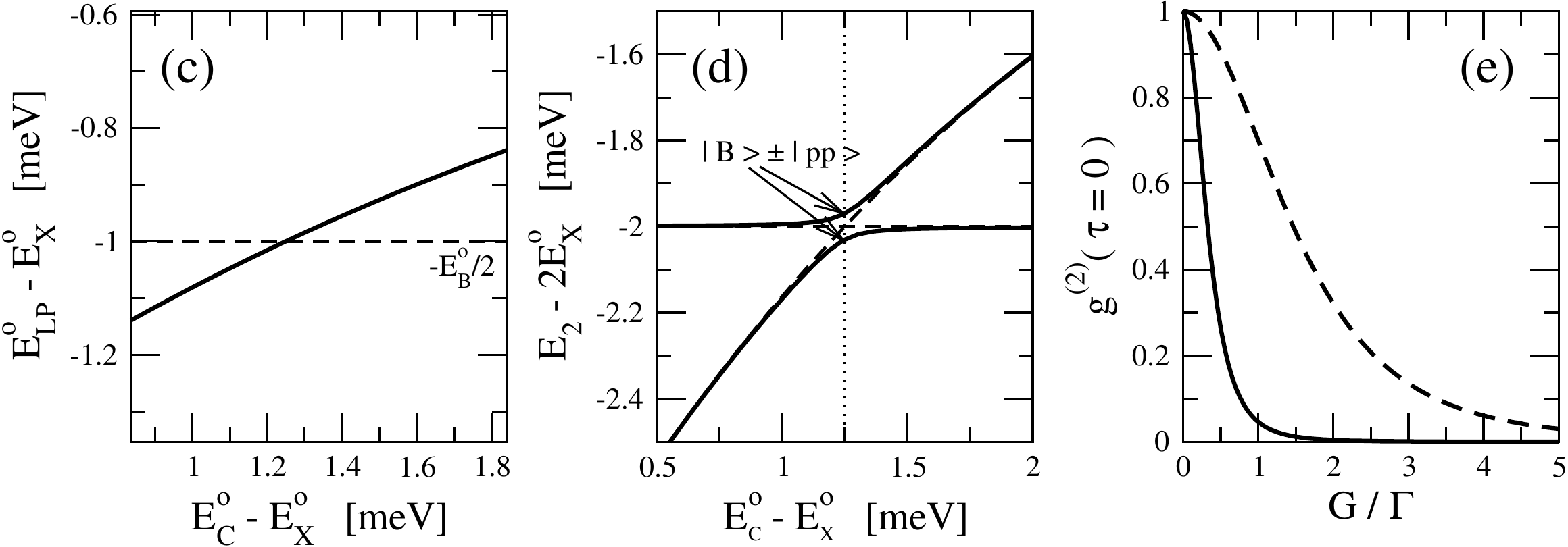}\hspace{0.5cm}
\end{center}
\caption{(a) Dispersion of the polariton frequency versus in-plane momentum; the exciton-cavity detuning is chosen such that the lower polariton is on (Feshbach) resonance with the biexciton energy (horizontal dot ted line).
Panel (b). Schematic diagram of the states belonging to the $N_{tot}=0,1,2$ manifolds exactly on Feshbach resonance
$2E_{LP}+V=E_B$. Panel (c). $k=0$ anticrossing of the lower
and upper polariton branches as a function of the bare cavity frequency
$E^o_C$. The horizontal dashed line indicates half the biexciton energy. Panel (d). Position of the different
states belonging to the $N_{tot}=2$ manifold as a function of the bare
cavity frequency $E_C^o$. Dashed line: bare $|pp\rangle$ and $|B\rangle$
states. Solid lines: states originating from the mixing of the biexciton
state $|B\rangle$ and the  $|pp\rangle$  state. The
vertical dotted line indicates the Feshbach resonance point. 
Rightmost panel (e). $\tau=0$ value of the second-order optical correlation function  $g^{(2)}(\tau)=\langle \phd(t)\, \phd(t+\tau)\, \ph(t+\tau)\, \ph(t) \rangle$ as a function of the exciton-biexciton coupling $G$ for a system on Feshbach resonance. Parameters:
$E_X^o=1.4$~meV; $\hbar \Omega_R=1.5$~meV; $G=0.03$~meV. 
Figure from~\cite{Carusotto:EPL2010}}
\label{fig:Feshbach}
\end{figure}

A great deal of the recent advances in  the field of strongly correlated atomic gases have been made possible by the so-called Feshbach resonance effect in atom-atom collisions~\cite{Chin:RMP2010}: the scattering cross section is dramatically enhanced when the energy of two colliding atoms is resonant with a long-lived quasi-bound molecular state. In typical experiments, an external magnetic field is used to tune the energy of the quasi-bound molecular state close to the energy zero of scattering states. At this point, the low-energy scattering amplitude diverges and a dilute ultra-cold atomic gas has the strongest possible interaction. Remarkable recent experiments with atomic Fermi gases have exploited such Feshbach resonances to demonstrate superfluid behavior at temperatures comparable to the Fermi temperature~\cite{Giorgini:2008RMP,Bloch:review2008}.

In the wake of the pioneering attempts~\cite{Savasta:SSC1999,Savasta:PRL2003}, a recent work~\cite{Wouters:PRB2007} has pointed out the possibility of exploiting an analogous Feshbach resonance effect on an intermediate biexciton state to enhance the optical nonlinearity of two-dimensional polaritons in planar microcavities. Biexcitons are two electron-two hole bound complexes analogous to hydrogen molecules in semiconducting materials and have been widely studied during the last decades in both bulk and confined geometries~\cite{Ivanov:PhysRep1998}. As it is described in Fig.\ref{fig:Feshbach}, a careful tuning of the length of a planar microcavity in the strong coupling regime allows to bring the energy of a pair of polaritons at the bottom of the lower polariton branch close to resonance with the biexciton state. This resonance is then responsible for a dramatic enhancement of the low-energy scattering cross-section for polariton-polariton collisions in the singlet channel. An analytic model of the biexciton Feshbach resonance was put forward in~\cite{Carusotto:EPL2010}, where a method inspired by the  atom-molecule approach to strongly interacting degenerate quantum gases~\cite{Heinzen:PRL2000,Kokkelmans:PRA2002} is used to obtain a approximate closed formula for the polariton scattering $T$ matrix element in the singlet channel,
\begin{equation}
T_{\uparrow\downarrow}(E)=\frac{|\bar{g}|^2}{E-E_B^o+\frac{m_{LP}}{4\pi\hbar^2}\,|\bar{g}|^2\,\log[\frac{E_{max}}{E}]+i\frac{m_{LP}}{4\hbar^2}\,|\bar{g}|^2}
\eqname{T}
\end{equation}
where $E$ is the collision energy, $E_B^o$ is the (bare) biexciton energy,  $\bar{g}$ is the polariton-biexciton coupling coefficient, and $E_{max}$ is a UV cut-off energy accounting for the microscopic details of the system. On physical grounds, the coupling coefficient $\bar{g}$ is expected to be related to the exciton-photon Rabi frequency $\Omega_R$ and the exciton Bohr radius (or, equivalently, the biexciton size) $a_B$ by $\bar{g}\approx \hbar \Omega_R\,a_B$. As usual in many-body theory~\cite{Petrov:PRL2000}, the biexciton contribution to the polariton-polariton interaction constant in a spatially extended two-dimensional geometry is obtained from the $T$-matrix \eq{T} as the amplitude for forward scattering at an energy $E$ determined by the oscillation frequency of the Bose field.

The same biexciton model~\cite{Carusotto:EPL2010} provides an estimation of the single-polariton nonlinearity coefficient in tightly confined geometries: the predicted efficiency of this {\em Feshbach blockade} effect to generate a non-classical stream of anti-bunched photons from realistic structures is illustrated in Fig.\ref{fig:Feshbach}(e). Exactly on Feshbach resonance, one can expect an effective detuning of the two-excitation state on the order of $\hbar \Omega_R a_B /L$ (note the different scaling on the characteristic spatial size $L$ of the cavity mode as compared to \eq{U_nonresonant}). Alternatively, additional decay channels of the biexciton state other than the decay into a pair of cavity polaritons can be exploited to obtain an efficient dissipative blockade effect.
As the biexciton mass is not renormalized by the coupling to the photon, a possible difficulty is the sizable inhomogeneous broadening of the biexciton energy due to short-distance fluctuations of the quantum well thickness.


\subsubsection{Unconventional photon blockade in photonic molecules}
\label{sec:bamba}

The discussion of the previous sections pointed out the occurrence of effective photon blockade effects as soon as the optical nonlinearity $U$ (or the vacuum Rabi frequency $g$ in a Jaynes-Cummings system) largely exceeds the loss rate $\Gamma$. This is a sufficient condition to observe strong antibunching in the transmitted light, but actually it is not strictly necessary.

This remarkable fact was first investigated in the field of cavity quantum electrodynamics~\cite{Carmicheal:PRL1985,Carmichael:OptCommun1991}: as a result of a peculiar quantum interference process, a sizable antibunching of transmitted light is predicted to be observed from cavities containing a large number of emitters with moderate values of the single emitter-cavity coupling. Experimental studies along these lines appeared in~\cite{Rempe:PRL1991,Foster:PRA2000}.

Very recently, \onlinecite{Liew:PRL2010} reported a numerical study of a driven-dissipative two-site Bose-Hubbard system, where a resonantly driven cavity is coupled by tunneling events to a non-pumped auxiliary cavity. The remarkable result of this work is that a strong antibunching  ($g^{(2)}(0) \ll 1$) can be obtained with a weak Kerr nonlinearity $U \ll \gamma$. The strongest antibunching is achieved for optimal finite values of the on-site repulsion energy $U$ and of the detuning between the pump and the cavity mode frequency. Independently from the previous work in cavity-QED, the origin of the strong antibunching was traced back to a sort of quantum interference effect.

\begin{figure}[t]
\begin{center}
\includegraphics[width=\columnwidth,angle=0,clip]{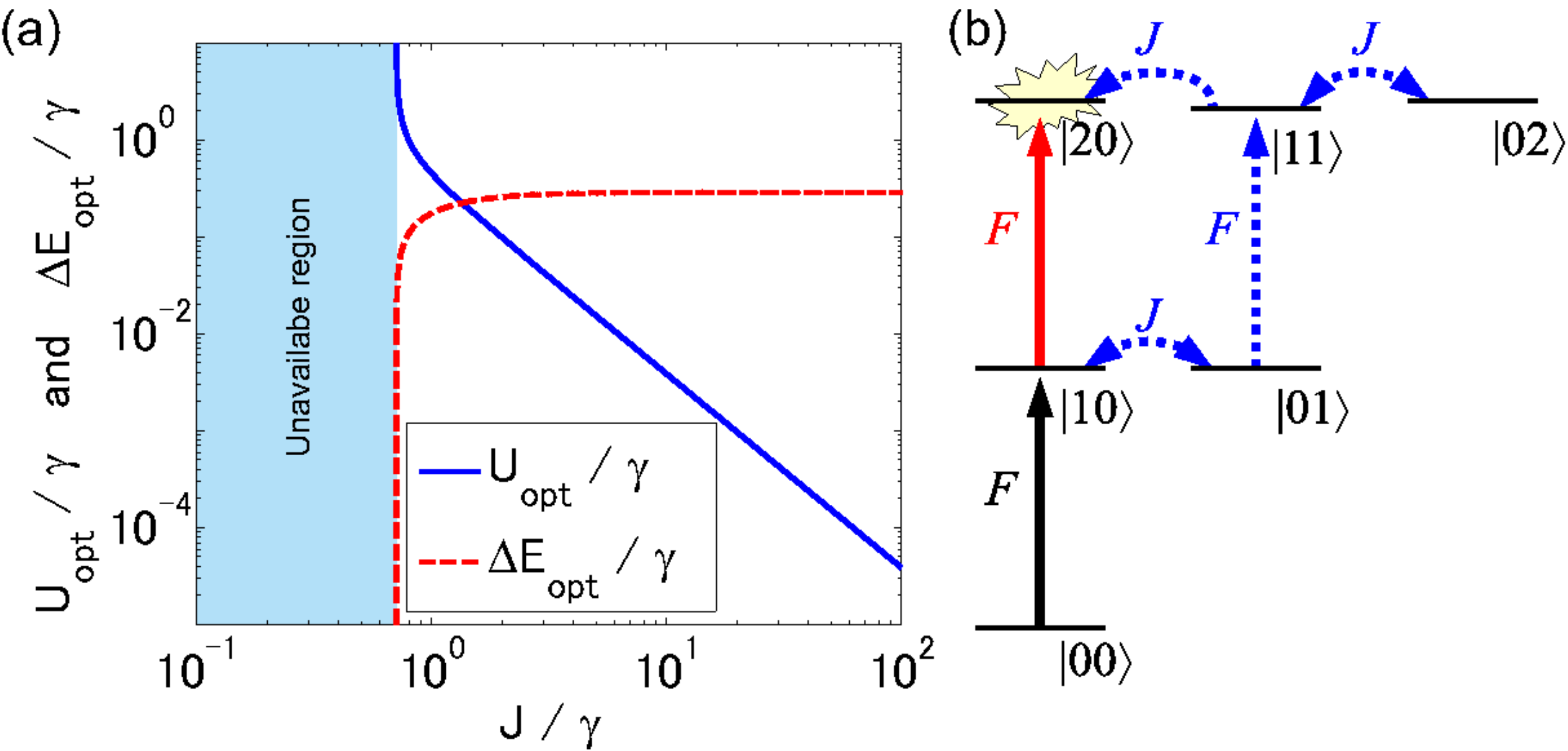} 
 \end{center}
\caption{(a) Optimal nonlinearity $U_{\text{opt}}$ and detuning $\Delta E_{\text{opt}}$ for
a driven-dissipative two-site Bose-Hubbard system are plotted
versus the inter-mode coupling strength $J$ normalized to mode broadening $\gamma$
($\gamma_1 = \gamma_2 = \gamma$ and $E_1 = E_2 = E$).
(b) Transition paths leading to the quantum interference responsible for the strong antibunching.
 From~\cite{Bamba:PRA2011}.}
\label{fig:bamba}
\end{figure}

The underlying physical mechanism was analytically unveiled in a following work~\cite{Bamba:PRA2011} using a wavefunction amplitude approach, which is exact in the limit of weak pump limit.
Such an approach is based on the following ansatz
\begin{multline}
\ket{\psi}
= C_{00}\ket{00} + \ee^{-\ii\wp t}\left( C_{10}\ket{10} + C_{01}\ket{01}\right)
+  \\ 
+\ee^{-\ii2\wp t}\left(C_{20}\ket{20} + C_{11}\ket{11} + C_{02}\ket{02}\right)
+ \ldots,
\end{multline}
to calculate the steady-state of the coupled cavity system, where $\omega_p$ is the pump frequency driving the first cavity. Here,
$\ket{mn}$ represents the Fock state with $m$ particles in the first cavity site and $n$ particles in the auxiliary cavity (second site). Under weak pumping conditions ($C_{00}
\gg C_{10}, C_{01} \gg C_{20}, C_{11}, C_{02}$), 
it is possible to determine the coefficients $C_{mn}$ in a iterative fashion
\cite{Carmichael:OptCommun1991}.

Remarkably, there exist values of the parameters for which one has $C_{20} = 0$, i.e. a vanishing probability of having $2$ photons in the first cavity: this is indeed what is required for an ideal photon blockade.
The analytical solutions of this condition have a particularly transparent form for equal energies and loss rates of the two cavity modes  ($E_1 = E_2 = E$; $\gamma_1 = \gamma_2 = \gamma$) and for $J \gg \gamma$. The condition $C_{20} = 0$ is satisfied when the on-site interaction $U$ and the pump detuning $\Delta E = E - \hbar \omega_p$ 
take the following optimal values:
\begin{equation}
\DE_{\text{opt}}  \simeq \frac{\gamma}{2\sqrt{3}} \hspace{0.7cm} \textrm{and} \hspace{0.7cm}
U_{\text{opt}} \simeq \frac{2}{3\sqrt{3}}\frac{\gamma^3}{J^2}.
\end{equation}
For $J\gg \gamma$, it directly implies that $U_{\text{opt}}/ \gamma \ll 1$, i.e., the optimal value of the nonlinear interaction energy is much smaller than the loss rate. 

It is worth mentioning that this dramatic release of the constraint on the $U/J$ parameter to observe photon blockade comes at the expense of fast oscillations of the photon correlation function $g^{(2)}(\tau)$ as a function of the delay time $\tau$: the characteristic scale of these oscillation is set by the tunneling rate $J\gg \Gamma$. As a result, photodetectors with a sufficient time resolution are required to observe the antibunching effect.

This unexpected photon blockade effect can be interpreted as arising from the destructive quantum interference between the amplitude of two transition paths having as initial state the vacuum in both cavity sites and $\ket{20}$ as final state, as shown in Fig.\ref{fig:bamba}. 
The interference is between the following two paths: (a) the direct
excitation from $\ket{10} \xrightarrow{F} \ket{20}$ (solid arrow), meaning that
two photons are sequentially injected by the resonant driving field in the first cavity.
(b) tunnel-coupling-mediated transition $\ket{10}
\overset{J}{\leftrightarrow} \ket{01} \xrightarrow{F} (\ket{11}
\overset{J}{\leftrightarrow} \ket{02}) \xrightarrow{J} \ket{20}$
(dotted arrows). For an optimal value of the nonlinearity and an optimal value of the pump frequency detuning, the two above mentioned transition paths have the same amplitude and opposite phases, leading to the destructive interference. 

It is worth noting that the nonlinearity $U_1$ of the pumped cavity mode is irrelevant for the antibunching. This implies that a finite, but weak nonlinearity is required only in the auxiliary (undriven) photonic mode to achieve the perfect destructive quantum interference that leads to $C_{20}=0$. In the same paper~\cite{Bamba:PRA2011} it was shown that an analogous effect also occurs for a two-site Jaynes-Cummings-Hubbard system: in contrast to the usual single-site photon blockade effect where a large vacuum Rabi frequency is needed \cite{Birnbaum:Nature2005,Lang:PRL2011}, the two-site geometry allows for a strong antibunching $g^{(2)}(0) \simeq 0$ even when $g/ \gamma \sim 1$. A related proposal has recently appeared in~\cite{Majumdar:PRL2012}.

This paradigm of photon blockade with weak nonlinearities has been extended to the case of cavities with polarization splittings in~\cite{Bamba:APL2011}. The additional degree of freedom provides a way to tailor such quantum interference photon blockade effects. In particular, using a polarized pump driving the first cavity, it is possible to get photon blockade for the counter-polarized mode in the auxiliary cavity, thus providing an efficient way to filter spurious pump photons through spatial and polarization filters. 

Given the novel physical mechanism underlying antibunching in weakly non-linear photonic molecules, a priori it is not  clear whether the predicted antibunching in a single cavity geometry may produce strong quantum correlations in an array of coupled photonic molecules. This important question was addressed in~\cite{Bamba:PRA2011}: a geometry consisting of ring of three photonic molecules whose the driven cavities are coupled with each 
other by a tunnel coupling. For a suitable choice of parameters, a nearly perfect antibunching could be revealed also in this case.

\subsection{Lossless many-photon gases at (quasi) equilibrium}

The first theoretical investigations of many-photon effects in arrays of coupled cavities appeared almost simultaneously in~\cite{Hartmann:NatPhys2006,Greentree:2006,Angelakis:PRA2007}. 
The basic idea of all these works closely follows classical approach of many-body physics with material particles~\cite{Fisher:PRB1989}, as recently anticipated and then observed with ultracold atoms~\cite{Jaksch:PRL98,Greiner:Nature2002}: depending on the ratio between tunneling $J$ and  interaction $U$ energies in the Bose-Hubbard Hamiltonian $\mathcal{H}_{BH}$ in \eq{BH}, the ground state of the system exhibits a quantum phase transition from a superfluid state \eq{SF} with spatially extended coherence to a Mott insulator state \eq{MI} with no coherence and suppressed number fluctuations on each site. 

In the first work~\cite{Hartmann:NatPhys2006}, the optical nonlinearity required for the blockade effect is provided by a gas of optically driven four-level atoms in an Electromagnetically Induced Transparency scheme inserted in the cavities as illustrated in Fig.\ref{fig:4level}; in the two others~\cite{Greentree:2006,Angelakis:PRA2007}, a Jaynes-Cummings-Hubbard model is considered with a simple two-level emitter strongly coupled to each cavity mode. 
The envisaged protocol is to tune the system in the deep Mott phase at large $U/J$ and prepare the system in a state with a single photon per site. Then, $U/J$ is adiabatically ramped down into the superfluid phase so that the system remains at all times in its ground state. A crucial advantage of the photon system over the atomic one is that the onset of the phase transition can be assessed {\em in situ} e.g. by measuring the number of photons in each cavity and its fluctuations. However, in all these works the photon gas is considered as a quasi-closed system, where loss processes are just an hindrance limiting the available time for the adiabatic ramp. 

A different geometry was considered in~\cite{Chang:NatPhys2008}: slow photons in a Electromagnetically Induced Transparency regime propagate along a one-dimensional fiber. By tuning the intensity of the EIT dressing beams, one can tune the strength of the effective repulsive interactions between the photons, as well as reduce their propagation speed along the fiber to zero. For sufficiently strong interactions, the stored photons within the fiber behave as a Tonks-Girardeau gas of impenetrable particles~\cite{Girardeau:1960}. As it was demonstrated in~\cite{Bajcsy:Nature2003}, the use of a pair of counterpropagating EIT beams allows to tune also the photon propagation speed along the fiber and to reduce it to zero. 

The experimental protocol that is proposed in~\cite{Chang:NatPhys2008} is then to inject a stream of fast and coherent photons into the fiber in a weakly interacting state. Once photons are loaded, the propagation speed is reduced to zero so that the gas of photons gets stored within the fiber. Simultaneously, the interaction constant is adiabatically ramped up, possibly achieving the Tonks-Girardeau regime. As in the previously quoted work, the ramps in the interaction constant and/or the photonic mass have to be performed slow enough for the gas to remain with good fidelity in the ground state. An exciting advantage of the fiber geometry is that the {\em in situ} density correlation function of the photon gas within the fiber can be measured by switching the photons again into a propagating state and detecting the arrival times on an external photodetector.

Following these pioneering proposals, a number of other works have appeared discussing the possibility of studying more complex many-body effects in photon gases, e.g. the mapping of spin models onto photonic systems~\cite{Hartmann:PRL2007}, the emergence of Bose glass phases due to the interplay of strong interactions with disorder~\cite{Rossini:PRL2007} (a different mean-field approach to Bose glass phases in weakly interacting polaritons systems was reported in~\onlinecite{Malpuech:PRL2007}), spin-charge separation effects in two-component photon gases (distinguished by their polarization state or their frequencies)~\cite{Angelakis:PRL2011}, a kind of BEC-BCS crossover~\cite{Huo:arXiv2011b}, fractional quantum Hall states~\cite{Cho:PRL2008}. A recent review of this rich literature can be found in~\cite{Hartmann:Laser2008,Angelakis:arXiv2012}. In all cases, a quasi-equilibrium picture was used, where the photon gas is modeled as a closed and conservative system and the standard concepts of many-body physics apply, as in liquid Helium, electronic systems and atomic gases.

\subsection{Non-equilibrium strongly correlated photon gases: Josephson interferometers and Tonks-Girardeau gases}
\label{sec:S-I-TG}

\begin{figure}[t!]
\begin{center}
\parbox[c]{3.51cm}{\includegraphics[width=3.cm,angle=0,clip]{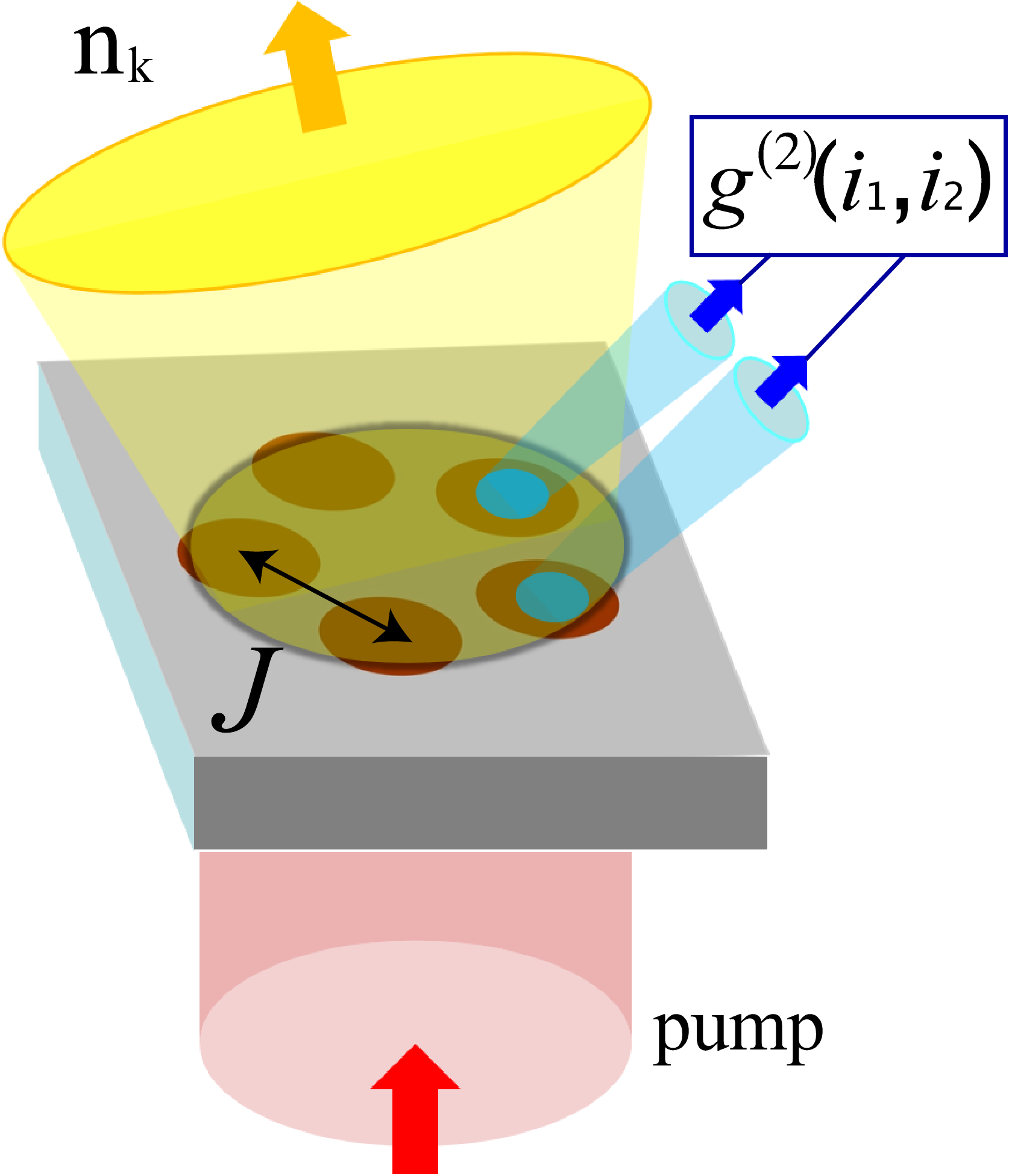}}
\parbox[c]{4.61cm}{\includegraphics[width=5cm,angle=0,clip]{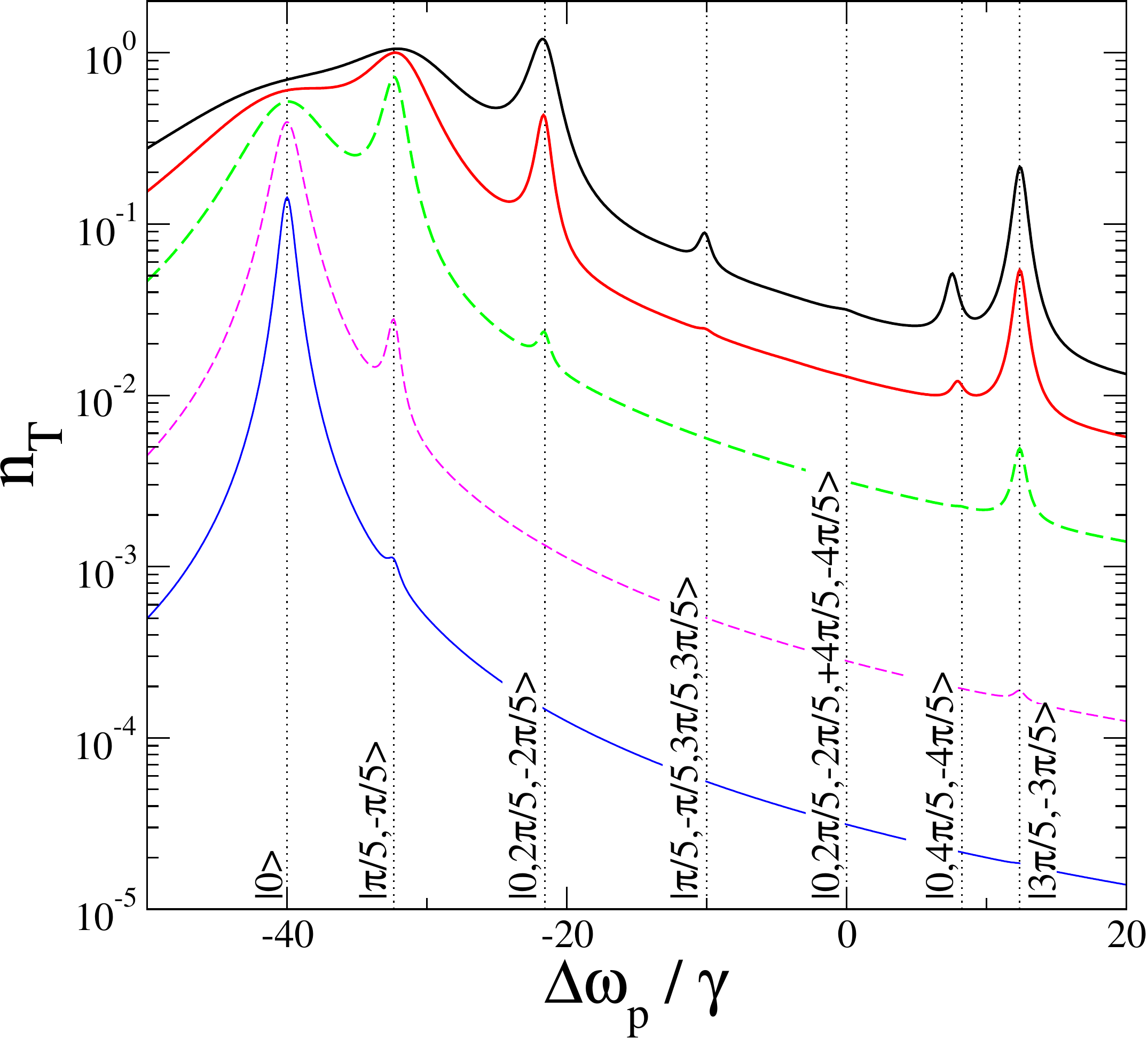}}\\
\includegraphics[width=0.95\columnwidth,angle=0,clip]{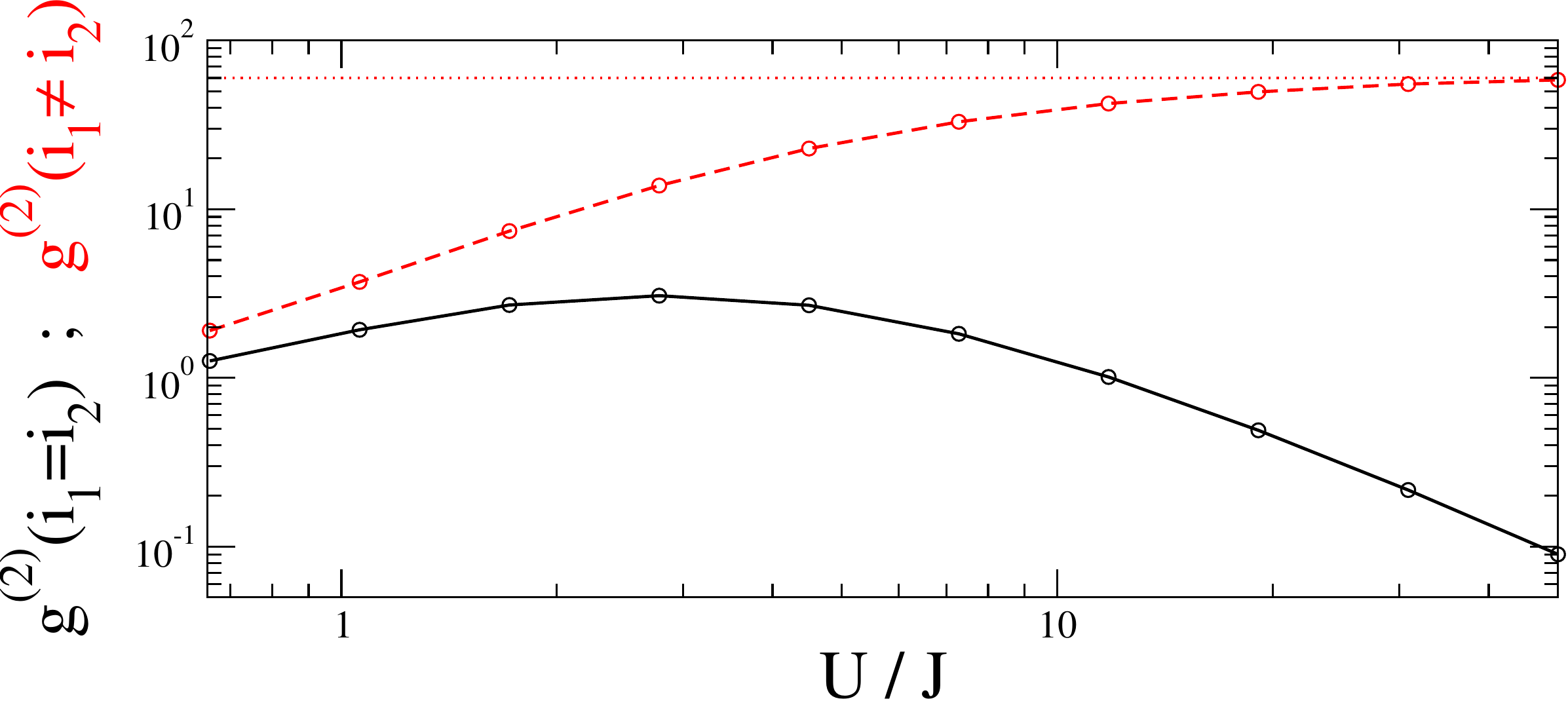}
 \end{center}
\caption{
Upper-left panel: a sketch of an array of $5$ nonlinear cavities in a ring geometry with periodic boundary conditions. The system is driven by a coherent laser and the properties of the strongly correlated photon system can be monitored, for example, by the spatial-dependent second-order coherence function $g^{2}$. Upper-right panel: steady-state mean number of photons as a function of the pump detuning $\Delta \omega_p = \omega_p-\omega_0$ in the impenetrable boson limit $U/J \to + \infty$. The different curves correspond to increasing values of the pump amplitude $F_p/\gamma = 0.1, 0.3, 1, 2, 3$. The vertical dotted lines indicate the spectral positions of the peaks predicted by the fermionization procedure with different number of photons.
The kets $\vert q_1 .... q_N \rangle$ represent the pseudo-momenta $q_1...q_N$ of the occupied fermionic orbitals according to Girardeau's the Bose-Fermi mapping~\cite{Girardeau:1960}.
Lower panel: two-photon correlation signal for light emission by a single site (black) and by different sites (red) when the interaction parameter is ramped from the weakly- to the strongly-interacting regime. For each value of $U/J$, the pump frequency is chosen on resonance with the lowest energy two-particle state.
Figure from \onlinecite{Carusotto:PRL2009}.}
\label{fig1PRL_TG}
\end{figure}

In the last decades, an intense activity has been devoted to the study of the physics of system far from thermodynamical equilibrium~\cite{Ruelle:2004}. In particular, a richer variety of many-body behaviors and phase transitions was predicted and in some case observed as a result of the interplay of driving, dissipation and interactions~\cite{Mukamel:1999}.
In this context, a special was recently devoted to the engineering of open systems of ultracold atoms and ions for quantum simulations of equilibrium and non-equilibrium quantum phases~\cite{Muller:arXiv2012}.

While the driven-dissipative nature of optical systems is naturally taken into account in the quantum optical literature~\cite{Carmicheal_book}, most of these works have focused their attention on few mode systems: in spite of the very restricted spatial dynamics, a rich wealth of quantum optical features was already predicted and observed. Exception are perhaps the pioneering literature on quantum solitons~\cite{Lai:PRA1989,Lai:PRA1989b,Kaertner:PRA1993,Drummond:Nature1993} and the on-going research on spatially multi-mode squeezing effects and the consequent possibility of sub-shot noise measurement of nano-displacements~\cite{Treps:Science2003,Treps:JOptB2004}.

The framework is even richer when the spatially extended geometry is combined with strong photon-photon interactions. A pioneering theoretical investigation of this physics appeared in~\cite{Gerace:NatPhys2009} for a simplest geometry consisting of a string of a few single-mode cavities coupled by tunneling processes: the strong on-site interactions result in strong photonic correlations in the non-equilibrium steady-state of the system, which are visible through the suppression of Josephson-like oscillations, as well as through the photon statistics of emitted light.

A different point of view on non-equilibrium strongly correlated photon gases was proposed in~\cite{Carusotto:PRL2009}, where an array of many cavities in the impenetrable photon regime under a coherent, continuous-wave pumping was considered, as sketched in the upper-left panel of Fig.\ref{fig1PRL_TG}.
Unambiguous signatures of the Tonks-Girardeau nature of the steady state were identified in observables as simple as the transmission or absorption spectrum of the device: each of the resonant peaks that are visible in the spectra in the upper-right panel of Fig.\ref{fig1PRL_TG} corresponds indeed to a different strongly correlated many-body state of the impenetrable photon gas. The labels on the vertical lines indicate the structure of each Tonks-Girardeau state in terms of the corresponding Fermi orbitals according to Girardeau's Bose-Fermi mapping~\cite{Girardeau:1960,Cazalilla:RMP2011}.

For a pump frequency on resonance with a given many-body eigenstate, the microscopic properties of the corresponding many-body wavefunction can be inferred from the coherence functions of the secondary emitted light. An example is shown in the lower panel of Fig.\ref{fig1PRL_TG} for a pump resonant on the lowest two-body state of a 3 sites lattice: depending on the ratio $U/J$, one moves from a weakly interacting regime where the two-body correlation functions on the same site and for different sites coincide, to a regime where strong on-site anti-bunching coexists with strong bunching on distinct sites. In~\cite{Umucalilar:PRL2012}, this same approach was applied to fractional quantum Hall states of impenetrable photons in the presence of an artificial gauge field.

Different aspects of the steady state of an array of nonlinear and lossy optical resonators driven by coherent lasers were addressed in~\cite{Hartmann:PRL2010}. In particular, for weak driving intensity the steady state was shown to be dominated by interactions in such a way that photon crystallize into dimers localized on neighboring sites and anticorrelations appear between distant sites. Much of these results were obtained by applying for the first time to the optical context the driven-dissipative version~\cite{Zwolak:PRL2004,Verstraete:PRL2004} of the novel numerical techniques based on the time evolving block decimation algorithm~\cite{Schollwock:RMP2005}. More extended studies along these lines have appeared in~\cite{Kiffner:NJP2011,Leib:NJP2010}. A detailed comparison of non-equilibrium many-photon physics in Bose-Hubbard and Jaynes-Cummings-Hubbard models was recently reported in~\cite{Grujic:arXiv2012} using sophisticated numerical techniques.

Quantum transport of strongly interacting photons along a continuous one-dimensional fiber was studied in~\cite{Hafezi:EPL2011,Hafezi:PRA2012}. In contrast to most previous work, photons are assumed to freely move along the fiber in the absence of any lattice potential: the theoretical description is carried out in a first-quantization picture based on a set of $n$-photon wavefunctions $\psi_n$. For each value $n$ of the total photon number, the corresponding wavefunction $\psi_n(x_1,\ldots,x_n)$ is a function of the $n$ photon coordinates $x_1,\ldots,x_n$ and is connected to the neighboring ones $\psi_{n\pm 1}$ by terms describing the photon injection and photon loss processes. In the present fiber geometry, these processes reduce to suitable boundary conditions at the two ends of the fiber.
The significant energy shift of the two particles states due to interactions is responsible for significant bunching and anti-bunching effects depending on the sign of interactions and the pump laser detuning. For attractive interactions, propagating two-photon bound states can be observed on the transmission resonances.

\subsection{Topological states and artificial gauge fields}
\label{sec:gauge}

The motion of charged particles in magnetic fields is a key paradigm of quantum mechanics and underlies a number of intriguing phenomena in very different contexts, from magnetohydrodynamics in astro- and geophysics to the fractional quantum Hall effect in solid-state physics~\cite{DasSarma,Yoshioka}: in all these systems, the charged particles are subject to a real magnetic field generated, e.g., by an electric current or by permanent magnets. An interesting development of the geometric ideas underlying the theoretical explanation of the quantum Hall effect has led to the discovery of a number of other materials with exotic properties due to the topological structure of the electron bands, the so-called topological insulator and topological superconductors~\cite{Qi:RMP11}.

Independently from these advances of solid state physics, a related concept of gauge structure was introduced in the concept of molecular physics~\cite{Mead:JChemPhys1979,Mead:RMP1992,Resta:JPhys2000}: the geometric phase that may appear in Born-Oppenheimer calculations as a result of an adiabatic motion across configuration space can be described in terms of an effective (abelian or non-abelian) gauge field. From a more abstract point of view, this is an example of the Berry phase (rotation) that is accumulated by a quantum system during an adiabatic sweep of some external parameter~\cite{Berry:PRSL1984} and can be described in terms of a $U(1)$ (or higher) gauge field~\cite{Wilczek:PRL1984}.

This idea was transposed to the field of atomic physics in~\cite{Dum:PRL1996}: when an atom is coherently dressed by a spatially and/or temporally varying optical and/or magnetic field, the adiabatic motion of its dark states experiences an effective gauge field due to the Berry phase. The high flexibility of atomic systems gave a novel twist to this research, as it became possible to engineer {\em artificial (or synthetic) gauge fields} for neutral atoms with almost arbitrary shapes~\cite{Dalibard:RMP2011}. An alternative way of studying this physics involves rapidly rotating atomic clouds~\cite{Cooper:AdvPhys2008}: modulo the centrifugal force, the Coriolis force that appears in the rotating frame of reference is in fact mathematically equivalent to a Lorentz force in a uniform magnetic field.

A striking experimental implementation of these ideas was reported in~\cite{Lin:Nature2009}, where the nucleation of a few quantized vortices in a Bose-Einstein condensate under the effect of an artificial gauge field was demonstrated. More recent experimental investigations have addressed the synthetic electric field generated by a time-dependent synthetic magnetic field~\cite{Lin:NatPhys2011}, the atomic bands in an optical lattice under a strong magnetic field~\cite{Aidelsburger:PRL2011}, a superfluid version of the Hall effect~\cite{LeBlanc:arXiv2012}, and a universal method to create artificial gauge fields by driving in time the lattice potential~\cite{Struck:PRL2012}. When the artificial gauge field is combined with strong atom-atom interactions, the atomic gases is expected to form strongly correlated gases that closely remind quantum Hall liquids, see e.g.~\cite{Cooper:PRL2001,Chang:PRA2005,Hafezi:PRA2007,Palmer:PRA2008,Umucalilar:PRA2007}.

\begin{figure*}[t]
\includegraphics[width=0.24\columnwidth,angle=0,clip]{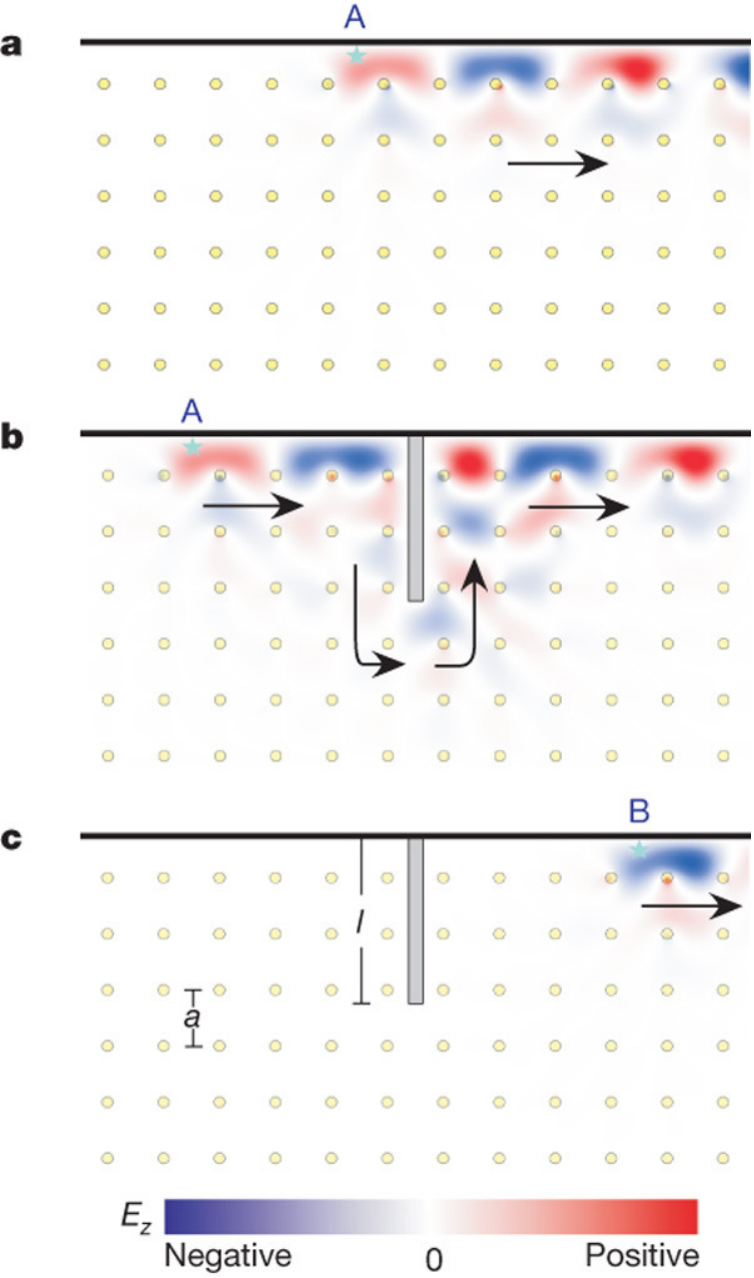}
\includegraphics[width=0.3\columnwidth,angle=0,clip]{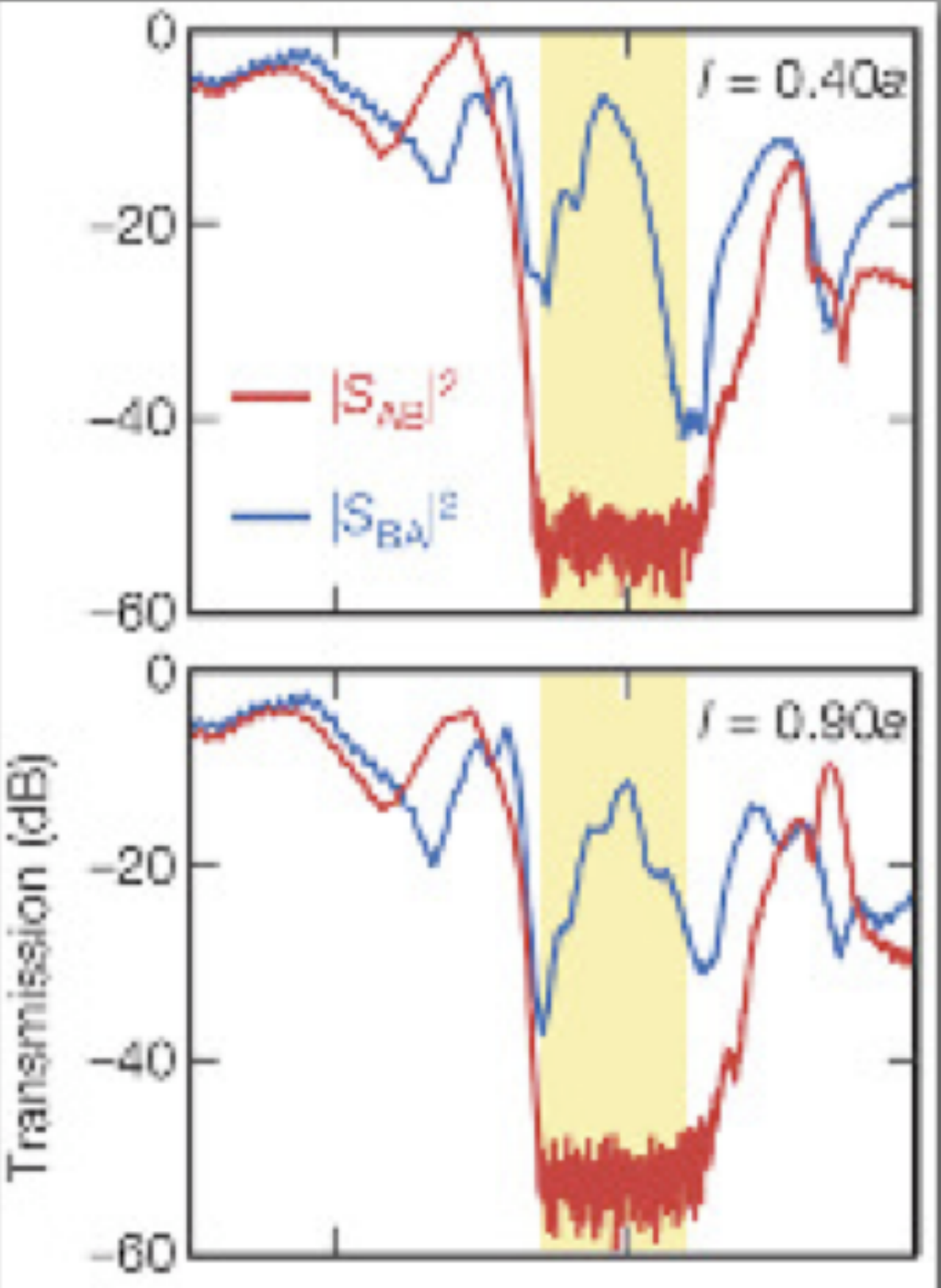}
\hspace{0.05\columnwidth}
\includegraphics[width=.50\columnwidth]{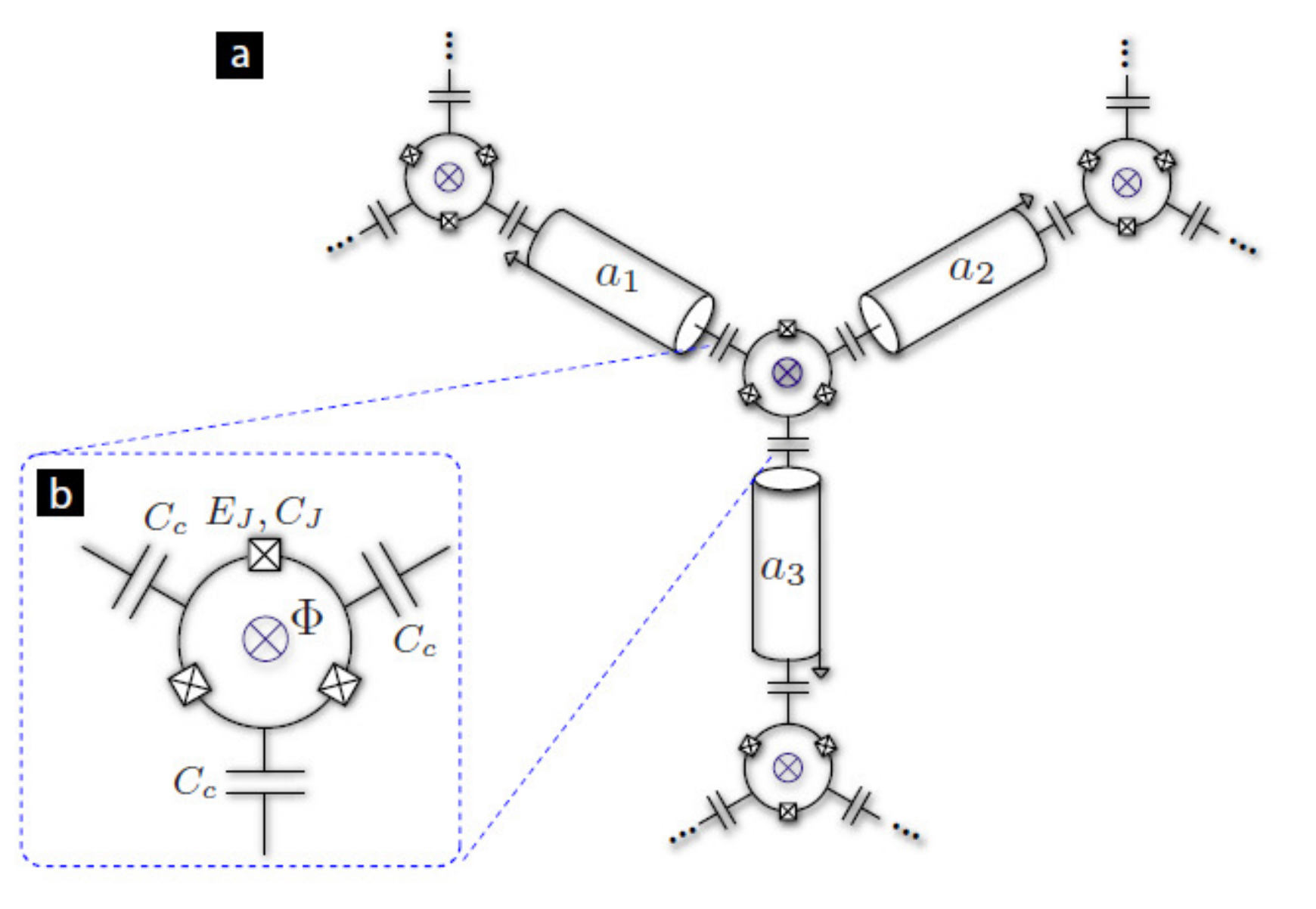}
\hspace{0.0\columnwidth}
\includegraphics[width=.45\columnwidth]{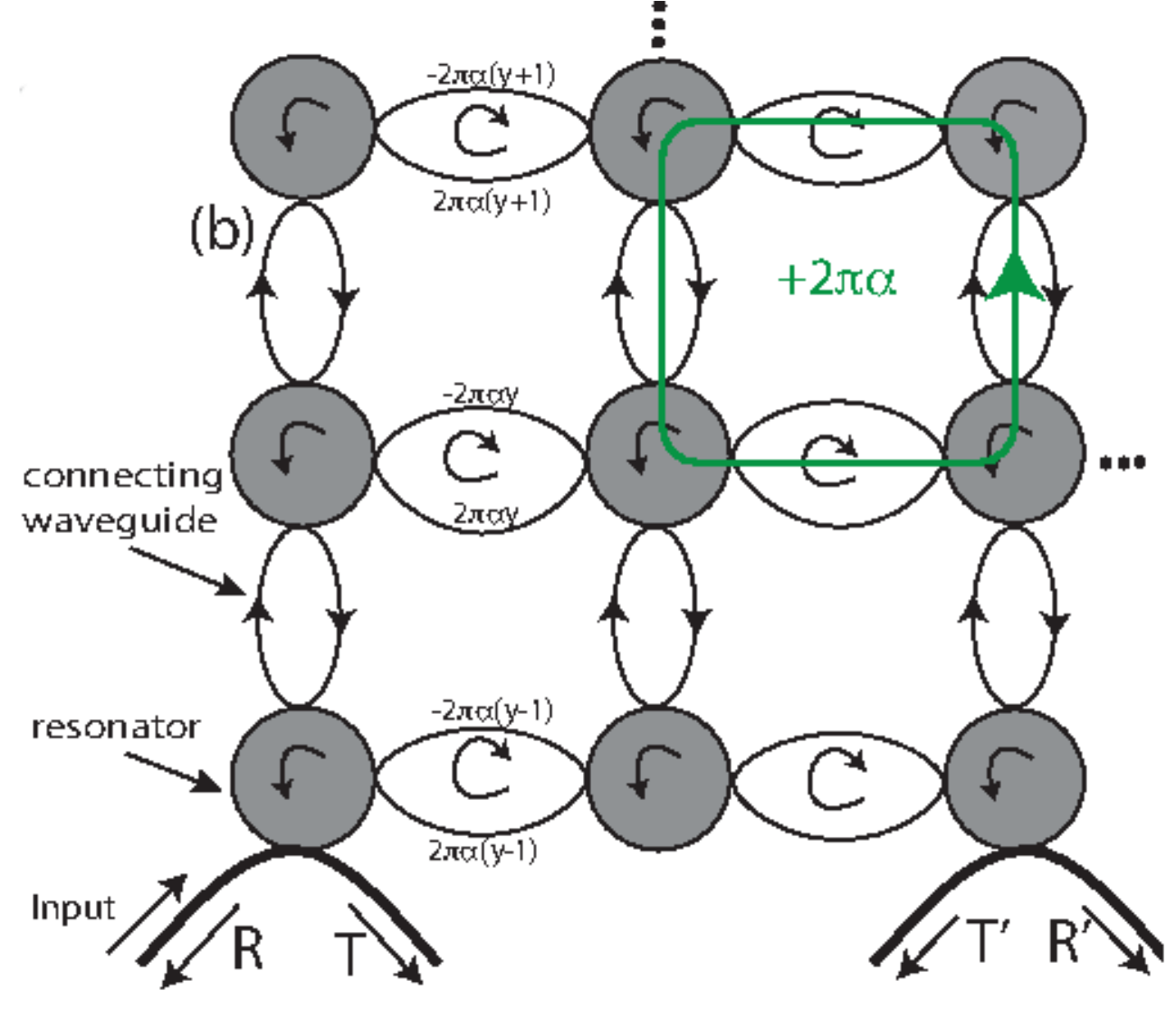}
\hspace{0.0\columnwidth}
\includegraphics[width=.38\columnwidth,clip]{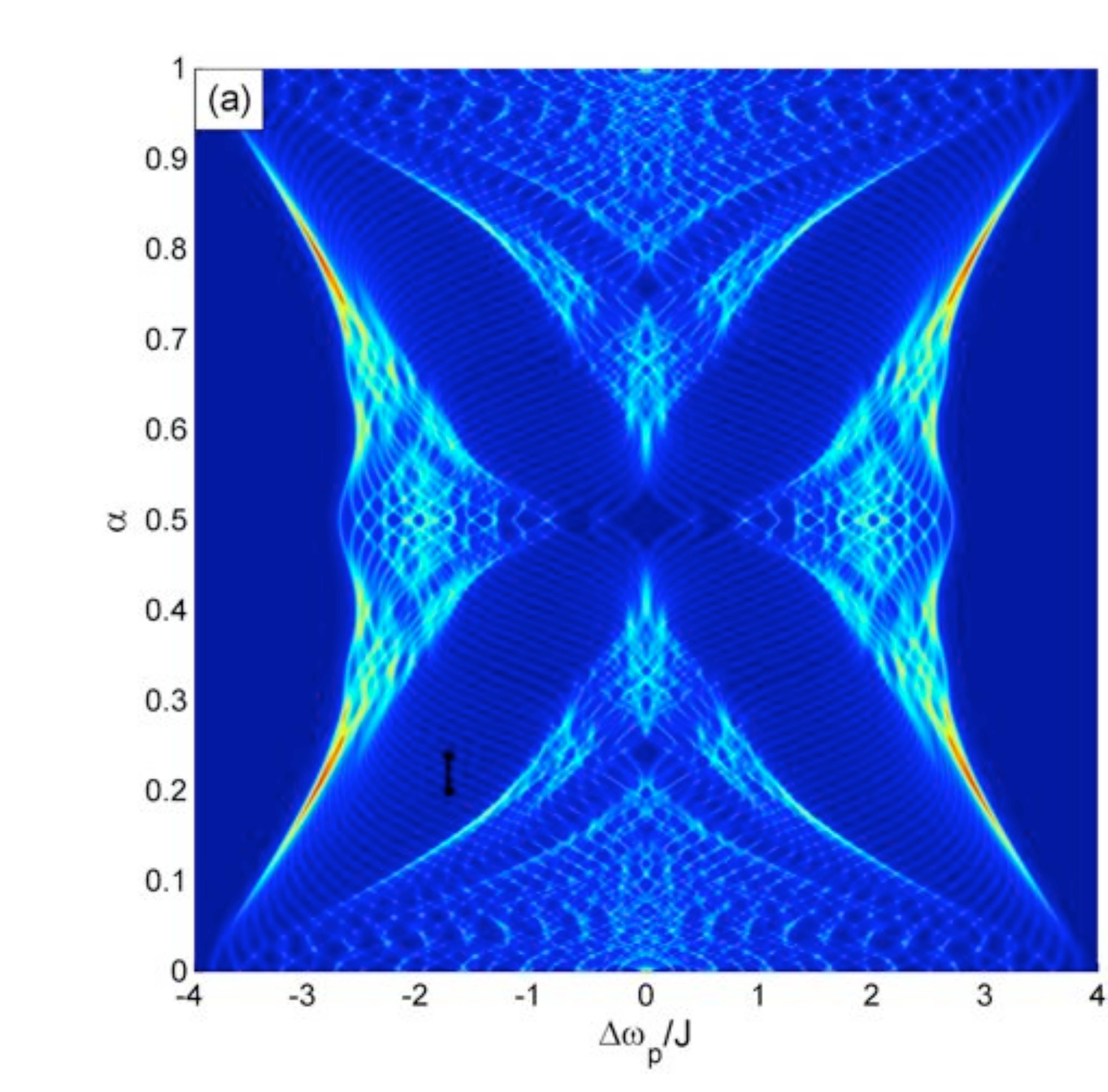}
\caption{Topological photonic states and artificial gauge fields for light. From left to right.
Two left-most panels: Photonic chiral edge states and effects of a large scatterer. Finite-element simulation of field propagation (left) and transmission spectra upon inclusion of the obstacle (right). Panels adapted from~\onlinecite{Wang:Nature2009}.
Third panel: Scheme of the circuit QED architecture to generate an artificial gauge field for microwave photons. Panel from~\cite{Koch:PRA2010}. 
Fourth panel: Sketch of the photonic device to generate an artificial gauge field for visible photons. Panel from \cite{Hafezi:NatPhys2011}.
Right-most panel: Hofstadter butterfly of states in the transmission spectrum of a $10\times 10$ two-dimensional array of cavities as a function of the pump frequency $\Delta\omega_p$ and of the magnetic flux per plaquette $\alpha$. Panel from \cite{Umucalilar:PRA2011}.}
\label{fig:gauge}
\end{figure*}

In parallel to these exciting advances, analogous studies have been undertaken in the photonic context: the basic idea was to look for photonic systems where the orbital motion of the photon is subject to an artificial gauge field. 
The first proposal in this direction~\cite{Haldane:PRL2008,Raghu:PRA2008} considered the two-dimensional photonic bands in an hexagonal array of dielectric rods showing a sizable Faraday effect breaking time-reversal invariance. Analogously to the electronic case, every photonic band is characterized by a topological invariant known as the Chern number: when two materials with different total Chern number are juxtaposed, a chiral state appears, which propagates along the interface with a unique directionality and is immune from scattering against disorder. In two dimensional electron gases, these states play an important role in the integer quantum Hall effect. In the photonic case, they form a sort of {\em one-way waveguide} along which light can propagate in one direction only, with no possibility of back-scattering at bends or imperfections. In particular, the absence of backwards propagating states and of bulk states at the same energy prevents light from being scattered by the obstacle, so that its net effect is limited to a phase shift. Other proposals to generate photonic topological insulators  without breaking time-reversal made use of arrays of toroidal microcavities~\cite{Hafezi:NatPhys2011} and and metamaterials with strong magneto-electric coupling~\cite{Khanikaev:arXiv2012}.

The idea of topologically protected one-way waveguides was experimentally implemented in~\cite{Wang:Nature2009} using a photonic crystal device in the microwave range with a square lattice geometry as proposed in~\cite{Wang:PRL2008}: time-reversal invariance was broken by the strong gyrotropic permeability of ferrite rods in the vicinity of the ferromagnetic resonance. As it is illustrated in Fig.\ref{fig:gauge}, one-way propagation of microwaves was assessed by measuring the transmission of light between two antennas possibly separated by a macroscopic obstacle (left-most panel). Transmission in the two directions (second panel from the left) differs by several orders of magnitudes and is almost unaffected by the presence of an obstacle: the microwaves are able to circumnavigate the obstacle without being scattered back, as it would instead happen in a standard waveguide. Soon later~\cite{Poo:PRL2011}, an important step forward was the demonstration of unidirectional propagation at the interface of a magnetic photonic crystal with air, without the requirement an ancillary cladding layer. Closely related studies of one-way waveguiding effects and non-reciprocal behaviors appeared in~\cite{Yu:PRL2008} using plasmon modes at the surface of a metal under a strong magnetic field and in~\cite{Lira:PRL2012} using a pair of single-mode waveguides under an AC-electrical drive. Other features due to non-reciprocal behavior in magneto-optical microwave devices in the presence of dissipation effects are discussed in the series of works~\cite{Dietz:PRL2007,Dietz:PRL2011}. 

Unfortunately, macroscopic photonic crystal devices as the ones in~\cite{Wang:Nature2009} seem unable to reach the regime of strong photon-photon interactions required for a fractional quantum Hall liquid of photons.
The first work that has addressed this question is~\cite{Cho:PRL2008}, where two-dimensional arrays of coupled optical cavities confining single atoms were theoretically considered. A gas of impenetrable bosons is encoded in the spin state of the atoms in their $s=1/2$ ground state; the artificial gauge field is obtained by suitably dressing the atoms with laser fields so to obtain a non-trivial hopping phase $\phi_{ij}$ in the generalized Bose-Hubbard model.
In particular, it was shown how to control $\phi_{ij}$ via the relative phase of the dressing beams. As an exciting application of this novel example of artificial gauge field, a viable protocol to prepare Laughlin states was also proposed in the presence of strongly repulsive interactions $U/J\to\infty$. A suitable adiabatic ramp of the system parameters (including an external potential) allows to bring the system from a ground state where each of the photons is localized on a separated site, to a different ground state which has an excellent overlap with a Laughlin state. Of course, this method of preparing the Laughlin state assumes a quasi-equilibrium regime in which the slow adiabatic following of the ground state is not disturbed by any loss process.

A different architecture to break time reversal symmetry and then obtain an artificial gauge field for photons was proposed in~\cite{Koch:PRA2010}. The idea is to use a lattice of cavities where tunnel coupling between neighbor sites occurs via a {\em circulator} element: an implementation of such a device in a circuit QED architecture with an array of coplanar waveguide resonators coupled by three junction Josephson rings connected in series and threaded by a static magnetic fields is sketched in the third panel of Fig.\ref{fig:gauge}. Signatures of the gauge field are predicted to be visible in the single-photon hopping dynamics in a few site geometry as well as in the photon bands in a spatially extended Kagom\'e lattice geometry. Further developments along these lines are summarized in~\cite{Petrescu:arXiv2012}. Another way of generating an artificial gauge field for photons by dynamically tuning the individual frequency of the each cavity was discussed in~\cite{Hayward:PRL2012} along the lines of proposals~\cite{Kolovsky:EPL2011,Eckardt:EPL2010} and an experiment~\cite{Struck:PRL2012} in the atomic physics context.

The interplay of the artificial gauge field with the driven-dissipative nature of photons was first theoretically investigated for a non-interacting photon system in~\cite{Hafezi:NatPhys2011}. The considered system was an array of toroidal microcavities coupled by tunnel events via the connecting waveguides as sketched in the fourth panel of Fig.\ref{fig:gauge}: in this configuration, a non-trivial tunneling phase $\phi_{ij}$ can be obtained by means of a suitable tayloring of the length of the connecting waveguides.
Alternative strategies to generate artificial gauge fields in photonic devices based on a planar microcavity architecture in the visible domain were proposed in~\cite{Umucalilar:PRA2011}. In these systems, a nontrivial tunneling phase between neighboring lattice sites can be induced by a suitable coupling of the orbital and polarization degrees of freedom in an optically active medium or by making the photon to perform a closed path in the space of polarizations during a tunneling event. This latter scheme generalizes to evanescent waves the geometric Pancharatnam phase that appear when photons propagate across a sequence of birefringent slabs~\cite{Pancharatnam:PIAS1956} or the Berry phase in twisted optical fibers~\cite{Tomita:PRL1986} and non-planar Mach-Zender interferometers~\cite{Chiao:PRL1988}. Its important advantage is the possibility of scaling the transverse patterning of the planar microcavity to a micrometer scale so to integrate the artificial gauge field with photon blockade.

In these two last proposals, the driven-dissipative nature of the photons is exploited to propose optical spectroscopic techniques to experimentally assess the effect of the artificial gauge field. Remarkably, a regime of large magnetic field appears to be under reach, where the loop integral $\sum_{\square} \phi_{ij}$ around a plaquette is of the order of $2\pi$. In standard electronic systems such a regime would correspond  to a magnetic flux per unit lattice cell of the order of the flux quantum $h/e$, which requires huge values of the magnetic field.
In such a strong magnetic field regime, the one-particle eigenstates of the conservative Hubbard model form the so-called Hofstadter butterfly of states~\cite{Hofstadter:PRB1976}: as it is illustrated in the right-most panel of Fig.\ref{fig:gauge} many features of this peculiar self-similar structure turn out to be observable in quantities as simple as reflection and transmission spectra in relatively small systems. Local access to the optical field provides full information on the spatial structure of the photon wavefunction.

In analogy to the earlier work on gyromagnetic photonic crystals~\cite{Wang:Nature2009}, the use of photonic lattices under strong magnetic fields to observe unidirectional propagation of light around the edge of a finite-sized structure was proposed in~\cite{Hafezi:NatPhys2011}. As a consequence of the nontrivial Chern number of the Hofstadter bands, edge states exist at energies within the gap and are topologically protected against imperfections and disorder: in the presence of a defect, light wraps around it and then keeps propagating in the same direction. These edge states are visible in the transmission spectra shown in the right-most panel of Fig.\ref{fig:gauge} as a comb of faint lines crossing the gaps.

The interplay of an artificial gauge field with sizable photon-photon interactions was first theoretically investigated in~\cite{Nunnenkamp:NJP2011} with a special eye to implementations in circuit QED devices.
In spite of the simplicity of the three-site geometry considered, an interesting Schr\"odinger cat structure was pointed out in the few photon eigenstates as well as novel two-photon resonances in the transmission spectra beyond linear regime, along lines very similar to~\cite{Carusotto:PRL2009}. Extension of this work in the direction of the observation of Laughlin states when strong artificial gauge fields are combined to photon blockade in a square lattice geometry has been recently undertaken in~\cite{Umucalilar:PRL2012}.

An alternative way of generating topologically non-trivial states of a photon gas was proposed in~\cite{Bardyn:arXiv2012} in analogy to the physics of $p$-wave superconductors: when an open chain of cavities in the impenetrable photon regime is pumped by parametric pumps injecting a squeezed vacuum in adjacent cavities (instead of coherent pumps as in~\onlinecite{Carusotto:PRL2009}), Majorana modes~\cite{Read:PRB2000,Alicea:arXiv2012} are expected to appear in the fermionized photon gas and can be detected via the mutual second-order coherence of the emitted light from the first and last cavities.

  \subsection{Emerging systems for strongly correlated photons}
\label{others}


In this last subsection, we complete the discussion on the prospects of strongly correlated photon gases by giving a few more details on two physical systems which are emerging as most prominent candidates to achieve strongly correlated photon phases in arrays of nonlinear cavities: semiconductor microstructures embedding quantum wells or quantum dots, and superconducting quantum circuits. Besides these solid-state systems, it is important to remind that a very recent experiment~\cite{Peyronel:Nature2012} has demonstrated the potential of atomic clouds in the Rydberg-EIT regime to obtain huge optical nonlinearities and induce strong correlations in photon streams.

\subsubsection{Semiconductor micro- and nano-resonators}
\begin{figure*}[t!]
\begin{center}
\includegraphics[width=0.55\columnwidth,angle=0,clip]{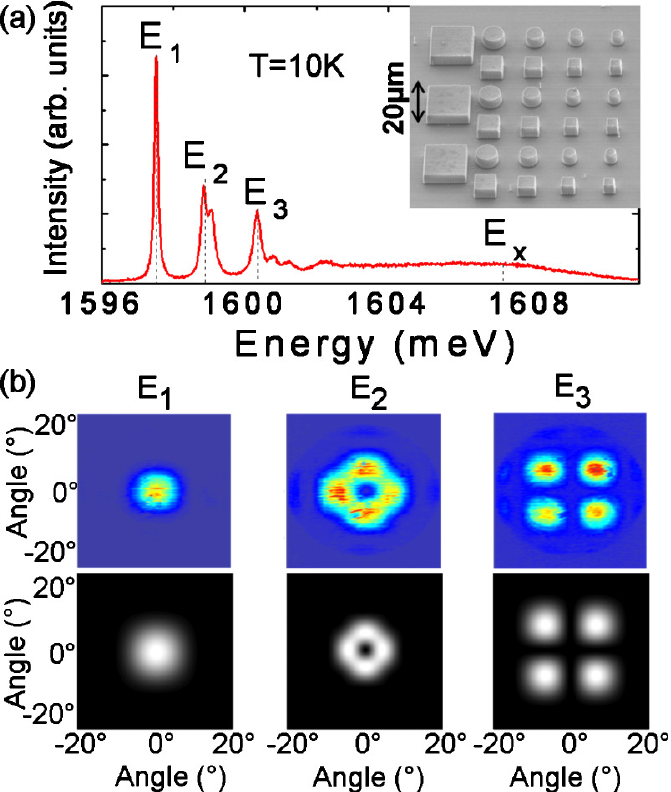}
\hspace{0.1\columnwidth}
\includegraphics[width=0.45\columnwidth,angle=0,clip]{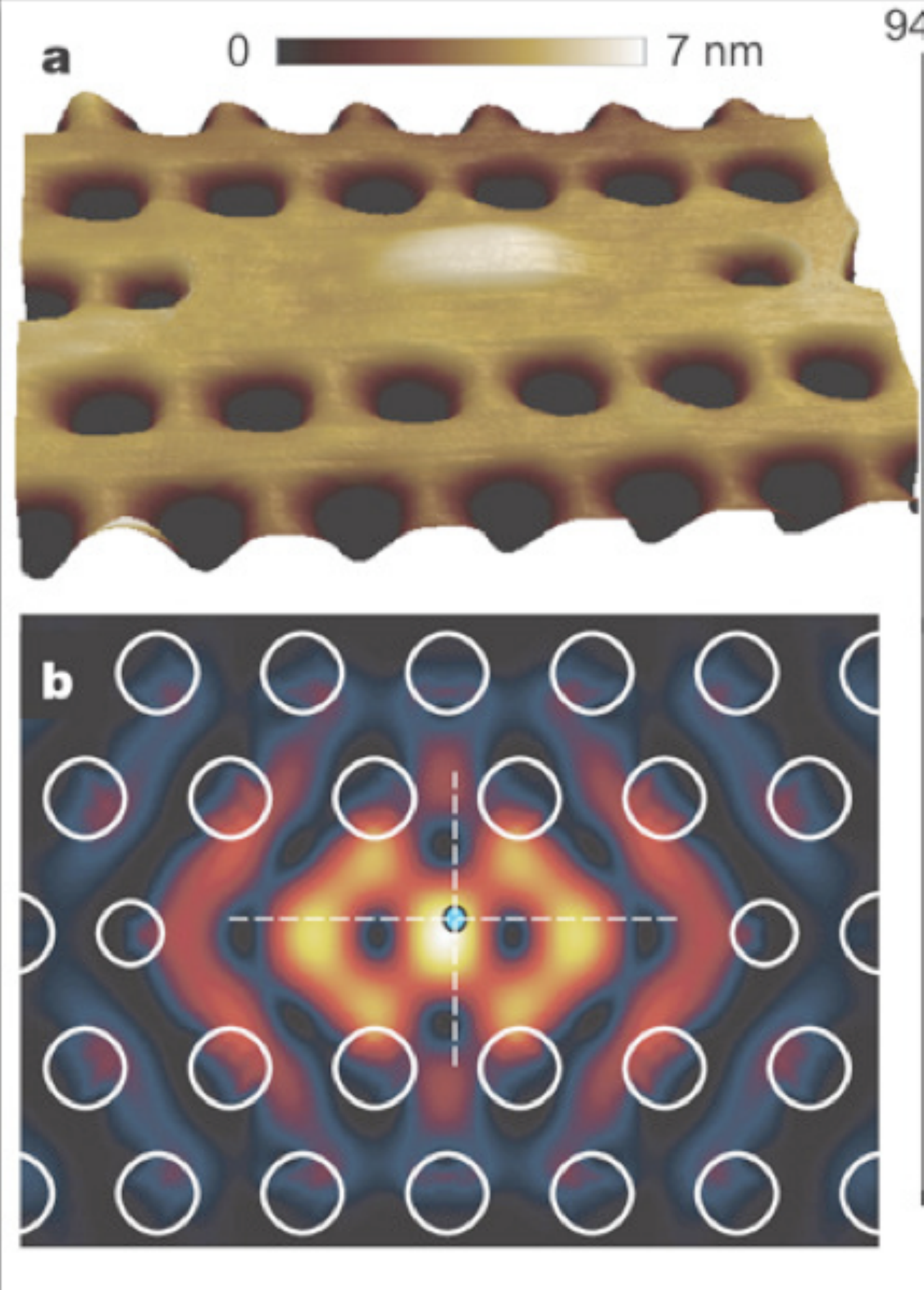}
\hspace{0.1\columnwidth}
\includegraphics[width=0.8\columnwidth,angle=0,clip]{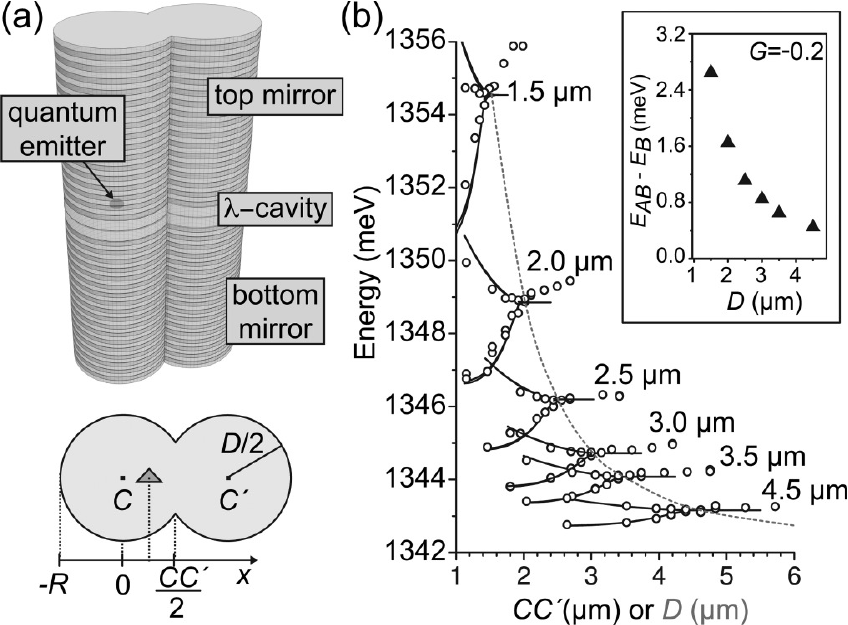}
 \end{center}
\caption{
Left panels: semiconductor resonators obtained by lithographic patterning of a semiconductor planar microcavity. The photoluminescence spectra (upper panel) are characterized by a series of resonances associated to the fully confined modes of the squared resonator indicated in the scanning electron micrograph shown in the inset. The bottom panel shows the measured and calculated far-field patterns of the photon modes. Figure from \cite{Ferrier:APL2010}. Central panels: photonic crystal nanocavity embedding a single quantum dot. Figure from \onlinecite{Hennessy:Nature2007}.
Right panels: (a) photonic molecule formed by etching a planar microcavity with a suitable mask. (b) the measured (symbols) and calculated (lines) energies of the first two optical modes of the photonic molecule for various diameters are shown. Figure from~\onlinecite{deVasconcellos:APL2011}.
}
\label{fig:coupled_nanocav}
\end{figure*}

In the first Sections of this article, we have reviewed the rich physics of dilute quantum fluids of light in spatially extended two-dimensional planar microcavity geometries. Several techniques have been developed to add some in-plane potential  to confine polaritons in all three dimensions or even to create arrays of single mode boxes. Some among the most advanced techniques are summarized in Sec.\ref{sec:external_pot}.
Photonic resonator obtained by laterally etching a planar semiconductor microcavity were pioneered in~\cite{Bloch:PhysE1998}. More recent samples fabricated along these lines are illustrated in the left panels of Fig. \ref{fig:coupled_nanocav}: a  direct image of the resulting pillar microcavities is shown in the inset, while the spatial structure of the lowest modes is highlighted in the lower panel. 

An alternative concept of photonic resonator in the infrared/visible range is obtained by introducing a defect in an otherwise periodic photonic crystal membrane: as it happens to electrons in solids, a isolated mode appear within the photonic gap, strongly localized in space in the vicinity of the defect. An example is shown in the central panels of Fig. \ref{fig:coupled_nanocav}. Most remarkable features of these {\em photonic crystal cavities} are the extremely high values of the $Q$-factor in the $10^6$ range and the extremely small mode volumes of the order of $(\lambda/2)^3$  that can be nowadays achieved~\cite{Joannopoulos:book}.

Optical coupling of neighboring photonic resonators via the overlap of the evanescent tails has been demonstrated for both systems~\cite{Bayer:PRL1998,Intonti:PRL2011,Dousse:Nature2010,deVasconcellos:APL2011}. A sketch and a review of the main spectroscopical properties of a pair of coupled pillar microcavities are illustrated in the right panels of Fig.\ref{fig:coupled_nanocav}. Some novel features related to Bose-Einstein condensation and Josephson dynamics in double cavity geometries have recently been experimentally addressed in~\cite{Galbiati:arXiv2011}.

In standard photonic devices based on semiconductor technology, the optical nonlinearity of the material medium used to confine the photon is moderate and the device operation is far from the strong interaction regime. The most common strategy to approach this regime is to embed in the cavity some active material showing a sharp material resonance in the spectral neighborhood of the photonic mode. In this way, the excitonic content of the resulting polariton modes introduces strong interactions that can eventually lead to a photon blockade behavior. 

A first successful route in this direction has been to insert an electronic quantum dot in a photonic crystal resonator: a quantum dot consists of a nanoscopic volume of a different material that is able to confine the electronic wavefunction in all three-dimensions~\cite{Bastard}: as a result, the electronic levels are quantized as in a standard atom and discrete optical transitions can be isolated. When one of these transitions is strongly coupled to the photonic mode, a Jaynes-Cummings system is recovered: provided the loss rate is weak enough, the optical nonlinearity that stems from the anharmonicity of the levels shown in the left panel of Fig.\ref{fig:JC} can result in an efficient photon blockade effect. 

First experimental evidences of photon blockade by a quantum dot embedded in a microcavity driven by a coherent pump was reported in \cite{Faraon:NatPhys2008}.
Later works have shown significantly clearer signatures of strong photon interactions~\cite{Reinhard:NatPhot2011}, as illustrated in the right panels of Fig.\ref{fig:JC}: (i) photon antibunching upon resonant excitation of the lowest-energy polariton state; (ii) photon bunching when the laser field is in two-photon resonance with the polariton eigenstates of the second Jaynes-Cummings manifold. A subsequent work~\cite{Volz:arXiv2011} has applied this giant optical nonlinearity to all-optical switching on a ultrafast time scale at the single photon level~\cite{Volz:arXiv2011}. 
{Other interesting experimental investigations of low-photon-number optical switching in microcavities embedding quantum dots have recently appeared in~\cite{Bose:PRL2012,Englund:PRL2012,Loo:arXiv2012}. A sophisticated four-wave mixing spectroscopy technique to study the nonlinear dynamics of these systems was implemented in~\cite{Kasprzak:NatMat2010}.}

In standard architectures with a single cavity mode coupled to a single quantum dot, an efficient photon blockade effect requires to be deeply in the strong emitter-cavity coupling regime, i.e., the vacuum Rabi frequency $g$ largely exceeding the loss rate. This means that the cavity Q-factor has to be kept as high as possible and, simultaneously, that the quantum dot is correctly positioned within the cavity mode profile. Experimentally, state-of-the-art samples display a coupling to loss rate of the order of 3, to be compared to 10 for single atoms in macroscopic optical cavities~\cite{Birnbaum:Nature2005} and 50 for superconducting circuits~\cite{Lang:PRL2011}. The strong coupling condition has been partially softened in~\cite{Bamba:PRA2011} where the use of a two-cavity geometry has been predicted to lead to efficient photon blockade even for loss rates comparable to the emitter-cavity coupling strength.

From the point of view of realizing strongly correlated states of photons in a many-cavity geometry, quantum dot-based architectures suffer from the significant drawback that neither the spatial position of individual dots nor the exact position of the electronic transitions can be controlled during the growth stage, as the different dots self-organize at random positions with random sizes. Indeed, in all quoted experiments, an ensemble of quantum dots was first prepared in the sample, then the photonic crystal cavity was fabricated in a careful way so to have the photonic mode located right on top of the quantum dot of interest with a frequency precisely tuned to the quantum dot exciton transition. A promising route to overcome this difficulty using spatially localized emitters may involve pyramidal quantum dots that can be grown on demand at specific spatial positions~\cite{Carron:APL2011} and a subsequent fine tuning of the exciton energy by Stark effect~\cite{Bennett:APL2010}. Other research lines that are active with similar objectives involve impurities in diamond crystals~\cite{Aharonovich:NatPhot2011} or single atoms~\cite{Dayan:Science2008} located on the surface of a monolithic microcavity and strongly coupled to a high-Q cavity mode.

An alternative strategy to realize a strong interaction condition is to use a quantum well as the active material~\cite{Verger:PRB2006}: excitons are then strongly confined only along the growth axis, while the patterning required for photon confinement affects their in-plane motion on a much longer micrometer scale. As the quantum well extends throughout the whole cavity and its thickness along the growth axis can be controlled at the level of a single atomic layer,  the experimental problem of correctly positioning the emitter is completely eliminated and the electronic contribution to inhomogeneous broadening can be dramatically suppressed. 

While the nonlinear dynamics of polaritons in confined systems embedding quantum wells has been very recently addressed in~\cite{Galbiati:arXiv2011}, it is still unclear whether the nonlinearity stemming from collisions between delocalized polaritons as described by \eq{V_LPLP} is strong enough to penetrate the $U/\gamma \gg 1 $.
A back-of-the-envelope approach based on the $U \propto 1/L^2$ scaling (here $L$ is the characteristic length of lateral confinement) would suggest that very tightly confined polariton boxes down to the sub-micron scale should allow to reach the photon blockade regime~\cite{Verger:PRB2006}. As the quality factor of the resonator may also get quenched when the spatial size of the resonator is reduced to sub-micron scales, full numerical calculations of the photonic mode profile in realistic devices are required. Alternative strategies to overcome this difficulty may be to take advantage of the reinforced interaction potential between the hybrid direct/indirect excitons first studied in~\cite{Cristofolini:Science2012} or of the quantum interference mechanism reviewed in Sec.\ref{sec:bamba}.

\subsubsection{Superconducting quantum circuits and circuit QED systems}

\begin{figure}[t]
\includegraphics[width=0.7\columnwidth,angle=0,clip]{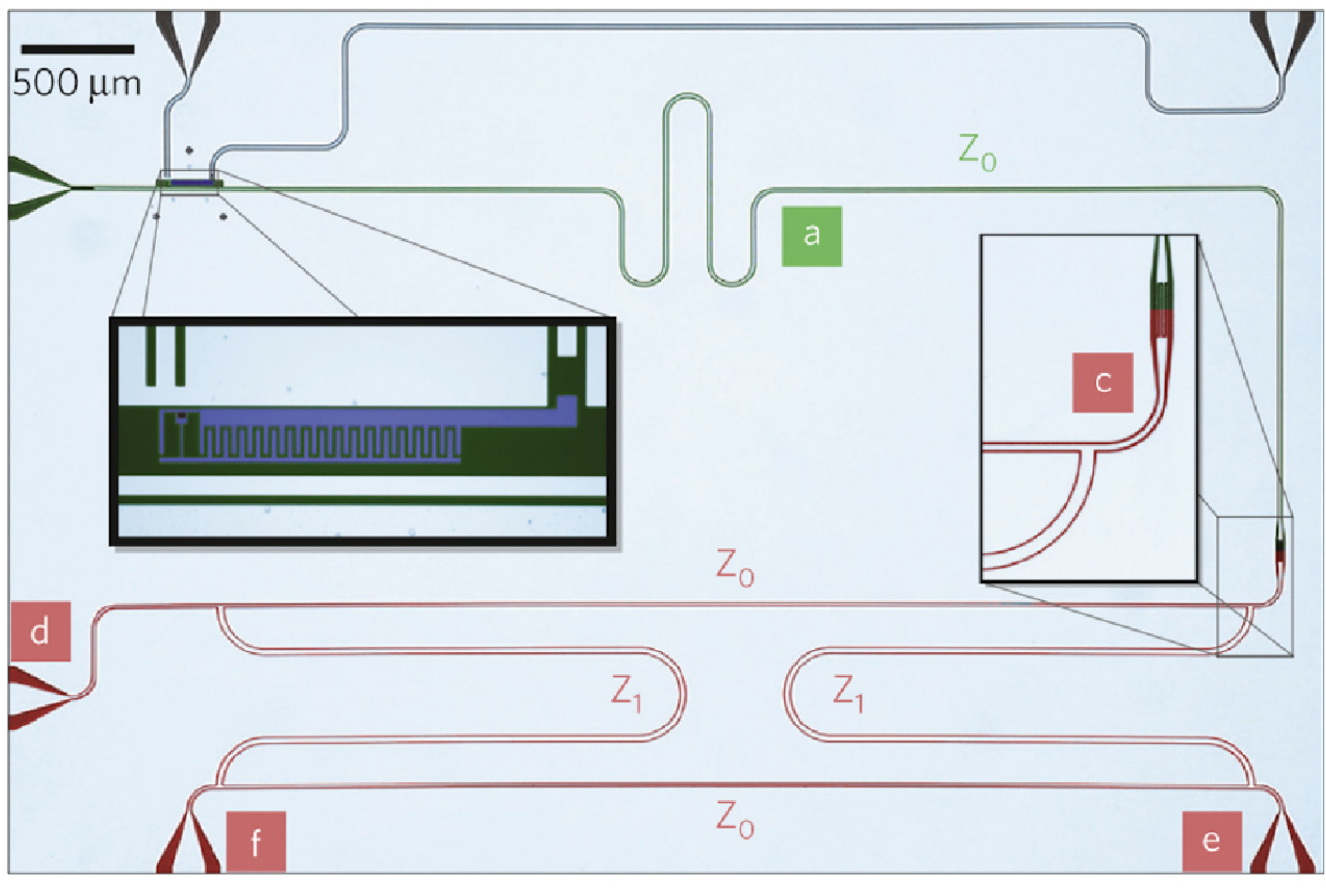}\\
\includegraphics[width=0.7\columnwidth,angle=0,clip]{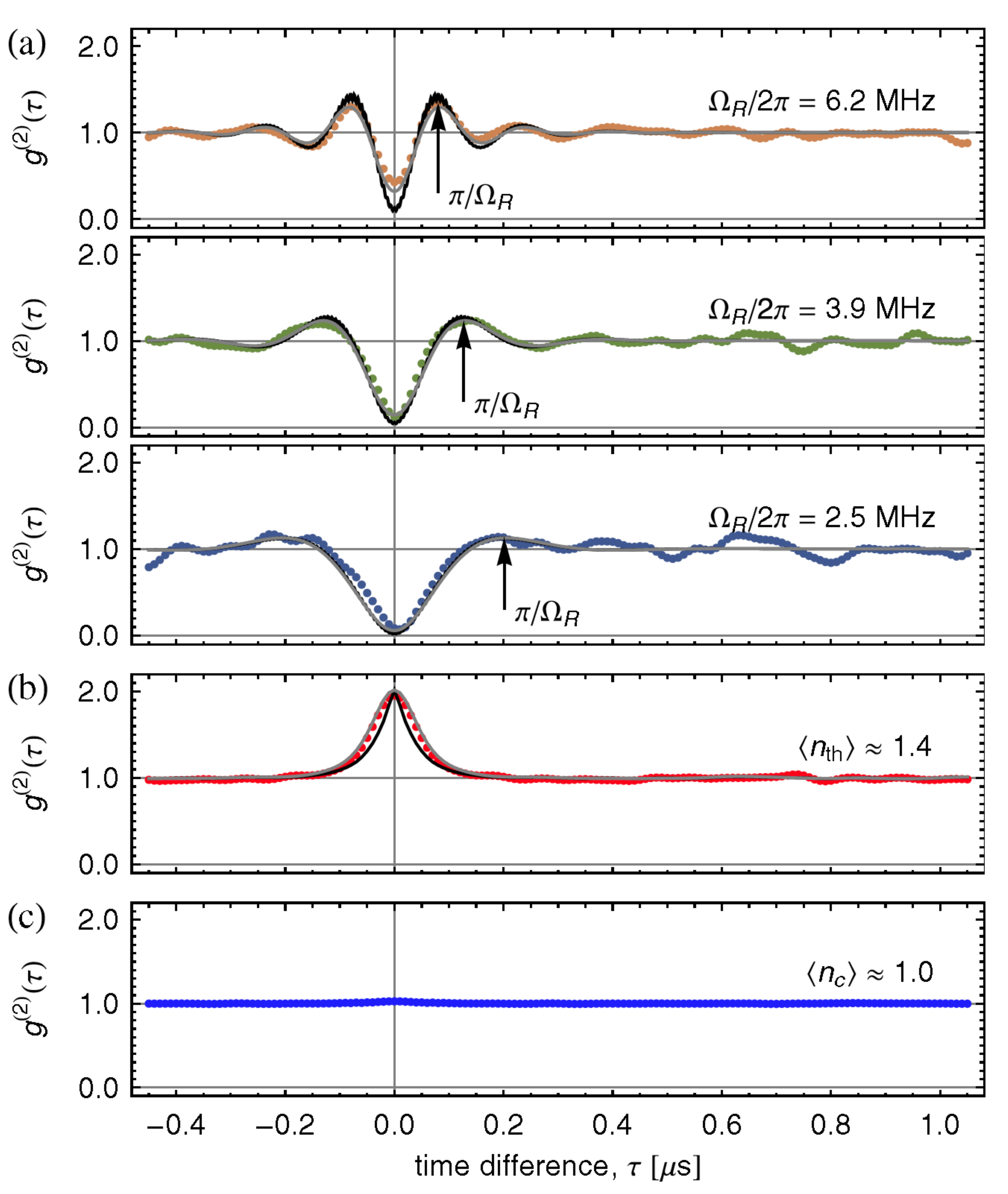}
\caption{Top panels: Sketch of a superconducting coplanar waveguide resonator (green) interacting with a Josephson transmon qubit (left inset) that can be biased with large-bandwidth flux and charge gate lines (top inputs). (b) Schematic of resonator with two-level system generating a photon in mode a, emitting it into mode c onto a beam splitter with modes d, e and f. 
Figure from \cite{Bozyigit:NatPhys2011}.
Bottom panels: Experimental demonstration of resonant photon blockade in  a circuit-QED system. (a)	Second-order correlation function measurements (dots) compared to theoretical calculations (lines) for different values of the drive amplitude. Same plot for a thermal incident field (b) and for a coherent drive (c). Figure from~\cite{Lang:PRL2011} 
}
\label{fig:circuit_QED}
\end{figure}

Superconducting quantum circuits based on Josephson junctions are another most promising playground where to investigate the physics of strongly correlated photons. Recent reviews of superconductor quantum circuits can be found in~\cite{Schoelkopf:Nature2008,You:Nature2011}. An example of architecture used in recent experiments is shown in the top panel of Fig.\ref{fig:circuit_QED}

A simple theoretical description of these systems is obtained by representing the macroscopic electromagnetic variables such as currents and voltages (or charges and fluxes) as non-commuting quantum operators and then describing the evolution of the system in terms of an Hamiltonian. In the simplest case of a $LC$ circuit, the Hamiltonian has the quadratic form of a harmonic oscillator. 

The addition of Josephson junctions to the circuit is essential for the emergence of quantum features and for quantum applications: thanks to its nonlinear current-voltage relationship, a Josephson junction introduces a non-dissipative nonlinearity into the circuit and produces a non-harmonic energy spectrum with non-equispaced energy levels.
As a result, the superconducting quantum circuit can be considered as an artificial atom with a discrete energy spectrum that depends on the specific connections and elements of the circuit. Under appropriate conditions, the two lowest states of the quantum circuit are much closer in energy than all higher states, so that this effectively behaves as a two-level artificial atom. By changing the type of circuit connections and the relative values of the capacitances, inductances and Josephson energies, a wide variety of artificial atoms can be obtained, among which the Cooper pair box\cite{Nakamura:Nature1999}, the quantronium\cite{Vion:Science2002}, the transmon~\cite{Koch:PRA2007} or the fluxonium~\cite{Manucharyan:Science2009}.

By inserting a Josephson two-level artificial atom in a superconducting transmission line resonator, it is possible to create a Jaynes-Cummings system where the artificial atom is coupled to the bosonic field of the microwave resonator. Given the analogy with usual cavity QED with real atoms~\cite{Raimond:RMP2001,cavityQED}, this exciting field of research field has been called circuit QED~\cite{Wallraff:Nature2004,Blais:PRA2004}.  
With respect to standard cavity QED implementations using atoms in optical or microwave cavities, the superconducting circuit QED offer better perspectives of scalability: lithographic techniques analogous to the ones used in semiconductor technology can be in fact used to create large arrays of resonators and Josephson artificial atoms on a single solid-state chip. In addition, circuit QED offer the possibility to have very large Rabi couplings comparable to the transition frequency~\cite{Devoret:Annalen2007} and to efficiently control {\em in-situ} and in a dynamical way the properties of the artificial atom with external fields. This latter property was indeed at the heart of the recent observation of an analog dynamical Casimir effect~\cite{Wilson:Nature2011}.

Compared to semiconductors, the main practical limitation of superconducting quantum circuits is that they must operate in the microwave frequency range and at mK temperatures of a dilution fridge in order to have $k_B T \ll \hbar \omega \ll \Delta$, where $\Delta$ is the superconducting gap energy and $\omega$ is the resonator frequency (typically in the GHz range). 
On the other hand, the figures of merit offered by these circuit QED systems are quite spectacular as compared to previous quantum optical systems: for instance, a superconducting Jaynes-Cummings system can display a value of the  $g/\gamma$ ratio between the vacuum Rabi frequency and the loss rate larger than $10^2$: a comparable value $g/\gamma\approx 300$ was obtained so far only with Rydberg atoms in superconducting cavities~\cite{Raimond:RMP2001}, but these systems are hardly scalable to many-cavity geometries. This remarkable large value of $g/\gamma$ was crucial in the recent experiments that have used Josephson atoms with tunable coupling to syntetie arbitrary quantum states of the coupled qubit-cavity field system and then perform a quantum tomography measurement of the state~\cite{Hofheinz:Nature2009}.

While single photon detectors in the microwave range are still in the exploratory phase~\cite{Romero:PRL2009,Chen:arXiv2010}, two-channel heterodyne schemes using linear amplitude detectors have allowed for accurate measurements of the quadratures of the microwave field down to the quantum limit~\cite{Gabelli:PRL2004,Bozyigit:NatPhys2011,Lang:PRL2011}: an example of application of this technique is illustrated in the bottom panel of Fig.\ref{fig:circuit_QED}, where we show how a strong photon anti-bunching and a marked sub-poissonian statistics can be obtained via a photon blockade mechanism under a coherent drive~\cite{Lang:PRL2011}.

The possibility of coupling in a controllable way several resonators has been demonstrated in~\cite{Mariantoni:PRB2008,Wang:PRL2011,Mariantoni:NatPhys2011}, paving the way to the study of arrays of resonators in the near future. These impressive experimental advances, together with the proposed implementations of artificial gauge fields for microwave photons~\cite{Koch:PRA2010,Nunnenkamp:NJP2011} puts superconducting circuits among the most promising candidates for the study of many-body physics in photon gases.

\section{Conclusions and perspectives} 
\label{conclusions}

In this review we have summarized recent developments in the theoretical and experimental study of gases of interacting photons in solid-state systems. An effective photon mass appears as a result of spatial confinement along the growth axis, while the nonlinear optical susceptibility of the material medium induces sizable binary interactions between photons. Strong coupling of the photon with some long-lived electronic excitation in the medium is a succesful strategy to reinforce binary interaction between the dressed bosonic particles that arise from the mixing of light with the matter excitation, the so-called {\em polaritons}. Recent experiments demonstrating Bose-Einstein condensation and superfluidity effects in luminous gases were indeed performed using polariton gases in semiconductor planar microcavities embedding quantum wells. 

In contrast to standard many-body systems such as liquid Helium, electron liquids in solid-state systems and ultracold atomic gases, polaritons have a finite lifetime as a consequence of radiative and non-radiative dissipative processes and some external pump is required to replenish the gas and compensate for the particle loss. As a result, the non-equilibrium steady state of the system is no longer determined by a standard thermodynamical equilibrium condition, rather by a dynamical balance of driving and dissipation. On one hand, the driven-dissipative nature of the photon fluid is responsible for novel phenomena in the collective dynamics; on the other hand it provides (ironically) a powerful tool for real-time diagnostics of the many-body system just by looking at the emitted light.

After a brief survey of the historical development of the concept of photon fluids, we started our presentation with a general theoretical summary of the quantum field description of photons and polaritons in solid-state devices: even if our focus was concentrated on planar geometries, the concepts are readily transferred to other geometries, including one-dimensional polariton wires, harmonic traps, and even periodic lattices. We have then illustrated the generalized non-equilibrium Gross-Pitaevskii equation that can be used to describe the steady-state and the collective dynamics of a dilute photon gas at the mean-field level. A solution of this equation is provided for the simplest geometries in the different pumping regimes. This general theoretical results are then used to explain recent experimental observations of the condensate shape and the momentum distribution.

Linearization of the generalized Gross-Pitaevskii equation around the steady state provides the spectrum of elementary excitations of the fluid. Based on the general features of this spectrum, we have reviewed the recent demonstration of superfluid hydrodynamic behaviors in polariton fluids. Under a coherent pumping, a variety of regimes was observed depending on the flow speed compared to the Landau critical velocity, from superfluid flow to the Bogoliubov-\u Cerenkov emission of phonons and single particle excitations in the wake of a weak defect; for large and strong defects, the strong local perturbation of the fluid has been shown to lead to the nucleation of topological excitations such as vortices and dark solitons. Intriguing and so far unexplained experimental observations of superfluid behaviors under a OPO pumping regime have been critically discussed.

In the last part of the review, we have presented an overview of the emerging field of strongly correlated photon gases. For strong photon-photon interactions, the mean-field description based on a Gross-Pitaevskii equation for the macroscopic wavefunction breaks down and the many-photon wavefunction starts displaying peculiar quantum correlations. 
Exception made for very simple single-mode cavities for which effective photon blockade has been already observed in a number of systems~\cite{Birnbaum:Nature2005,Faraon:NatPhys2008,Reinhard:NatPhot2011,Lang:PRL2011}, the first experimental claim of strongly interacting photons in spatially extended geometry has appeared very recently using an atomic gas in the Rydberg-EIT configuration~\cite{Peyronel:Nature2012}. In the meanwhile, an intense activity is devoted to the theoretical study of systems where quantum states of photon matter can be generated, for instance Mott insulators, Tonks-Girardeau gases, Laughlin states of quantum Hall physics. Also in this case, the driven-dissipative nature of the photon gas is responsible for a wealth of unexplored features due to the interplay of non-equilibrium statistical mechanics, quantum optics, many-body physics, possibly with long-term applications for all-optical quantum information processing.

In spite of the great achievements that the research on photon gases has experienced in the last few years, an even larger number of questions are still awaiting experimental and/or theoretical answer. From the experimental point of view, the big challenge is in our opinion the development of devices in the visible and/or in the microwave range of frequencies where strong photon-photon interactions can be associated with a non-trivial spatial dynamics, either in free space geometries or in lattice ones. A specific discussion of two systems that we believe most promising in this direction is given in the last section of the article, namely semiconductor micro- and nano-resonators for visible or infrared light and circuit-QED systems for microwaves.

Important questions are still open also in the dilute gas regime. In addition to a complete explanation of observations, more experimental work to fully characterize the elementary excitation spectrum of a dilute photon gas in the different pumping regimes is required and is expected to shine light on novel aspects of superfluidity. Of course, all this discussion is not restricted to systems in the strong coupling regime where the elementary excitations have a polaritonic nature: for instance, vertical cavity surface emitting laser devices are promising alternative candidates for the study of the collective dynamics of the photon gas. In turn, a better understanding of these many-body features of the non-equilibrium Bose-Einstein transition will shine new light on the laser threshold of spatially extended devices such as VCSELs: in particular, very little appear to be known yet on the critical fluctuations in the vicinity of the transition point.

Another research direction that is expected to have a tremendous impact on both fundamental science and on applications is the one on the generation of artificial gauge fields for photons: on one hand, quantum mechanics of particles in the presence of gauge fields has been predicted to show a number of novel fascinating effects. On the other hand, we expect that their full exploitation in practical photonic devices will allow to take advantage of topological features to elaborate optical information with unprecedented robustness towards experimental imperfections. Inclusion of optical nonlinearities in the arena is an almost unexplored field that may considerably enlarge the spectrum of possibilities with a number of novel unexpected effects and possibly a bright future for the physics of luminous quantum fluids.

\section*{Acknowledgments}
This review article could never be completed without the continuous stimulating discussions with the many friends and colleagues with whom we had the chance of collaborating during all these years on the fascinating subject of quantum fluids of light, in particular A. Amo, M. Bamba, J. Bloch, A. Baas, A. Bramati, S. De Liberato, J. W. Fleischer, D. Gerace, E. Giacobino, M. Hafezi, A. Imamo\u glu, G. C. La Rocca, S. Pigeon, M. Richard, D. Sanvitto, D. Sarchi, V. Savona, A. Smerzi, J. Tignon, H. T\"ureci, R. O. Umucal\i lar, T. Volz, M. Wouters. \\
I. C. acknowledges partial financial support from ERC via the QGBE grant. 
C. C. is member of Institut Universitaire de France and acknowledges support from ANR via the QPOL and QUANDYDE grants. 

\bibliographystyle{apsrmp}

\bibliography{carusotto_final}

\newpage


\end{document}